\documentclass[a4paper,twoside]{mythesis}

\usepackage{amsmath}
\usepackage{amssymb}
\usepackage{graphicx}% Include figure files
\usepackage{dcolumn}% Align table columns on decimal point
\usepackage{bm}% bold math
\def\ov#1{{\overline{#1}}}

\def\wt#1{{\widetilde{#1}}}
\def\vb#1{{\mbox{\boldmath$#1$}}}
\def\pd#1#2{\frac{\partial #1}{\partial #2}}
\def\pds#1#2{{\partial _{#2} #1}}

\def\wh#1{\widehat{#1}}
\def\bdot{\,\vb{\cdot}\,}

\def\grad{{\boldsymbol{\nabla}}}

\def\dotp#1#2{{\boldsymbol{#1}\bdot\boldsymbol{#2}}}
\def\crop#1#2{{\boldsymbol{#1}\times\boldsymbol{#2}}}
\def\half{{\frac{1}{2}}}

\def\dfrac#1#2{\frac{#1}{#2}}

\def\stackp#1#2{\begin{array}{c}	\mathrm{#1}\\ #2 \end{array}}

\def\onlinecite{\cite}

\begin{document}

\begin{titlepage}
\begin{center}
%\begin{singlespace}
\vspace*{1.0cm}
\begin{tabular}{c c c}
\includegraphics[height=2.0cm]{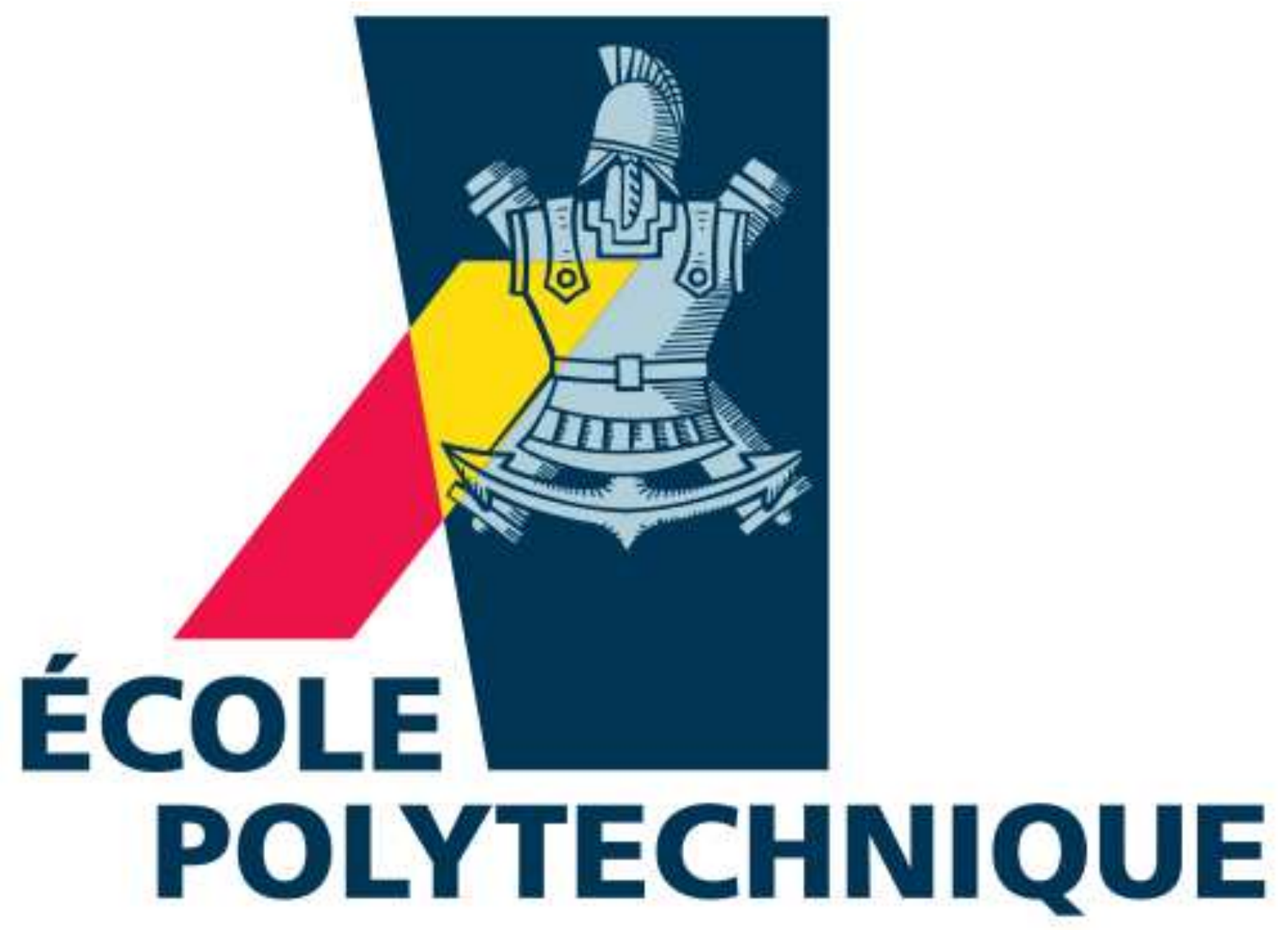} & \includegraphics[height=2.0cm]{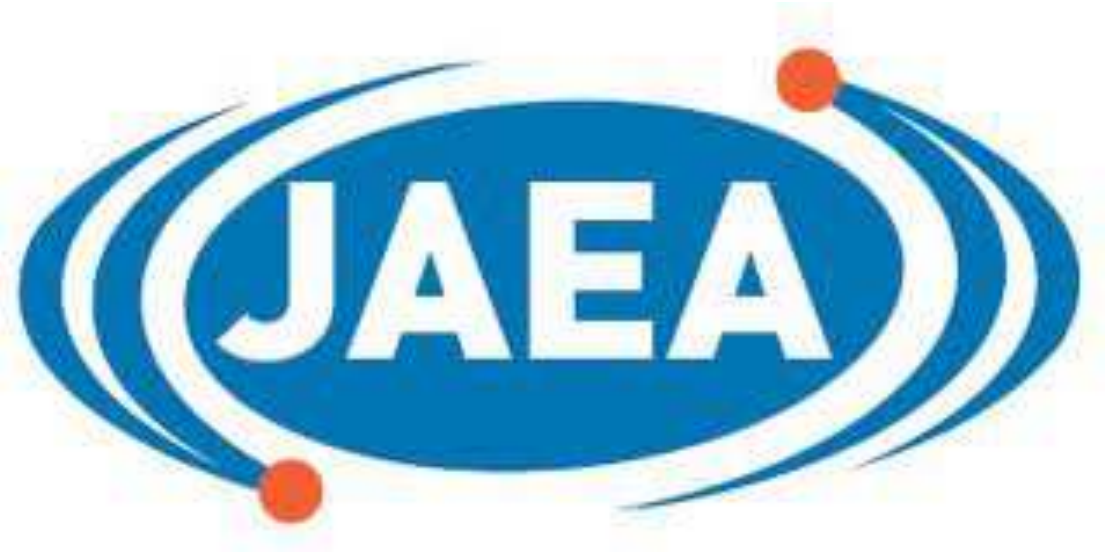} & \includegraphics[height=2.0cm]{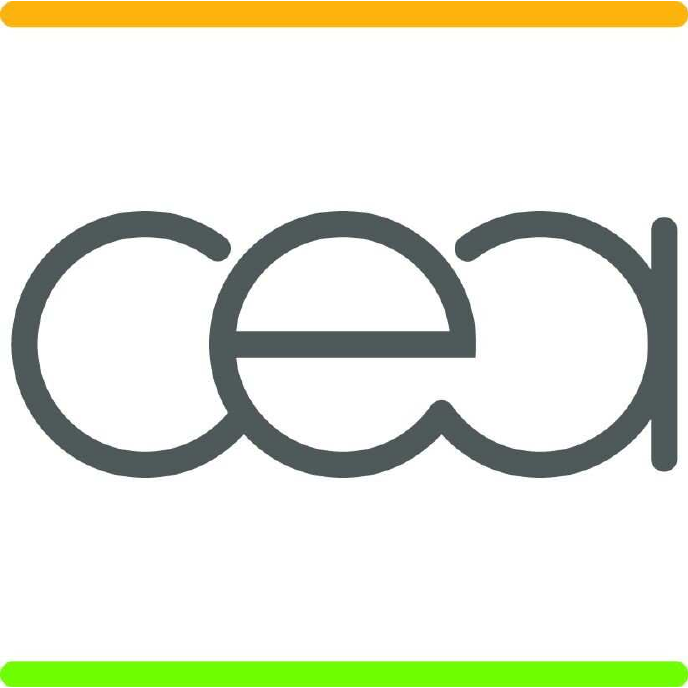}\\
%\textsf Ecole Doctorale de l'Ecole Polytechnique & Japan Atomic Energy Agency & IRFM, Association Euratom-CEA\\
\textsf Ecole Doctorale de & Japan Atomic & IRFM, Association\\
l'Ecole Polytechnique & Energy Agency & Euratom-CEA\\
Route de Saclay & 6-9-3 Higashi-Ueno Taito-ku & Centre de Cadarache\\
F-91128 Palaiseau, France. & 110-0015 Tokyo, Japan & F-13108 St Paul-Lez-D., France.\\
\end{tabular}\\
\vspace{2.0cm}
{\huge \textsf{The Berk-Breizman Model as a Paradigm for\\
\vspace{0.2cm}Energetic Particle-driven Alfv\'en Eigenmodes}}
\par
\vspace{1.0cm}
{\large by}
\par
\vspace{0.5cm}
{\Large \textbf{Maxime~Lesur}}
\par
\vspace{1.5cm}
{\large A dissertation submitted for the degree of}
\par
\vspace{1.0cm}
{\large \textbf{Doctor of Philosophy}}
\par
in
\par
{\large \textbf{Plasma Physics\\
\vspace{0.1cm}Ecole Doctorale de l'Ecole Polytechnique}}
\par
\vfill
\begin{tabular}{l l}
\hline \\
Dr.~Saddrudin Benkadda & CNRS Research Director\\
Dr.~Xavier Garbet & CEA Research Director - Advisor\\
Dr.~Virginie Grandgirard & CEA Research Scientist\\
Dr.~Yasuhiro Idomura & JAEA Research Scientist - Advisor\\
Dr.~Jean-Marcel Rax & Professor at Ecole Polytechnique - PhD Advisor\\
Dr.~Laurent Villard & Professor at EPFL - Referee\\
Dr.~Fulvio Zonca & Senior Researcher at ENEA - Referee\\
\par \\
\hline
\end{tabular}
\par
\vspace{1.5cm}
\textsf{Sep 2007 - Dec 2010}
%\end{singlespace}
\end{center}
\end{titlepage}

%\pagenumbering{roman}

% Skipping one page
\newpage
$\quad$
\thispagestyle{empty} %no page number
\clearpage
\newpage

\newpage
\vspace*{5.0cm}
\includegraphics{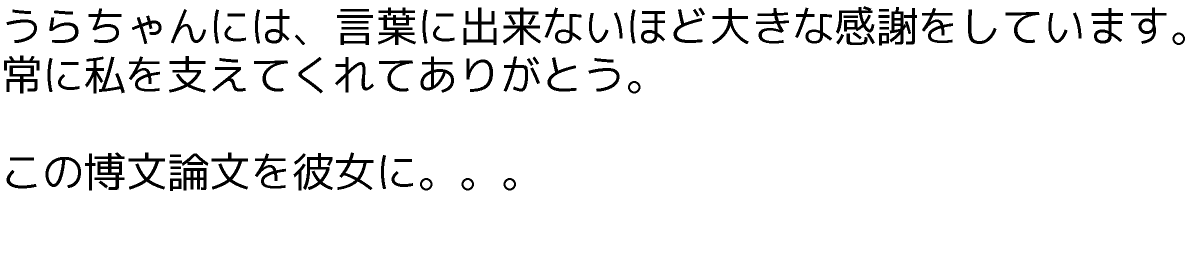}

\thispagestyle{empty} %no page number

\chapter*{Acknowledgements}

\thispagestyle{empty} %no page number

A PhD thesis is like a marathon in some sense. More than reaching the finish line, the point is what is learned in the process, from the first training session to the last spark. But, unlike a marathon, which, with enough inner strength, can be achieved without external help, the success of a PhD hinges on the qualities of surrounding people. A few lines in this manuscript is the least I can do to thank those who contributed to the quality of this work, or the quality of my PhD student's life. And since it is human nature to read the Acknowledgment section only, to check if one's name is included, I hope I do not forget anybody...\\

My first thanks naturally go to my two supervisors, Yasuhiro Idomura and Xavier Garbet, who both showed incredible pedagogy and patience, while keeping relationships simple and friendly.

Y.~I.~has been a supervisor like every PhD student should dream of. He formulated many ideas that are developed in this manuscript, helped me structure my thoughts and writings, instilling his analytic way of thinking, and proofread most of my output with the utmost care, while giving me complete freedom in my final choices. I am very honored to have been his first PhD student, and hope this was not my last opportunity to work with him. I am also grateful to him for helping me set up in Japan and explaining parts of Japanese culture.

X.~G.~always had time to spare for me, despite incredibly tight agenda. While we had less occasions to interact because of the distance, these interactions were extremely precious to the present work and to me. Since he magically has the answer for any problem, he was like a trump card when I was stuck. I am grateful for his teaching the importance of advertisement and diplomacy in research.

Jean-Marcel Rax kindly accepted to take the formal role of thesis director, and was the eye of the doctoral school of Ecole Polytechnique while I was away.\\

I sincerely thank Shinji Tokuda for profound lessons on Landau damping and MHD, which were given with extreme pedagogy.
Koji Shinohara gave me invaluable help in understanding experimental problematics and diagnostic issues.
The quality of this work was greatly improved by eye-opening discussions with Yasushi Todo, Boris Breizman, Matthew Lilley, Simon Pinches, Fulvio Zonca, and Patrick Diamond.
I am grateful for fruitful collaborations and nice memories with Christine Nguyen and Nicolas Dubuit.

I want to thank the friends I found in Japan, who made me feel like I was leaving home after three years, while I was coming back to my country.
Warm thanks go to my \textit{sempai} S{\'e}bastien Jolliet, for his friendship. I am indebted to him for answering my questions on many topics, and for sharing some of my best memories of Japan.
Thanks to my \textit{kouhai} Miho Janvier for listening to my patronizing speeches and for her words of motivation.
Thanks to Yann Camenen for his advices on planning the after-thesis and for his relaxed attitude to which I take example.
I also benefited from interactions with Ken Kawaoto and Remy Cardinet, since teaching is often even more formative for the teacher than for the student.
I thank all the people of CCSE in Inarich{\^o}, especially Masahiko Machida for always having something to chat about. I thank the theory group and the JT-60 team in Naka, Nobuyuki Aiba for frequent rides from the station, Masatoshi Yagi, Yasutomo Ishii, Naoaki Miyato, Makoto Hirota, Junya Shiraishi and Mitsuru Honda for memorable welcoming and farewell parties. I thank Toshio Hirayama and the late Toshihide Tsunematsu for approving my stay.

In France, I owe many thanks to my long-time friends, Audrey, Julie, Christophe, Edouard, and Mathieu, whose trust made me keep looking ahead and reaching forward. Witnessing their friendship, untouched despite the distance, was a real comfort.
I also thank my mother who dealt with personal issues in France while I was in Japan, and my whole family for their constant support.
I thank Virginie Grandgirard for insightful discussions on numerical issues and for helping me set up in the legendary Hameau ; Guillaume Latu for introducing me to OpenMP and helping me set up on Cadarache super-calculators. I thank my office-mates, Farah Hariri who really brightened my working days and weekends in the Hameau, and David Zarzoso who was so nice with me even when I was using all our computing resources. Thanks to J{\'e}r{\'e}mie Abiteboul for helping me send the manuscript in time, and for his communicative cheerfulness. Thanks to Shimpei Futatani for visiting our office many mornings. Thanks to the other students or ex-students for removing any residual stress, Guilhem Dif-Pradalier, Patrick Tamain, Eric Nardon, Stanislas Pamela, Antoine Merle, Nicolas Mellet, Thimoth{\'e}e Nicolas, Didier Vezinet, Dmytro Meshcheriakov, Hugo Bufferand, Gr{\'e}goire Hornung, Francois-Xavier Duthoit, Sylvain Rauch, and Jonathan Jacquot. Thanks to the regulars of the 9$:$45 AM coffee break for making me laugh every day, Gloria Falchetto, Marina B{\'e}coulet, Chantal Passeron, Yanick Sarazin, Patrick Maget, Joan Decker, R{\'e}mi Dumont, Philippe Ghendrih, Simon Allfrey, Guido Huysmans, and Maurizio Ottaviani. I thank Tuong Hoang, Alain B{\'e}coulet, and Xavier Litaudon for their support. 

I thank the people of NIFS, CRPP, and NFRI who hosted my visits. I also thank people I met in conferences or seminars, with whom I had insightful discussions, Guo-Yong Fu, Kazuo Toi, Rowdy Vann, Choong-Seock Chang, Chang-Mo Ryu, Raffi Nazikian, Kenji Imadera, and many others. Thanks to the organizers and participants of Festival de Th{\'e}orie 2007 and 2009, where I learned so much.% and ITPA meetings.%I also want to thank Kenji Imadera, Raffi Nazikian for diverse discussions in conferences.

Warm thanks to Anthony Genot, who just obtained a PhD from Merton college, and remained an invaluable friend throughout my thesis. He put me back on track when metaphysical doubts were seizing me. He also experimented and reported the various phases of a thesis a few months before me so that I would always be prepared...
% Working on nanotechnologies, he also gave me an insight on particularities of research in the plasma community.\\

For taking care of administrative procedures, I thank Kazutoyo Onozaki, St\'ephanie Villechavrolle, and Valerie Icard, who also organized the defense.

Finally I thank the members of my jury, for taking the time to read my manuscript and to attend my defense.

This work was performed within the frame of a co-thesis agreement between the doctoral school of the Ecole Polytechnique, the Japan Atomic Energy Agency, and the Commissariat \`a l'Energie Atomique. It was supported by both the JAEA Foreign Researcher Inviting Program and the European Communities under the contract of Association between EURATOM and CEA, and by the MEXT, Grant No.~22866086. Financial support to attend the IAEA-FEC 2010 conference in Daejeon was kindly provided by NFRI-WCI Center for Fusion Theory. The views and opinions expressed herein do not necessarily reflect those of the European Commission. Computations were performed on Altix3700 and BX900 systems at JAEA, and on Norma system at CEA.\\

\thispagestyle{empty} %no page number

%Finally, no word of acknowledgement can make justice to the unfaltering support of Ula-chan, who gives a meaning to my efforts.\\

%I dedicate this work to her.\\

\newpage
\thispagestyle{empty}
\chapter*{Abstract}

%\selectlanguage{english}
%\begin{abstract}
The achievement of sustained nuclear fusion in magnetically confined plasma relies on efficient confinement of alpha particles, which are high-energy ions produced by the fusion reaction. Such particles can excite instabilities in the frequency range of Alfv\'en Eigenmodes (AEs), which significantly degrade their confinement and threatens the vacuum vessel of future reactors.
%Consequent ejection of alpha particles before they could give their energy to the core to sustain the reaction, and potential damages on the vacuum vessel are major concerns for future reactors.
In order to develop diagnostics and control schemes, a better understanding of linear and nonlinear features of resonant interactions between plasma waves and high-energy particles, which is the aim of this thesis, is required.
%In particular, quantitative linear predictions of the Toroidicity-induced Alfv\'en Eigenmode (TAE), which results from the toroidal coupling of two poloidal modes in Tokamak devices, remain elusive in the general case. However, for an isolated single resonance, it is possible to reduce the description of wave-particle interactions to a simple harmonic oscillator. 
In the case of an isolated single resonance, the description of AE destabilization by high-energy ions is homothetic to the so-called Berk-Breizman (BB) problem, which is an extension of the classic bump-on-tail electrostatic problem, including external damping to a thermal plasma, and collisions. 
A semi-Lagrangian simulation code, COBBLES, is developed to solve the initial-value BB problem in both perturbative ($\delta f$) and self-consistent (full-$f$) approaches. Two collision models are considered, namely a Krook model, and a model that includes dynamical friction (drag) and velocity-space diffusion.
The nonlinear behavior of instabilities in experimentally-relevant conditions is categorized into steady-state, periodic, chaotic, and frequency-sweeping (chirping) regimes, depending on external damping rate and collision frequency. The chaotic regime is shown to extend into a linearly stable region, and a mechanism that solves the paradox formed by the existence of such subcritical instabilities is proposed.
Analytic and semi-empirical laws for nonlinear chirping characteristics, such as sweeping-rate, lifetime, and asymmetry, are developed and validated. Long-time simulations demonstrate the existence of a quasi-periodic chirping regime. Although the existence of such regime stands for both collision models, drag and diffusion are essential to reproduce the alternation between major chirping events and quiescent phases, which is observed in experiments.
Based on these findings, a new method for analyzing fundamental kinetic plasma parameters, such as linear drive and external damping rate, is developed. The method, which consists of fitting procedures between COBBLES simulations and quasi-periodic chirping AE experiments, does not require any internal diagnostics. This approach is applied to Toroidicity-induced AEs (TAEs) on JT-60 Upgrade and Mega-Amp Spherical Tokamak devices, which yields estimations of local kinetic parameters and suggests the existence of TAEs relatively far from marginal stability. The results are validated by recovering measured growth and decay of perturbation amplitude, and by estimating collision frequencies from experimental equilibrium data.

\thispagestyle{empty}
\chapter*{R{\'e}sum{\'e}}

%\selectlanguage{french}
%\begin{abstract}
Le succ{\`e}s de la fusion nucl{\'e}aire par confinement magn{\'e}tique repose sur un confinement efficace des particules alpha, qui sont des ions hautement {\'e}nerg{\'e}tiques produits par les r{\'e}actions de fusion. De telles particules peuvent exciter des instabilit{\'e}s dans le domaine de fr{\'e}quence des modes d'Alfv\'en (AEs) qui d{\'e}gradent leur confinement et risquent d'endommager l'enceinte {\`a} vide de r{\'e}acteurs futurs.
Afin de d{\'e}velopper des diagnostiques et moyens de contr{\^o}le, une meilleure compr{\'e}hension des comportements lin{\'e}aire et non-lin{\'e}aire des interactions r{\'e}sonantes entre ondes plasma et particules {\'e}nerg{\'e}tiques, qui constitue le but de cette th{\`e}se, est requise.
Dans le cas d'une r{\'e}sonance unique et isol{\'e}e, la description de la d{\'e}stabilisation des AEs par des ions {\'e}nerg{\'e}tiques est homoth{\'e}tique au probl{\`e}me de Berk-Breizman (BB), qui est une extension du probl{\`e}me classique de l'instabilit{\'e} faisceau, incluant un amortissement externe vers un plasma thermique, et des collisions.
Un code semi-Lagrangien, COBBLES, est d{\'e}velopp{\'e} pour r{\'e}soudre le probl{\`e}me aux valeurs initiales de BB selon deux approches, perturbative ($\delta f$) et auto-coh{\'e}rente (full-$f$). Deux mod{\`e}les de collisions sont consid{\'e}r{\'e}s, {\`a} savoir un mod{\`e}le de Krook, et un mod{\`e}le qui inclue la friction dynamique et la diffusion dans l'espace des vitesses.
Le comportement non-lin{\'e}aire de ces instabilit{\'e}s dans des conditions correspondantes aux exp{\'e}riences est cat{\'e}goris{\'e} en r{\'e}gimes stable, p{\'e}riodique, chaotique, et de balayage en fr{\'e}quence (sifflet), selon le taux d'amortissement externe et la fr{\'e}quence de collision. On montre que le r{\'e}gime chaotique d{\'e}borde dans une r{\'e}gion lin{\'e}airement stable, et l'on propose un m{\'e}canisme qui r{\'e}soud le paradoxe que constitue l'existence de telles instabilit{\'e}s sous-critiques.
On d{\'e}veloppe et valide des lois analytiques et semi-empiriques r{\'e}gissant les caract{\'e}ristiques non-lin{\'e}aires de sifflet, telles que la vitesse de balayage, la dur{\'e}e de vie, et l'asym{\'e}trie. Des simulations de longue dur{\'e}e d{\'e}montrent l'existence d'un r{\'e}gime de sifflets quasi-p{\'e}riodiques. Bien que ce r{\'e}gime existe quel que soit l'un des deux mod{\`e}les de collision, la friction et la diffusion sont essentielles pour reproduire l'alternance entre sifflets et p{\'e}riodes de repos, telle qu'observ{\'e}e exp{\'e}rimentalement.
Gr{\^a}ce {\`a} ces d{\'e}couvertes, on d{\'e}veloppe une nouvelle m{\'e}thode pour analyser des param{\`e}tres cin{\'e}tiques fondamentaux du plasma, tels que le taux de croissance lin{\'e}aire et le taux d'amortissement externe. Cette m{\'e}thode, qui consiste {\`a} faire correspondre les simulations de COBBLES avec des exp{\'e}riences d'AEs qui pr{\'e}sentent des sifflets quasi-p{\'e}riodiques, ne requiert aucun diagnostique interne. Cette approche est appliqu{\'e}e {\`a} des AEs induits par la toroidicit{\'e} (TAEs) sur les machines JT-60 Upgrade et Mega-Amp Spherical Tokamak. On obtient des estimations de param{\`e}tres cin{\'e}tiques locaux qui sugg{\`e}rent l'existence de TAEs relativement loin de la stabilit{\'e} marginale. Les r{\'e}sultats sont valid{\'e}s en recouvrant la croissance et d{\'e}croissance de l'amplitude des perturbations mesur{\'e}es, et en estimant les fr{\'e}quences de collision {\`a} partir des donn{\'e}es exp{\'e}rimentales d'{\'e}quilibre.
%\end{abstract}
%\selectlanguage{english}

\pagenumbering{roman}

\tableofcontents
%\listoffigures
%\listoftables

\chapter*{Abbreviations}

\noindent
BB: Berk-Breizman.\\
EP: Energetic Particles.\\
MHD: Magneto-HydroDynamics.\\
SAW: Shear-Alfv\'en Wave.\\
AE: Alfv\'en Eigenmode.\\
TAE: Toroidicity-induced Alfv\'en Eigenmode.\\
BAE: Beta-induced Alfv\'en Eigenmode.\\
KAW: Kinetic Alfv\'en Wave.\\
EPM: Energetic Particle Mode.\\
1D $\ldots$ 6D: one dimension $\ldots$ six dimensions. Real space dimensions, unless explicitly specified to be phase-space dimensions.\\
NBI: Neutral Beam Injection.\\
ICRH: Ion Cyclotron Resonant Heating.\\
DCE: Displacement Current Equation.\\
BGK: Bernstein-Greene-Kruskal.\\
CIP: Constrained Interpolation Profile.\\
R-CIP: Rational interpolation version of CIP.\\
CIP-CSL: Conservative Semi-Lagrangian version of CIP.\\
ITER: International Thermonuclear Experimental Reactor (Cadarache, France).\\
JT-60U: JT-60 Upgrade (Naka, Japan).\\
DIII-D: Doublet III-D (San Diego, USA).\\
MAST: Mega Amp Spherical Tokamak (Culham, UK).\\
START: Small Tight Aspect Ratio Tokamak (formerly Culham, UK).\\
NSTX: National Spherical Torus Experiment (Princeton, USA).\\
LHD: Large Helical Device (Toki, Japan).\\
CHS: Compact Helical System (Toki, Japan).\\
COBBLES: COnservative Berk-Breizman semi-Lagrangian Extended Solver.\\
HAGIS: Hamiltonian Guiding Center System.\\

\addtocounter{page}{-4}

\pagenumbering{arabic}

\chapter{Introduction}
\label{ch:intro}

This manuscript has been written in an effort toward pedagogy, privileging clarity over details, physical pictures over mathematical developments. We hope it can be used as an introduction to the Berk-Breizman problem and its numerical computation. However we cannot pretend being able to introduce notions such as nuclear fusion, magnetic plasma confinement, Magneto-HydroDynamics (MHD), kinetic theory, tokamak geometry, or particle orbits, with as much clarity as in well-known textbooks. For this reason, the reader is assumed to be familiar with the basis of the above fields.
In this introduction we expose the background and motivation for our work.

\section{Energetic particles-driven AEs}

In an ignited tokamak operating with a deuterium-tritium mix, the confinement of $\alpha$-particles is critical to prevent damages on the first-wall and to achieve break-even. The reason is that these high energy particles must be kept long enough in the plasma core to allow enough of their energy to heat thermal populations by slowing-down processes. A major concern is that high energy ions can excite plasma instabilities in the frequency range of Alfv\'en Eigenmodes (AEs), which significantly enhance their transport. 
Ever since the recognition of this issue in the 1970's, considerable progress has been made in the theoretical understanding of the principal Alfv\'enic instabilities. However, the estimation of the mode growth rate is complex, and the question of their stability in ITER \cite{machine_ITER} remains to be clarified. In addition, estimation of the effect on transport and development of diagnostics and control schemes require significant progress on our understanding of nonlinear behaviors.

We limit our framework to the tokamak configuration. In this work, we focus our interest on the Toroidicity induced Alfv\'en Eigenmode (TAE) \cite{cheng85}, which is a shear-Alfv\'en wave (SAW) located in a gap of the SAW continuum that is created by toroidal coupling of two successive poloidal modes, and which can be destabilized by resonant interactions with high-energy ions. TAEs have been observed to be driven by $\alpha$-particles in burning plamas \cite{wong96,nazikian97}, Ion-Cyclotron-Resonance Heating (ICRH) \cite{wong94}, and Neutral Beam Injection (NBI) \cite{wong91,heidbrink91}. In this work we consider only the latter situation (NBI-driven TAEs).
%Most present-day devices operate with pure deuterium plasma, to avoid complications linked with the handling of tritium. In the absence of $\alpha$-particles as a heat source, several heating schemes are used, including NBI, which is the source of high-energy ions we consider in this work.

\section{The BB problem as a paradigm for TAEs}

In general, TAEs are described in a three-dimensional (3D) configuration space. However, near a phase-space surface where the resonance condition is satisfied (resonant phase-space surface), it is possible to obtain a new set of variables in which the perturbation is described by a one-dimensional (1D) effective Hamiltonian in 2 conjugated variables \cite{berkbreizman97ppr,berkbreizman97pop,wong98,garbet08}, if we assume an isolated single mode. In this sense, the problem of kinetic interactions between TAEs and fast-particles is homothetic to a simple 1D single mode bump-on-tail instability. The so-called Berk-Breizman (BB) problem \cite{berkbreizman93,berkbreizman97ppr,berkbreizman97pop,berkbreizman96} is a generalization of the bump-on-tail problem, where we take into account an external wave damping accounting for background dissipative mechanisms, and a collision operator. Observed qualitative and quantitative similarities between BB nonlinear theory and both global TAE simulations \cite{wong98,pinches04S47} and experiments \cite{fasoli98,heeter00} are an indication of the validity of this reduction of dimensionality.
Similar arguments exist for other Alfv\'en wave instabilities such as the geodesic acoustic mode (GAM) \cite{berk06}, and the beta Alfv\'en eigenmode (BAE) \cite{nguyen_thesis}. These analogies enable more understanding of fully nonlinear problems in complex geometries by using a model that is analytically and numerically tractable. This approach is complementary to heavier 3D analysis.

\section{Frequency sweeping}

A feature of the nonlinear evolution of AEs, the frequency sweeping (chirping) of the resonant frequency by 10-30$\%$ on a timescale much faster than the equilibrium evolution, has been observed in the plasma core region of tokamaks JT-60 Upgrade (JT-60U) \cite{kusama99}, DIII-D \cite{heidbrink95}, the Small Tight Aspect Ratio Tokamak (START) \cite{mcclements99}, the Mega Amp Spherical Tokamak (MAST) \cite{pinches04S47}, the National Spherical Torus Experiment (NSTX) \cite{fredrickson06}, and in stellerators such as the Large Helical Device (LHD) \cite{osakabe06}, and the Compact Helical Stellerator (CHS) \cite{takechi99}. In general, two branches coexist, with their frequency sweeping downwardly (down-chirping) for one, upwardly (up-chirping) for the other. In many experiments, asymmetric chirping has been observed, with the amplitude of down-chirping branches significantly dominating up-chirping ones.
Chirping TAEs have been reproduced in 3D simulations with a hybrid MHD/drift-kinetic code \cite{todojpfr03}, and with a drift-kinetic perturbative code \cite{pinches04S47}.
Qualitatively similar chirping modes are spontaneously generated by the BB model, and have been shown to correspond to the evolution of holes and clumps in the velocity distribution. Theory relates the time evolution of the frequency shift with the linear drive $\gamma _L$ and the external damping rate $\gamma _d$ \cite{berkbreizman97pla}, when the evolution of holes and clumps is adiabatic.
The idea of using chirping velocity (or sweeping rate) as a diagnostic for growth rates looks promising. If we assume that the mode is close to marginal stability, $\gamma _L \approx \gamma _d$, and if we further assume that holes and clumps are in the adiabatic regime, then the growth rates are simply obtained by fitting analytic prediction to experimental measurement of chirping velocity.
However, these two assumptions are not verified in general, which limits the applicability of this simple approach.
In most experiments, chirping events are quasi-periodic, with a period in the order of the millisecond. On the one hand, the long-time simulation of repetitive chirping with 3D codes is a heavy task, which has not been undertaken yet. On the other hand, simulations of the BB model with many chirping events have been performed \cite{vann07,lilley10}, but such solutions do not feature any quasi-periodicity. In this sense, repetitive chirping that qualitatively agrees with an experiment has not been reproduced, neither with 1D nor 3D simulations.

\section{Aim of this thesis}

Our main goals are to improve our understanding of wave-particle nonlinear resonant interactions, develop diagnostics and identify control parameters for AEs. From these backgrounds, a straight-forward plan is to:
\begin{itemize}
\item clarify the link between BB model and AEs,
\item extend BB theory where it is insufficient to explain experimental observations,
\item analyze experimental data by applying the BB model to AEs.
\end{itemize}

\section{Outline}

In brief, we reduce the problem of TAEs to a numerically and analytically tractable paradigm, the BB problem. We make a survey of linear and nonlinear physics of the BB model, and work on improving our understanding, by extending theory and by using systematic numerical observations, focusing on the frequency sweeping regime. Armed with new findings and robust numerical tools, we finally go back to the original problem of TAEs, explaining quantitative features of experimental observations, and developing a new method for accurate linear predictions.

In Chap.~\ref{ch:tae}, we review the physics of the TAE, and simplify the description of the problem from 6D to 2D in phase-space, around a single resonant phase-space surface. The first step in this procedure is to isolate the gyromotion, which is decoupled from the wave for typical TAEs with $f\sim 100$ kHz, by changing variables to the guiding-center coordinates. This change of variables is based on the so-called Lie transform perturbation theory, which we review in order to get a better grasp of involved hypothesis. Then we show how to reduce the guiding-center Hamiltonian, as well as collision operator and background mechanisms, and discuss limitations of this simplified description. 
In Chap.~\ref{ch:bbmodel}, we review basic nonlinear physics of the BB problem. We develop and verify numerical tools that we use in following chapters. In particular, we develop and verify a kinetic code to solve the initial-value problem. We show that numerical simulations in experimentally relevant conditions, with cold-bulk and weak, warm-beam distributions, require a careful choice of advection scheme. Among the family of so-called Constrained Interpolation Profile (CIP) schemes, a locally conservative version (CIP-CSL) displays the best convergence properties. As an intermediate summary, we cast the analogies between BB model and 1D description of TAEs.
In Chap.~\ref{ch:nonlinearities}, we investigate kinetic features of self-consistent interactions between an energetic particle beam and a weakly unstable electrostatic wave in 1D plasma, within the BB framework, with an emphasis on chirping regime and subcritical instabilities. We show that the nonlinear time-evolution of chirping in 1D simulations can be used to retrieve information about linear input parameters with good precision. We identify a regime where chirping events are quasi-periodical. This regime exists whether the collision model is annihilation/creation type, or takes into account dynamical friction and velocity-space diffusion.
Based on these findings, in Chap.~\ref{ch:experiment}, we propose a new method to estimate local linear drive, external damping, and collision frequencies from the spectrogram of magnetic field variations measured by a Mirnov coil at the edge of the plasma. This method, which relies on a fitting of normalized chirping characteristics between the experiment and BB simulations, is described and applied to TAE experiments in MAST and JT-60U. We show that the BB model can successfully reproduce features observed in the experiment only if the collision operator includes drag and diffusion terms. We find a quantitative agreement for the diffusion frequency, and a qualitative agreement for the drag frequency, between the values obtained with our fitting procedure, and an independent estimation obtained from experimental equilibrium measurements.
In the conclusion (Chap.~\ref{ch:conclusion}), we summarize our findings, and we suggest numerical and experimental investigations these findings enable.

\chapter{Energetic particle-induced TAEs}
\label{ch:tae}

The TAE is a shear-Alfv\'en wave located in a gap of the SAW continuum that is created by toroidal coupling of two successive poloidal modes, which can be destabilized by energetic particles. This chapter deals with the modelization at several levels of complexity of an isolated single-mode TAE. In Sec.~\ref{sec:review_tae}, we give a short review of the basic TAE physics, and of the state-of-the-art of its linear stability analysis, emphasizing a need for improved accuracy. When the TAE is an isolated single-mode, it is possible to reduce the problem to a simple harmonic oscillator, by expanding the perturbed Hamiltonian around a resonant phase-space surface. This reduction from 3D to 1D becomes particularly transparent in angle-action variables ($\alpha _i$, $J_i$), once the perturbation has been put in the form $\exp (\imath l_i\dot\alpha _i-\imath \omega t)$, where $l_i$ are integers. In Sec.~\ref{sec:gyrokinetic_TAE}, we show how to obtain the latter form. The first step is to separate the gyromotion, which does not resonate with typical TAEs, by changing variables to guiding-center coordinates. This procedure can be done while conserving the Hamiltonian properties by making use of Lie transform perturbation theory, which we review thoroughly. In Sec.~\ref{sec:3D_to_1D}, we show how to reduce the Hamiltonian, collision operator, and background damping mechanisms from their 3D description to a 1D model.

\section{Review \label{sec:review_tae}}

This short review is intended to introduce notions and notations that we use when applying the BB model to the TAE, and to motivate our linear analysis of TAEs. For a more comprehensive review of energetic-particle driven AEs, see Ref.~\cite{heidbrink08}.

\subsection{Toroidal geometry}

Hereafter, we make use of magnetic flux coordinates \cite{hazeltinemeiss}, also called as straight field-line coordinates, ($r$, $\theta$, $\zeta$), where $r$, $\theta$ and $\zeta$ are minor radius, poloidal and toroidal angle, respectively, to describe the nested poloidal flux surfaces of the equilibrium magnetic field $\boldsymbol{B_0}$. In these coordinates, $\boldsymbol{B_0}$ takes the following form,
\begin{equation}
\boldsymbol{B_0} \;=\; \grad{\chi} \, \boldsymbol{\times} \, \grad{(q \theta-\zeta)},
\end{equation}
where $\chi (r)$ is the poloidal flux (divided by $2 \pi$), and the safety factor, defined by
\begin{equation}
q(r) \;\equiv\; \frac{\boldsymbol{B_0}\bdot \grad{\zeta}}{\boldsymbol{B_0}\bdot \grad{\theta}},
\end{equation}
is the flux surface-averaged number of toroidal rotation that a field line undergoes in one poloidal rotation.

\subsection{Physics of the TAE}

\begin{figure}
\begin{center}
\includegraphics{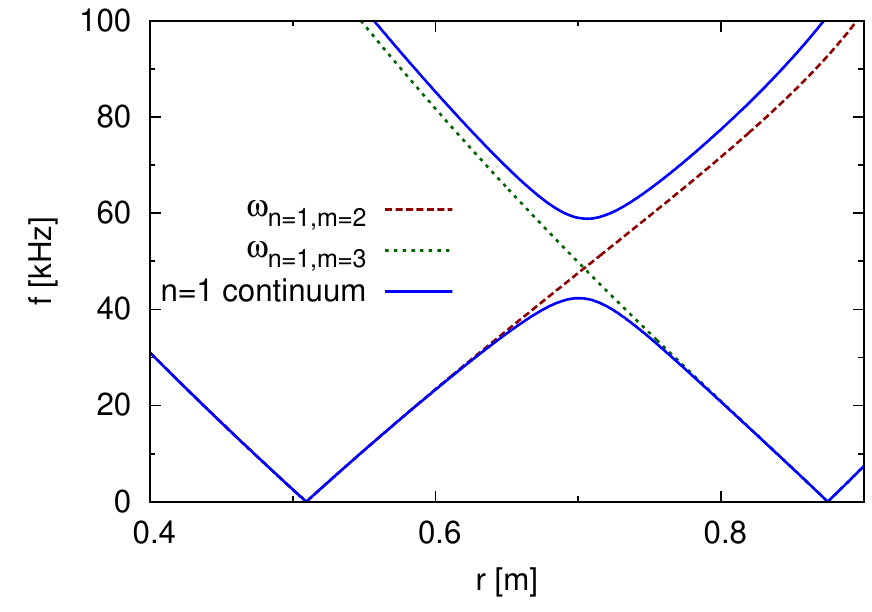}
\caption{Alfv\'en continuum for $n=1$, with and without coupling between $m=2$ and $m=3$ poloidal modes. The q profile has been modeled by $q(r)=1.2+2.1\,(r/a)^{1.5}$, and the electronic density by $n_e(r)=0.11+1.57\,(1-r^2/a^2)^{0.3}$ [$10^{19}$ m$^{-3}$]. Other parameters are $B_0=1.2$ T, $R_0=3.3$ m, and $a=0.96$ m. Note that the discrepancy, relatively far from resonance, between the upper branch of coupled continuum and the uncoupled $m=2$ branch, is accounted by errors of second-order in the aspect ratio. Similar discrepancy is observed in Fig.~1 of Ref.~\cite{fu89} for example.}
\label{fig:continuum}
\end{center}
\end{figure}
%The measured TAE frequency, and extent of chirping are represented by an horizontal line and a vertical arrow, respectively. 

In an homogeneous magnetized plasma, linear ideal-MHD arguments show the existence of a shear-Alfv\'en wave of frequency $\omega _\mathrm{A}$ with the dispersion relation
\begin{equation}
\omega _\mathrm{A}^2 \;=\; k_{\|}^2 \, v_\mathrm{A}^2, \label{eq:SAW_disp_rela}
\end{equation}
where
\begin{equation}
v_\mathrm{A} \;=\; \frac{B_0}{\sqrt{\mu _0 \sum_i n_i m_i}}
\end{equation}
is the Alfv\'en velocity, and $k_{\|}$ is the wave number in the direction of the equilibrium magnetic field $\boldsymbol{B_0}$.
Let us consider axisymmetric toroidal plasmas. In the cylindrical limit, the periodicities of the system require that there exists two integers, a toroidal mode number $n$ and a poloidal mode number $m$, such that
\begin{equation}
k_{\|} \;=\; \frac{n\,-\,m / q(r)}{R_0},
\end{equation}
where $R_0$ is the distance from the symmetry axis of the tokamak to the magnetic axis. In a non-homogeneous plasma in a sheared magnetic field, both $k_{\|}$ and $v_\mathrm{A}$ are functions of $r$. The simple dispersion relation Eq.~(\ref{eq:SAW_disp_rela}) is still valid in this configuration \cite{cheng86}, and it is called the Alfv\'en continuum. Since phase velocity is a function of radius, a wave packet with finite radial extent would suffer from phase-mixing, the so-called continuum damping. Except for energetic particle modes, which are outside of the scope of this work, resonant drive by fast particles is not enough to overcome this damping.
However, a toroidal coupling of two successive poloidal modes $m$ and $m+1$ breaks up the continuous spectrum. This is illustrated in Fig.~\ref{fig:continuum}, which shows the Alfv\'en continuum for $n=1$, $m=2$ and $m+1=3$, in cylindrical geometry, where two poloidal continuum are decoupled, and in toroidal geometry, with a two-mode coupling model \cite{cheng86,fu89}. The latter is obtained with equilibrium plasma parameters corresponding to JT-60U shot E32359 at $t=4.2$s, which we analyze in Chap.~\ref{ch:experiment}, assuming concentric circular magnetic flux surfaces, retaining toroidicity effects in the first order in inverse aspect ratio. Though we show only the $\omega >0$ half-plane, the continuum is symmetric with respect to $\omega$. Coupled modes are ($n$, $m$) and ($-n$, $-m-1$) for $\omega>0$, and ($n$, $m+1$) and ($-n$, $-m$) for $\omega<0$. The gap is centered at a radius $r_\mathrm{A}$ such that $q(r_\mathrm{A})=(m+1/2)/n$, where the two continuous spectra would cross in the absence of coupling, and where $\left| k_{\|} \right| = 1/2qR_0$. The resulting discrete eigenmode is a TAE, at a frequency $\omega _\mathrm{A}=v_\mathrm{A} / 2 q R_0$.

For a deuterium plasma with typical magnetic field $B_0 \sim 1 T$ and density $n_i \sim 10^{20} m^{-3}$, the Alfv\'enic energy is $E_\mathrm{A} \equiv m_i v_\mathrm{A}^2 /2 \sim 10 keV$, which is in the range of passing particles induced by NBI. For ITER parameters, $E_\mathrm{A} \sim 1 MeV$, which is in the range of passing $\alpha$-particles born from fusion reactions. In both cases, TAEs can be driven unstable by resonance with energetic particles.
For far-passing particles, the resonance condition is $\Omega = \omega _\mathrm{A}$, where
\begin{equation}
\Omega \;=\; n\,\omega _\zeta + l \,\omega _\theta, \label{eq:tae_resonance_condition}
\end{equation}
where $\omega _\zeta = v_\| / R_0$ and $\omega _\theta = v_\| / q R_0$ are frequencies of toroidal motion and poloidal motion, respectively, and $l=-m$ for co-passing particles, $l=m$ for counter-passing particles \cite{todo98}. Since we analyze TAEs driven by co-injected ions, we can simplify following discussions by considering only co-passing particles. Then, the resonance condition is
\begin{equation}
\omega _\mathrm{A} \;-\; n\,\frac{v_\|}{R_0} \;+\; m\,\frac{v_\|}{q \,R_0} \;=\; 0. \label{eq:tae_resonance_condition_co}
\end{equation}

%where now $v_{A}$ is the Alfv\'en velocity at the magnetic axis, since we take an orbit-averaged point-of-view.

\subsection{Linear stability}

In theory, the linear growth rate $\gamma\approx \gamma _L - \gamma _d$ is positive when the drive by fast particles overcomes damping processes to background plasma. The growth rate can be estimated either by linear stability codes, such as PENN \cite{jaun95}, TASK/WM \cite{fukuyama02}, NOVA-K \cite{cheng92}, or CASTOR-K \cite{borba02} ; or by gyro- or drift- kinetic perturbative nonlinear initial value codes, such as FAC \cite{candy97} or HAGIS \cite{pinches98}. The analysis requires internal diagnostics that are not always available.

The linear drive $\gamma _L$ depends on several factors such as spatial and energy gradients of EP distribution, and alignment between particle orbit and the eigenmode. In the limit of large aspect ratio, analytic theory \cite{fu89} yields successful estimations of $\gamma _L$ for well-defined numerical models \cite{todo95}. However, in the general case, complicated factors cited above forbid accurate analytic estimation, as yet. Moreover, improvements in measurement of EP distributions are needed to allow estimation of $\gamma _L$ for experimental TAEs.

The external damping rate $\gamma_d$ involves complicated mechanisms, which include continuum damping (since parts of the mode extend spatially into the continuum) \cite{zonca92}, radiative damping \cite{mett92}, Landau damping with thermal species \cite{betti92,zonca96}, and collisional damping by trapped electrons \cite{gorelenkov92}. Most of these mechanisms are still under investigation, and cannot be quantified by existing theory. Experimentally, $\gamma _d$ can be estimated by active measurements of externally injected perturbations \cite{fasoli95,fasoli00}. However, the applicability of this technique is limited to dedicated experiments.

Moreover, if the system is close to marginal stability, $\gamma$ is sensitive to small variations of driving and damping terms, and is very sensitive to plasma parameters such as the $q$ profile.
In addition, the existence of unstable AEs in a regime where linear theory predicts $\gamma < 0$, or subcritical AEs, has not been ruled out. To access the subcritical regime, nonlinear analysis is necessary.
Therefore, accurate linear stability analysis requires significant theoretical and experimental improvements, or a new approach.

%amplitudes of poloidal harmonics of TAE and drift velocity,

\section{Angle-action description} \label{sec:gyrokinetic_TAE}

\subsection{Kinetic description}

Since the kinetic equation is at the center of interest of this thesis, it is useful to get a deep understanding of its meaning, by reviewing its derivation from fundamental principles. The steps we summarize here are detailed in Ref.~\cite{kralltrivelpiece} for example.\\

A many-body system is completely described by the microscopic distribution $N(\boldsymbol{z},t)=\sum \delta(\boldsymbol{z}-\boldsymbol{z_i})$, where $\boldsymbol{z_i}\equiv(\boldsymbol{x_i},\boldsymbol{v_i})$, $\boldsymbol{x_i}$ and $\boldsymbol{v_i}$ are the position and velocity of a particle indexed as $i$, and the sum is taken over all particles. To simplify the present discussion, we consider a single-species system of $N_p$ particles, without external forces, and normalize the total phase-space volume, assumed constant, to $1$. Substituting Newton equations of motion into the partial time derivative of $N$ yields the so-called Klimontovich-Dupree equation,
\begin{equation}
\pd{N}{t} \;+\; \boldsymbol{v}\bdot \pd{N}{\boldsymbol{x}} \;+\; \boldsymbol{a^\mathrm{micro}}\bdot \pd{N}{\boldsymbol{v}} \;=\; 0, \label{eq:klimontovich_dupree}
\end{equation}
where $\boldsymbol{a^\mathrm{micro}}(\boldsymbol{z})$ is the acceleration due to microscopic electromagnetic fields, at the exception of those due to a particle located at $\boldsymbol{z}$, if such a particle exists. The microscopic electromagnetic fields obey Maxwell equations, where density and current are velocity moments of $N$.

Since it is impossible to reproduce any many-body experiment at the microscopic level, it is much more efficient to take an ensemble point-of-view, where distributions and fields are smooth functions of phase-space. The statistical properties are completely determined by the distribution $F_N(\boldsymbol{z'_1},\boldsymbol{z'_2},\ldots, \boldsymbol{z'_{N_p}},t)$, where $F_N \mathrm{d} V_1 \ldots \mathrm{d} V_{N_p}$ is the probability of finding, at a time $t$, particle 1 within $\mathrm{d} V_1$, particle 2 within $\mathrm{d} V_2$, $\ldots$ and particle $N_p$ within $\mathrm{d} V_{N_p}$, and $\mathrm{d} V_i$ is an infinitesimal phase-space volume at the neighborhood of $\boldsymbol{z}=\boldsymbol{z'_i}$. By integrating $F_N$ over all $\boldsymbol{z'_i}$ but $\boldsymbol{z'_1}$, we can define the one-particle distribution function $f_1$, where $f_1(\boldsymbol{z'_1},t) \mathrm{d} V_1$ is the probability of finding one particle within $\mathrm{d} V_1$ at $t$. $f_1$ is related to the microscopic distribution by
\begin{equation}
f_1(\boldsymbol{z_1},t)\;=\; \frac{\left\langle N(\boldsymbol{z_1},t) \right\rangle}{N_p},
\end{equation}
where $\left\langle N \right\rangle$ is the ensemble average $\int F_N N \mathrm{d} \boldsymbol{z'_1} \ldots \mathrm{d} \boldsymbol{z'_{N_p}}$.
Similarly, by integrating $F_N$ over all $\boldsymbol{z'_i}$ but $\boldsymbol{z'_1}$ and $\boldsymbol{z'_2}$, we can define the two-body distribution function $f_2(\boldsymbol{z'_1},\boldsymbol{z'_2},t)$, which is related to the microscopic distribution by
\begin{equation}
f_2(\boldsymbol{z_1},\boldsymbol{z_2},t)\;=\; \frac{\left\langle N(\boldsymbol{z_1},t) N(\boldsymbol{z_2},t) \right\rangle}{N_p^2} \;-\; \frac{\delta(\boldsymbol{z_1}-\boldsymbol{z_2})}{N_p} f_1(\boldsymbol{z_1},t),
\end{equation}
where we used a large particle number approximation, $N_p \gg 1$.
In the absence of atomic and nuclear processes, $N_p$ is a constant, then the ensemble average of Eq.~(\ref{eq:klimontovich_dupree}) yields
\begin{equation}
\pd{f_1}{t} \;+\; \boldsymbol{v}\bdot\pd{f_1}{\boldsymbol{x}} \;+\; \boldsymbol{a}\bdot \pd{f_1}{\boldsymbol{v}} \;=\; \left.\dfrac{\mathrm{d} f_1}{\mathrm{d} t}\right| _{\mathrm{coll.}}, \label{eq:averaged_klimontovich_dupree}
\end{equation}
where the average acceleration $\boldsymbol{a}\equiv\left\langle \boldsymbol{a^\mathrm{micro}} \right\rangle$ is given by electromagnetic fields that obey Maxwell equations, where density and current are velocity moments of $f_1$. The collision term,
\begin{equation}
\left.\dfrac{\mathrm{d} f_1}{\mathrm{d} t}\right| _{\mathrm{coll.}} \;\equiv\; \boldsymbol{a}\bdot \pd{f_1}{\boldsymbol{v}} \;-\; \frac{1}{N_p}\, \left\langle \boldsymbol{a^\mathrm{micro}}\bdot \pd{N}{\boldsymbol{v}} \right\rangle,
\end{equation}
is shown to vanish in the absence of particle interactions, yielding Vlasov equation, which is valid on a time-scale much shorter than a collisional time-scale. Eq.~(\ref{eq:averaged_klimontovich_dupree}) is not a closed equation, because the second part of the collision term involves expressions of the form $\left\langle N N \right\rangle$. Under the Coulomb approximation, which forbids any retardation effect, and which is valid if the thermal velocity is much slower than the speed of light, we can reduce the latter term as a function of $f_2$. Similarly, the equation which gives the evolution of $f_2$ involves terms of the form $\left\langle N N N \right\rangle$, and so on. Altogether, we have a chain of equations for which we need a closure.
A collision operator is an approximative statistical operator that accounts for particle interactions, which provides such closure. A clear review of collision operators is given in Ref.~\cite{helandersigmar}. In this thesis, we focus on a Fokker-Planck collision operator, which is based on the fact that, in a Tokamak plasma, collisions are dominated by binary Coulomb collisions with small-angle deflection.
%Although it can be demonstrated rigorously, we can easily understand intuitively that, in a low-density plasma, particle interactions are dominated by binary Coulomb collisions with

In the following, we write $f_1$ simply as $f$. The kinetic equation, Eq.~(\ref{eq:averaged_klimontovich_dupree}), can be put in Hamiltonian form,
\begin{equation}
\frac{\partial f}{\partial t} \;-\; \left\{ h,\,f \right\} \;=\; \left.\dfrac{\mathrm{d} f}{\mathrm{d} t}\right| _{\mathrm{coll.}}, \label{eq:3d_kinetic}
\end{equation}
where $h$ is the Hamiltonian, and $\{\}$ are Poisson brackets. The l.h.s.~is the variation of $f$ following particle orbits, or so-called Lagrangian derivative of $f$, noted $\mathrm{D}_t f$.

\begin{figure}
\begin{center}
\includegraphics{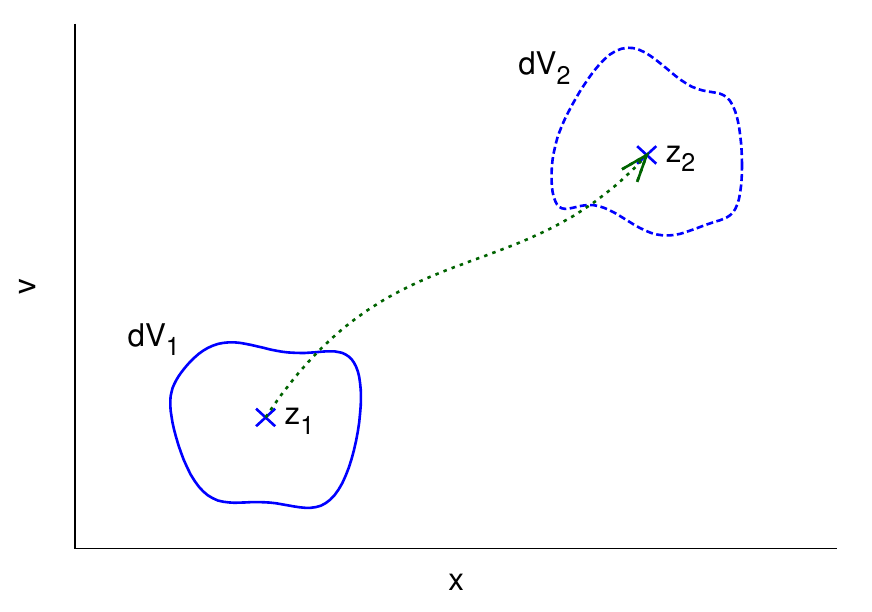}
\caption{Illustration of Liouville's theorem. Time-evolution of an infinitesimal volume in a 2D phase-space ($x$, $v$), delimited by a solid curve at $t=t_1$ and by a dashed curve at $t=t_2$.  The phase-space volume is constant, $\mathrm{d} V_1=\mathrm{d} V_2$, and, in the absence of collision, the number of particles inside the volume is constant too.}
\label{fig:liouville}
\end{center}
\end{figure}

The fact that, in the absence of collisions, $f$ is conserved along particle orbits, can be seen as a direct consequence of Liouville's theorem, which states that the density in phase-space is constant along particle orbits. Let us consider an infinitesimal volume of phase-space $\mathrm{d} V_1$ surrounding a particle at $z_1$ at $t=t_1$, as illustrated in Fig.~\ref{fig:liouville}. The particle and all points of $\mathrm{d} V_1$ evolve following the equations of motion until a time $t=t_2$ where the particle is at $z_2$, and $\mathrm{d} V_1$ is changed to a volume $\mathrm{d} V_2$. Liouville's theorem arises from the following properties,
\begin{itemize}
\item Since the points of the boundary of $\mathrm{d} V$ follow the equations of motion, in the absence of collisions, the number of particles inside $\mathrm{d} V$ is constant;
\item Time-evolution can be seen as a series of infinitesimal canonical transformations generated by the Hamiltonian;
\item Poincar{\'e}'s invariant: the volume element in phase-space is a canonical invariant.\\
\end{itemize}

%We are interested in the interactions between a TAE and energetic ions, whose distribution $f$ evolves, in the collisionless limit, following a Vlasov equation in 6D (3D position, 3D velocity) phase-space $z$,

%Gyrokinetic description of the interactions between trapped particles and the TAE. A comparison of Alfven velocity with resonant velocity for trapped particle vs passing particle for NNBI and ICRH justifies considering passing particles for NNBI and trapped particles for ICRH.

\subsection{Review of Lie transform theory}

When working in canonical variables, it is easy to exploit the properties of a Hamiltonian system. However, for some systems it is difficult to find canonical variables, and the most convenient variables may not be canonical. Moreover, if some variables are canonical for the unperturbed system, they may be non-canonical for the perturbed one. By applying the Lie near-identity transformation theory to the phase-space Lagrangian, one can change any Lagrangian into a simpler form in coordinates that reveal the symmetries of the system, and use this formulation to study the properties of a perturbed Hamiltonian system in any arbitrary noncanonical variables \cite{cary83}.

\subsubsection{\label{subsubsec:noncanHamMech}Noncanonical Hamiltonian Mechanics}

\def\PoissonTensor{\mathsf{J}}

Consider a Hamiltonian system with $N$ degrees of freedom. Hamilton equations are given by the application of a variational principle to the scalar Lagrangian $L$, in some arbitrary coordinates $z^\mu$ on the $2N+1$-dimensional extended phase space, ($t$, $z^\mu$),
\begin{equation}
\delta \int L(z^\mu (\lambda),\lambda) \, \mathsf{d}\lambda \;=\; 0,
\label{eq:varprcple}
\end{equation}
where $\lambda$ is an arbitrary parameter. Hereafter, Greek indices $\alpha$, $\mu$ and $\nu$ run from $0$ to $2N$, whereas Latin indices $a$ and $b$ run from $1$ to $N$, and $i$, $j$, $k$ from $1$ to $2N$.
Thus, any Hamiltonian system is characterized by its Lagrangian
\begin{equation}
L \;=\; \gamma _\mu \dfrac{\mathsf{d} z^\mu}{\mathsf{d}\lambda},
\end{equation}
or equivalently by its fundamental one-form, $\gamma _\mu \mathsf{d} z_\mu$.
In the canonical extended phase-space $z^\mu = (t,q^a,p^a)$, when $z^0 \equiv t$ is the time coordinate,
\begin{eqnarray}
\gamma _0 &=& -h(z^\mu ,t),\\
\gamma _a &=& p_a,\\
\gamma _{a+N} &=& 0,
\end{eqnarray}
where $h$ is the Hamiltonian.
However, in noncanonical variables, all the $\gamma _\mu$ may be nonzero.

From Eq.~(\ref{eq:varprcple}), which has the same form in any new coordinate system $Z^\mu$ with $\gamma _\mu \mathrm{d} z_\mu = \Gamma _\mu \mathrm{d} Z_\mu$, Euler-Lagrange equations are obtained as
\begin{equation}
\omega _{\mu \nu} \dfrac{\mathrm{d} z^\nu }{d\lambda} \;=\; 0,
\end{equation}
where $\omega$ is a tensor defined by
\begin{equation}
\omega _{\mu \nu} \;=\;  \pd{\gamma _\nu }{z^\mu} \,-\, \pd{\gamma _\mu }{z^\nu} ,
\end{equation}
and its restriction to the symplectic coordinates $\wh{\omega _{ij}}$ is the Lagrange tensor.
Thus the flow $\mathrm{d} _{\lambda}z^\mu$ is an eigenvector of $\omega _{\mu \nu}$ with eigenvalue $0$. The solution is unique only after a normalization. In physical terms, we can take $\lambda = t$. Since the Jacobian of the coordinate transformation $\gamma _{\mu} (z^\mu) = (-h,p_a,0) $ is nonsingular, we can invert the Lagrange tensor. The Poisson tensor $\PoissonTensor=\wh{\omega} ^{-1}$ then yields the equations of motion,
\begin{equation}
\dfrac{\mathrm{d} z^i }{\mathrm{d} z^0} \;=\; \PoissonTensor ^{i j} \left( \pd{\gamma _j}{z^0} \,-\, \pd{\gamma _0}{z^i} \right). \label{eq:unpEqMotion}
\end{equation}

\subsubsection{\label{subsubsec:pertTh}Lie transform technique}

We consider a fundamental one-form $\gamma = \gamma _0 + \epsilon \gamma _1$ which consists of an equilibrium part $\gamma _0$ for which the flow is well-known, and a small perturbation $\gamma _1$. We want to study the effect of the perturbation on the flow.
The strategy is to search for a near-identity transformation that will reveal the symmetries of the perturbed system. Indeed, if the fundamental one-form is independent of some coordinate $z^\alpha $, then, as an application of Noether's theorem,
\begin{equation}
\dfrac{\mathrm{d} z^\alpha }{\mathrm{d} t} \;=\; \pd{\gamma _\alpha }{z^\mu} \dfrac{\mathrm{d} z^\mu }{\mathrm{d} t} \;=\; -\omega _{\alpha \mu} \dfrac{\mathrm{d} z^\mu}{\mathrm{d} t} \;=\; 0,
\end{equation}
revealing an exact invariant.
The general form of a near-identity transformation with a small parameter $\epsilon$ is
\begin{equation}
Z^\mu \;=\; z^\mu \,+\, \epsilon Z_{1f}^\mu (z) \,+\, \epsilon ^2 Z_{2f}^\mu (z) \,+\, \ldots
\end{equation}
Rather than an expansion in $\epsilon$ which is difficult to invert, we express the transformation in operator form. We denote a forward transformation $ Z^\mu \;=\; Z_{f}^\mu (z,\epsilon) $, and a backward transformation $ z^\mu \;=\; Z_{b}^\mu (Z,\epsilon)  $.
In the Lie transform technique, the coordinate transformation is specified by a generator $g^\mu$ such that
\begin{equation}
\pd{Z_{f}^\mu}{\epsilon} (z,\epsilon) \;=\; g^\mu \left( Z_{f}(z,\epsilon) \right),
\end{equation}
and $Z_{f}^\mu (z,0) \,=\, z^\mu$.

The forward transformation of a scalar $f(z)$ into $F(Z,\epsilon)$ is given by
\begin{eqnarray}
F &=& e^{-\epsilon L_g} f, \\
f &=& e^{+\epsilon L_g} F,
\end{eqnarray}
where $L_g$ is defined by its action on a scalar $L_g f \;=\; g^\mu \partial _\mu f$, and its action on a one-form $\left( L_g \gamma \right)_\mu \;=\; g^\nu \left( \partial _\nu \gamma _\mu \,-\, \partial _\mu \gamma _\nu \right)$.
The transformation of coordinates is just a special case of scalar transformation,
\begin{equation}
Z_b^\alpha \;=\; e^{-\epsilon L_g} I^\alpha ,
\end{equation}
where $I^\alpha (z) \equiv z^\alpha = Z_b^\alpha (Z)$ is the coordinate function.
The transformation of a one-form $\gamma(z)$ into $\Gamma(Z,\epsilon)$ is given by the functional relationships,
\begin{eqnarray}
\Gamma &=& e^{-\epsilon L_g} \gamma \,+\, \mathrm{d} S, \label{eq:pushForwOneForm}\\
\gamma &=& e^{+\epsilon L_g} \Gamma \,+\, \mathrm{d} s,
\end{eqnarray}
where $S$ and $s$ are scalar functions.

\subsubsection{Higher order perturbation theory}

We now consider a fundamental one-form in the form of an expansion $\gamma = \gamma _0 + \epsilon \gamma _1 + \epsilon ^2 \gamma _2 + \ldots $.
We introduce a push-forward transformation operator $T = \ldots T_3 T_2 T_1$, where each $T_n = e^{-\epsilon ^n L_n}$ is a Lie transform operator, and $L_n$ is a short for $L_{g_n}$.
Substituting these new definitions into Eq.~(\ref{eq:pushForwOneForm}), we have $\Gamma = T \gamma + \mathrm{d} S$, which yields for each successive order, $\Gamma _0 \,=\, \gamma _0$, and
\begin{eqnarray}
\Gamma _1 &=& \mathrm{d} S_1 \,-\, L_1 \gamma _0 \,+\, \gamma _1 \nonumber \\
\Gamma _2 &=& \mathrm{d} S_2 \,-\, L_2 \gamma _0 \,+\, \gamma _2 \,-\, L_1 \gamma _1 \,+\, \frac{1}{2} L_1^2 \gamma _0 \nonumber \\
\ldots \nonumber \\
\Gamma _n &=& \mathrm{d} S_n \,-\, L_n \gamma _0 \,+\, C_n, \;\;\;\;\;\;\;\;\;\; \mathrm{for}\;\; n\; \neq\; 0. 
\end{eqnarray}

Let us recall that our aim is to simplify the fundamental one-form in the new coordinates $\Gamma$. To this aim we have to solve successively the latter equations for the gauge scalar $S_n$ and the generating vector $g_n^\mu$ (Each $C_n$ is given by $\gamma _n$ and the result of the preceding order $n-1$).
In these equations, the generating vector appears only in the one-form $ \left( L_n \gamma _0 \right)_\mu \;=\; g_n^\nu \omega _{0 \mu \nu}$. We already discussed that $\omega _{0 \mu \nu}$ has a unique null eigenvector. Then we can add any multiple of this eigenvector to $g_n^\mu$ without changing the equations we are now trying to solve. Let us suppose this eigenvector has a nonzero component in the time direction, then we can set
\begin{equation}
g_n^0\;=\;0, 
\end{equation}
so that the time-coordinate doesn't change ($Z^0 = z^0 = t$).
At this point, we are left with $2N+1$ unknowns, namely $2N$ generating functions $g_n^i$ and one scalar gauge $S_n$. A priori we should be able to bring the $2N+1$ components of $\Gamma _n$ into a simpler form. A good strategy is to make all its symplectic components $\Gamma _{n i}$ vanish by choosing
\begin{equation}
g_n^j \;=\; \left( \partial _i S_n \,+\, C_{n i}  \right) \, \PoissonTensor _0^{i j}.
\end{equation}

Let us now focus on the time component of the new one-form,
\begin{equation}
\Gamma _{n 0} \;=\; -H_n \;=\; \partial _t S_n \;-\; g_n^j \wh{\omega _{0 j 0}} \;+\; C_{n 0}         .
\end{equation}
It is convenient to introduce the lowest order velocity vector, defined as the time derivative along the unperturbed orbits of the coordinates :
\begin{equation}
V_0^\mu \;\equiv\; \left. \dfrac{\mathrm{d} z^\mu}{\mathrm{d} t}\right| _0.
\end{equation}
Substituting the unperturbed equations of motion (\ref{eq:unpEqMotion}), we find that the scalar gauge is given by its total time derivative over the unperturbed orbit,
\begin{equation}
\left. \dfrac{\mathrm{D} S_n}{\mathrm{D} t} \right| _0 \;\equiv\; V_0^\mu \pd{S_n}{z^\mu} \;=\; \Gamma _{n 0} \;-\;  V_0^\mu C_{n \mu}        .
\end{equation}
In integrating the latter equation, we want to avoid any secularity effect. Then we should remove any secular perturbation by taking
\begin{equation}
\Gamma _{n 0} \;=\; \left<  V_0^\mu  C_{n \mu}  \right> _0,
\end{equation}
where the average is to be taken over the unperturbed orbits.\\

Finally, in the new coordinates $Z^\mu = \left[ e^{+\epsilon L_g} I^\mu \right] z$, the new one-form is given by
\begin{equation}
\Gamma _{n} \;=\; \left<  V_0^\mu  C_{n \mu}  \right> _0 \, dt , \label{eq:GiveNewOneForm}
\end{equation}
if we choose the generating functions
\begin{equation}
g_n^j \;=\; \left( \partial _i S_n \,+\, C_{n i}  \right) \, \PoissonTensor _0^{i j}, \label{eq:GiveGeneratingFunctions}
\end{equation}
where the gauge scalar is given by
\begin{equation}
S_n \;=\; -\, \oint _0 \wt{ V_0^\mu C_{n \mu} }, \label{eq:GiveGaugeScalar}
\end{equation}
and a tilde represents the oscillatory part.

To illustrate the benefits of Lie transform theory compared to classical perturbation theory, and to quantify its validity limit, we apply it to a simple mathematical problem in App.~\ref{app:lie}.

%\begin{itemize}
%\item Results of Brizard's paper
%\item Validity limit of gyrokinetics for a single particle orbit in slab geometry (details in the Appendix \ref{app:slab_gyrokinetics})
%\end{itemize}

\subsection{Guiding-center Lagrangian}

Guiding-center theory provides reduced equations of motion of a charged particle in a slowly-varying (in time and space) electromagnetic field, where the fast gyromotion is decoupled from the relatively slow drifting motion of the guiding-center. Guiding-center theory is based on the closeness to the limit of a fixed and uniform magnetic field, where the orbit of a particle in a frame following its gyration center is a circle. The perturbation from this situation is quantified by a small parameter $\epsilon$, if we assume the following ordering,
\begin{equation}
\epsilon \;\sim\; \omega / \Omega _c \;\sim\; \rho / L \;\ll\; 1,
\end{equation}
where $\Omega _c = eB/m$ is the cyclotronic frequency (or gyrofrequency), $\rho$ is the Larmor radius (or gyroradius), $\omega$ is a characteristic frequency of interest, and $L$ is the scalelength of field variation.

On the one hand, the standard derivation of guiding-center equations of motion is based on an averaging procedure \cite{northrop63}, which removes important properties of the Hamiltonian formulation.
On the other hand, the modern derivation \cite{littlejohn83} is based on Lie-transform perturbation theory. This approach preserves the validity of Noether's theorem and the validity of Liouville's theorem, to each order in $\epsilon$. In addition, expansion to arbitrary order is straightforward. Moreover, since we keep the Hamiltonian formulation, further reductions of dimensionality are still possible with Lie transforms.

The starting point is the one-form in canonical cartesian phase-space ($\boldsymbol{x}$, $\boldsymbol{p}$),
\begin{equation}
\gamma \;=\; \boldsymbol{p} \bdot \boldsymbol{\mathrm{d} x} \,-\, h(\boldsymbol{x},\boldsymbol{p},t) \, \mathrm{d} t,
\end{equation}
where $\boldsymbol{p} = m \boldsymbol{v} + e \boldsymbol{A}$, and $\boldsymbol{A}$ is the vector potential. The one-form can be expressed in noncanonical phase-space $(\boldsymbol{x},\boldsymbol{v})$, and written as an expansion in $\epsilon$,
\begin{equation}
\gamma \;=\; \gamma _0 \,+\, \epsilon \, \gamma _1 \,+\, \ldots ,
\end{equation}
where
\begin{eqnarray}
\gamma _0 &=&  e \boldsymbol{A}(\boldsymbol{x},t) \bdot \boldsymbol{\mathrm{d} x} \,-\, e \varphi _0 (\boldsymbol{x},t),\\
\gamma _1 &=& m \boldsymbol{v} \bdot \boldsymbol{\mathrm{d} x} \,-\, \left(e \varphi _1 (\boldsymbol{x},t) \,+\, \frac{m}{2} v^2 \right) \mathrm{d} t ,
\end{eqnarray}
and where $\varphi$ is the scalar potential.

In Ref.~\cite{littlejohn83}, Lie-transform theory is applied to change variables to 
new coordinates, where the new one-form $\Gamma$ does not depend on the gyroangle. When this procedure is carried up to the second order, we obtain the guiding-center one-form,
\begin{equation}
\Gamma \;=\; \left(m U \boldsymbol{b}(\boldsymbol{R},t) \,+\, e \boldsymbol{A}(\boldsymbol{R},t) \right) \bdot \boldsymbol{\mathrm{d} R} \,+\, M \mathrm{d} \xi \,-\, H \mathrm{d} t, \label{gc_oneform}
\end{equation}
where
\begin{equation}
H \;=\; e \varphi (\boldsymbol{R},t) \,+\, \frac{m}{2} U^2 \,+\, (e/m) M B(\boldsymbol{R},t) ,
\end{equation}
and the guiding-center coordinates $\boldsymbol{Z} \equiv (\boldsymbol{R},U,M,\xi)$ as
\begin{eqnarray}
\boldsymbol{R} & \equiv & \boldsymbol{x} \; - \; \epsilon \, \dfrac{v_{\bot}}{\Omega _{ci}} \boldsymbol{a},\\
U & \equiv & \dotp{v}{b} = v_{\|},\\
M & \equiv & \dfrac{mv_{\bot}^2}{2\Omega _{ci}},\\
\xi & \equiv & \tan ^{-1} \left( \dfrac{\dotp{v}{e_1}}{\dotp{v}{e_2}} \right),
\end{eqnarray}
where $(\boldsymbol{e_1},\boldsymbol{e_2},\boldsymbol{b})$ is an orthogonal unit vectors system, $\boldsymbol{a} \equiv \cos (\xi) \boldsymbol{e_1} - \sin (\xi) \boldsymbol{e_2}$, and $\boldsymbol{c} \equiv \crop{a}{b}$.

In Eq.~(\ref{gc_oneform}), the angle-action variables of the gyromotion, ($\xi$, $M$), appear in canonical form in the symplectic part, and $H$ does not depend on $\xi$. Thus, since the gyromotion is irrelevant to fast-ion/TAE interactions, we can drop the term $M \mathrm{d} \xi$ in the guiding-center one-form.

\subsection{Guiding-center action-angle variables}

In Ref.~\cite{meiss90}, the zeroth order guiding-center one-form $\Gamma _0$ is put in the form,
\begin{equation}
\Gamma _0 \;=\; P_\theta \mathrm{d} \theta \,+\, P_\zeta \mathrm{d} \zeta \,+\, M \mathrm{d} \xi - H(P_\theta ,P_\zeta ,\theta ,M) \mathrm{d} t,
\end{equation}
where $P_{\zeta}$ is the toroidal canonical angular momentum,
\begin{equation}
P_{\zeta} \,\equiv \, -e \chi (r) \,+\, m_b R_0 v_{\|},
\end{equation}
and $P_\theta$ is the poloidal canonical angular momentum.
The latter expression can be used as a starting point to develop an action-angle formalism where the perturbed Hamiltonian takes a standard form. In Ref.~\cite{berkbreizman95nf}, a canonical transformation is performed to obtain
\begin{equation}
\Gamma _0 \;=\; J_\theta \mathrm{d} \widetilde{\theta} \,+\, J_\zeta \mathrm{d} \widetilde{\zeta} \,+\, J_\xi \mathrm{d} \widetilde{\xi} - H(J_\theta ,J_\zeta ,J_\xi) \mathrm{d} t,
\end{equation}
where $\dot{\widetilde{\xi}}=\Omega _c$, $\dot{\widetilde{\theta}}=\omega _\theta$, and $\dot{\widetilde{\zeta}}=\omega _\zeta$ are the unperturbed frequencies of the gyromotion, poloidal motion, and toroidal motion, respectively. $\widetilde{\theta}$ and $\widetilde{\zeta}$ reduce to the geometric angles $\theta$ and $\zeta$ if we neglect finite aspect ratio effects (But we do not neglect them, since we would remove the toroidicity from which the TAE originates).

%\subsection{Action-angle variables of the bounce motion}
%
%Lie transform theory can be applied to the guiding-center bounce motion to construct a set of canonical angle-action variables \cite{littlejohn82}. The first couple, ($\xi$, $\mu$), where $\xi$ is the cyclotronic angle and $\mu$ is the magnetic moment, describes the gyromotion. The second couple, ($\alpha$, $\beta$), is such that $\boldsymbol{B_0}=\grad{\beta}\boldsymbol{\times} \grad{\alpha}$ (where we inverted the notations for $\alpha$ and $\beta$ from the reference). We take $\alpha =q \theta-\zeta$ and $\beta = \chi$. The third couple, ($\psi$, $J$), corresponds to the bounce-motion.
%
%[Redaction needed: summarize the derivation of bounce-motion angle-action variables from \cite{littlejohn82}. (This part is my excuse for including Lie-transforms in the manuscript)]

%\begin{itemize}
%\item Results of White and Chance's papers
%\item Results of Littlejohn's paper and application to $(r,\theta,\zeta)$ in axisymmetric systems
%\end{itemize}
%\subsection{Hamiltonian in action-angle variables} \label{subsec:aa_hamiltonian}
%\begin{itemize}
%\item Hamiltonian expressed in action-angle variables, where the perpendicular magnetic vector-potential perturbation has a fixed spatial structure (Calculations in Appendix \ref{app:action_angle_hamiltonian})
%\item In the large aspect ratio, we should recover [Berk Breizman Pekker NF(1995)] for passing particles, and Wong$\&$Berk for well-trapped particles.
%\end{itemize}

\subsection{Application to TAE}

Although TAEs resonate mainly with passing particles, when the source of high-energy ions is isotropic, a large fraction of the energy transfer may be accounted by resonance with the bounce-motion (or banana motion) of toroidally trapped particles \cite{todo98}. However, for tangential NBI ions, to which we confine our analysis, it is sufficient to describe resonance with far passing particles.

In a gauge where the perturbed scalar potential is zero, the TAE can be described by a perturbation Hamiltonian,
\begin{equation}
H _1 \;=\; -e\, \boldsymbol{A_{1\perp}}\bdot \boldsymbol{v_\mathrm{gc}}, \label{eq:perturbed_lag}
\end{equation}
where $\boldsymbol{v_\mathrm{gc}}$ is the guiding-center velocity. Here, $\boldsymbol{A_{1\perp}}$ is the perpendicular part of the perturbed vector potential, and we have neglected a second order term $\boldsymbol{A_{1\perp}}\bdot \boldsymbol{A_{1\perp}}$. In a small plasma pressure (small $\beta$) limit, we can neglect parallel gradients, then $\boldsymbol{A_{1\perp}}$ can be split into magnetic compression,
\begin{equation}
\boldsymbol{A_{1\perp}^\mathrm{c}} \;=\; \boldsymbol{b_0} \times \grad{\Pi},
\end{equation}
and magnetic shear,
\begin{equation}
\boldsymbol{A_{1\perp}^\mathrm{s}} \;=\; \grad{\Phi} \;-\; (\boldsymbol{b_0}\bdot\grad{\Phi}) \boldsymbol{b_0}, \label{eq:shear_part}
\end{equation}
where $\boldsymbol{b_0} \equiv \boldsymbol{B_0}/B_0$.
For the TAE, which is a shear Alfv{\'e}n wave, the latter part only is relevant, hence the excitation is described by a single scalar function $\Phi$.

In Ref.~\cite{berkbreizman95nf} the perturbed Hamiltonian is put in the form
\begin{equation}
H_1 \;=\; V(\boldsymbol{J}) \, \cos (\boldsymbol{l}\bdot\boldsymbol{\alpha} \,-\, \omega t ), \label{eq:perturbation_in_aa}
\end{equation}
in arbitrary tokamak geometry, where ($\boldsymbol{\alpha}$, $\boldsymbol{J}$) are angle-action variables for the solvable unperturbed Hamiltonian $H_0(\boldsymbol{J})$, and $\boldsymbol{l} = (l_1,l_2,l_3)$ is a triplet of three integers.
Substituting Eq.~(\ref{eq:shear_part}) into Eq.~(\ref{eq:perturbed_lag}) yields
\begin{eqnarray}
H _1 &=& -e\,\boldsymbol{v_\mathrm{gc}} \bdot \grad{\Phi} \;+\; e\, (\boldsymbol{b_0}\bdot\grad{\Phi})\, \boldsymbol{b_0}\bdot\boldsymbol{v_{\mathrm{gc}}} \, \\
  &=& -e\, \left. \frac{\mathrm{d} \Phi}{\mathrm{d} t}\right| _{0} \;+\; e\, \pd{\Phi}{t} \;+\; e\, (\boldsymbol{b_0}\bdot\grad{\Phi})\, v_{\|}, \label{eq:perturbed_lag2}
\end{eqnarray}
where $\left. \mathrm{d} \Phi / \mathrm{d} t \right| _{0}$ is the Lagrangian derivative of $\Phi$ along unperturbed particle orbit, which can be removed from the Lagrangian without altering Euler equations.

For a single $n/m$ mode of frequency $\omega$, the eigenfunction takes the following form,
\begin{equation}
\Phi (r,\theta,\zeta) \;=\; C(t) e^{-\imath \omega t -\imath \varphi (t)} \sum _{l=m}^{m+1} \Phi _l (r) e^{\imath n \zeta - \imath l \theta} \;+\; \mathrm{c.c.},
\end{equation}
where $C$ and $\varphi$ are the amplitude and phase of the wave.
Substituting the eigenfunction and the expression $\boldsymbol{B_0} \;=\; (\partial \chi / \partial r) \grad{r} \, \boldsymbol{\times} \, \grad{(q \theta-\zeta)}$ into Eq.~(\ref{eq:perturbed_lag2}), the TAE excitation is reduced to
\begin{equation}
H _1 \;=\; -\imath e\, C(t) e^{-\imath \omega t -\imath \varphi (t)} \sum _{l=m}^{m+1} \Phi _l (r) e^{\imath n \zeta - \imath l \theta} \left[ \omega\,-\, v_{\|} \boldsymbol{b_0} \bdot \left( n\grad{\zeta} - m\grad{\theta} \right) \right] \;+\; \mathrm{c.c.},
\end{equation}
where we have neglected time-derivation of slowly varying phase and amplitude of the wave.

Then we change the variables to the canonical angle-actions ($\widetilde{\theta}$, $\widetilde{\zeta}$, $J_\theta$, $J_\zeta$), in order to express the perturbed Hamiltonian as a Fourier series in $\widetilde{\theta}$,
\begin{equation}
H_1 \;=\; -\imath e\, C(t) e^{-\imath \omega t -\imath \varphi (t)} e^{\imath n \widetilde{\zeta}} \sum _{p=-\infty}^{+\infty} V _p (J_\theta, J_\zeta) e^{\imath p \widetilde{\theta}}  \;+\; \mathrm{c.c.}, \label{eq:tae_perturb_flux_coord}
\end{equation}
where
\begin{equation}
V_p = \int \frac{\mathrm{d} \widetilde{\theta}}{2\pi} e^{-\imath p \widetilde{\theta} -\imath n \widetilde{\zeta}} \sum _{l=m}^{m+1} \Phi _l (r) e^{\imath n \zeta - \imath l \theta} \left[ \omega\,-\, v_{\|} \boldsymbol{b_0} \bdot \left( n\grad{\zeta} - m\grad{\theta} \right) \right]. \label{eq:Vp_berk}
\end{equation}
Formally, the problem is expressed in the desired form of Eq.~(\ref{eq:perturbation_in_aa}), but the numerical computation of the Fourier components $V_p$, which is needed for quantitative comparison of absolute physical quantities between 3D and 1D model, requires to relate geometric angles and canonical angles, which may be complicated depending on the equilibrium configuration.
However, since each particle of the resonant phase-space surface interact with the wave in the same way at the same frequency (though with different strength), comparison of quantities that are normalized to the mode frequency is possible, even without evaluating $V_p$ coefficients.

\section{Reduction to a one dimensional problem} \label{sec:3D_to_1D}

%From the resonance condition Eq.(\ref{eq:tae_resonance_condition_co}), writing $\Delta r \,q'=q(r_m)-q(r_{m+1})$, the distance $\Delta r$ between the resonant surfaces of two neighbouring $m$ modes is estimated as
%\begin{equation}
%\Delta r \;\sim\; \dfrac{1/q'}{n-\omega _\mathrm{A} R_0 / v_\|} \;\sim\; \dfrac{1}{n q'},
%\end{equation}
If we consider only small toroidal mode numbers $n$, $n$ and $n+1$ modes are isolated. Let us consider a single toroidal mode number. On the one hand, since on a flux surface $r=r_m$ where a poloidal mode number $m$ is centered, the safety factor is $q(r_m)=(2m+1)/(2n)$, then we can estimate the distance $\Delta r = r_{m+1}-r_m$ between two neighbouring $m$ modes by writing $\Delta r \,q'\approx q(r_m)-q(r_{m+1})$, as
\begin{equation}
\Delta r \;\approx\; \dfrac{1}{n q'}.
\end{equation}
On the other hand, the characteristic width of TAE modes $\delta r$ is of the order of $\delta r \sim r_m^2/nqR_0$ \cite{cheng85}. Hence, for typical parameters, $\delta r / \Delta r \sim (r_m/R_0)S \ll 1$, where $S\equiv r q'/q$ is the magnetic shear. Thus, TAEs have a two-scale radial structure, the larger scale corresponding to the enveloppe of the TAE. In our analysis, we assume that the number of $m$ harmonics involved is small enough to consider resonances one by one, as isolated $n$, $m$ mode. The latter hypothesis is reasonable for sufficiently core-localised, low-$n$ TAEs.
We must keep in mind, though, that high-$n$ TAEs are likely to be destabilized in future devices such as ITER, in which case it may be necessary to take into account multiple-wave resonances.

The evolution of the distribution $f(\boldsymbol{x},\boldsymbol{v},t)$ of energetic ions is described in 3D configuration space by the kinetic equation (\ref{eq:3d_kinetic}), with the perturbed Hamiltonian Eq.~(\ref{eq:perturbation_in_aa}).
In the following, we reduce the problem to a 1D Hamiltonian system, by considering a single $n$, $m$ mode.

\subsection{Reduction of the Hamiltonian} \label{subsec:1D_hamiltonian}

%Keeping a single mode in Eq.[Equation number needed], the Hamiltonian is in the form
%\begin{equation}
%H \;=\; H_0(\boldsymbol{J}) \;+\; H_1(\boldsymbol{\alpha},\boldsymbol{J}),
%\end{equation}
%where ($\boldsymbol{\alpha}$, $\boldsymbol{J}$) are the angle-action variables for the solvable unperturbed Hamiltonian $H_0$, and $H_1$ is periodic,
%\begin{equation}
%H_1 \;=\; A(\boldsymbol{J}) \, \cos (\boldsymbol{l}\bdot\boldsymbol{\alpha} \,-\, \omega t ),
%\end{equation}
%with $\boldsymbol{l} = (l_1,l_2,l_3)$.

Formally, the resonant phase-space surface, $\boldsymbol{J}=\left\{ \boldsymbol{J}_R \text{ such that } J_{R3} = F(J_{R1},J_{R2}) \right\}$, is defined by a function $F$. The resonance condition,
\begin{equation}
\boldsymbol{l} \bdot \boldsymbol{\Omega} (\boldsymbol{J}_R) \;=\; \omega,
\end{equation}
where $\boldsymbol{\Omega} (\boldsymbol{J}) \equiv \pd{H_0}{\boldsymbol{J}} (\boldsymbol{J})$, is satisfied on the resonant phase-space surface.

Once the perturbed Hamiltonian has been put in the form of Eq.~(\ref{eq:perturbation_in_aa}), we can reduce the problem to one action and one angle \cite{lichtenberg,garbet08}, by performing a canonical transformation $\boldsymbol{J}\bdot \mathrm{d}\boldsymbol{\alpha} - H \mathrm{d} t = \boldsymbol{I}\bdot \mathrm{d}\boldsymbol{\psi} - H' \mathrm{d} t + \mathrm{d} S$ with the generating function
\begin{equation}
S \;=\; -\dotp{I}{\psi} \;+\; I_3 (\boldsymbol{l}\bdot\boldsymbol{\alpha} \,-\, \omega t) \;+\; I_1 \alpha _1 \;+\; I_2 \alpha _2 \;+\; F(I_1,I_2) \alpha _3.
\end{equation}
This procedure yields,
\begin{center}
\begin{tabular}{r l r l}
$J_1 \;=$ & $I_1 \;+\; l_1 \, I_3$ & $\psi _1 \;=$ & $\alpha _1 \;+\; \alpha _3 \, \pds{F}{I_1}$ \\
$J_2 \;=$ & $I_2 \;+\; l_2 \, I_3$ & $\psi _2 \;=$ & $\alpha _2 \;+\; \alpha _3 \, \pds{F}{I_2}$ \\
$J_3 \;=$ & $F(I_1,I_2) \;+\; l_3 \, I_3 \ \ \ \ \ \ \ \ \ \ \ \  $ & $\psi _3 \;=$ & $\boldsymbol{l}\bdot\boldsymbol{\alpha} \;-\; \omega t$,
\end{tabular}
\end{center}
and $H \;=\; H' \;+\; \omega \, I_3$. Thus, near the resonant phase-space surface, $\boldsymbol{J}\;=\;\boldsymbol{J}_R \,+\, I_3 \, \boldsymbol{l}$, and we can expand the new Hamiltonian around this surface,
\begin{eqnarray}
H'(\boldsymbol{\psi},\boldsymbol{I}) &=& H_0(\boldsymbol{J}_R \,+\, I_3 \, \boldsymbol{l}) \;+\; V (\boldsymbol{J}_R \,+\, I_3 \, \boldsymbol{l}) \cos \psi _3 \;-\; I_3 \, \omega  \\
 &=& H_0(\boldsymbol{J}_R) \;+\; I_3 \, \left( \boldsymbol{l} \bdot \boldsymbol{\Omega}(\boldsymbol{J}_R) \,-\, \omega \right) \;+\; \dfrac{1}{2} D \, I_3 ^2  \nonumber \\
& & \qquad \qquad \qquad \qquad \;+\; V (\boldsymbol{J}_R \,+\, I_3 \, \boldsymbol{l}) \cos \psi _3,
\end{eqnarray}
with $D(\boldsymbol{J}_R) \equiv l_i l_j \pds{}{J_i}\pds{}{J_j} H_0 (\boldsymbol{J}_R)$.

If the variations of $H(\boldsymbol{J})$ are small around $\boldsymbol{J}_R$, we can replace $V (\boldsymbol{J}_R \,+\, I_3 \, \boldsymbol{n})$ by $V (\boldsymbol{J}_R)$ in the latter expression, and obtain the new Hamiltonian $H' \;=\; H_0 (\boldsymbol{J}_R) \;+\; H_{1,\boldsymbol{J}_R} (\psi _3 , I_3)$, with
\begin{equation}
H_{1,\boldsymbol{J}_R} (\psi , I) \;\equiv \; \dfrac{1}{2} \, D \, I ^2 \;+\; V \cos \psi. \label{eq:TAE1DHamiltonian}
\end{equation}
Thus, the problem has been reduced to a 1D Hamiltonian problem for the angle-action variables ($\psi$, $I$)$\equiv$($\psi _3$, $I_3$).

Substituting the expression of the TAE perturbation, Eq.~(\ref{eq:tae_perturb_flux_coord}) into Eq.~(\ref{eq:perturbation_in_aa}), we obtain
\begin{eqnarray}
\psi &=& p \widetilde{\theta} \,+\, n \widetilde{\zeta} \,-\, \omega t,\\
I &=& \frac{J_\zeta - J_{\zeta \mathrm{R}} }{n} \;\approx \; -e\, \frac{\chi - \chi _\mathrm{R} }{n} ,\\
D &\approx & n^2 \, \frac{\partial ^2 H_0}{\partial J_\zeta ^2} (\boldsymbol{J _\mathrm{R}}), \\
V &=& -\imath e \, C(t) V _p (J_{\theta \mathrm{R}}, J_{\zeta \mathrm{R}}),
\end{eqnarray}
where the subscript $\mathrm{R}$ means that the quantity is evaluated at the resonance, and with $\boldsymbol{\alpha}=(\widetilde{\xi},\widetilde{\theta},\widetilde{\zeta})$, $\boldsymbol{J}=(J_\xi,J _\theta, J_\zeta)$, and $\boldsymbol{l}=(0,p,n)$.

%[Work needed: Estimation of the terms $\partial \Omega /\partial P$, $A$]

\subsection{Reduction of the collision, source and sink terms} \label{subs:3d_to_1d_collisions}

A first, simple model is obtained by reducing the effects of collisions to the recovery of an equilibrium energetic particle distribution, with a recovery rate $\nu _a (v)$.

%We see that the drag term is a model for the slowing-down of energetic particles, which is mainly due to collisions with electrons, while the diffusion term is a model for pitch-angle scattering and parallel velocity diffusion, which originate mainly from collisions with ions.

A more rigorous treatment of collision processes is obtained if we project a collision operator that describes Coulomb collisions perceived by energetic ions, on the resonant phase-space surface.
We consider collisions on energetic particles by thermal electrons ($s=e$), ions ($s=i$), and carbon impurities ($s=c$), and describe them by a Fokker-Planck collision operator \cite{helandersigmar} that acts on the distribution of energetic particles ($s=b$).
%\begin{equation}
%\left.\dfrac{df}{dt}\right| _{\mathrm{coll.}} \;=\; \sum _s \mathcal{C}_{bs}(f), \label{eq:collision_operator_3d}
%\end{equation}
%where the contribution of each species $s$,
%\begin{eqnarray}
%\mathcal{C}_{bs}(f) \;=& \;\gamma _{bs}\,\frac{\partial}{\partial v_{\alpha}}\,\left[ -\left(\frac{1}{m_b} + \frac{1}{m_s} \right)\, \frac{\partial H_s}{\partial v_{\alpha}} f \right. \nonumber \\
%&\left. +\,\frac{1}{2 m_b}\, \frac{\partial ^2 G_s}{\partial v_{\alpha} \partial v_{\beta}} \frac{\partial f}{\partial v_{\beta}} \right], \label{eq:cbs}
%\end{eqnarray}
%involves the Rothenbluth potentials $G_s(\boldsymbol{v})$ and $H_s(\boldsymbol{v})$ \cite{rosenbluth57}.
%This collision operator includes pitch-angle scattering, slowing down, and parallel velocity diffusion.
In spherical coordinates ($v$,$\Theta$), where $\Theta$ is the angle between $\boldsymbol{v}$ and $\boldsymbol{b}$, neglecting gyroangle dependency,
\begin{equation}
\left.\dfrac{\mathrm{d} f}{\mathrm{d} t}\right| _{\mathrm{coll.}} \;=\; \nu _\mathrm{defl}\, \dfrac{1}{2} \dfrac{1}{\sin{\Theta}} \dfrac{\partial}{\partial \Theta} \left( \sin{\Theta} \dfrac{\partial f}{\partial \Theta} \right) \;+\; \dfrac{1}{v^2} \dfrac{\partial}{\partial v} \left[ v^3 \left( \nu_\mathrm{slow} f \,+\, \dfrac{1}{2} \nu_{\|} v \dfrac{\partial f}{\partial v} \right) \right],
\end{equation}
where $\nu _\mathrm{defl}$, $\nu_\mathrm{slow}$ and $\nu_{\|}$ are pitch-angle scattering, slowing-down, and parallel velocity diffusion rates, respectively, $v_\| = v \cos{\Theta}$ is the parallel velocity of energetic particles.

We consider a TAE with toroidal mode number $n$, resulting from the coupling of $m$ and $m+1$ poloidal modes. To simplify the following discussion, we consider strongly co-passing beam particles that resonate with the TAE at a velocity $v\approx v_\| = v_\mathrm{A}$. Then the resonance condition is given by $\Omega = \omega _\mathrm{A}$, where $\Omega$ is given by Eq.~(\ref{eq:tae_resonance_condition}).
To project the Fokker-Planck operator on the resonant phase-space surface, we follow the procedure described in Refs.~\cite{berkbreizman97ppr,lilley09}. We replace $\partial _{\boldsymbol{v}} f$ by $\mathcal{J} \boldsymbol{b_0}\, \partial _{\Omega} f$, where $\mathcal{J}$ is the Jacobian of the coordinate transformation from $v_\|$ to $\Omega$,
\begin{equation}
\mathcal{J} \;=\; \frac{\partial P_{\zeta}}{\partial v_\|} \, \left. \frac{\partial \Omega}{\partial P_{\zeta}} \right| _{v_\|} \;=\; \frac{m S m_b v_\|}{2 r^2 e_b B_0},
\end{equation}
and $S$ is the magnetic shear. Here $e_s$ and $m_s$ are charge and mass of a species $s$, respectively, and $b$ stands for beam particles.
This procedure yields 
\begin{equation}
\left.\dfrac{\mathrm{d} f}{\mathrm{d} t}\right| _{\mathrm{coll.}} \;=\; \nu _f^2 \, \frac{\partial f}{\partial \Omega} \,+\, \nu _d^3 \, \frac{\partial ^2 f}{\partial \Omega ^2}, \label{eq:total_collisions}
\end{equation}
with
\begin{eqnarray}
\nu _f^2 &=& v_\| \,\mathcal{J}\, \left( 2\nu _\| \,+\, \nu _\mathrm{slow} \,-\, \nu _\mathrm{defl} \right) ,\\
\nu _d^3 &=& \frac{v^2}{2} \,\mathcal{J}^2\, \left( \nu _\| \cos{\Theta} \,+\, \nu _\mathrm{defl} \sin{\Theta} \right).
\end{eqnarray}

We assume Maxwellian background distributions, with a same temperature $T_0$. Typical experiments satisfy the following ordering of thermal velocities, $v_{Tc}<v_{Ti}\ll v_\mathrm{A} \ll v_{Te}$, while the beam energy $E_b$ is much larger than $T_0$.
Within these assumptions, around the resonance,
\begin{eqnarray}
\nu _f^2 &=& \frac{v_\| \mathcal{J}}{v^3} \, \sum _s \frac{n_s \gamma _{bs}}{m_s} \, \left[ \mathrm{erf}\,\eta _s \,-\, \frac{2\eta _s}{\sqrt{\pi}}\,e^{-\eta _s^2} \right] , \label{eq:nuf} \\
\nu _d^3 &=& \frac{\mathcal{J} ^2}{2v^3}\, \sum _s \frac{n_s \gamma _{bs}}{2 m_b \eta _s ^2}\, \left[ \left( (2\eta _s^2-1) v_{\perp}^2+2v_\|^2 \right) \,\mathrm{erf}\,\eta _s \, \right. \nonumber \\
& & \qquad \qquad \qquad \qquad \qquad \left. + \, \frac{2\eta _s}{\sqrt{\pi}}\,(v^2-3v_\|^2)\,e^{-\eta _s^2} \right],\label{eq:nud}
\end{eqnarray}
where $\eta _s\equiv v / v_{Ts}$, $v_{\perp}=v \sin \Theta$, $v_\|=v_\mathrm{A}$,
\begin{equation}
\gamma _{bs} \;=\; \frac{e_b^2 e_s^2 \log{\Lambda}}{\epsilon _0 ^2 m_b},
\end{equation}
$\epsilon _0$ is the vacuum permittivity, and $\log{\Lambda}$ is the Coulomb logarithm. Since the magnetic moment is an invariant of the motion of injected beam ions from deposition to resonant phase-space surface, $v_\perp^2=v_b^2(1-R_T^2/R_0^2)$, where $v_b$ is the velocity of beam particles, and $R_T$ is the tangential radius of the beam.

The equivalent collision operator in the BB model is obtained by substituting $\Omega = k v$ in Eq.~(\ref{eq:total_collisions}).

%\begin{equation}
%\mathcal{J} \;=\; m_b R_0\, \left.\frac{\partial \Omega}{\partial P_{\zeta}}\right| _{E'}.
%\end{equation}
%and derivatives are taken at constant $E'$, with $E'=E-\omega P_{\zeta} /n$.
%For typical parameters, $e \psi \gg m_b R_0 v_\mathrm{A}$, thus $P_{\zeta}(r,v_\|)\approx P_{\zeta}(r)$, then $\left. \partial _{P_{\zeta}} \Omega \right| _{E'}=\left. \partial _{P_{\zeta}} \Omega \right| _{v_\|}=\partial _{r} \Omega / \partial _r P_{\zeta}$, which yields

%It is somewhat more transparent to directly project the collision operator expressed in Cartesian coordinates. Substituting $f=f[\boldsymbol{x},\Omega(r,v_\|),t]$ in Eq.~(\ref{eq:cbs}) yields
%\begin{eqnarray}
%\mathcal{C}_{bs}(f) \;=& 4\pi \gamma _{bs}\, \left(\frac{1}{m_b}+\frac{1}{m_s}\right)\,f_s\,f \nonumber \\
%&-\, \frac{\gamma _{bs}}{m_s} \, \frac{\partial H_s}{\partial v_\|}\, \mathcal{J} \, \frac{\partial f}{\partial \Omega} \nonumber \\
%&+\,\frac{\gamma _{bs}}{2 m_b}\, \frac{\partial ^2 G_s}{\partial v_\| ^2} \, \mathcal{J} ^2 \, \frac{\partial ^2 f}{\partial \Omega ^2}, \label{eq:collisions_with_omega}
%\end{eqnarray}
%where the first term is a source term, which depends on the distribution of species $s$, $f_s$.
%Assuming that $\partial _{v_\|} f \sim f/v_\mathrm{A}$, the source term in Eq.~(\ref{eq:collisions_with_omega}) is then negligible.

\subsection{Reduction of the background damping mechanisms}

Since the time-scale of fast-particle evolution is much faster than background thermal populations evolution, these two dynamics are decoupled. Hence we can reasonably treat the effects of background damping in an extrinsic way. We assume that all background damping mechanisms affect linearly the wave energy $\mathcal{W}$,
\begin{equation}
\frac{\mathrm{d} \mathcal{W}}{\mathrm{d} t} \;=\; -\,2\,\gamma _d\, \mathcal{W}(t).
\end{equation}
Damping includes mechanisms such as radiative damping, which strength depends on the frequency \cite{mett92}. Hence, in a rigorous model, $\gamma _d$ should be a function of $\omega$. However, theory needs to be developed before this complex interplay can be taken into account.
Thus, we limit our framework to cases where $\gamma _d$ can be treated as a constant. This framework is consistent with a fixed mode frequency.

%[Reading needed: Is time-scale of background dampings the MHD time-scale ?][Redaction needed: Validity of this model]

\subsection{Limitations}

%Since the frequency and growth rate of TAEs are very sensible to the $q$ profile, our assumption of fixed mode structure, implying fixed Magneto-Hydrodynamic (MHD) equilibrium, is not necessarily satisfied even on a time-scale much shorter than a time-scale of MHD equilibrium evolution, which is of the order of the second on large devices such as JT-60U. This can be seen for example in the spectrogram of magnetic fluctuations for JT-60U shot number E36378 \cite{shinohara01}, where the frequency of a TAE is changed by $20\%$ between $t=3.65$s and $3.7$s. In practice, if the TAE frequency observed in experiments stays nearly constant during a certain time-window, we infer that the fixed mode structure assumption is satisfied.

We assumed an isolated single mode, which is reasonable for sufficiently core-localized, low-$n$ TAEs. However, for future devices with higher-$n$, an other model that includes multiple-modes interactions has to be developed. % [Work needed: what if there is co- and counter-injection simultaneously ?]

Since it assumes a fixed mode structure, implying a fixed Magneto-Hydrodynamic (MHD) equilibrium, the above reduced model looses its validity on a time-scale of MHD equilibrium evolution, which is of the order of the second on large devices such as JT-60U. We must also require that wave amplitude is low enough, such that nonlinear redistribution of energetic particles does not significantly alter the mode structure and frequency. In practice, since the frequency and growth rate of TAEs are very sensitive to the $q$ profile, if the frequency of a low-amplitude TAE observed in experiments stays nearly constant during a certain time-window, we infer that the fixed-mode-structure assumption is satisfied for this time-window.

In the case of frequency sweeping, which is the case we apply to experiments, it is sometimes argued that since the frequency is changed, so must be the mode structure. However, we must distinguish at least three classes of frequency sweeping, namely, slow frequency sweeping (slow-FS), fast frequency sweeping (fast-FS), and so-called abrupt large-amplitude events (ALE) \cite{shinohara01}, although it is not clear for the latter if the frequency does sweep. In the case of JT-60U shot number E32359, which is analyzed in Chap.~\ref{ch:experiment}, slow-FS have a timescale of $100-200$ ms, and their frequency is correlated with bulk equilibrium variations, therefore they are out of the scope of the above reduce model.
Fast-FS have a timescale of $1-5$ ms, and the associated redistribution of energetic ions is relatively small \cite{shinohara02}, therefore are consistent with a fixed-mode-structure hypothesis.
Although the occurrence of fast-FS and ALEs seems to be linked, ALEs are identified as so-called Energetic Particle-driven Modes (EPMs) \cite{briguglio07}, have significantly larger amplitude and shorter timescale ($200-400$ $\mu$s), and induce significant loss of energetic ions, and are out of the scope of this work since we assume a weak drive and a constant density of energetic ions in the BB model.

Finally, modeling all background damping mechanisms as an extrinsic, fixed linear damping on the wave is a strong assumption, whose validation requires more understanding of these mechanisms. We must assume that $\gamma _d$ does not depend neither on the wave amplitude, nor on the energetic population. In the case of frequency sweeping, the assumption is clearly violated if the nonlinear modification of frequency is of the order of the linear frequency, especially if a chirping phase-space structure approaches the SAW continuum, where damping rate depends largely on the frequency.

Overall, the above reduced model is suitable for describing resonant interactions between energetic particles and a weakly driven, isolated TAE, even for slightly-chirping modes, as long as
\begin{itemize}
\item phase-space structures are well confined within the continuum gap ;
\item redistribution of energetic population is negligible as far as wave dispersiveness and damping mechanisms are concerned ;
\item we look at time-scales much smaller than the equilibrium evolution time-scale.
\end{itemize}

\chapter{The Berk-Breizman model}
\label{ch:bbmodel}

The BB model, describing the fully nonlinear wave-particle interactions between energetic particles and an electrostatic wave in a 1D plasma, is an extension of the Vlasov-Poisson system, where we take into account collisions and external wave damping. In this system, we consider a bump-on-tail velocity distribution comprising a Maxwellian bulk and a beam of energetic particles, and we apply a small electrostatic perturbation. The apparent simplicity of the corresponding equation system hides surprisingly rich physics. Depending on the parameters of the model, the perturbation may be damped or amplified due to a transfer of energy between resonant particles and the wave. In the stable case, when the perturbation is small, linear theory predicts exponential decay of the wave amplitude, which in the absence of collisions and external damping is known as Landau damping \cite{landau46}. In the unstable case, on the contrary, linear theory predicts exponential growth of the wave amplitude. Then, trapping of resonant particles significantly modifies the distribution function and an island structure appears. Saturation of the instability and following nonlinear evolution are determined by a competition among the drive by resonant particles, the external damping, the particle relaxation which tends to recover the initial positive slope in the distribution function, and particle trapping that tends to smooth it.
It has been predicted \cite{berkbreizman97ppr,berkbreizman97pop,berkbreizman96} and observed \cite{vann03} that three kinds of behavior emerge, namely steady-state, periodic, or chaotic responses, depending on the strength of each factor. In addition, chaotic solutions can display significant shifting of the mode frequency (chirping), both upwardly and downwardly, as holes and clumps are formed in the distribution \cite{berkbreizman97pla,berkbreizman98erratum,berkbreizman99}.

In Sec.~\ref{sec:bb_eqns}, we recall the equations of the BB model, in both full-$f$ and $\delta f$ approaches.
In Sec.~\ref{sec:bblin}, we show the linear dispersion relation, and present tools that we use for accurate linear analysis.
For our purposes of validating and extending BB theory, and of applying it to TAEs, we develop a kinetic code based on the Constrained Interpolation Profile - Conservative Semi-Lagrangian (CIP-CSL) scheme \cite{nakamura01} for solving the initial value problem. In Sec.~\ref{sec:cobbles}, we present the main principles of our code, which we name COBBLES. In the full-$f$ case, we show that a locally conservative implementation is a key point for robust simulations in experimentally-relevant conditions, which are particularly stringent in a numerical point-of-view.
In Sec.~\ref{sec:verification}, we verify nonlinear capabilities of COBBLES. In the collisionless limit without external damping, we are able to solve the simpler Vlasov-Poisson system and recover saturation level, relative oscillation amplitude, and the so-called Bernstein-Greene-Kruskal (BGK) steady-state solution \cite{bgk57}. Then, we analyze the conservation and convergence properties for a system with finite $\gamma _d$ and collision frequencies. Further, we benchmark COBBLES against a parameter scan given in former works by Vann \cite{vann03}. A drag and diffusion collision operator is verified by recovering qualitative steady-state distributions predicted by theory. Finally, we consider multiple-waves interaction and estimate a particle diffusion coefficient which agrees with quasi-linear theory.
In Sec.~\ref{sec:analogies}, we summarize the analogies between BB model and 1D model of TAE.

\section{Basic equations and physics} \label{sec:bb_eqns}

Depending on the application, it may be convenient to cast the BB model either in a self-consistent form (full-$f$) or in a perturbative form ($\delta f$). 

\subsection{Normalization}

For the sake of concision in this thesis, and to avoid numerical treatment of too large and too small numbers in simulations, we adopt the normalization shown in Tab.~\ref{tab:normalization}, where the plasma frequency $\omega _p$ is defined by $\omega _p^2 = n_0 e^2 / (\epsilon _0 m)$, $e$, $m$, and $n_0$ are the charge, mass, and total density, respectively, of the evolving species, $\lambda _\mathrm{D} = v_{th}/\omega _{p}$ is the Debye length, and $v_{th}$ is a typical thermal velocity.\\

\begin{table*} \begin{center}
\caption{Correspondence between physical and normalized quantities.}
\begin{tabular}{c c}
&  \\[-5pt]
\hline \hline
&  \\[-10pt]
Physical quantity & Normalization constant  \\[3pt]
\hline
&  \\[-8pt]
Time & $\omega_p^{-1}$ \\[8pt]
Length & $\lambda _\mathrm{D}$ \\[8pt]
Velocity & $v_{th}$ \\[8pt]
Density & $n_0$ \\[8pt]
Distribution $f$ & $n_0/v_{th}$ \\[8pt]
Electric field & $m v_{th}^2 / (e\lambda _\mathrm{D})$ \\[8pt]
Energy, Hamiltonian & $m v_{th}^2$ \\[8pt]
Power & $m \lambda _\mathrm{D} n_0 v_{th}^2 \omega_p$\\[5pt]
\hline  \hline
\end{tabular} 
\label{tab:normalization}
\end{center} \end{table*}

\subsection{Full-\texorpdfstring{$f$}{f} BB model} \label{subs:ff_bb}

\begin{figure}
\begin{center}
\includegraphics{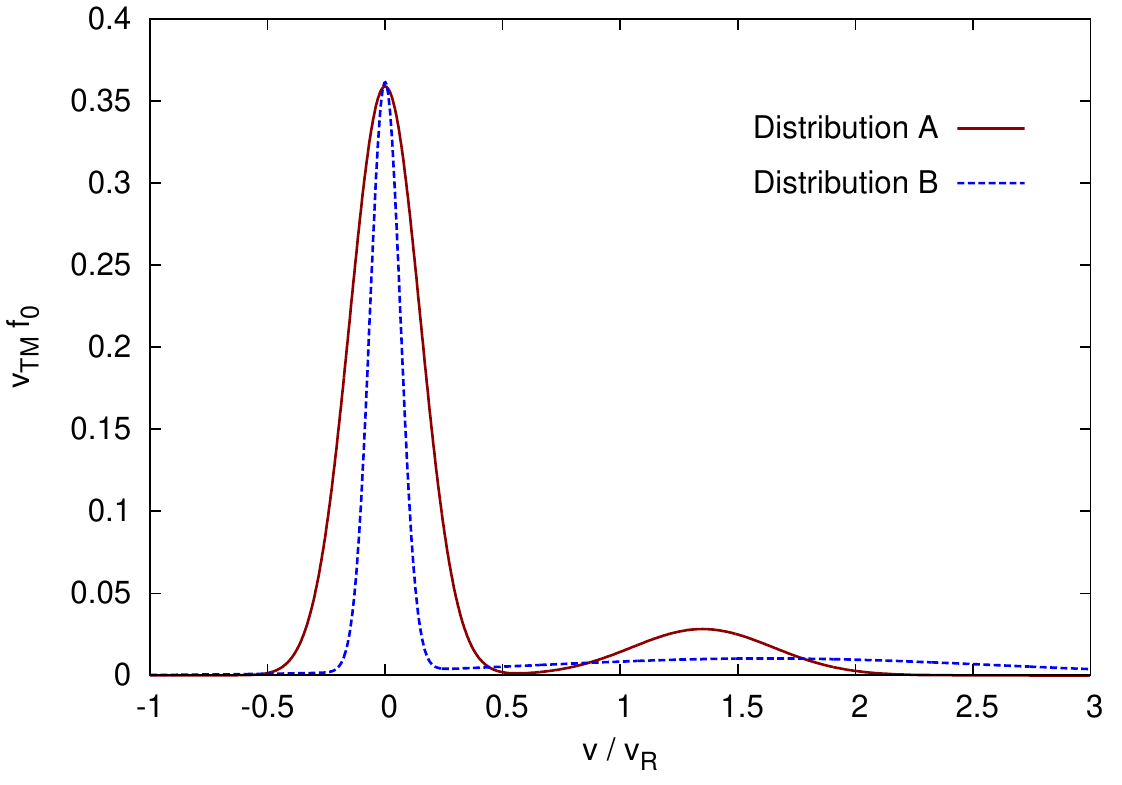}
\caption{Initial distribution function $f_0$. Velocity is normalized to the resonant velocity $v_R$. Full curve is distribution A, broken curve is distribution B.}
\label{fig:full-f_f0}
\end{center}
\end{figure}

We consider a 1D plasma with a distribution function $f(x,v,t)$. In the initial condition, the velocity distribution,
\begin{equation}
f_0(v) \;\equiv\; \ov{f}(v,t=0) \;=\; f_0^M(v) + f_0^B(v),
\end{equation}
where $\ov{f}$ is the spatial average of $f$, comprises a Maxwellian bulk,
\begin{equation}
f_0^M(v) \;=\; \dfrac{n_M}{v_{TM} \, \sqrt{2\pi}} \, e^{-\frac{1}{2}\left(\frac{v}{v_{TM}}\right)^2},
\end{equation}
and a beam of high-energy particles,
\begin{equation}
f_0^B(v) \;=\; \dfrac{n_B}{v_{TB} \, \sqrt{2\pi}} \, e^{-\frac{1}{2}\left(\frac{v-v_B}{v_{TB}}\right)^2},
\end{equation}
where $n_M$ and $n_B$ are bulk and beam densities, which verify $n_M+n_B=1$, $v_{TM}$ and $v_{TB}$ are thermal velocities of bulk and beam particles, and $v_B$ is the beam drift velocity. To ensure charge neutrality, we assume a fixed background population of the opposite charge with a distribution function $f_0(v)$. Fig.~\ref{fig:full-f_f0} shows two typical initial distribution functions, with a cold bulk and a weak, warm beam. The first one, to which we refer in the following as \textit{distribution A}, with $n_B=0.1$, $v_{TM}=0.5$, $v_{TB}=1.0$, and $v_B=4.5$, is used to benchmark our code in \ref{subs:benchmark}. The second one, \textit{distribution B}, with $n_B=0.1$, $v_{TM}=0.2$, $v_{TB}=3.0$, and $v_B=5.0$, hence a warmer beam and a colder bulk, is used to validate and develop some aspects of BB theory. For both distributions, we will always choose a system size $L=2\pi /k$ with $k=0.3$.\\

The evolution of the distribution is given by the kinetic equation
\begin{equation}
\pd{f}{t} \,+\, v \, \pd{f}{x} \,+\, E \, \pd{f}{v} \;=\; \mathcal{C}(f-f_0), \label{model:kinetic}
\end{equation}
where $E$ is the electric field, and $\mathcal{C}(f-f_0)$ is a collision operator.

In this work, we consider either one of the following two collision models.
On the one hand, a large part of existing theory for the BB-model deals with collisions in the form of a Krook operator \cite{bgk54},
\begin{equation}
\mathcal{C}_\mathrm{K}(f-f_0) \;=\; -\nu _a (v) \left( f - f_0 \right), \label{model:krook}
\end{equation}
which is a simple model for collisional processes that tend to recover the initial distribution at a rate $\nu _a$, including both source and sink of energetic particles. If we assume cold and adiabatic bulk plasmas, $\nu _a (v)$ acts only on the beam. Reflecting this situation, we design the velocity dependency of $\nu _a (v)$ as
\begin{equation}
\nu _a (v) \;=\; \left \{ \begin{array}{l r}
\nu _a  &  \mbox{if $v\,>\,v_{\nu}$} \\
0  &  \mbox{else}
\end{array}   \right. , \label{model:nua}
\end{equation}
where $v_{\nu}$ satisfies $f_0^M(v_{\nu})/f_0^M(0) = \epsilon _{\nu}$, and we choose $\epsilon _{\nu} = 10^{-3}$. Hence, $\nu _a (v)$ is constant in the beam region, and zero in the bulk region, except for the benchmark in Sec.~\ref{subs:benchmark} where it is explicitly stated to be constant everywhere.

On the other hand, a more realistic collision operator, the one-dimensional projection of a Fokker-Plank operator, Eq.~(\ref{eq:total_collisions}), includes a dynamical friction (drag) term and a velocity-space diffusion term,
\begin{equation}
\mathcal{C}_\mathrm{FP}(f-f_0) \;=\; \dfrac{\nu _f^2(v)}{k} \dfrac{\partial \left( f - f_0 \right)}{\partial v} \;+\; \dfrac{\nu _d^3(v)}{k^2} \dfrac{\partial ^2 \left( f - f_0 \right)}{\partial v^2}, \label{model:diffdrag}
\end{equation}
where $k$ is the wave number for the resonance under investigation, and with similar velocity-dependence for $\nu _f$ and $\nu _d$.
An other large part of existing theory deals with the latter operator in the absence of drag ($\nu_f=0$). Investigations of the effects of dynamical friction are fairly recent \cite{lilley09,lang10}.

We define the effective collision frequency as $\nu _\mathrm{eff}\equiv \nu _a$ in the Krook case and $\nu _\mathrm{eff}\equiv \nu _d^3/\gamma _{L0}^2$ in the case with diffusion. A dimensional analysis gives a typical lifetime of phase-space structures as $\nu _\mathrm{eff}^{-1}$.

In the expression of the electric field,
\begin{equation}
E(x,t) \,=\, \hat{E_k} (t) e^{\imath k x} \,+\, c.c. ,\label{model:wave}
\end{equation}
we assume a single mode of wave number $k$, reflecting the situation of an isolated single mode AE.
The displacement current equation (DCE),
\begin{equation}
\pd{E}{t} \;=\; -\int v \left( f - f_0 \right) \mathrm{d} v \;-\; 2 \, \gamma _d \, E, \label{model:dce}
\end{equation}
yields the time evolution of the wave. In the initial condition we apply a small perturbation, $f(x,v,t=0) = f_0(v) (1 + \epsilon \cos{k x})$, and the initial electric field is given by solving Poisson's equation. In Eq.~(\ref{model:dce}), an external wave damping has been added to model all linear dissipation mechanisms of the wave energy to the background plasma that are not included in the previous equations \cite{berkbreizman93}.
The presence of a factor $2$ in front of $\gamma _d$ is consistent with Berk and Breizman's literature and will be justified in Sec.~\ref{sec:bblin}.\\

\subsubsection{Conservation properties}

Before deriving the conservation properties of this model, it is useful to note the following property. If $f(x,v,t)$, $g[f(x,v,t),t]$ and $h(x,v,t)$ are arbitrary functions, analytic in a phase-space $\Gamma \equiv (x,v)$, then
\begin{eqnarray}
\int \left\{ h,f \right\}_{\Gamma} \, g(f,t) \, \mathrm{d} x \, \mathrm{d} v &=& \int _0^L \mathrm{d} x \left[ f \, \frac{\partial (g\,h)}{\partial x} \right]_{-\infty}^{\infty} \;-\; \int _{-\infty}^{\infty} \mathrm{d} v \left[ f \, \frac{\partial (g\,h)}{\partial v} \right]_{0}^{L}\\
& = & 0 \qquad \text{for usual boundary conditions,}
\end{eqnarray}
where integration in the l.h.s.~is over the whole phase-space surface.

In the ideal situation, the model presented above ensures conservation of total particle number $N(t) \equiv \int f \mathrm{d} x \mathrm{d} v$.
This is proven by taking the integral over the whole phase-space of the kinetic equation, which can be written as
\begin{equation}
\pds{f}{t} \,-\, \left\{ h,f \right\} _{x,v} \;=\; \mathcal{C}(f-f_0),
\end{equation}
where $h$ is the Hamiltonian,
\begin{equation}
h(x,v,t)\;=\;v^2/2 \;+\; \varphi(x,t), \label{eq:bb_hamiltonian}
\end{equation}
and $\varphi$ is the electrostatic potential. Thus we obtain
\begin{equation}
\dfrac{\mathrm{d} N}{\mathrm{d} t} \;=\; \int \mathcal{C}(f-f_0) \,\mathrm{d} x \,\mathrm{d} v. \label{step:particle_conservation}
\end{equation}
With Krook operator, if $\nu _a (v) = \nu _a$ is taken as constant, the latter equation can be written as
\begin{equation}
\dfrac{\mathrm{d}  \Delta N}{\mathrm{d} t} \;+\; \nu _a \Delta N \;=\; 0, \label{eq:exact_particle_conservation}
\end{equation}
where $\Delta N (t) \equiv N(t)-N(0)$. Since $\Delta N(0)=0$, Eq.~(\ref{eq:exact_particle_conservation}) yields the conservation of total particle number, $\mathrm{d} \Delta N/\mathrm{d} t=0$. However, in numerical simulations, some spurious leakage of particles from velocity boundaries induces a small error in this conservation.
When $\nu _a (v)$ has the velocity dependence of Eq.~(\ref{model:nua}), Eq.~(\ref{eq:exact_particle_conservation}) is changed to
\begin{equation}
\dfrac{\mathrm{d}  \Delta N}{\mathrm{d} t} \;+\; \nu _a \Delta N \;=\; \nu _a \,L \, \int _{-\infty}^{v_{\nu}} \left( \ov{f} - f_0 \right) \, \mathrm{d} v.
\end{equation}
In the bulk part $v<v_{\nu}$, we assume that the variation of the distribution is negligible, $\left| \ov{f}(v, t) - f_0(v) \right|$ $\ll$ $f_0(v)$, and we show the approximative conservation of total particle number,
\begin{equation}
\left| \dfrac{\mathrm{d}  \Delta N}{\mathrm{d} t} \;+\; \nu _a \Delta N \right| \;\ll\; \nu _a \,N(0). \label{result:particle_conservation}
\end{equation}
With Fokker-Plank operator, $\mathrm{d} N / \mathrm{d} t=0$ is immediate from Eq.~(\ref{step:particle_conservation}).

Let us now derive an energy equation to relate power transfers between wave, particles, and external damping. Detailing each term is useful to clarify different possible decomposition of the power transfer, corresponding to different point of view.
Taking the integral over phase space of the product of the Hamiltonian with the kinetic equation, and dividing it by the system size yields
\begin{equation}
\dfrac{1}{L} \, \int h \, \pd{f}{t} \, \mathrm{d}\Gamma \;=\;  \int \dfrac{v^2}{2} \, \pd{f}{t} \, \dfrac{\mathrm{d} x}{L} \, \mathrm{d} v \;+\; \int \varphi \, \pd{f}{t} \, \dfrac{\mathrm{d} x}{L} \, \mathrm{d} v \;=\; P_{\nu}, \label{step:power_transfer}
\end{equation}
where the right hand side shows the collisional power transfer $P_{\nu} \equiv P_{\nu}^{\mathcal{T}} + P_{\nu}^{\varphi}$,
\begin{eqnarray}
P_{\nu}^{\mathcal{T}}(t) &\equiv & \int \mathcal{C}(f-f_0) \dfrac{v^2}{2} \, \dfrac{\mathrm{d} x}{L} \, \mathrm{d} v, \\
P_{\nu}^{\varphi}(t) &\equiv & \int \mathcal{C}(f-f_0) \varphi \, \dfrac{\mathrm{d} x}{L} \, \mathrm{d} v.
\end{eqnarray}
The left integral of the left hand side is the kinetic part $\mathrm{d}\mathcal{T}/\mathrm{d} t$, where $\mathcal{T}$ is the total particle kinetic energy density,
\begin{equation}
\mathcal{T}(t) \;\equiv\; \int \dfrac{v^2}{2} \ov{f} \mathrm{d} v.
\end{equation}
Substituting the kinetic equation into the right integral of the left hand side, we find out that the field part is $-P_h + P_{\nu}^{\varphi}$, where we define the particle power transfer as
\begin{equation}
P_h(t) \;\equiv\; \int v \, E \, f \, \dfrac{\mathrm{d} x}{L} \, \mathrm{d} v. 
\end{equation}

There are two ways of decomposing $P_h$. The substitution of DCE (\ref{model:dce}) into the expression of $P_h$ yields
\begin{equation}
P_h \;=\; -\dfrac{\mathrm{d}\mathcal{E}}{\mathrm{d} t} - P_d,
\end{equation}
where the electric field energy density is defined by
\begin{equation}
\mathcal{E}(t) \;\equiv\; \int \dfrac{E^2}{2}  \dfrac{\mathrm{d} x}{L},
\end{equation}
and the external damping power transfer by
\begin{equation}
P_d(t) \;\equiv\; 4 \, \gamma _d \, \mathcal{E}.
\end{equation}
On the other hand, by substituting the kinetic equation (\ref{model:kinetic}) into the expression of $\mathrm{d}\mathcal{T}/\mathrm{d} t$ we obtain
\begin{equation}
P_h \;=\; -P_{\nu}^{\mathcal{T}} \;+\; \dfrac{\mathrm{d}\mathcal{T}}{\mathrm{d} t}.
\end{equation}

We can further separate the response of the particules into resonant and non-resonant pieces if we decompose the distribution into $f \equiv f^\mathrm{R} + f^\mathrm{NR}$ and define $\mathcal{T}\equiv \mathcal{T}^\mathrm{R} + \mathcal{T}^\mathrm{NR}$. The non-resonant part of the kinetic energy is the sloshing energy,
\begin{equation}
\mathcal{T}^\mathrm{NR} \;\equiv\; \int \dfrac{v^2}{2} \ov{f^\mathrm{NR}} \mathrm{d} v
\end{equation}
For non-resonant particles, the velocity is oscillatory and we can replace the amplitude of its oscillation by the linear response $E_0/\omega$, where $E_0=2|\hat{E_k}|$, and we obtain $\mathcal{T}^\mathrm{NR}=\mathcal{E}$. The wave energy $\mathcal{W}$ is composed of the field energy $\mathcal{E}$ and the sloshing energy $\mathcal{T}^\mathrm{NR}$ which supports the wave,
\begin{equation}
\mathcal{W} \;\equiv\; \mathcal{T}^\mathrm{NR} \;+\; \mathcal{E} \;=\; 2 \, \mathcal{E}.
\end{equation}

Finally, we can express the power transfer equation in two equivalent ways, depending on whether we consider the non-resonant kinetic energy as part of the total kinetic energy or as part of the total wave energy. In the electric field point of view,
\begin{equation}
\dfrac{\mathrm{d}\mathcal{E}}{\mathrm{d} t} \;+\; P_h \;+\; 4 \, \gamma _d \, \mathcal{E} \;=\; 0, \label{result:power_transfer_E}
\end{equation}
showing the balance between the field and all the particles.
In the wave-as-quasi-particles point of view,
\begin{equation}
\dfrac{d\mathcal{W}}{dt} \;+\; P_h^\mathrm{R} \;+\; 2 \, \gamma _d \, \mathcal{W} \;=\; 0, \label{result:power_transfer_W}
\end{equation}
where $P_h ^\mathrm{R}$ is the resonant power transfer,
\begin{equation}
P_h ^\mathrm{R} \;\equiv\; \int v \, E \, f^\mathrm{R} \, \frac{\mathrm{d} x}{L}\,\mathrm{d} v \;=\; P_h \;-\; \dfrac{\mathrm{d}\mathcal{T}^\mathrm{NR}}{\mathrm{d} t}, \label{def:resonant_power}
\end{equation}
showing the balance between wave and resonant particles.

%A straightforward analysis of the above model yields a power balance equation. Let us define the electric field energy,
%\begin{equation}
%\mathcal{E}(t) \;\equiv\; \int \dfrac{E^2}{8\pi} \mathrm{d} x. \label{def:electricEnergy}
%\end{equation}
%The power transferred from the perturbed electric field to both bulk and beam particles, not including sloshing energy, is given by
%\begin{equation}
%P_h(t) \;\equiv\; \int v \, E \, f \, \mathrm{d} x \, \mathrm{d} v. 
%\end{equation}
%We show the power balance equation,
%\begin{equation}
%P_E \;+\; P_h \;+\; P_d \;=\; 0, \label{result:power_transfer_E}
%\end{equation}
%where $P_E \equiv \mathrm{d} _t \mathcal{E}$ is the electric field power transfer, and $P_d \equiv 4 \gamma _d \, \mathcal{E}$ is the power transferred to the background plasma by dissipative mechanisms.

%[Additions needed: Relation between direct current equation and poisson equation ; Discussions on Krook operator ; Discussions on background damping. Note that the system looses its self-consistency ; Entropy conservation ?]

\subsection{\texorpdfstring{$\delta f$}{df} BB model}

If the bulk particles interact adiabatically with the wave, their contribution to the Lagrangian can be expressed as part of the electric field. Then it is possible to adopt a perturbative approach, and to cast the BB model in a reduced form that describes the time evolution of beam particles only \cite{berkbreizman95,cary93}.
The evolution of the beam distribution, $f^B(x,v,t)$, is given by the kinetic equation
\begin{equation}
\pd{f^B}{t} \,+\, v \, \pd{f^B}{x} \,+\, \tilde{E} \, \pd{f^B}{v} \;=\; \mathcal{C}\left(f^B - f^B_0\right), \label{eq:df_kinetic}
\end{equation}
where the pseudo-electric field $\tilde{E}$ is defined as
\begin{equation}
\tilde{E}(x,t) \; \equiv \; Q(t) \, \cos (\psi) \,-\, P(t) \, \sin (\psi), \label{eq:df_reduced_wave}
\end{equation}
where $\psi \equiv kx-\omega t$. In this model, the real frequency of the wave is imposed as $\omega = 1$. This restriction does not forbid nonlinear phenomena like frequency sweeping, since both amplitude and phase of the wave are time-dependent. In this thesis, we renormalize physical quantities for the $\delta f$ model so that they do not depend on $k$. In practice, we choose $k=1$.
In the collision operators, $\nu _a$, $\nu _f$ and $\nu _d$ are taken as constants, since, with the $\delta f$ description, velocity dependency is not needed to avoid affecting bulk plasma with collisions.

The evolution of the pseudo-electric field is given by
\begin{eqnarray}
\dfrac{\mathrm{d} Q}{\mathrm{d} t} \;=& -\dfrac{1}{2\pi} \, \int f^B(x,v,t) \, \cos (\psi) \, \mathrm{d}x \, \mathrm{d}v &-\; \gamma _d \, Q, \label{model:reduced_Q}\\
\dfrac{\mathrm{d} P}{\mathrm{d} t} \;=& \dfrac{1}{2\pi} \, \int f^B(x,v,t) \, \sin (\psi) \, \mathrm{d}x \, \mathrm{d}v &-\; \gamma _d \, P. \label{model:reduced_P}
\end{eqnarray}
The initial values of $Q$ and $P$ are given by solving Poisson's equation. Note that the latter equations, without factor $2$ in front of $\gamma _d$, are consistent with Eq.~(\ref{model:dce}).\\

\begin{figure} \begin{center}
\includegraphics{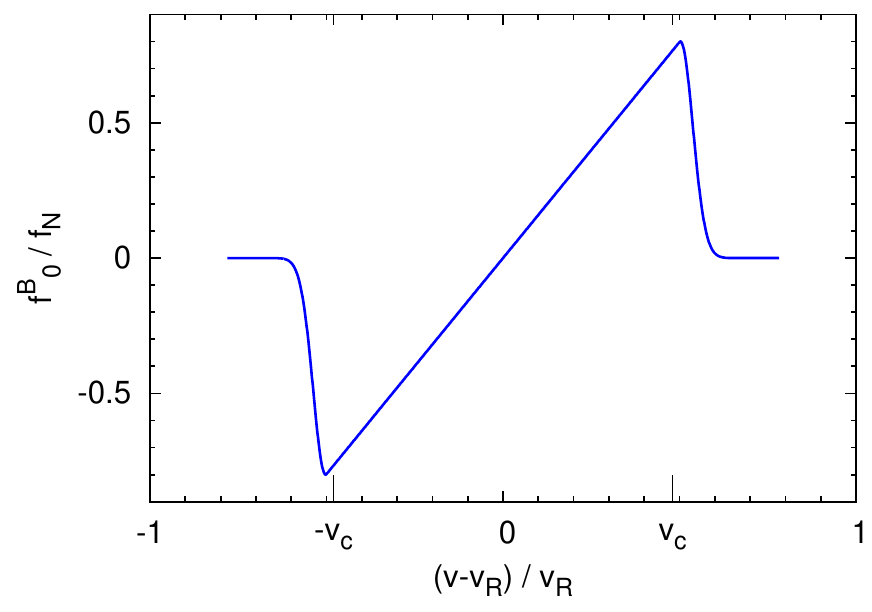}
\caption{Initial velocity distribution function, normalized to $f_N\equiv\gamma _{L0} / (\sqrt{2\pi} v_R)$.}
\label{fig:df_f0}
\end{center} \end{figure}

In the collisionless case, one can see from the linear dispersion relation Eq.~(\ref{eq:df_disp_rela}) that $\omega = 1$ only if $f^B_0$ is symmetric around the resonant velocity, $v_R \equiv \omega / k$. Since we assumed $\omega = 1$ from the start, we consider only such distributions, for the model to be self-consistent.
The velocity distribution of beam particles in the initial condition is shown in Fig.~\ref{fig:df_f0}. A constant slope is imposed between $v=-v_c$ and $v=v_c$, where $v_c$ is an arbitrary cut-off velocity. The zero average ensures that the plasma frequency is not perturbed by the beam density. Smooth joins between the constant gradient region and the large velocity regions are necessary to prevent numerical oscillations at $v\approx \pm v_c$. Since we always choose $v_c$ large enough so that border effects are negligible, an initial distribution is fully characterized by its slope at resonant velocity, in other words by $\gamma _{L0}$.\\

\subsubsection{Conservation properties}

Arguments similar to those for full-$f$ model yield the conservation of total particle number.
The power balance is changed to,
\begin{equation}
P_h \;+\; P_E \;+\; P_d \;=\; 0,
\end{equation}
where $P_h$ is the kinetic power transfer,
\begin{equation}
P_h \;\equiv\; \int v\,E\,f^B\,\mathrm{d} x\,\mathrm{d} v,
\end{equation}
$P_E$ is the electric field power transfer,
\begin{equation}
P_E \;\equiv\; \dfrac{2\pi}{k} \, \dfrac{\mathrm{d}}{\mathrm{d} t} \left( P\dot{Q}-Q\dot{P} \,+\, \dfrac{Q^2+P^2}{2} \right),
\end{equation}
and $P_d$ is the power transfer due to external damping and collisions. In the Krook case,
\begin{equation}
P_d \;\equiv\; \dfrac{2\pi}{k} \, \left[ (\gamma_d + \nu_a)\, \left( P\dot{Q}-Q\dot{P} \right) \,+\, \gamma_d \, \left( Q^2+P^2 \right) \right], \label{eq:external_power}
\end{equation}
while in the Fokker-Planck case, collisions do not contribute to this latter power transfer, thus $P_d$ is obtained by substituting $\nu _a=0$ in Eq.~(\ref{eq:external_power}).

Compared to the full-$f$ model, the $\delta f$ model does not take into account effects of time-evolution of bulk particles, which is a caveat when assessing limit of theory that breaks-up when phase-space structures approach the bulk, but it has an advantage in the application to experiment, where we assume fixed mode structure, hence fixed background plasma. Moreover, the velocity range required to simulate a similar resonant region can be significantly reduced with the $\delta f$ model, saving computation time.
%The full-$f$ model with Krook collisions is considered in Vann's literature, while the $\delta f$ model with either Krook or diffusion-only collisions has been studied extensively in Berk and Breizman's literature. In this thesis, we use the full-$f$ model with Krook collisions, the $\delta f$ model with Krook
Since, in this thesis, we often refer to literature by Berk and Breizman, by Vann, or by Lilley, we clarify which approaches and which collision operators have been studied by these authors, in Tab.~\ref{tab:who_did_what}.

%\begin{table*} \begin{center}
%\caption{Non exhaustive list of approaches and collision operators in the literature. "BB" refers to Berk, Breizman and coworkers, "Lesur" refers to this thesis.}
%\begin{tabular}{l c c c c}
%\hline \hline 
%& & & & \\[-10pt]
%           & $\nu_a$ & $\nu _a(v)$ & $\nu _d$ & $\nu _f$ and $\nu _d$  \\[3pt]
%\hline 
%& & & & \\[-8pt]
%full-$f$   & Vann & Lesur &  &   \\[8pt]
%$\delta f$ & BB, Lilley and Lesur & - & BB, Lilley and Lesur & Lilley and Lesur  \\[5pt]
%\hline  \hline
%\end{tabular} 
%\label{tab:who_did_what}
%\end{center} \end{table*}

\begin{table*} \begin{center}
\caption{Non exhaustive list of approaches and collision operators in the literature. "BB" refers to Berk, Breizman and coworkers, "Lesur" refers to this thesis.}
\begin{tabular}{l c c}
&  \\[-5pt]
\hline \hline 
& &  \\[-10pt]
Author  & Approach & Collisions  \\[3pt]
\hline 
BB     & $\delta f$            & Krook / Diffusion \\[8pt]
Vann   & full-$f$              & Krook \\[8pt]
Lilley & $\delta f$            & Krook / Diffusion / Diffusion+Drag \\[8pt]
Lesur  & $\delta f$ / full-$f$ & Krook / Diffusion / Diffusion+Drag \\[5pt]
\hline  \hline
\end{tabular} 
\label{tab:who_did_what}
\end{center} \end{table*}

\section{Linear analysis} \label{sec:bblin}

When the perturbation is small, linear theory predicts exponential growth or decay \cite{landau46} of the wave amplitude. For the full-$f$ model with Krook collisions, the linear dispersion relation,
\begin{equation}
\gamma \;+\; 2\, \gamma _d \;-\; \imath \, \omega \;=\; \int _{\Gamma} \dfrac{v\, \pds{f_0}{v}}{(\gamma \,+\, \nu _a) \,+\, \imath \, (k\,v \,-\, \omega)} \, \mathrm{d} v, \label{eq:ff_disp_rela}
\end{equation}
where $\Gamma$ is the appropriate Landau contour \cite{landau46}, yields the linear growth rate $\gamma$, and the real frequency $\omega$ of the wave. We implemented an algorithm to solve the latter equation, applying a method of residue for locating the zeros of an analytic function in the complex plane \cite{davies86}. We refer to this algorithm as Davies solver.
In the following, $\gamma _L$ is defined as the linear growth rate for $\gamma _d = \nu _a = 0$.

\begin{figure} \begin{center}
\includegraphics{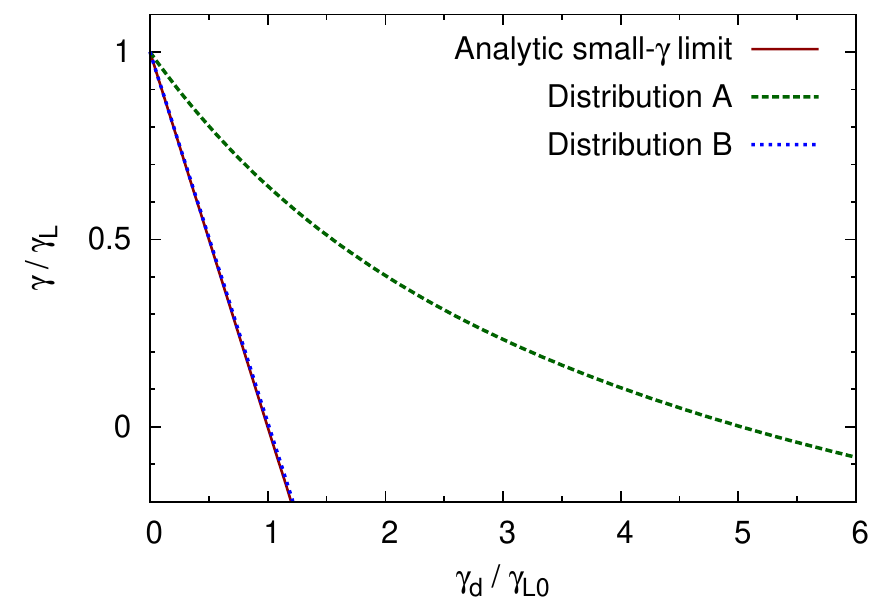}
\caption{Growth rate in the collisionless limit. Solid line corresponds to Eq.~(\ref{eq:collisionless_gamma}), dashed line corresponds to distribution A, for which $\gamma _L=0.1981$, and dotted line corresponds to distribution B, for which $\gamma_L=0.0324$.}
\label{fig:collisionless_linear}
\end{center} \end{figure}

In the collisionless limit, if we assume a small perturbation and a linear growth rate $\gamma$ much smaller than the real frequency $\omega$, the dispersion relation reduces to
\begin{equation}
\gamma \;=\; \gamma _L \;-\; \dfrac{2\,\gamma _d}{\omega \, \left. \partial_\Omega \mathcal{R}(D_L) \right\vert _{\Omega = \omega}},
\end{equation}
where
\begin{equation}
\mathcal{R}(D_L) \;=\; 1 \;-\; \frac{1}{\Omega} \, \mathcal{P} \int \frac{v \partial _v f_0}{k v-\Omega} \mathrm{d} v,
\end{equation}
and $\mathcal{P}$ is a notation for Cauchy's principal value. In the cold Maxwellian limit, $\gamma _L=\gamma _{L0}$, where $\gamma _{L0}$ is a measure of the slope of initial distribution at resonant velocity,
\begin{equation}
\gamma _{L0} \;\equiv\; \dfrac{\pi}{2k^2} \, \left. \pd{f_0}{v} \right| _{v=v_R}.
\end{equation}
In this limit, a simple relation, $\gamma \,=\, \gamma _L - \gamma _d$, stands. However, with our choice of distribution function, we must keep in mind that there is some discrepancy between $\gamma$ and $\gamma _L - \gamma _d$,
\begin{equation}
\gamma \;=\; \gamma _L \, \left( 1\,-\, \frac{\gamma _d}{\gamma _{L0}} \right) \quad \mathrm{for } \;\nu _a=0,\; \gamma \ll \omega. \label{eq:collisionless_gamma}
\end{equation}
Fig.~\ref{fig:collisionless_linear} shows the growth rate estimated by Davies solver as a function of external damping, in the collisionless limit, for both initial distributions A and B. Eq.~(\ref{eq:collisionless_gamma}) is recovered in the limit of small $\gamma$.

With Fokker-Planck collisions, the kinetic equation in Fourier-Laplace space is a second order differential equation in $v$, which prevents a similar treatment. However, we can take another approach, which is also valid in the Krook case, where we search for solutions of the form $\exp (p t)$, where $p\equiv\gamma - \imath \omega$. Writing $f_k(v,t)=f_p(v) e^{pt}$ and $E_k(t)=E_p e^{pt}$ the Fourier component of $f-f_0$ and $E$, respectively, we obtain a linear equation system,
\begin{eqnarray}
(p\,+\,\imath k v) f_p \;+\; E_p \, \pd{f_0}{v} &=& -\nu _a \, f_p \;+\; \frac{\nu _f^2}{k}\,\pd{f_p}{v} \;+\; \frac{\nu _d^3}{k^2}\,\frac{\partial ^2 f_p}{\partial v^2} ,\\
(p\,+\,2\gamma _d) E_p &=& - \int v \,f_p\, \mathrm{d} v.
\end{eqnarray}
Discretizing the velocity space as $v_j=j \Delta v$ for $j=1\cdots N_v$, the latter system is approximated to first order accuracy in $\Delta v$ by
\begin{eqnarray}
(p\,+\,\imath k v_j) f_j \;+\; E_p \, \pd{f_0}{v}(v_j) &=& -\nu _a \, f_j \;+\; \frac{\nu _f^2}{2k\Delta v}\,(f_{j+1}-f_{j-1}) \nonumber \\
& &  \qquad  \quad  \;+\; \frac{\nu _d^3}{k^2\Delta v^2}\,(f_{j+1}-2f_j+f_{j-1}),\\
(p\,+\,2\gamma _d) E_p &=& - \Delta v \sum _{j=1}^{N_v} v_j \,f_j,
\end{eqnarray}
where $f_j\equiv f_p(v_j)$, and boundary conditions are $f_0=f_{N_v+1}=0$.
This system of $N_v+1$ equations can be put in matrix form, 
\begin{equation}
\mathsf{M} \cdot \boldsymbol{F} \;=\;p \,\boldsymbol{F},
\end{equation}
where $\boldsymbol{F}$ is a $N_v+1$ dimension vector defined by
\begin{eqnarray}
F _j &=& f_j \quad (j=1\cdots N_v),\\
F _{N_v +1} &=& E_p,
\end{eqnarray}
and $\mathsf{M}$ is a $N_v+1$ dimension square matrix defined by
\begin{eqnarray}
\mathsf{M} _{j,j} &=& -\imath k v_j \;-\; \nu _a \;-\; 2\frac{\nu_d^3}{k^2\Delta v^2} \quad (j=1\cdots N_v),\\
\mathsf{M} _{j+1,j} &=& \frac{\nu_d^3}{k^2\Delta v^2} \;+\; \frac{\nu_f^2}{2k\Delta v}   \quad (j=1\cdots N_v-1),\\
\mathsf{M} _{j,j+1} &=& \frac{\nu_d^3}{k^2\Delta v^2} \;-\; \frac{\nu_f^2}{2k\Delta v}   \quad (j=1\cdots N_v-1),\\
\mathsf{M} _{N_v +1,j} &=& -\partial _v f_0(v_j)  \quad (j=1\cdots N_v),\\
\mathsf{M} _{j,N_v +1} &=& -v_j \Delta v  \quad (j=1\cdots N_v),\\
\mathsf{M} _{N_v +1,N_v +1} &=& -2\gamma _d,
\end{eqnarray}
where $\mathsf{M} _{i,j}$ is the element of column $i$, line $j$.
\begin{figure} \begin{center}
\includegraphics{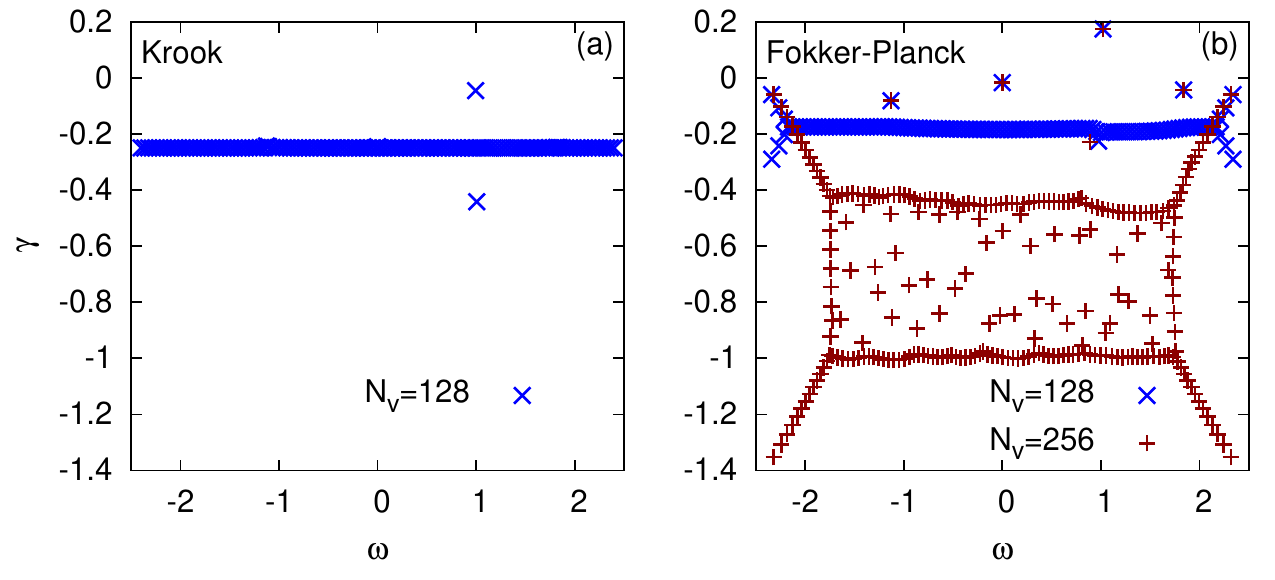}
\caption{Eigenvalues for initial distribution A, with $\gamma_d=0.1$. (a) Krook case, with $\nu_a=0.25$. (b) Fokker-Planck case, with $\nu_f=0.02$ and $\nu_d=0.05$.}
\label{fig:evs}
\end{center} \end{figure}
We solve the above eigenvalue problem using LAPACK library.
In the Krook case, $\omega +\imath \gamma=kv-\imath \nu _a$ constitutes a continuum of trivial solutions with $E=0$. This eigenvalue method does not yield solutions with $\gamma < -\nu _a$. Moreover, as $\gamma$ approaches $-\nu _a$, increasing number of grid points are needed to accurately estimate the growth rate, since continuum solutions tend to perturb nontrivial solutions. Eigenvalues found by this method are shown in Fig.~\ref{fig:evs}(a), for initial distribution A, for which $\gamma _L=0.1981$, and with $\gamma_d=0.1$ and $\nu_a=0.25$. The least stable solution is $\omega=0.9953$, $\gamma =-0.04551$, which is confirmed with Davies solver (see Table \ref{tab:linear_check}). Thus, damped solutions can be found, if $|\gamma | < \nu _a$.
In the Fokker-Planck case, the continuum is topologically changed, and depends on both distribution function and grid points number. Thus, when estimating growth rates with this method, we must be careful to use values that are converged with $N_v$. This is illustrated in Fig.~\ref{fig:evs}(b), which shows eigenvalues for distribution A, with $\gamma_d=0.1$, $\nu_f=0.02$ and $\nu_d=0.05$. The most unstable solution is $\omega=1.0204$, $\gamma =0.1737$, which is in agreement with full-$f$ simulation (see Table \ref{tab:linear_check}).

For the $\delta f$ model with Krook collisions, the linear dispersion relation is changed to
\begin{equation}
\gamma \;+\; \gamma _d \;-\; \imath \, (\omega -1) \;=\; \frac{1}{2k} \int _{\Gamma} \frac{\pds{f^B_0}{v}}{(\gamma \,+\, \nu _a) \,+\, \imath \, (k\,v \,-\, \omega)} \, \mathrm{d} v. \label{eq:df_disp_rela}
\end{equation}
As announced before, we have $\omega = 1$ only if the imaginary part of the right-hand side vanishes, in other words if $f^B_0$ is anti-symmetric around the resonant velocity.
In the collisionless case, for $\gamma / \omega \ll 1$, Eq.~(\ref{eq:df_disp_rela}) yields
\begin{equation}
\gamma \;=\; \gamma _{L0} \;-\; \gamma _d. \label{eq:gaml_minus_gamd}
\end{equation}

If we search for solutions of the form $\exp (p t)$, as $f_k(v,t)=f_p(v) e^{pt}$ and $Z(t)=Z_p e^{pt}$, where $Z(t)\equiv [Q(t)+\imath P(t)] \exp (-\imath t)$, we obtain a linear equation system,
\begin{eqnarray}
(p\,+\,\imath k v) f_p \;+\; \frac{Z_p}{2} \, \pd{f_0}{v} &=& -\nu _a \, f_p \;+\; \frac{\nu _f^2}{k}\,\pd{f_p}{v} \;+\; \frac{\nu _d^3}{k^2}\,\frac{\partial ^2 f_p}{\partial v^2} ,\\
(p\,+\,2\gamma _d+\imath) E_p &=& - \frac{1}{k}\,\int f_p\, \mathrm{d} v.
\end{eqnarray}
The discretized version of the latter system can easily be put in the form of an eigenvalue matrix problem, and solved in a way similar to the full-$f$ case.

%[Work needed: check and include hybrid method]

\section{The kinetic code COBBLES\label{sec:cobbles}}

\subsection{Numerical implementation}

Let us recall that the BB model is an extension of the Vlasov-Poisson system, which is recovered in the collisionless, closed system ($\gamma _d = 0$) limit. In a previous work \cite{lesur07}, we developed a 1D semi-Lagrangian full-$f$ Vlasov code, based on the Cubic-Interpolated-Propagation (CIP) scheme \cite{nakamura99} and the splitting method \cite{cheng76}, which enabled accurate simulations of the Vlasov-Poisson system. In this thesis, we extend our code to include distribution relaxation and extrinsic dissipation, and develop a $\delta f$ version. We refer to these codes as full-$f$ COBBLES and $\delta f$ COBBLES, respectively, COBBLES standing for COnservative Berk-Breizman semi-Lagrangian Extended Solver.

In both codes we solve DCE instead of Poisson equation. Looking at the spatial average of Eq.~(\ref{model:dce}),
\begin{equation}
%\frac{\mathrm{d}\ov{E}}{\mathrm{d} t}\;=\;-(\mathsf{P}-\mathsf{P}_0)\;-\; 2\,\gamma _d \ov{E},
\frac{\mathrm{d}\ov{E}}{\mathrm{d} t}\;=\;- \int v \left( \ov{f} - f_0 \right) \mathrm{d} v \;-\; 2\,\gamma _d \ov{E}, \label{eq:average_field}
\end{equation}
we see that a small deviation from a constant total momentum can be the source of a systematic error in the average electric field. Such deviation arises when Krook collisions are included, or can be caused by numerical error. To avoid this problem, we replace $\int v f_0 \mathrm{d} v$ by $\int v \ov{f} \mathrm{d} v$ in the DCE \cite{vann_thesis}. Then Eq.~(\ref{eq:average_field}) is changed to $\mathrm{d} _t \ov{E}+2\gamma _d \ov{E}=0$, which ensures a zero average electric field, since $\left. \ov{E} \right| _{t=0} = 0$.

Let us now describe the main points of our algorithm. All quantities like $f$ are sampled on uniform Eulerian grids with $N_x$ and $N_v$ grid points in the $x$ and $v$ directions, respectively, within the computational domain $\{ (x,v)$ $|$ $0\leq x < L,$ $v_{\mathrm{min}}\leq v \leq v_{\mathrm{max}} \}$. For distribution A, cut-off velocities are always chosen as $v_\mathrm{min}=-8$ and $v_\mathrm{max}=8$. For distribution B, $v_\mathrm{min}=-10$ and $v_\mathrm{max}=18$. We define the Courant-Friedrichs-Lewy number $\mathrm{CFL}$ $\equiv$ $v_{\mathrm{max}}\, \Delta t \, N_x$ $/$ $(2 L)$ as a measure of the time-step width $\Delta t$. We use the Strang splitting \cite{strang68} formula to obtain a second-order accuracy in time \cite{vann03}. For each time-step, we perform the following steps,
\begin{enumerate}%[1.]
\item Advect $\pds{f}{t} \,+\, v \, \pds{f}{x} \;=\; 0$ for a time $\Delta t / 2$
\item \label{step:collisions} Solve $\pds{f}{t} \;=\; -\, \nu _a \, \left( f - f_0 \right) \;-\; (\nu _f^2/k) \,\partial _v f_0 \;+\; (\nu _d^3/k^2) \,\partial _v^2 (f-f_0)$ for a time $\Delta t / 2$
\item \label{step:dce} Solve DCE for a time $\Delta t / 2$
\item \label{step:v} Advect $\pds{f}{t} \,+\, (E(x) \,-\, \nu _f^2/k) \, \pds{f}{v} \;=\; 0$ for a time $\Delta t$
\item Repeat the step 3.
\item Repeat the step 2.
\item Repeat the step 1.
\end{enumerate}

Numerically, step \ref{step:dce} is performed by a forward Euler scheme. Note that the implementation of friction is split into steps \ref{step:collisions} and \ref{step:v}. In step \ref{step:collisions}, $f$ is replaced by $f_0+\exp (-\nu_a \Delta t/2)(f-f_0)\,-\,(\nu _f^2/k)\partial _v f_0 \Delta t /2$, then the diffusion equation is solved by the Crank-Nicolson method \cite{crank47}.
The remaining problem, corresponding to steps 1 and 4, is the advection of a 1D hyperbolic equation,
\begin{equation}
\pds{F}{t} \;+\; u \, \pds{F}{\lambda} \;=\; 0, \label{eq:1Dadvec}
\end{equation}
where $u$ is constant in the $\lambda$ direction, $\lambda$ is a generalized advection coordinate, and $F$ is a general function of $\lambda$ and $t$. We aim at long-time accurate simulations in the whole ($\gamma _d$,\, $\nu _a$) space. The choice of advection scheme is of crucial importance to reach this goal. In Appendix \ref{app:numerical_implementation}, we recall the CIP-CSL algorithm, which we use for solving Eq.(\ref{eq:1Dadvec}), and its extension to the position-velocity phase-space, as presented in Ref.~\cite{nakamura01}. The key idea is that in addition to the distribution function, we advect its integrated value $\rho$ to keep a flux balance between neighboring grids. We justify this choice in the following section.
Boundary conditions are periodic in configuration space, and zero-flux at velocity boundaries.

\begin{figure}
\begin{center}
\includegraphics{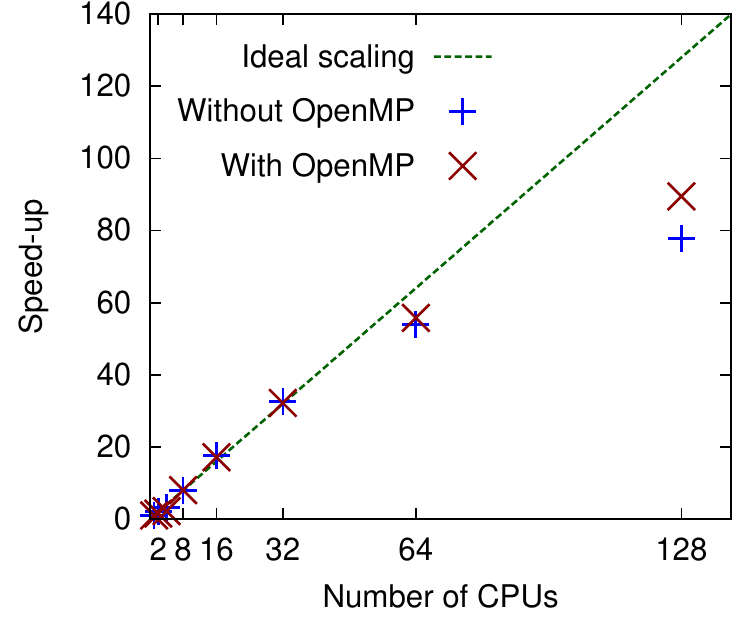}
\caption{Strong-scaling for the COBBLES code, with and without OpenMP enabled.}
\label{fig:strong_scaling}
\end{center}
\end{figure}

COBBLES is coded in Fortran 90 language. It is parallelized in a hybrid fashion, using MPI in the velocity direction and OpenMP. Fig.~\ref{fig:strong_scaling} shows the speed-up on JAEA's BX900 systems as the number of Central Processing Units (CPUs) is increased, at fixed grid-points number $N_x \times N_v \,=\, 64 \times 4096$. Although there is room for optimization, the observed scaling properties are sufficient for our purposes. The use of OpenMP in addition to MPI provides a significant speed-up for 128 CPUs.

Differences between $\delta f$ and full-$f$ versions are in the initialization, which defines $f^B$ instead of $f$, and in the DCE, which is replaced by Eqs.~(\ref{model:reduced_Q}) and (\ref{model:reduced_P}).

Note that we also implemented the treatment of two species, which was used to benchmark simulations of Ref.~\cite{nguyen10prl}, though this feature is not used in this thesis.

\subsection{Comparison of advection schemes} \label{subs:schemes}

Compared to the benchmark of \ref{subs:benchmark}, our choices of parameters for testing theory, in Chap.~\ref{ch:nonlinearities}, constitute difficult conditions for numerical stability of full-$f$ simulations and drastically increases computational cost. As a consequence, we must take special care in choosing an advection scheme that minimizes computational time. Therefore we discuss the relevancy of the CIP-CSL advection scheme, and present a comparison with four other advection schemes.

In choosing the advection scheme, we focus on stability and convergence properties, which are estimated with severe benchmark parameters relevant for analysis in Sec.~\ref{subs:bif00327}, where we use distribution B (with a cold bulk and a weak, warm beam).
Compared to distribution A, which is used as initial condition in the following benchmark (Sec.~\ref{subs:benchmark}), simulations with initial distribution B are more sensitive to numerical errors such as numerical diffusion. Firstly, the colder the bulk, the less grid points are available in the bulk, leading to artificial heating. Secondly, the weak warm beam induces weaker linear instabilities, which produce narrower islands in phase space. To resolve such a narrow island, increased grid resolution is needed. Furthermore, for steady-state solutions, when the island is narrower we observe unphysical drive after nonlinear saturation, which suggests that the region of flattening acquire spurious gradient by influence of surrounding distribution.
In this work, we aim at producing a numerical scan of nonlinear behavior in the whole parameter space. Near marginal stability, the linear growth rate $\gamma$ is very small (we limit the investigation range to $\left| \gamma  \right| > 10^{-6}$ to avoid excessive computation cost) and long-time computations ($t \sim 10^5$) are required. For this reason, we cannot afford too much grid points, and we have to take utmost care in choosing a robust and quickly converging numerical scheme.\\

\begin{figure}
\begin{center}
\includegraphics{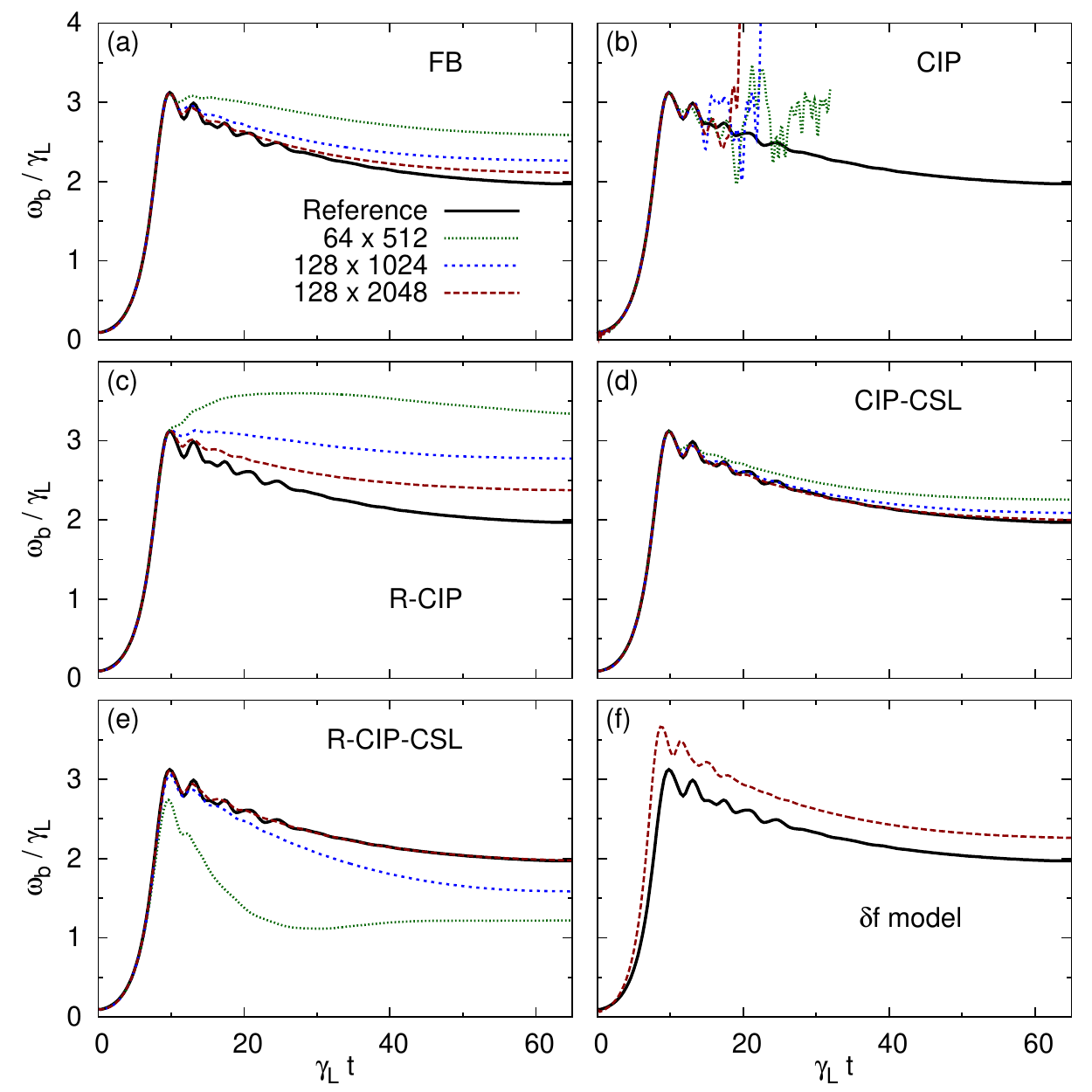}
\caption{Time evolution of the normalized bounce frequency for different advection schemes. Solutions are shown for grid resolution $N_x \times N_v$ of $64 \times 512$, $128 \times 1024$, and $128 \times 2048$, and for a reference run described in the text. (a-e) Full-$f$ simulations with initial distribution B, $\nu _a(v>v_{\nu})=0.002$, $\gamma _d=0.002$, and $\mathrm{CFL}=0.9$. Each sub-figure corresponds to one of five advection schemes. (f) $\delta f$ simulation with $\gamma _{L0}=0.037$ (such that $\gamma _L$ is the same as for full-$f$), $\nu _a=0.002$, $\gamma _d=0.002$, and $\mathrm{CFL}=0.9$.}
\label{fig:convergenceTest}
\end{center}
\end{figure}

A comparison of several advection schemes for one of the case of Fig.~\ref{fig:bifurcation00327} (with distribution B) is shown in Fig.~\ref{fig:convergenceTest}. The time evolution of a beam instability with a low dissipation $\nu _a(v>v_{\nu})=0.002$, and a small external damping $\gamma _d=0.002$, for increasing grid resolution, is compared to a reference run for each of five schemes : Flux-Balance (FB) \cite{fijalkow99}, CIP \cite{nakamura99}, CIP with rational function interpolation (R-CIP) \cite{xiao96}, CIP-CSL, and Rational - CIP-CSL (R-CIP-CSL) scheme \cite{xiao02}. The reference run is obtained with a high resolution $N_x \times N_v \,=\, 256 \times 4096$ using CIP-CSL.
%The reference run is chosen as the average of the solutions given by FB, R-CIP, CIP-CSL and R-CIP-CSL, with a high resolution $N_x \times N_v \,=\, 256 \times 4096$.
The CIP scheme is a low-diffusion and stable scheme, and is implemented in a way that exactly conserves the total mass. However, it is not locally conservative. After several amplitude oscillations in the nonlinear phase, we observe the apparition of numerical oscillations in the velocity direction in a large gradient region of the distribution, which appears between a cold bulk and a beam. While, in this test case, numerical divergence eventually occurs even for very high resolution with the CIP scheme, the other schemes show convergence to a same solution. The FB scheme is only second-order accurate, so that convergence is slow compared to the CIP-based schemes, which are third-order accurate in general \cite{nakamura99}. Rational function interpolation aims at preventing numerical oscillations by preserving convex-concave and monotonic properties, but at the expense of this property, numerical diffusion produces spurious drive leading to higher saturation levels. R-CIP-CSL produces less numerical diffusion than R-CIP, but convergence is slower than with CIP-CSL. Finally, the CIP-CSL scheme shows quick convergence without unfavorable numerical oscillations, and therefore, we use this scheme in the following simulations.

As an illustration of $\delta f$ model, we include in Fig.~\ref{fig:convergenceTest} the time-evolution of beam instability obtained by $\delta f$ COBBLES (with the CIP-CSL scheme), for similar parameters as those of the full-$f$ simulations.

\section{Verification of COBBLES\label{sec:verification}}

For concision, we present only the verification of full-$f$ COBBLES, except for conservation properties, for which it is revealing to compare full-$f$ and $\delta f$ approaches, and for verification of drag and diffusion collision operator, since reference material is based on hypothesis of $\delta f$ model. As a preliminary test, we compared linear growth rate and real frequency measured in COBBLES simulations with those obtained with Davies solver in the Krook case, or by solving the eigenvalue problem in the Fokker-Planck case, and found good quantitative agreement, which is illustrated in Table \ref{tab:linear_check}.

%\begin{table*} \begin{center}
%\caption{Linear frequency and growth rate, obtained by solving the eigenvalue problem, or using Davies solver, or by fitting an exponential and performing a Fourier analysis of electric field time-series in full-$f$ COBBLES simulations. In both cases, the initial distribution is distribution A, with $\gamma _L=0.1981$, and $\gamma_d=0.1$.}
%\begin{tabular}{l l c c c}
%\hline \hline 
%& &  \\[-10pt]
%Collision operator & Collision frequencies & Eigenvalue & Davies solver & COBBLES simulation \\[3pt]
%\hline 
%Krook & $\nu_a=0.25$ & $\omega=0.9953$, $\gamma =-0.04551$ & $\omega=0.9953$, $\gamma =-0.04551$ & $\omega=0.9948$, $\gamma =-0.04549$ \\[8pt]
%Diffusion+Drag & $\nu_f=0.02$, $\nu_d=0.05$ & $\omega=1.0204$, $\gamma =0.1737$ & - & $\omega=1.0176$, $\gamma=0.1743$ \\[5pt]
%\hline  \hline
%\end{tabular} 
%\label{tab:linear_check}
%\end{center} \end{table*}

\begin{table*} \begin{center}
\caption{Linear frequency and growth rate, obtained by solving the eigenvalue problem, or using Davies solver, or by fitting an exponential and performing a Fourier analysis of electric field time-series in full-$f$ COBBLES simulations. In both cases, the initial distribution is A, with $\gamma _L=0.1981$, and $\gamma_d=0.1$.}
\begin{tabular}{l c c}
&  \\[-5pt]
\hline \hline 
& &  \\[-10pt]
Collision operator & Krook & Diffusion+Drag \\[3pt]
Collision frequencies & $\nu_a=0.25$ & $\nu_f=0.02$, $\nu_d=0.05$ \\[3pt]
\hline \\[-5pt]
Eigenvalue & $\omega=0.9953$, $\gamma =-0.04551$ & $\omega=1.0204$, $\gamma =0.1737$   \\[8pt]
Davies solver & $\omega=0.9953$, $\gamma =-0.04551$ & $\ldots$ \\[8pt]
COBBLES simulation & $\omega=0.9948$, $\gamma =-0.04549$ & $\omega=1.0176$, $\gamma=0.1743$ \\[5pt]
\hline  \hline
\end{tabular} 
\label{tab:linear_check}
\end{center} \end{table*}

\subsection{Collisionless closed system\texorpdfstring{ ($\gamma _d$ $=$ $\nu _{a,f,d}$ $=$ $0$)}{}}

Our purpose is to test nonlinear capabilities of COBBLES. Let us consider the simpler collisionless Vlasov-Poisson model without external damping, corresponding to the BB model without any collision nor extrinsic dissipation.
In the unstable case, linear growth goes on until a significant number of resonant particle trajectories are modified by electrostatic trapping. In the nonlinear phase, the distribution develops an island structure in phase-space, and becomes flat on average in the resonant velocity region. The instability saturates and linear theory breaks down. As a measure of the electric field amplitude $E_0$, we use $\omega _b$, the bounce frequency of particles that are deeply trapped in the electrostatic potential, which is defined by $\omega _b ^2 \equiv k E_0$. O'Neil extended the theory of collisionless wave-particle interaction in the nonlinear phase \cite{oneil65}, within the assumptions $\gamma _L / \omega _b \ll 1$ and $\omega / \omega _b \gg 1$. He obtained an analytic estimation of the evolution of wave amplitude. In the small-time limit, $\omega _b t \ll 1$, the electric field amplitude is estimated as
\begin{equation}
\dfrac{\omega _b(t)}{\omega _b(0)}\;=\; \exp \dfrac{\gamma _L}{\pi \omega _b} \int _0 ^1 \mathrm{d} \kappa \left( 1-\cos \dfrac{2 \omega _b t}{\kappa} \right). \label{result:oneil}
\end{equation}

\begin{figure}
\begin{center}
\includegraphics{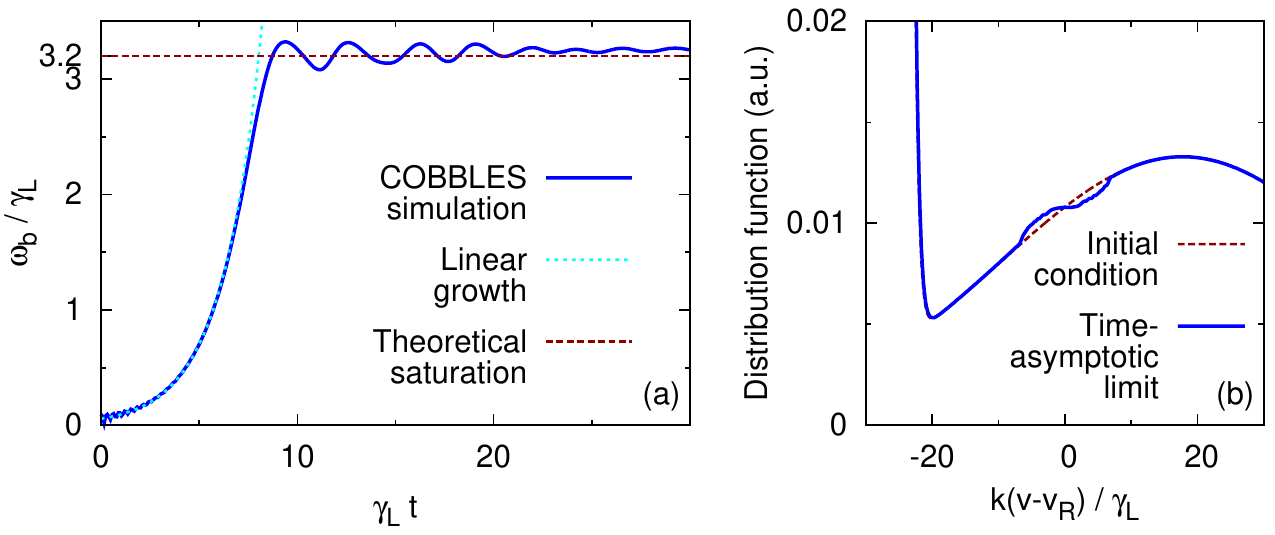}
\caption{Full-$f$ COBBLES simulation with $\gamma _d = \nu _{a,f,d} = 0$, for initial distribution B, with $N_x \times N_v = 256 \times 2048$ grid points. (a) Nonlinear evolution of normalized bounce frequency. (b) Snapshots of distribution function.}
\label{fig:oneil}
\end{center}
\end{figure}

Fig.~\ref{fig:oneil} shows the evolution of normalized bounce frequency $\omega _b / \gamma _L$, along with snapshots of the distribution function, for initial distribution B. We recover the linear growth rate obtained from Davies solver within $1\%$ error. The nonlinear evolution of the wave is in qualitative agreement with the analytic estimation (\ref{result:oneil}) in its validity limit (for the first few amplitude oscillations). Although it is impossible to quantitatively compare all the features of this analytic solution because of an ambiguity in the initial time in Eq.(\ref{result:oneil}), we observe a good agreement for the amplitude oscillations frequency, and for the relative amplitude of these oscillations compared to the saturation level. Furthermore, the saturation level is close to the value $\omega _b / \gamma _L \sim 3.2$, which was numerically obtained in Refs.~\cite{cary93,oneil71} with the $\delta f$ BB model.

\begin{figure}
\begin{center}
\includegraphics{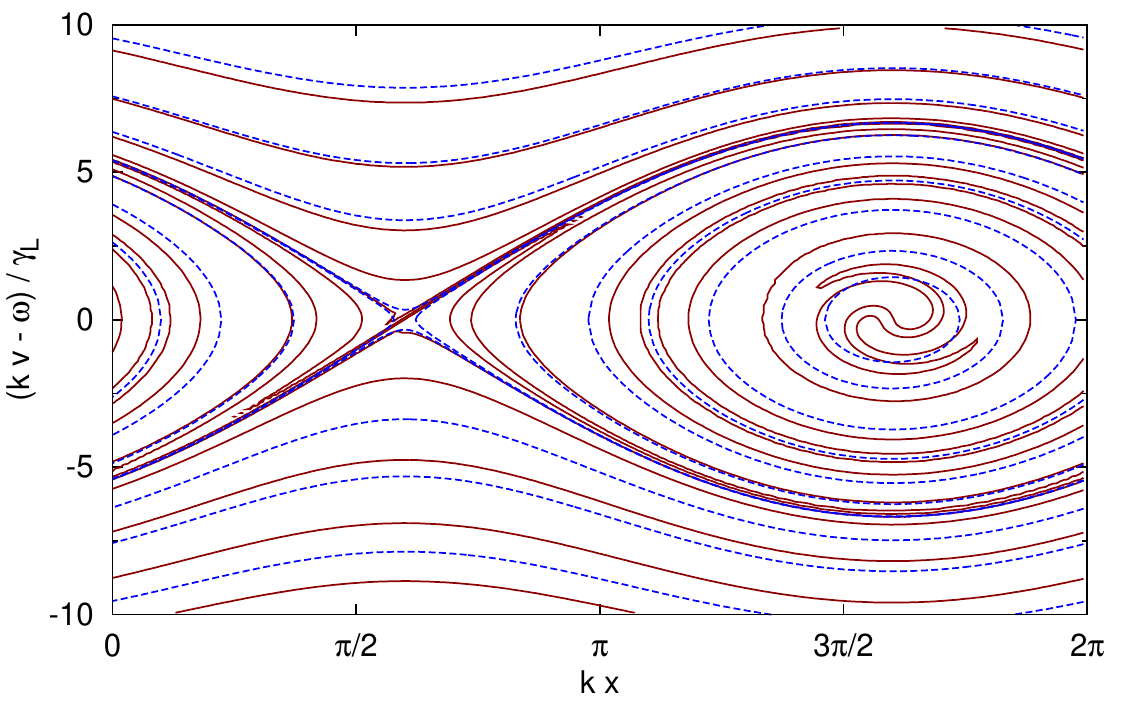}
\caption{Contour plot of the time-asymptotic distribution function (solid curves) and constant energy curves (dashed curves) for the simulation of Fig.~\ref{fig:oneil}.}
\label{fig:bgk}
\end{center}
\end{figure}

In the time-asymptotic limit, assuming some infinitesimal amount of collision, the steady-state solution of the Vlasov-Maxwell system is a distribution given as a function of the energy only. This BGK solution is consistent with a non-zero electric field. Fig.~\ref{fig:bgk} is a contour plot of the distribution function in the time-asymptotic limit of numerical simulation, on which several constant energy curves are superposed. We clearly observe an island structure, which agrees with the BGK solution. This island is topologically different from the initial condition, thus some collisions are needed to violate Liouville's theorem and obtain the BGK solution. In numerical simulations, finite numerical dissipation, which is due to interpolation on a discretized grid, smears out fine-scale structures near the separatrix, allowing a reconnection of contour lines of $f$.

\subsection{Conserved quantities}

\begin{figure}
\begin{center}
\includegraphics{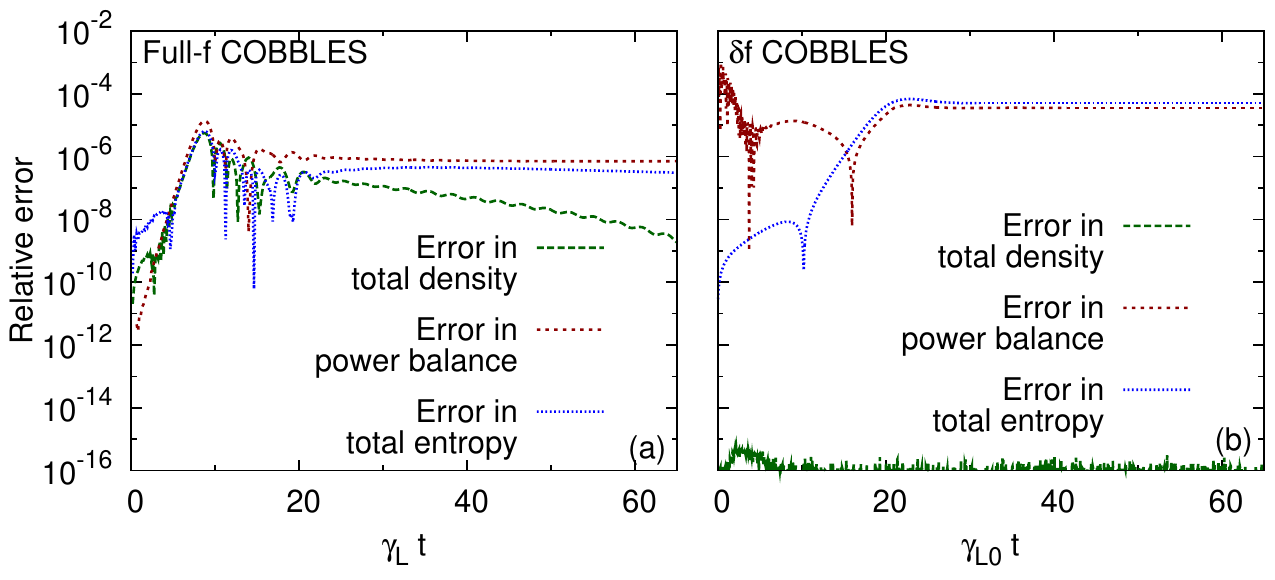}
\caption{Conservation properties of COBBLES. Time-evolution of the relative error in total particle number calculated with from $f$, in total particle number calculated from $\rho$ ("total density"), in power balance, and in total entropy, for steady-state simulations. (a) Full-$f$ simulation with initial distribution B, $\nu _a(v>v_{\nu})=0.002$, $\gamma _d=0.002$, $N_x \times N_v = 64 \times 512$ grids, and $\mathrm{CFL}=0.9$. (b) $\delta f$ simulation with $\gamma_{L0}=0.1$, $\gamma_d=0.05$, $\nu_a=0.05$, $N_x \times N_v = 64 \times 512$ grids, and $\mathrm{CFL}=1.1$.}
\label{fig:conservation_properties}
\end{center}
\end{figure}

Total particle number in simulations is calculated by taking the sum over the computation domain of the integrated value of the electronic distribution, $N(t_n)=\sum _{i,j} \rho ^n _{i,j}$. When $\nu _a$ is a constant, the relative error in particle conservation is, as expected from a locally conservative scheme, of the order of machine precision (We are working with 64 bits double-precision variables, which use 8 bits for the exponent and 56 bits for the precision, so that $2^{56} \sim 10^{16}$ is the minimum numerical error). Even when $\nu _a (v)$ has the velocity dependence of the equation (\ref{model:nua}), the relative error is negligible ($<10^{-9}\,\%$), as shown in Fig.~\ref{fig:conservation_properties}(a).

\begin{figure}
\begin{center}
\includegraphics{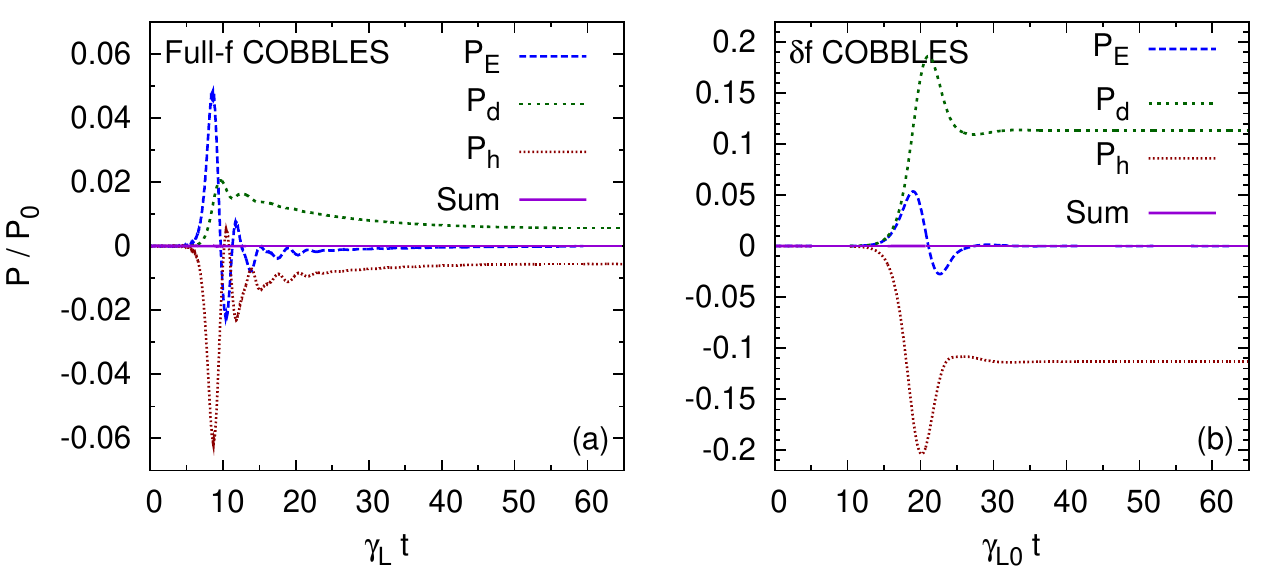}
\caption{Time-evolution of the power balance. The power transfer is normalized to $P_0 \equiv \pi v_R \gamma _L^4 / k^2$. $P_E$, $P_d$ and $P_h$ are the electric field, dissipative, and particle power transfers, respectively. The solid curve is the sum of these three power transfers. (a) Power balance in a full-$f$ simulation with same parameters as Fig.~\ref{fig:conservation_properties}(a). (b) Power balance in a $\delta f$ simulation with same parameters as Fig.~\ref{fig:conservation_properties}(b).}
\label{fig:powerBalance}
\end{center}
\end{figure}

In both cases, numerical simulations show good fidelity to the power balance, even for a relatively small number of grid points. The relative error in power balance, $\left| P_E + P_d + P_h \right| / (|P_E|+|P_d|+|P_h|)$, is included in Fig.~\ref{fig:conservation_properties}. A direct comparison between $\delta f$ and full-$f$ is not really meaningful, since simulation parameters, and definitions of $P_E$ and $P_d$ are different. Fig.~\ref{fig:powerBalance} illustrates how the different power transfers (normalized to $P_0 \equiv \pi v_R \gamma _L^4 / k^2$) compensate with each others.

%Note that, since the example we took in the $\delta f$ case is not a steady-state solution, $|P_E|+|P_d|+|P_h|$ vanishes at many times, thus we redefine the relative error in power balance as $\left| P_E + P_d + P_h \right|/\left< |P_E|+|P_d|+|P_h| \right>_t$, where $\left< \right>_t$ is a time average on $\Delta t=50$.

Entropy can be seen as a measure of numerical dissipation, since it grows as small structures dissipate. In full-$f$ simulations we define entropy as a sum over grid points of $f \log f$, and we check that $f$ is always strictly positive. In $\delta f$ simulations, we arbitrarily assume a bulk distribution as $f^{M}= 2 |f^B_\mathrm{min}|$ in the resonant region, where $f^B_\mathrm{min}$ is the minimum of $f^B_0$. The factor $2$ is to ensure that $f=f^{M}+f^B$ does not take negative values because of perturbations in $f^B$ near its minimum. The relative error in the total entropy is included in Fig.~\ref{fig:conservation_properties}.

%\subsection{Convergence properties}
%
%\begin{itemize}
%\item Relative error in power balance or E(t) itself as a function of $N_x$, $N_v$ and $dt$.
%\end{itemize}

\subsection{Benchmark} \label{subs:benchmark}

We consider five kinds of behavior for the time-evolution of the instability in the Krook case, and produce the behavior bifurcation diagram in the ($\gamma _d$,\, $\nu _a$) space. These behaviors are illustrated in Chap.~\ref{ch:nonlinearities} (Fig.~\ref{fig:nonlinear_behaviors}). The category is obtained by an analysis of the electric field energy density $\mathcal{E}(t)$ and of the spectrogram of electric field. A numerical solution is defined as
\begin{enumerate}
\item Damped: if the asymptotic-time limit of $\mathcal{E}(t)$ is zero;
\item Steady-state: if the asymptotic-time limit of $\mathcal{E}(t)$ is finite;
\item Periodic: if for large enough $t$ there is a period $\tau$ for which $\mathcal{E}(t+\tau)\rightarrow \mathcal{E}(t)$;
\item Chirping: if there is a spectral component whose frequency significantly shifts in time.
\item Chaotic: if $\mathcal{E}(t)$ is bounded, but does not satisfy one of the previous conditions.
\end{enumerate}
The categories 1., 2., 3.~and 5.~are defined in the same way as Vann \cite{vann03}, and we added a new diagnosis for the characterization of chirping solutions. Each numerical solution is systematically categorized by an algorithm based on a decision tree which is based on the one developed by Vann. We describe this algorithm in Appendix \ref{app:categorization_algo}.

\begin{figure}
\begin{center}
\includegraphics{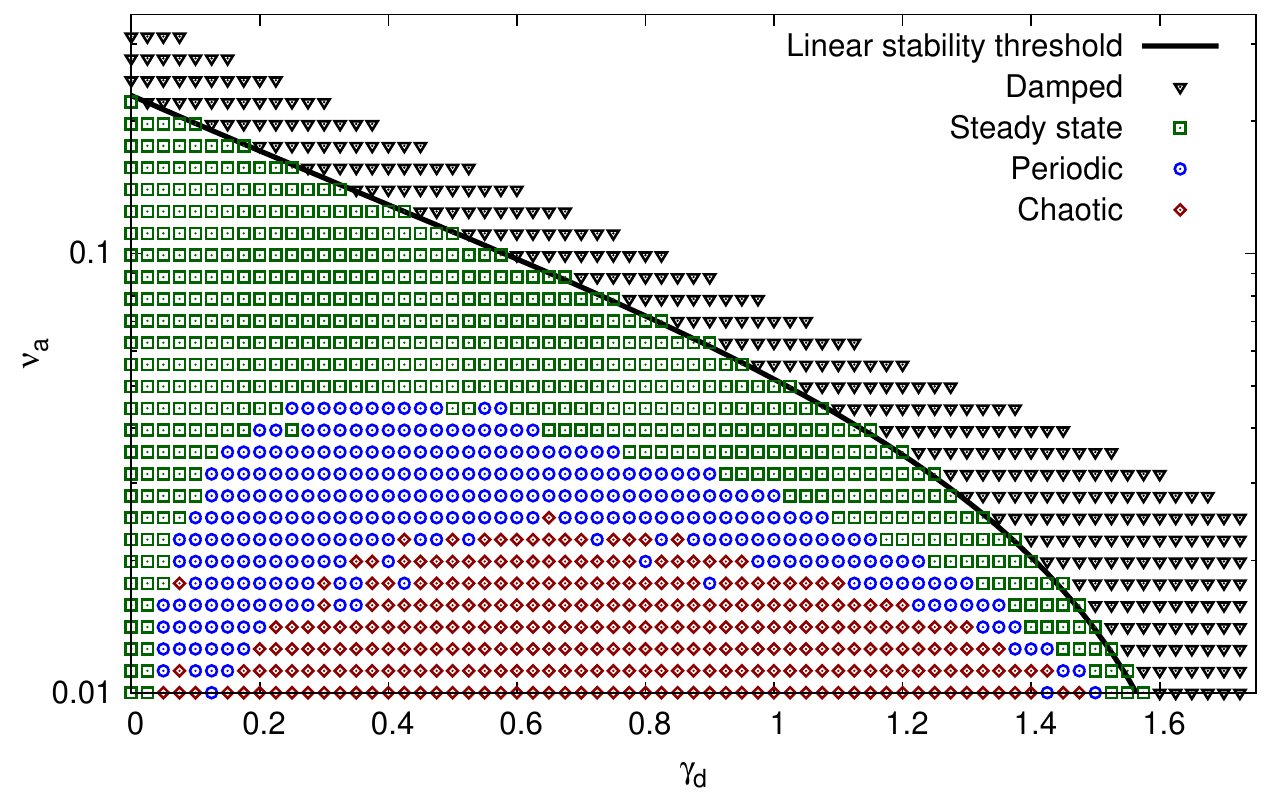}
\caption{Behavior bifurcation diagram. The classification of each solution is plotted in the ($\gamma _d$,\, $\nu _a$) parameter space. The solid curve is the linear stability threshold obtained by solving the linear dispersion relation numerically. The parameters of these simulations are $N_x \times N_v = 64 \times 512$ and $\mathrm{CFL}=1.2$.}
\label{fig:bifurcation01981}
\end{center}
\end{figure}

As a benchmark of both COBBLES code and our categorization algorithm, we reproduce results presented in Fig.~3.~of Ref.~\cite{vann03} (Note that our definition of $\gamma _d$ is consistent with Berk and Breizman's literature, and differs from Vann's article by a factor 2). The initial distribution is A, for which $\gamma _L = 0.1981$. The field energy of the initial perturbation is $2\times 10^{-4}$ of the total energy, which corresponds to $\omega _b / \gamma _L = 0.05$ at $t=0$. We perform a series of simulations in the parameter space
%, $\left\lbrace (\gamma _d , \nu _a) \;|\; 0\leq \gamma _d / \omega _0 < 1.75,\, 0.01\leq \nu _a / \omega _0 \leq 0.35\,e^{-1.5\gamma _d} \right\rbrace$
$(\gamma _d ,\, \nu _a)$
, where $\nu _a$ is chosen as velocity-independent. We set the time-duration of each simulation to $t_{\mathrm{max}}=3000$. In the categorization algorithm, we choose $t_{\mathrm{min}}=1000$, $\epsilon _1 = 10^{-12}$, $\epsilon _2 = 0.05$, $\epsilon _3 = 0.01$, $\epsilon _4 = 10^{-9}$, and $\epsilon _5 = 0.25$. The resulting behavior bifurcation diagram is shown in Fig.~\ref{fig:bifurcation01981}.
The 1416 simulations used for this plot took approximately 115 CPU hours on an Altix3700Bx2 array of Intel Itanium2 processors. The categorization of $92 \,\%$ of these time-series is in agreement with the reference, most of the difference coming from a different way of sorting out chaotic from periodic solutions. This result is a further indication of the validity of both COBBLES and categorization algorithms.\\

\subsection{Steady-state with drag and diffusion}

\begin{figure} \begin{center}
\includegraphics{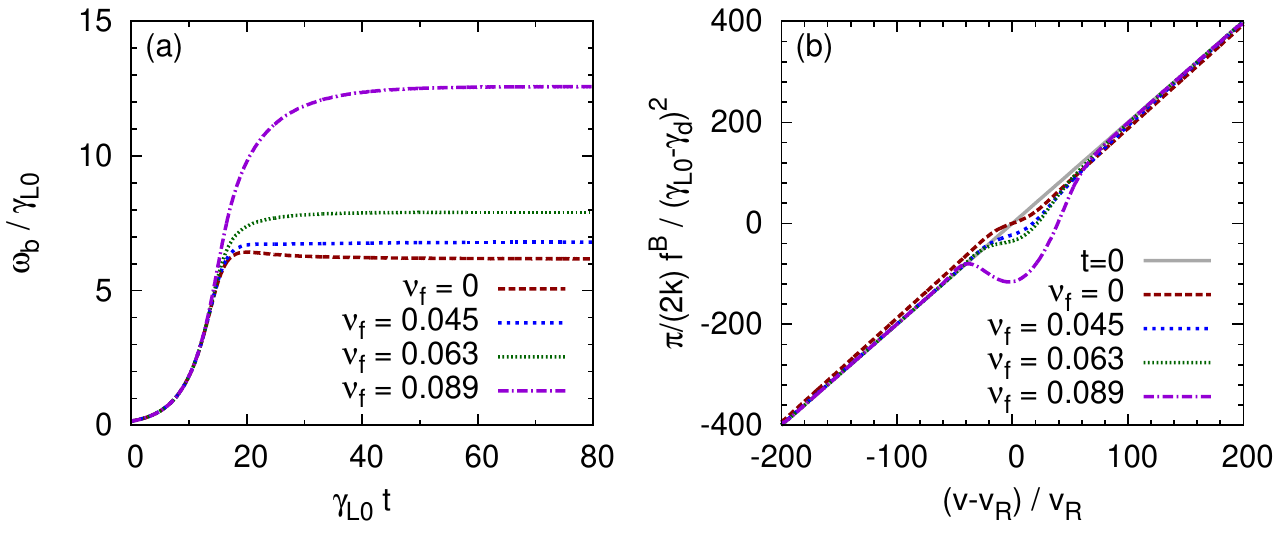}
\caption{Steady-state saturation with drag and diffusion. $\delta f$ simulation with $\gamma _{L0} = 0.02$, $\gamma _d = 0.01$, $\nu _d = 0.1$, and $\nu _f$ given in the legend. (a) Saturation level. (b) Saturated distribution function.}
\label{fig:steady_diffdrag}
\end{center} \end{figure}

To verify our implementation of collision operator with drag and diffusion, we confirmed that a Gaussian perturbation in the velocity distribution follows the analytic solution of the diffusion equation in the absence of electric field and drag, and is simply advected at a rate $\dot{v}=\nu _f^2/k$ in the absence of electric field and diffusion.\\

As an additional test, we compare nonlinear steady-state solutions between $\delta f$-COBBLES and analytic predictions derived in Ref.~\cite{lilley_thesis}. Fig.~\ref{fig:steady_diffdrag} shows steady-states in $\delta f$-COBBLES simulations with different collision frequencies. Fig.~\ref{fig:steady_diffdrag}(b) in this manuscript, and Fig.~6.13 in Ref.~\cite{lilley_thesis}, which share the same normalization, can be directly compared. We confirm quantitative agreement with theory.

%[Work needed: recover and include analytic theory in Fig.~\ref{fig:steady_diffdrag}]

\subsection{Multiple-modes interaction}

\begin{figure}
\begin{center}
\includegraphics{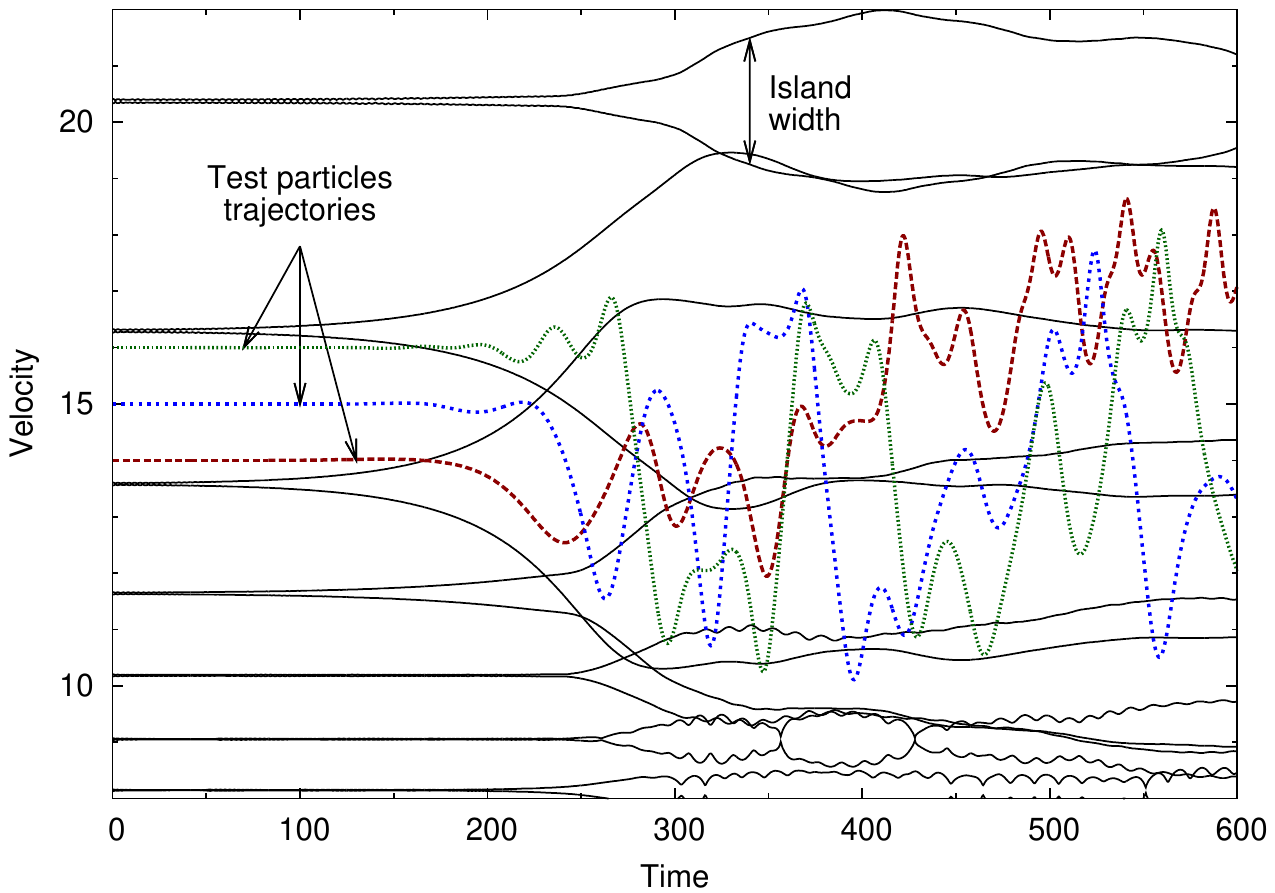}
\caption{Evolution and overlapping of multiple islands. Dashed and dotted lines are trajectories of three test-particles.}
\label{fig:islands_width}
\end{center}
\end{figure}

Though this feature is not used in this thesis, we also test multiple-waves capabilities of COBBLES.
When many electrostatic waves are excited, the amplitude of each wave grows exponentially until nonlinear saturation occurs, and each wave develops an island structure in phase-space. If the width of each island is much smaller than the distance between the phase velocities of two neighbouring waves, we can treat the problem as a superposition of the former single wave-particle problem. However, if island structures overlap each others, particle trajectories are not integrable. We consider a situation where there exists a velocity interval within which the phase velocities of many waves are close enough and their islands overlap.
We perform a full-$f$ COBBLES simulation, without collisions nor external damping, with $n_B=0.05$, $v_{TM}=v_{TB}=4.0$, $v_B=16.0$, $v_\mathrm{max}=-v_\mathrm{min}=30$, $L=512$, $N_x \times N_v = 512 \times 64$, initializing 10 waves with wave numbers $k_m\equiv 2\pi m /L$ and a random phase for each wave $m$. Fig.~\ref{fig:islands_width} shows the position and width of each island, and trajectories in the velocity direction of three test particles evolving within the resonant region. We observe overlapping of islands, and resonant particles seem to undergo Brownian motion in the velocity direction.

When particle diffusion time is much longer than correlation time, quasi-linear theory \cite{sagdeevgaleev} predicts velocity diffusion of particles in the resonant region, leading to a flattening of the distribution, as we observe in numerical simulations. In the resonant region, the quasi-linear diffusion coefficient $D_\mathrm{QL}$ can be estimated as
\begin{equation}
D_\mathrm{QL} \;=\; \dfrac{\pi \, \sum _m  \left| E _{k_m} \right| ^2 /{k_m}}{\Delta v_R},
\end{equation}
where $E _{k_m}$ is the Fourier component for the wave number $k_m$ of the electric field, and $\Delta v_R$ is the width of the whole resonant region.

Another way of estimating the diffusion coefficient involves the variance of the displacement in velocity of a large number of test particles. For any time interval $\Delta t$ larger than the correlation time, but smaller than the distribution relaxation time, this estimated coefficient $D_\mathrm{P}$ is given by
\begin{equation}
D_\mathrm{P} \;=\; \dfrac{\left< \left[  v(t_0 + \Delta t) - v(t_0)  \right]^2 \right> }{2 \Delta t},
\end{equation}
where angle brackets represent an ensemble average.

\begin{figure}
\begin{center}
\includegraphics{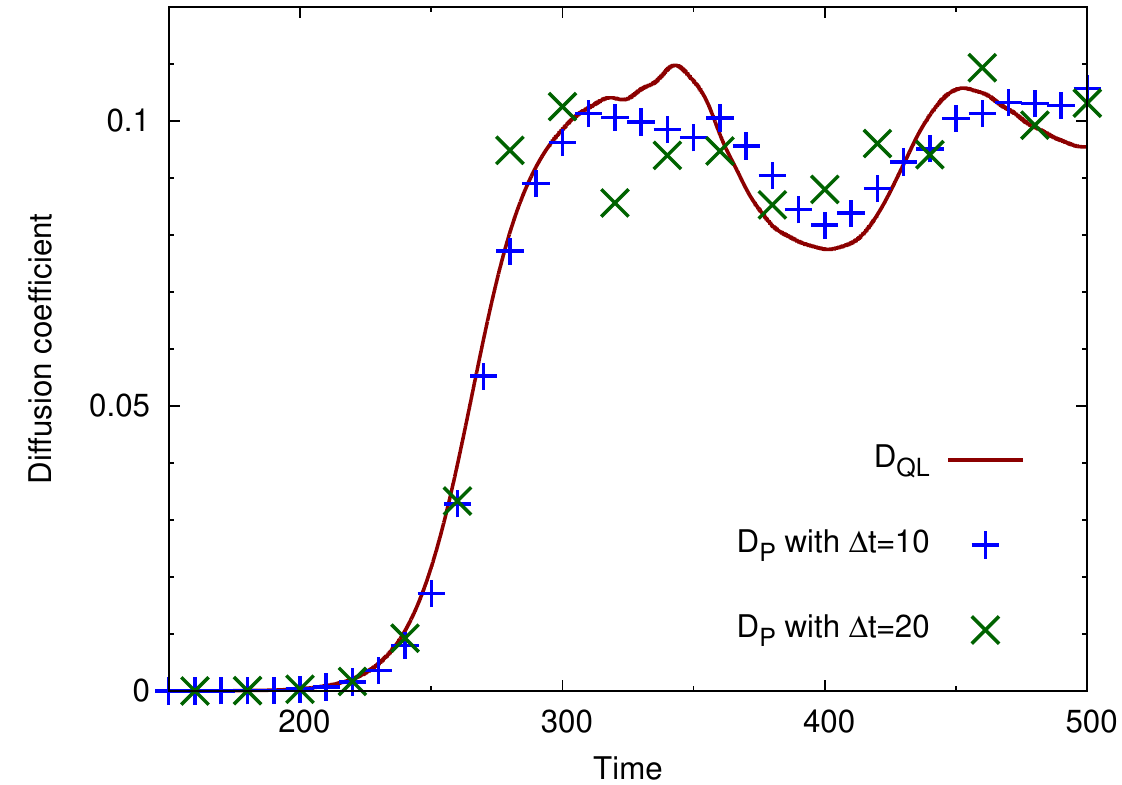}
\caption{Time-evolution of the quasi-linear diffusion coefficient estimated by quasi-linear theory (solid line), and by following test particles trajectories (crosses).}
\label{fig:diffusion_coeff}
\end{center}
\end{figure}

In our simulation, we estimate $D_\mathrm{P}$ by following the trajectories of $3\times 10^5$ test particles, which are initialized with an uniform distribution over the resonant region, and a random position. Fig.~\ref{fig:diffusion_coeff} shows that $D_\mathrm{QL}$ matches $D_\mathrm{P}$ in simulation, even as we double $\Delta t$.

%\begin{itemize}
%\item Validity ? See [Y.~Kominis, A.~K.~Ram and K.~Hizanidis, PRL 104, 235001 (2010)]
%\item Typical trajectory of a test particle for trapped, passing and near separatrix.
%\item $2\pi$-time-advance mapping of a particle trajectory
%\item Diffusion coefficient: Comparison of quasilinear estimation and test particle estimation
%\end{itemize}

\section{The BB model as a paradigm for the TAE\label{sec:analogies}}

The BB problem can be put in Hamiltonian form with the Hamiltonian given in Eq.~(\ref{eq:bb_hamiltonian}). Let us make a canonical transformation with the generating function $S\equiv v_\mathrm{R} \left( x - v_\mathrm{R} t / 2 \right)$ to a moving-frame coordinate set, ($\psi$, $I$), where $\psi\equiv kx-\omega t$ and $I\equiv (v-v_\mathrm{R})/k$. The new Hamiltonian,
\begin{equation}
h_\mathrm{eff} \;=\; h \;-\; I\,\omega \;-\; \dfrac{\omega ^2}{2k^2} \;=\; \dfrac{k^2}{2} I^2 \;+\; \dfrac{\omega _b ^2}{k^2} \,\cos \psi,
\end{equation}
takes a standard form, which is shared with the effective Hamiltonian of the TAE, Eq.~(\ref{eq:TAE1DHamiltonian}). Therefore, the BB problem is a simple 1D model that is homothetic to a whole class of instabilities, including EP-driven TAEs.

\begin{figure}
\begin{center}
\includegraphics[scale=0.6,angle=-90]{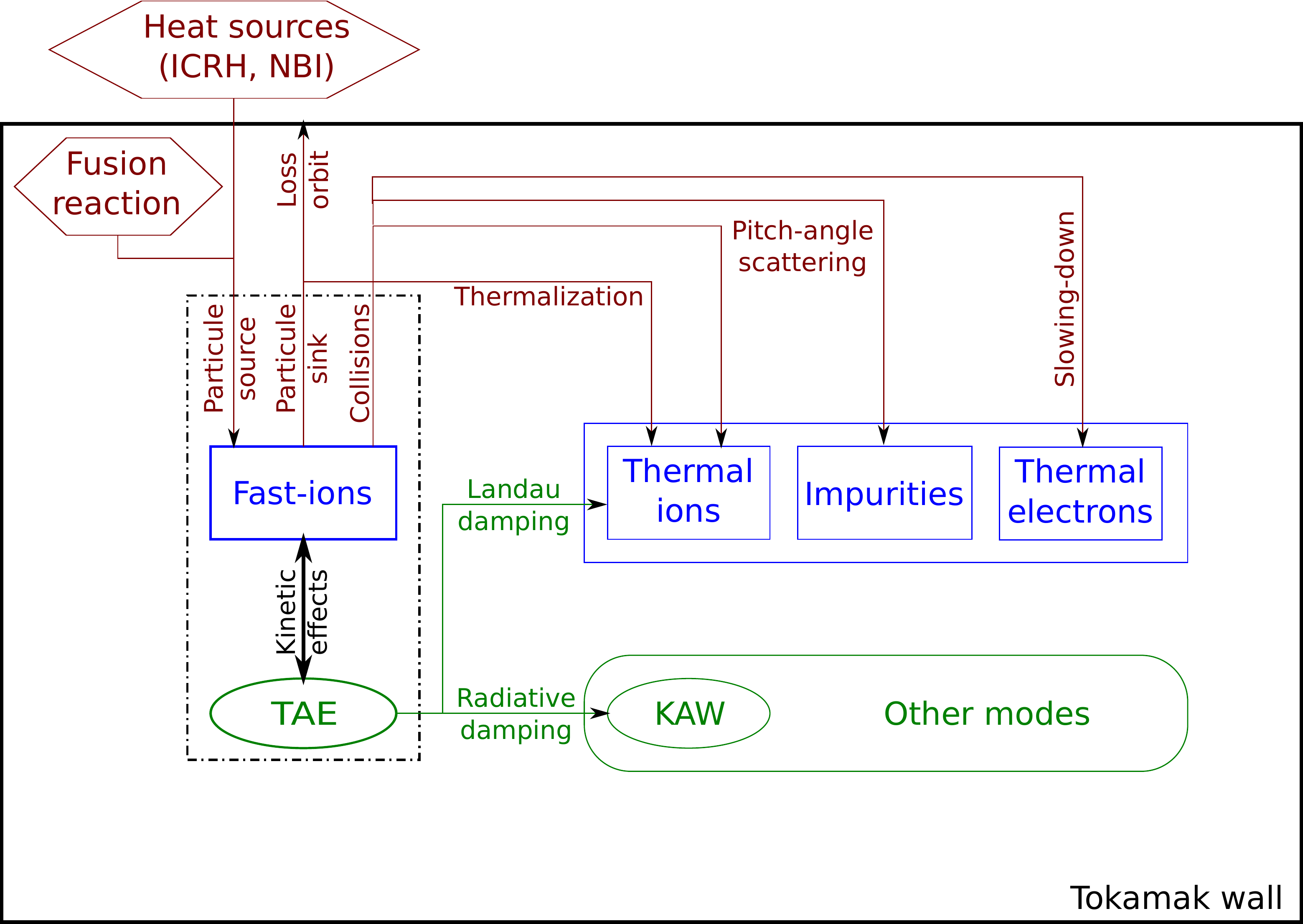}
\caption{Schematic representation of wave-particle interactions relevant to the TAE. The dotted-dashed rectangle represents the limit of the BB representation of TAEs.}
\label{fig:bbmodel}
\end{center}
\end{figure}

Fig.~\ref{fig:bbmodel} is a schematic representation of wave-particle interactions relevant to TAEs. A dotted-dashed rectangle represents the physics encompassed by the BB representation of TAEs. All physics outside this rectangle are treated as input parameters in the BB model.

Table \ref{tab:analogies} summarizes the parallels between the BB model and the reduced model for the TAE.
\begin{table*} \begin{center}
\caption{Analogies between the BB model and the reduced model for the TAE.}
\begin{tabular}{c c c}
&  \\[-5pt]
\hline \hline
Harmonic oscillator & BB model & 1D model for TAEs  \\[3pt]
\hline
$\stackp{Hamiltonian}{\frac{1}{2}D\,I^2 \,+\, V\,\cos{\psi}}$ & $\stackp{Hamiltonian}{\frac{1}{2} (v-v_\mathrm{R})^2 \,+\, \varphi(x,t)}$ & $\stackp{Hamiltonian}{\frac{1}{2}D\,I^2 \,+\, V\,\cos{\psi}}$ \\[10pt]
$\stackp{Action}{I}$ & $\stackp{Velocity\; in\; the\; wave\; frame}{I(v) = (v-v_\mathrm{R})/k}$ & $\stackp{Deviation\; from\; resonant\; surface}{I = e (\chi _\mathrm{R}-\chi)/n}$ \\[10pt]
$\stackp{Angle}{\psi}$ & $\stackp{Position\; in\; the\; wave\; frame}{\psi(x,t) = kx-\omega t}$ & $\psi = p \widetilde{\theta} \,+\, n \widetilde{\zeta} \,-\, \omega t$ \\[10pt]
$\stackp{Effective\; mass}{D}$ & $D=k^2$ & $D\approx n^2\left| \frac{\partial ^2 H_0}{\partial \chi ^2} \right|$ \\[10pt]
$\stackp{Oscillations\; amplitude}{V}$ & $\stackp{Electric\; field\; amplitude}{V=\omega _b ^2 / k^2}$ & $\stackp{Magnetic\; field\; amplitude}{V=-\imath e \, C(t) V _p (J_{\theta \mathrm{R}}, J_{\zeta \mathrm{R}}) \sim \delta \boldsymbol{A_\perp}}$ \\[10pt]
\hline  \hline
\end{tabular} 
\label{tab:analogies}
\end{center} \end{table*}

\chapter{Kinetic nonlinearities}
\label{ch:nonlinearities}

To different nonlinear behaviors of the BB model correspond different regimes of resonant energy transfer. In terms of alpha-particle issues, there is an interplay between two effects. On the one hand, energy transfer from particles to the wave, which is supported by the thermal plasma, is favorable since the energy from fusion reactions must be channeled to heat the bulk plasma. On the other hand, a large wave amplitude is unfavorable, since it is associated with transport and energetic-particle ejection.
A survey of kinetic nonlinearities provides important insight into the optimum regime.
%In the previous chapter, we have seen that complex nonlinear behaviors emerge from the BB model in the unstable case.
% State of research and general goal
%In this chapter, we investigate nonlinear interactions in various regimes.

In Sec.~\ref{sec:nonlinear_regimes}, we perform a systematic parameter scan in ($\gamma _d$, $\nu _a$) in the full-$f$, Krook case. We confirm the validity of available theory, and show limits of $\delta f$ approach.
In Sec.~\ref{sec:chirping}, we investigate nonlinear chirping features, since they can provide precious information about the state of the plasma. Existing quantitative predictions of these features are verified for both collision operators, and we extend theory by including the effects of beam distribution shape, finite collision frequency, and drag.
In Sec.~\ref{sec:subcritical}, we investigate instabilities that arise in a regime where linear theory predicts wave damping, provided that initial perturbation is large enough. We propose a mechanism to explain the apparent contradiction between linear theory and the behavior of subcritical instabilities observed in simulations, and perform a numerical investigation of an initial amplitude threshold.

\section{Nonlinear regimes\label{sec:nonlinear_regimes}}

Theories \cite{berkbreizman90,berkbreizman96,berkbreizman97pla,berkbreizman98erratum,berkbreizman99} have been developed by Berk, Breizman, and coworkers, to quantitavely predict nonlinear behaviors in various parameter regimes and to explain underlying mechanisms. 
On the one hand, some of these theories have been validated by numerical simulations based on the $\delta f$ model. A concern with this perturbative approach is that, as instability grows, the resonant region may expand and ultimately include a significant portion of bulk particles. An other concern is that, when chirping occurs, corresponding resonant velocity may propagate into the bulk. In such situations, kinetic effects of bulk plasma should also be taken into account.
On the other hand, full-$f$ simulations have been performed by Vann and coworkers \cite{vann03}. However, as we show in \ref{subs:schemes}, this approach shows some difficulty in simulating situations considered in the aforementioned theories, which assume a plasma near marginal stability with a cold bulk and a weak beam. In fact, these theories have not been validated with this approach.
Filling the gap between these two fronts of the current state of research, with quantitative comparisons between available theory and full-$f$ model, is the aim of this section. We investigate the validity of analytic theories for the following nonlinear features,
\begin{itemize}
\item saturation level in a parameter regime above marginal stability;
\item saturation level and bifurcation criterion between steady-state and periodic solutions near marginal stability;
\item time-evolution of a frequency shifting mode.
\end{itemize}
We choose initial bump-on-tail distribution B, which was introduced in \ref{subs:ff_bb}. The parameters of distribution B were actually chosen so that we stay within the validity limit of these theories, with sufficiently cold bulk and sufficiently weak warm beam. We recall these parameters as $n_B = 0.1$, $v_{TP} = 0.2$, $v_{TB} = 3.0$, $v_B = 5.0$, which give, for $k=0.3$, $\gamma _L = 0.0324$ and $\omega = 0.925$. As mentioned before, $\nu _a$ is a function of the velocity such that collisions affect only the beam particles. The field energy of initial perturbation is $2\times 10^{-8}$ of total energy, or $\omega _b / \gamma _L = 0.3$.

\subsection{Nonlinear saturation}

\begin{figure} \begin{center}
\includegraphics{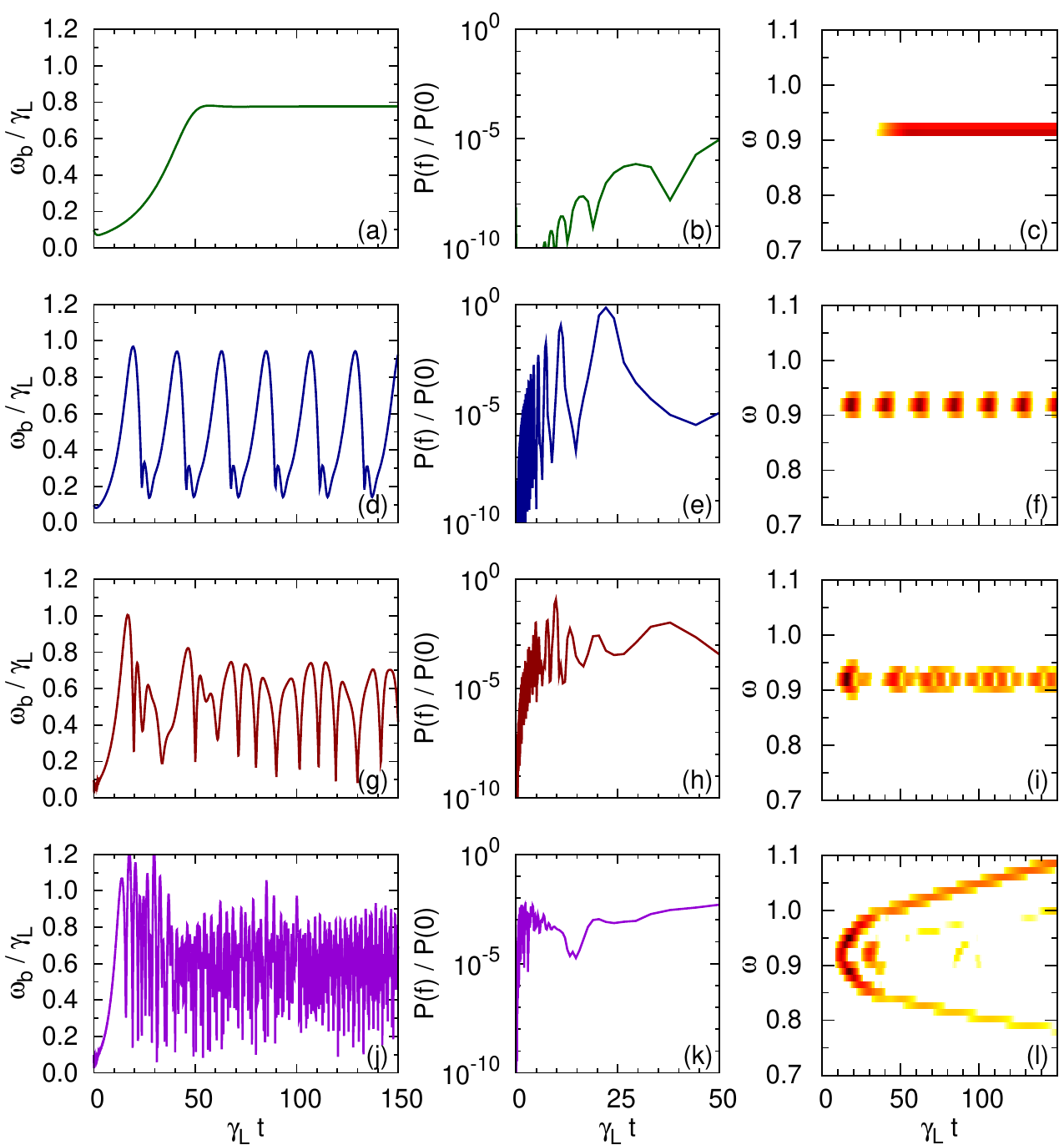}
\caption{Typical nonlinear behavior in steady-state (a-c), periodic (d-f), chaotic (g-i) and chirping (j-l) cases. Full-$f$ simulations with initial distribution B, $\gamma _d=0.03$, and values for $\nu _a$ are $0.02$, $0.008$, $0.005$, and $0.00002$, respectively. Left: Time-evolution of electric field amplitude. Center: Fourier spectrum of electric field amplitude after $t=3000$, with a time-window $
\Delta t=10^5$. Right: Spectrogram of electric field at $x=0$, with a moving Fourier-window of width $\Delta t=400$.}
\label{fig:nonlinear_behaviors}
\end{center} \end{figure}

Fig.~\ref{fig:nonlinear_behaviors} shows four examples of nonlinear saturation in the unstable case, corresponding to steady-state, periodic, chaotic, and chirping behaviors, obtained by varying $\nu _a$ at fixed $\gamma _d$.
Spectrums are obtained by applying Fast Fourier Transform to time-series, which are filtered by a Hann window \cite{NumericalRecipes}. Included are both spectrum of electric field amplitude, where we consider only times after nonlinear saturation, and spectrogram of the electric field measured at some arbitrary point in configuration space.

%[Redaction needed: physical interpretation]

\subsection{Scan in the (\texorpdfstring{$\gamma _d$, $\nu _a$}{external damping rate, collision frequency}) space} \label{subs:bif00327}

\begin{figure}
\begin{center}
\includegraphics{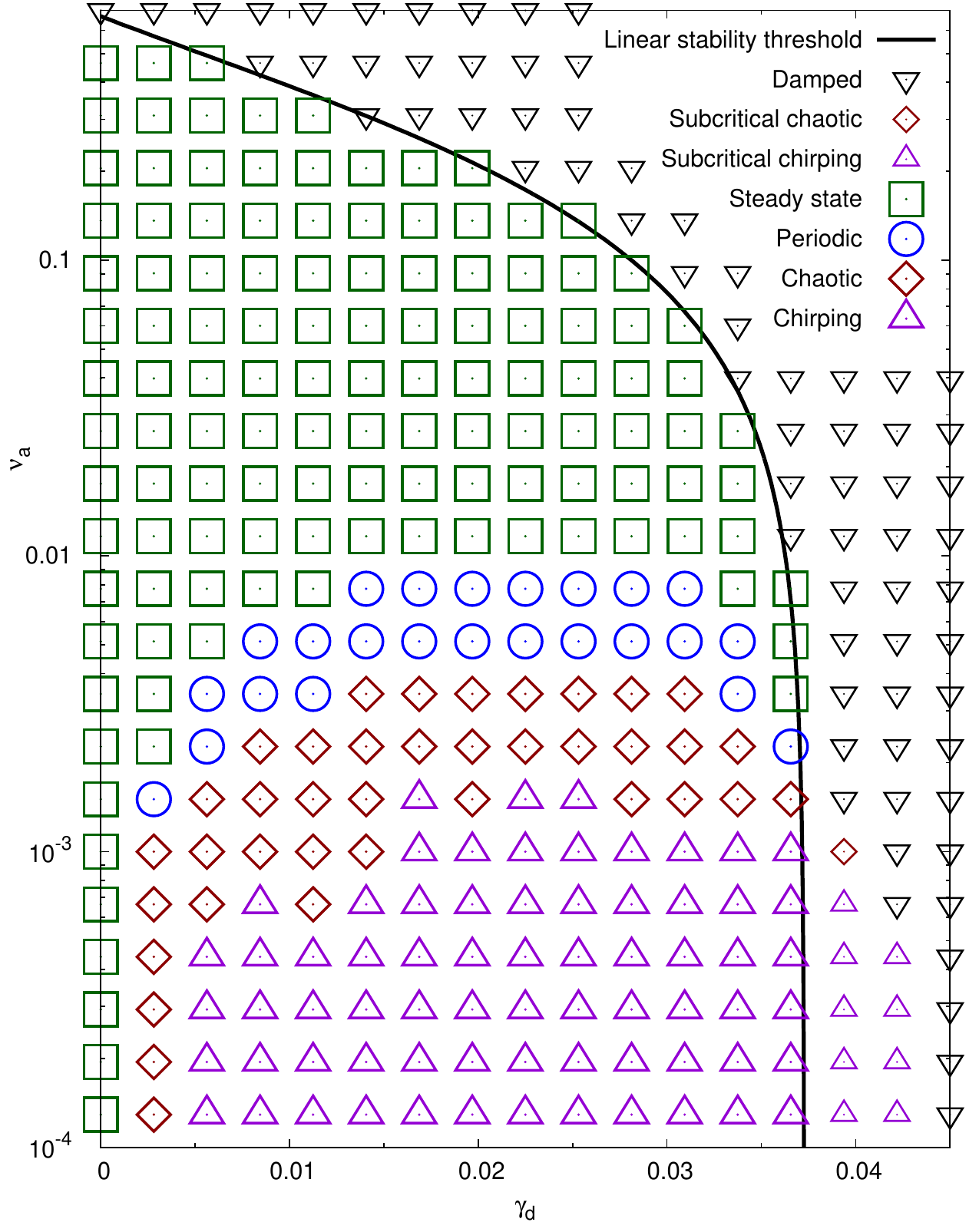}
\caption{Behavior bifurcation diagram for a cold bulk, weak warm beam distribution. The classification of each solution is plotted in the ($\gamma _d$,\, $\nu _a$) parameter space. The solid curve is the linear stability threshold obtained by Davies solver. The parameters of these full-$f$ simulations are $\mathrm{CFL}=3.0$, $v_{\mathrm{min}}= -10$, $v_{\mathrm{max}}=18$, $N_v=2048$ and $N_x$ ranges from $128$ to $256$. Smaller diamonds and triangles on the right of the linear stability threshold represent subcritical instabilities.}
\label{fig:bifurcation00327}
\end{center}
\end{figure}

For the benchmark in \ref{subs:benchmark}, we set a same value for the maximum time of every numerical simulations. However, we must now take into account computational cost, which is much larger because of a decrease of $\gamma _L$ by one order of magnitude. As we approach marginal stability, the time window must be increasingly large to successfully capture the nonlinear behavior. To reduce computational cost, we choose a time-window size as a function of $\gamma$, as
\begin{equation}
t_{\mathrm{max}}  \;=\;  20\,\dfrac{2\pi}{| \gamma |}.
\end{equation}
The frequency of amplitude oscillations is of the order of $\omega _b$, which is empirically of the order of $\gamma$ after the transient phase, so that such time windows contain at least a few amplitude oscillations, enough to sort steady-state, periodic and chaotic responses.
In the categorization algorithm described in App.~\ref{app:categorization_algo}, we choose $t_{\mathrm{min}}=t_{\mathrm{max}}/2$, and each time series is sampled every $\Delta t_s$ $=$ $20$. $\epsilon _0 = 10^{-14}$ is chosen as a free-streaming criterion, $\epsilon _6 = 0.05$, $\epsilon _7 = 0.05$ are used to sort out chirping solutions, and the other $\epsilon$-thresholds are the same as above ($\epsilon _1 = 10^{-12}$, $\epsilon _2 = 0.05$, $\epsilon _3 = 0.01$, $\epsilon _4 = 10^{-9}$, and $\epsilon _5 = 0.25$).
The behavior of wave amplitude time-series obtained by full-$f$ COBBLES is characterized in Fig.~\ref{fig:bifurcation00327}.
The 391 simulations used for this plot required approximately 15000 CPU hours on an Altix3700Bx2 array of Intel Itanium2 processors.

Agreement between linear stability threshold and the boundary between linearly stable and unstable simulations confirms that the problem of recurrence is taken care of by the free streaming test in our categorization algorithm.
When $\gamma _d \ll \gamma _L$ and $\nu _a \ll \gamma _d$, a bursty behavior, characterized by a succession of bursts with characteristic growth and decay rates of $\gamma _L$ and $\gamma _d$, respectively, and with a quiescent phase in between that lasts a time $1/\nu _a$, as described in \cite{berkbreizman92}, is expected. A few solutions in the chaotic region appear to follow this picture. However, most of the chaotic solutions do not feature a significantly quiescent phase. Consequently, an attempt at sorting out a pulsating regime from the chaotic region seems vain.
For small collision rates, we observe instabilities in the linearly stable region, which suggests the possibility of subcritical instabilities. This effect is discussed in Sec.~\ref{sec:subcritical}. The chirping regime is discussed in details in Sec.~\ref{sec:chirping}. The physics of several other regions of this diagram is discussed in the remaining of this section.

%[Figure needed: Plot of the long-time spectrum vs $\nu _a$ to illustrate bifurcations]

\subsection{\label{subs:aboveMargStab}Steady-state above marginal stability\texorpdfstring{ ($\gamma _d$ $\sim$ $\nu _a$ $\ll$ $\gamma _L$)}{}}

When external damping and distribution relaxation are of the same order and both are small compared to the linear drive, we expect and observe the saturation of wave amplitude to a steady-state in the time-asymptotic limit. To estimate a saturation level, we assume a rate of annihilation of beam particles much smaller than the saturated bounce frequency, $\nu _a \ll \omega _b$ at $t\rightarrow \infty$. We also assume that the resonant region is narrow compared to the resonant velocity, $4 \, \omega _b / k \ll \omega / k$, so that we can assume that the contribution to resonant power transfer comes from a narrow region around $v_R$.
Berk and Breizman derived a relation yielding the saturation level in this situation \cite{berkbreizman90},
\begin{equation}
\omega _b \;=\; 1.96 \, \dfrac{\nu _a}{\gamma _d} \, \gamma _L. \label{result:above_threshold_saturation_level}
\end{equation}
Thus, if we re-normalize all quantities to the linear growth rate, then within the aforementioned assumptions, the saturation level depends only on the ratio of $\nu _a$ to $\gamma _d$.

\begin{figure}
\begin{center}
\includegraphics{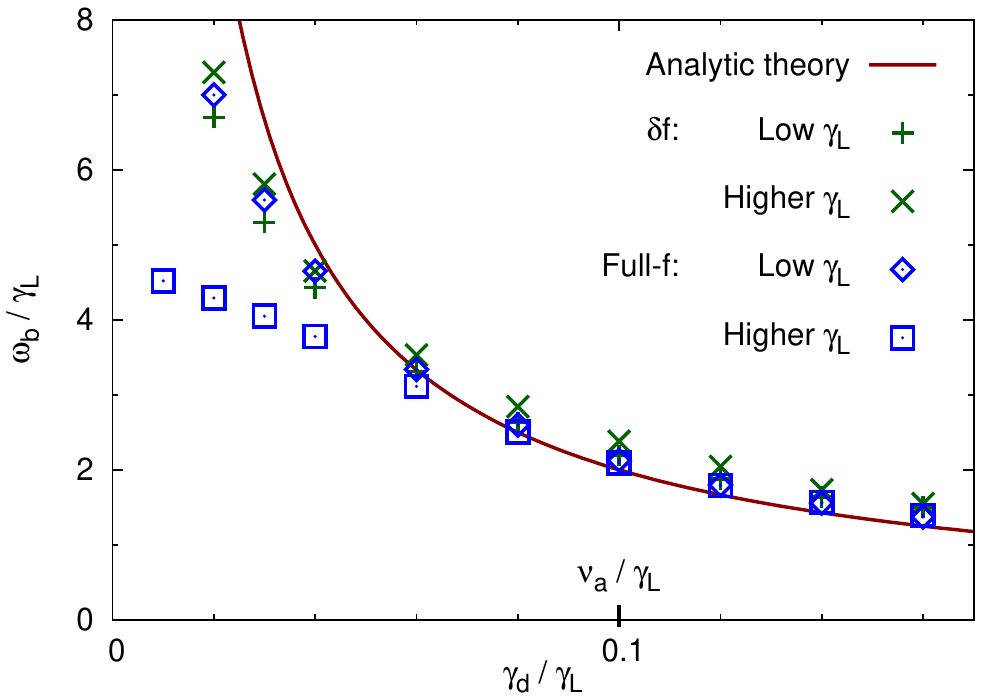}
\caption{Saturation level at a given collision frequency $\nu _a / \gamma _L = 0.1$ for distribution B (Low-$\gamma _L$ case), and for higher beam density and lower beam temperature (Higher-$\gamma _L$ case). The parameters of full-$f$ simulations are those of Fig.~\ref{fig:bifurcation00327}. Parameters of $\delta f$ simulations are such that $\gamma _L$ take the same two values.}
\label{fig:above_threshold_saturation_level}
\end{center}
\end{figure}

We investigate the validity of this theory by numerically computing the scaling law for the saturation level at a given normalized relaxation rate $\nu _a / \gamma _L = 0.1$. In a previous work \cite{berkbreizman95}, such a scan has been done using a $\delta f$ particle code, and the results showed good agreement with analytic prediction in a region where $\gamma _d \sim \nu _a$.
Fig.~\ref{fig:above_threshold_saturation_level} shows the saturation level obtained from theory, Eq.(\ref{result:above_threshold_saturation_level}), and from both $\delta f$ and full-$f$ COBBLES simulations. When the initial distribution is distribution B, we observe quantitative agreement between theory and both $\delta f$ and full-$f$ simulations in the parameter region $\gamma _d \sim \nu _a$.

To reveal some limitations of $\delta f$ model, the same computation is done for a distribution with a slightly higher beam density, $n_B = 0.15$, and a slightly lower beam temperature, $v_B = 2.5$, giving $\gamma _L = 0.067$ instead of $0.032$. When we use $\delta f$ model, the scaling law is roughly independent of $\gamma _L$ and is in agreement with theory in a parameter region $\gamma _d \sim \nu _a$, in agreement with aforementioned work. On the other hand, when we take into account the evolution of bulk plasma, we observe a significant dependency of the saturation amplitude on the linear growth rate. For larger $\gamma _L$, we find some discrepancy with theory in the low $\gamma _d$ region, because the island width $\Delta v$ becomes of the order of the resonant velocity ($\Delta v / v_R = 0.14\, \omega _b / \gamma _L$ in the low-$\gamma _L$ case, and $\Delta v / v_R = 0.29\, \omega _b / \gamma _L$ in the higher-$\gamma _L$ case). This result shows that it is necessary to take into account the effect of bulk particles to accurately discuss the validity limit of this theory.

\subsection{\label{subs:nearMargStab}Near-marginal steady-state and periodicity\texorpdfstring{ ($\gamma \,\approx\, \gamma _L -\gamma _d \,\ll\, \gamma _L$)}{}}

When $\gamma \ll \gamma _L$, a reduced integral equation for the time evolution of electric field amplitude has been developed using an extension based on the closeness to marginal stability \cite{berkbreizman96}. Within the assumption $\omega _b / \gamma \ll 1$,
%\begin{eqnarray}
%\dfrac{d\omega _b ^2}{dt} & =\; (\gamma _L ^* - \gamma _d)\, \omega _b ^2 \;-\; \dfrac{\gamma _L ^*}{2} \, \int ^{t} _{t/2} \mathrm{d}t_1 \, \int ^{t_1} _{t-t_1} \mathrm{d}t_2 (t-t_1)^2 \nonumber \\
%&\times \, e^{-\nu _a (2t-t_1-t_2)} \, \omega _b ^2 (t_1)\, \omega _b ^2 (t_2)\, \omega _b ^2 (t+t_2-t_1), \label{eq:bb_integral_equation}
%\end{eqnarray}
\begin{eqnarray}
\dfrac{d\omega _b ^2}{dt} &=& (\gamma _{L0} - \gamma _d)\, \omega _b ^2 \;-\; \dfrac{\gamma _{L0}}{2} \, \int ^{t} _{t/2} \mathrm{d}t_1 \, \int ^{t_1} _{t-t_1} \mathrm{d}t_2 (t-t_1)^2 \nonumber \\
& & \qquad \qquad \qquad e^{-\nu _a (2t-t_1-t_2)} \, \omega _b ^2 (t_1)\, \omega _b ^2 (t_2)\, \omega _b ^2 (t+t_2-t_1). \label{eq:bb_integral_equation}
\end{eqnarray}
For a cold bulk, warm beam distribution, in the collisionless limit, as we approach marginal stability, Eq.~\ref{eq:collisionless_gamma} reduces to
\begin{equation}
\gamma \;\approx\; \gamma _{L0} \;-\; \gamma _d, \label{eq:gamma_for_Cbulk_Wbeam}
\end{equation}
which agrees with the linear part of the latter integral equation (\ref{eq:bb_integral_equation}). In Ref.~\cite{berkbreizman96}, the analytic treatment is carried on by normalizing time by $\gamma _{L0} - \gamma _d$.

We observe that the relation (\ref{eq:gamma_for_Cbulk_Wbeam}) is a good approximation in most of the parameter space. However, as we get closer to the linear stability threshold, the relative error $|\gamma _{L0} - \gamma _d - \gamma|/(|\gamma _{L0} - \gamma _d|+|\gamma|)$ approaches unity for finite collisions. In addition, for our choice of distribution, there is a $14\%$ discrepancy of $\gamma _{L0} = 0.0368$ compared to $\gamma _L = 0.0324$. We infer that we can replace $\gamma _{L0} - \gamma _d$ by $\gamma$ in the integral equation (\ref{eq:bb_integral_equation}) and use $\gamma$ itself as the relevant choice of normalization parameter.

% $ \gamma _L ^* / \omega _0 = 0.0372640791038408789 $

\begin{figure}
\begin{center}
\includegraphics{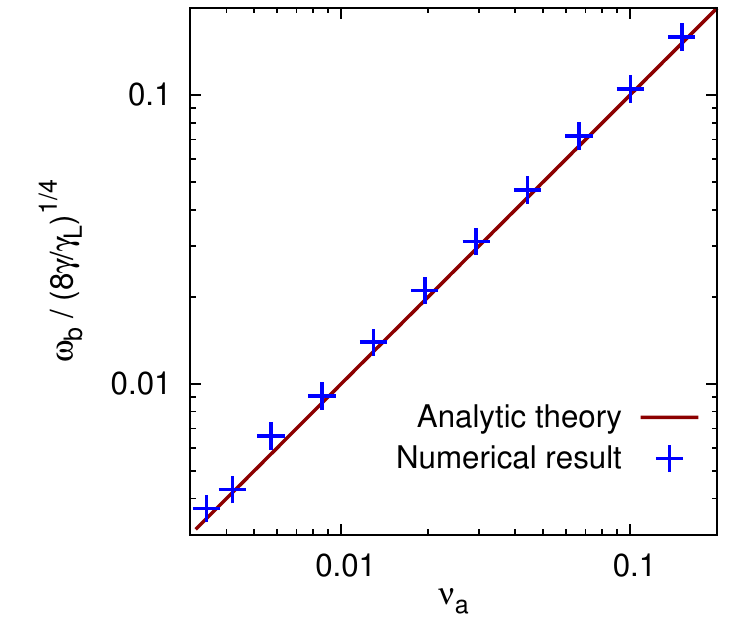}
\caption{Saturation level near marginal stability. Measured in full-$f$ simulations for the distribution and the numerical parameters of Fig.~\ref{fig:bifurcation00327}.}
\label{fig:near_threshold_saturation_level}
\end{center}
\end{figure}

This procedure yields a steady solution,
\begin{equation}
\omega _b ^2 \;=\; 2\sqrt{2} \, \nu _a ^2 \, \sqrt{\dfrac{\gamma}{\gamma _{L0}}}. \label{eq:sat_level_near_marginal}
\end{equation}
A series of simulations near marginal stability ($0.005$ $< \gamma / \gamma _L$ $< 0.02$), for $\nu _a$ spanning 2 orders of magnitude, confirms the validity of the latter expression. Fig.~\ref{fig:near_threshold_saturation_level} shows quantitative agreement with the saturation level of numerical solutions.

\begin{figure}
\begin{center}
\includegraphics{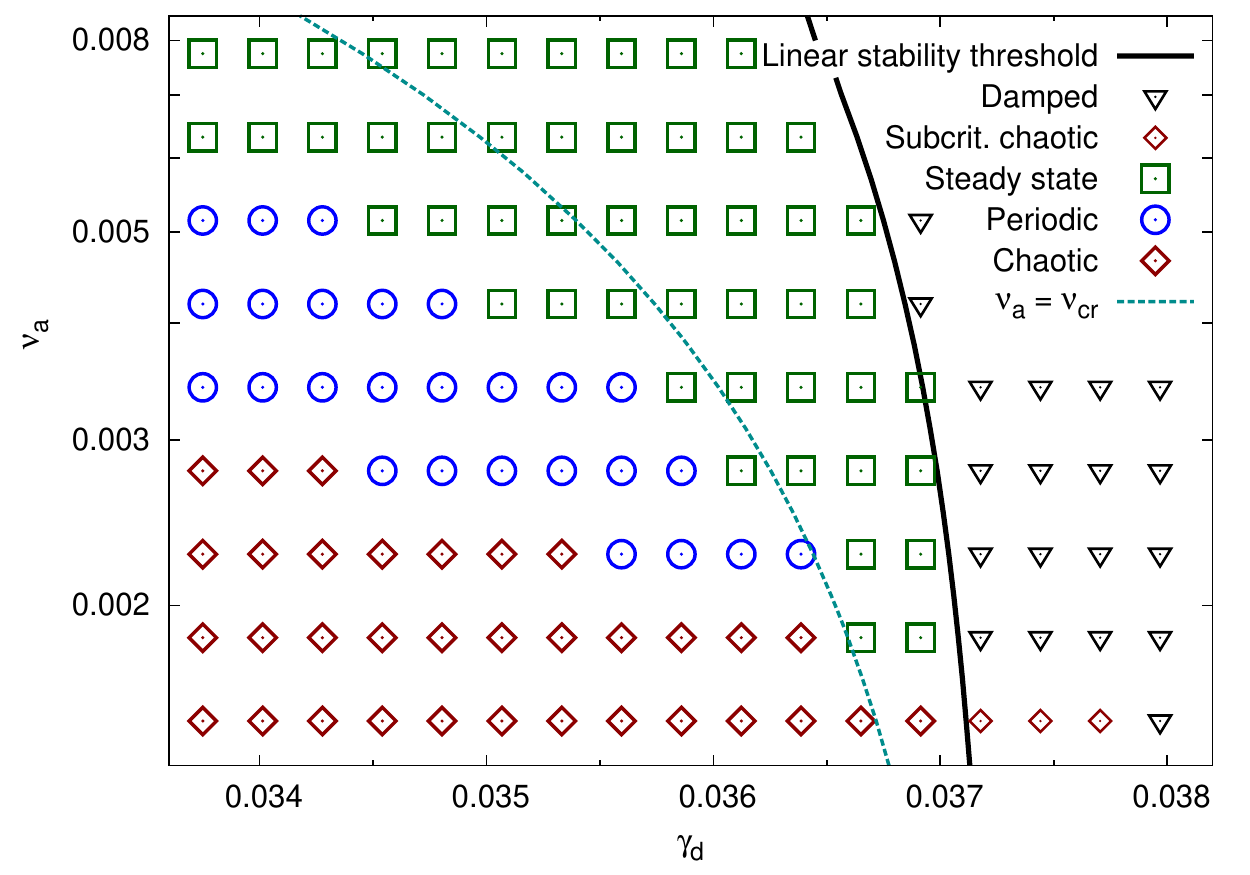}
\caption{Zoom in the behavior bifurcation diagram (Fig.~\ref{fig:bifurcation00327}) in a region near marginal stability where steady-periodic bifurcation occurs. Dashed line is the critical distribution relaxation. Smaller diamonds on the right of the linear stability threshold correspond to subcritical instabilities.}
\label{fig:near_threshold_bifurcation}
\end{center}
\end{figure}

\begin{figure}
\begin{center}
\includegraphics{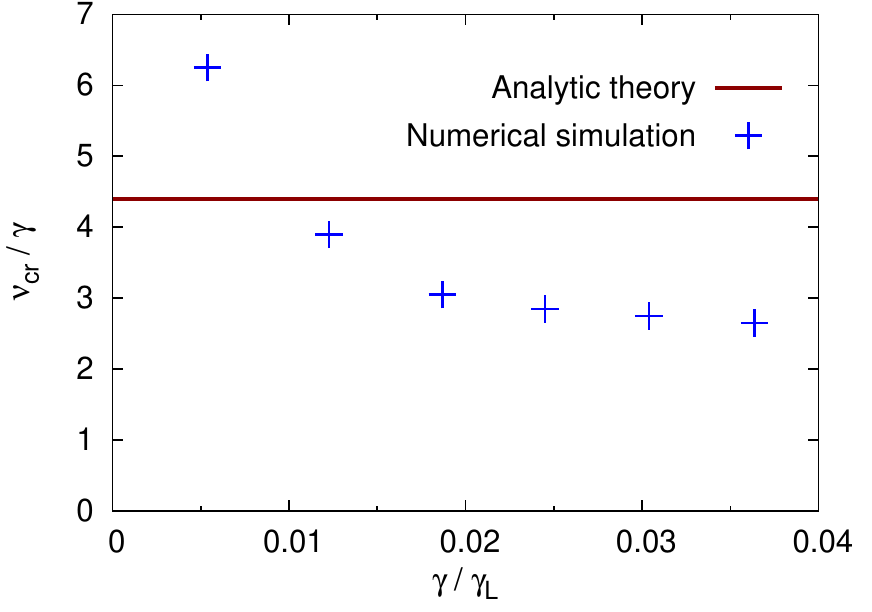}
\caption{Critical distribution relaxation for steady - periodic bifurcation near marginal stability obtained from full-$f$ simulations with numerical parameters of Fig.~\ref{fig:bifurcation00327}.}
\label{fig:critical_nu}
\end{center} \end{figure}

Nonlinear stability analysis reveals that the steady solution (\ref{eq:sat_level_near_marginal}) is unstable when $\nu _a < \nu _\mathrm{cr}$, with $\nu _\mathrm{cr} = 4.4 \gamma$. To assess this criterion for the bifurcation from steady-state to periodic solutions, a zoom in the behavior bifurcation diagram (Fig.~\ref{fig:bifurcation00327}) in a region near marginal stability where this bifurcation occurs is presented in Fig.~\ref{fig:near_threshold_bifurcation}. We observe a qualitative agreement between the steady-periodic boundary and $\nu _\mathrm{cr}$. However, when $\gamma / \gamma _L < 0.01$, chaotic solutions appear for $\nu _a\gg \nu _\mathrm{cr}$. This discrepancy is explained by the existence of nonlinear excitations. As we approach marginal stability, the nonlinear behavior becomes sensible to the initial perturbation level. To prove this point, we perform a series of simulations in the vicinity of the bifurcation with an initial amplitude reduced from $\omega _b / \gamma _L = 0.3$ to $\omega _b / \gamma _L = 3\times 10^{-7}$. Fig.~\ref{fig:critical_nu} shows the values of $\nu / \gamma$ for the bifurcation between steady-state and periodic solutions. The bifurcation occurs somewhere in between. We confirm that $\nu _\mathrm{cr} / \gamma$ stays close to the predicted value of $4.4$ for smaller values of $\gamma$.

%However, when $\gamma$ is too small compared to $d_t \omega _b$, the nature of the reduced integral equation (\ref{eq:bb_integral_equation}) is changed and theory breaks down. This explains the discrepancy at $0.0365 < \gamma _d / \omega _0 < 0.0373$, where we numerically observe $\gamma / d_t \omega _b \ll 1$, and chaotic solutions appear above $\nu = \nu _\mathrm{cr}$.

\section{Nonlinear features of chirping\label{sec:chirping}}

In the collisionless limit, when $\nu _a \,<\, \gamma \,\ll\, \gamma _L$, the integral equation (\ref{eq:bb_integral_equation}) is consistent with explosive solutions that diverge in a finite time, which suggests that the mode energy is partitioned into several spectral components. The resulting sideband frequencies have been observed to shift both upwardly and downwardly \cite{berkbreizman97pop}, the frequency shift $\delta \omega (t)$ increasing in time.

These chirping solutions arise when hole and clump structures \cite{berkbreizman97ppr} are formed in phase-space. They belong to a chaotic regime, and each chirping event is slightly different. In this section, we are interested in the nonlinear chirping characteristics, averaged over a significant number of chirping events. In particular, in our simulations, the first chirping event is observed to stand out from the statistics, with a larger extent of chirping - up to twice as much as any other one of the following series of repetitive chirping. This may be due to the fact that the first chirping benefits from a perfectly constant velocity-slope, while following events suffer from the interference of phase-space structures that remain from previous chirping events. Since the latter condition seems more experimentally-relevant, the first chirping is ignored in the present analysis, unless stated otherwise.

Since we want to use chirping features as experimental diagnostics, it is necessary to validate and develop corresponding theory. These features are quantified from raw simulation results in this section, and from experimental data in Chap.~\ref{ch:experiment}, using an algorithm described in Appendix \ref{app:chirping_algo}.

\subsection{Holes and clumps}

\begin{figure} \begin{center}
\includegraphics{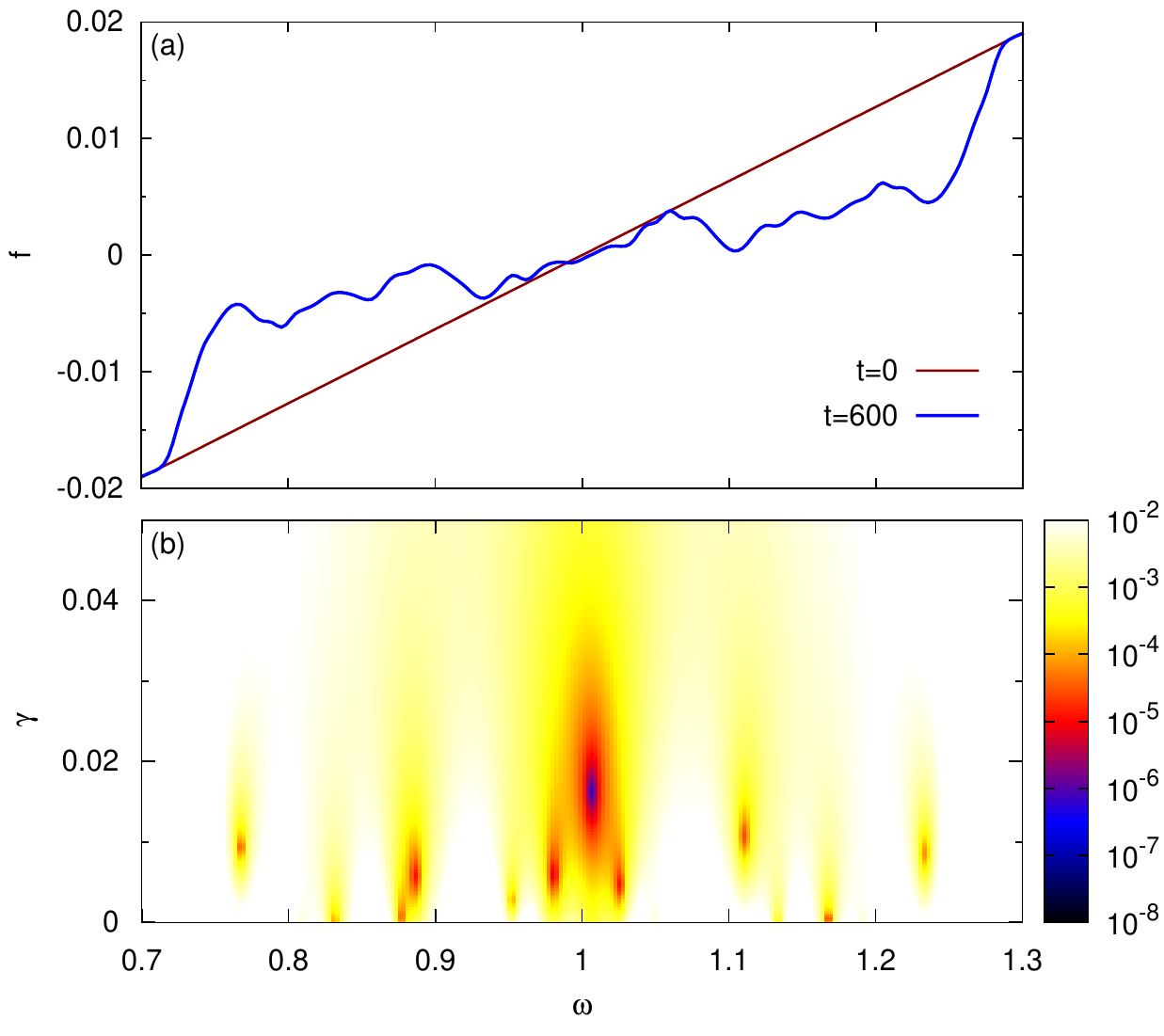}
\caption{Hole and clumps. (a) Snapshot of velocity distribution for a $\delta f$ simulation with $\gamma _{L0}=0.1$, $\gamma _d=0.05$, $\nu _a=0.001$, $N_x \times N_v=64\times 2048$, and $\mathrm{CFL}=1.1$. (b) Determinant of the linear dispersion relation, where the distribution function is taken from the simulation at $t=600$. Logarithmic color scale spanning 6 orders of magnitude. The center of each dark area corresponds to a linear mode.}
\label{fig:holes_clumps}
\end{center} \end{figure}

Holes and clumps are nonlinear coherent structures with time-dependant velocities. In general, several holes and several clumps, with different amplitude, coexist, as shown in Fig.~\ref {fig:holes_clumps}(a), which is a snapshot of the velocity distribution for a Krook $\delta f$ simulation. In Fig.~\ref {fig:holes_clumps}(b), we plot $D$ as a function of the real frequency $\omega$ and the growth rate $\gamma$, where
\begin{equation}
D(\omega,\gamma)\;\equiv\; \gamma \;+\; \gamma _d \;-\; \imath \, (\omega -1) \;-\; \frac{1}{2k} \int _{\Gamma} \frac{\pds{f^B_0}{v}}{(\gamma \,+\, \nu _a) \,+\, \imath \, (k\,v \,-\, \omega)} \, \mathrm{d} v,
\end{equation}
so that $D=0$ gives the linear dispersion relation Eq.~(\ref{eq:df_disp_rela}). This plot suggests that to each hole/clump corresponds an eigenmode, with frequency and growth rate both functions of time.

%[Redaction needed: plot snapshots of phase-space][Work needed: test particle to see if a hole chirps with its particles][Work needed: Momentum conservation \textit{a la} P.~Diamond to explain the growth of a hole/clump]

%The velocity distribution after nonlinear saturation shown in Fig. illustrates the fact that several holes and clumps with different amplitudes can co-exist. 

\subsection{Chirping velocity (or sweeping rate)}

\subsubsection{Available theory}

Ref.~\cite{berkbreizman97pla} shows how one can isolate one spectral component and model it by a BGK wave to obtain the time-evolution of one chirping event. This theory is based on the following assumptions:
\begin{itemize}
\item The resonant velocity of a hole/clump evolves slowly enough for trapped particle orbits to keep their coherency, $\dot{\delta \omega} / \omega _b^2$, $\ddot{\delta \omega} / \omega _b^3$ $\ll 1$;
\item The width of a hole/clump evolves slowly enough for trapped particle orbits to keep their coherency, $\dot{\omega _b} / \omega _b^2 \ll 1$;
\item Holes and clumps are narrow enough that they don't overlay each others, $\omega _b / \delta \omega \ll 1$.
\end{itemize}
Within the above assumptions, the perturbation of passing particle distribution is negligible, and a bounce-average treatment of trapped particle distribution yields the frequency shift, in the collisionless limit, as
%In Ref.~\onlinecite{berkbreizman97pla}, the frequency shift $\delta \omega$ is obtained in the collisionless limit as 
\begin{equation}
\delta \omega (t) \;=\; \alpha \, \gamma _{L0} \, \sqrt{\gamma _d \, t}, \label{eq:chirping_velocity}
\end{equation}
with $\alpha \approx 0.44$ ; and a saturation level as
\begin{equation}
\omega _b \;\approx\; 0.54 \, \gamma _{L0} . \label{eq:chirping_steadystate}
\end{equation}
These analytic expressions have been found to agree with 1D $\delta f$ particle simulations, \cite{berkbreizman97pla}, with both Krook and diffusion-only collision operators, and with 3D HAGIS simulations \cite{pinches04S47}.

\subsubsection{Numerical validation}

\begin{figure}
\begin{center}
\includegraphics{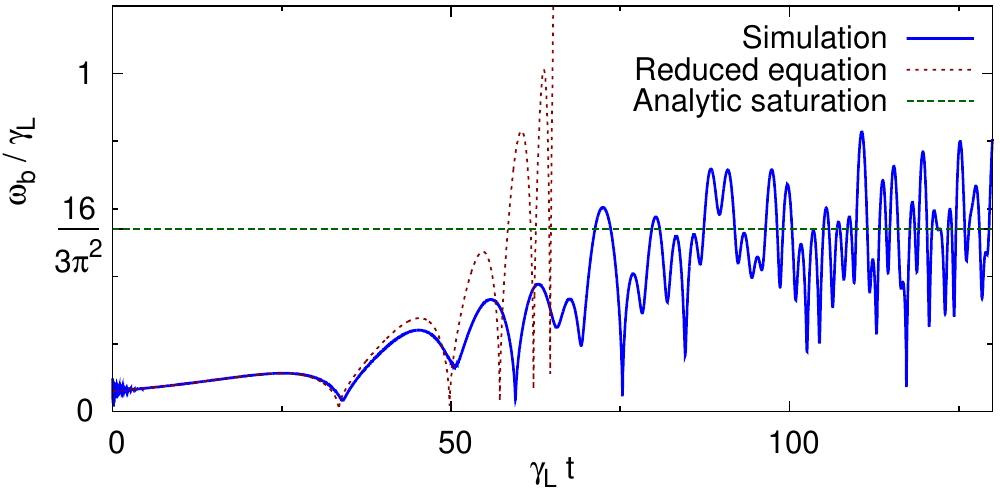}
\caption{Time evolution of the normalized field amplitude, obtained from full-$f$ simulation and from the reduced equation (\ref{eq:bb_integral_equation}), with initial distribution B, $\gamma _d = 0.035$, and $\nu _a = 10^{-4}$.}
\label{fig:chirping_burst}
\end{center}
\end{figure}

When $\gamma _d$ is finite and $\nu _a$ is small enough, we observe such chirping solutions in full-$f$ COBBLES simulations. Fig.~\ref{fig:chirping_burst} shows the time evolution of field amplitude, with initial distribution B, when $\gamma _d = 0.035$, and $\nu _a = 10^{-4}$, so that $\gamma = 0.05\,\gamma _{L0}$. The simulation result agrees with a numerical solution of reduced equation (\ref{eq:bb_integral_equation}), until the field amplitude approaches the applicability limit. After saturation, the solution is close to analytic prediction Eq.~(\ref{eq:chirping_steadystate}).

\begin{figure}
\begin{center}
\includegraphics{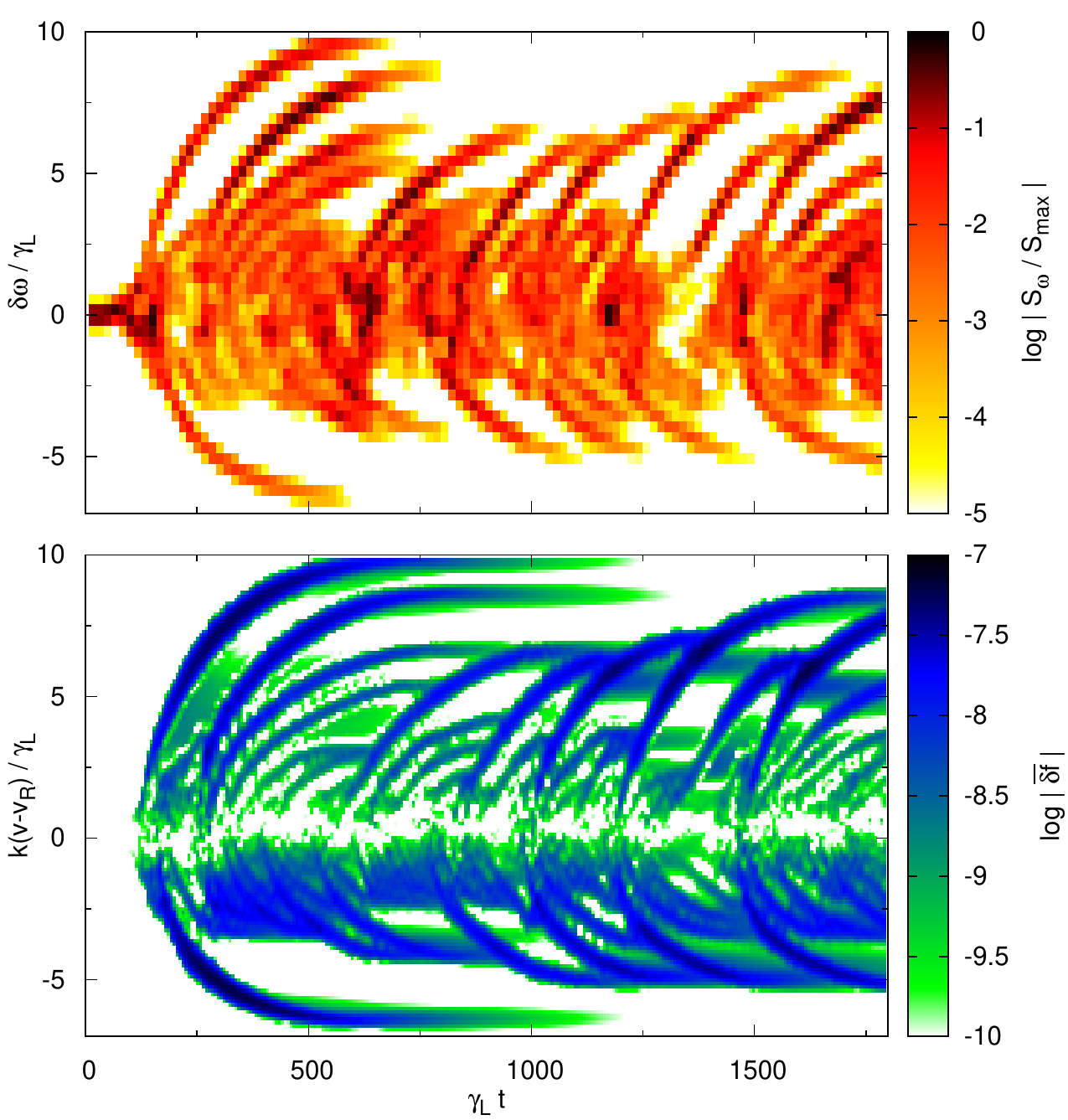}
\caption{(a) Logarithmic scale plot of the time evolution of the frequency power spectrum. At each time, the spectrum is normalized to its maximum value. (b) Logarithmic scale plot of the time evolution of $\ov{\delta f} \equiv \ov{f}-f_0$.}
\label{fig:chirping}
\end{center}
\end{figure}

\begin{figure}
\begin{center}
\includegraphics{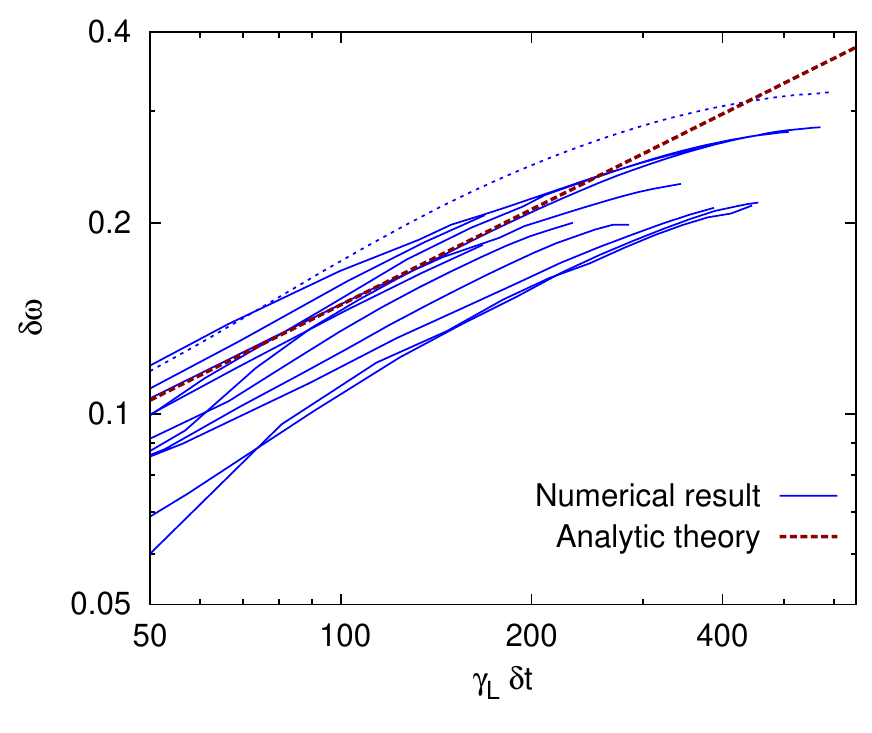}
\caption{Time-evolution of the 12 largest upshifting branches in logarithmic scale. Dotted line is the first chirping event. Dashed line is the analytic prediction Eq.~(\ref{eq:chirping_velocity}).}
\label{fig:chirping_exponent}
\end{center}
\end{figure}

Fig.~\ref{fig:chirping}(a) shows the spectrogram of electric field. Frequency sweeping events occur repetitively as the slope of the distribution is successively recovered and flattened.
The time evolution of each chirping event is slightly variable. To quantify the agreement with theory, we extract the 12 largest upshifting branches. In Fig.~\ref{fig:chirping_exponent}, we show these branches shifted to the same initial time, where each point is obtained by interpolating local maxima from the discrete frequency spectrum. From this plot, we conclude that, before border effects occur, chirping evolution follows a square-root law in time, as expected from theory.
This suggests a possibility to recover the product $\gamma _{L0} \, \sqrt{\gamma _d}$ from such power spectrum. By fitting a square-root law to the branches of Fig.~\ref{fig:chirping_exponent}, we obtain an average value of $\gamma _{L0} \, \sqrt{\gamma _d} = 0.0057$, with a standard deviation of $14\,\%$, when the input value is $\gamma _{L0} \, \sqrt{\gamma _d} = 0.0061$. When a plasma is close to marginal stability, we can assume $\gamma_{L0}\sim\gamma_d$ ($\gamma_{L0}\sqrt{\gamma_d}\sim\gamma_d^{3/2}$), and this chirping diagnostics may be used as a rough estimation of the extrinsic damping rate of the bulk plasma $\gamma_d$. The agreement of the expression Eq.~(\ref{eq:chirping_velocity}) with chirping found in experimental devices \cite{pinches04S47} supports this claim.

\begin{figure}
\begin{center}
\includegraphics{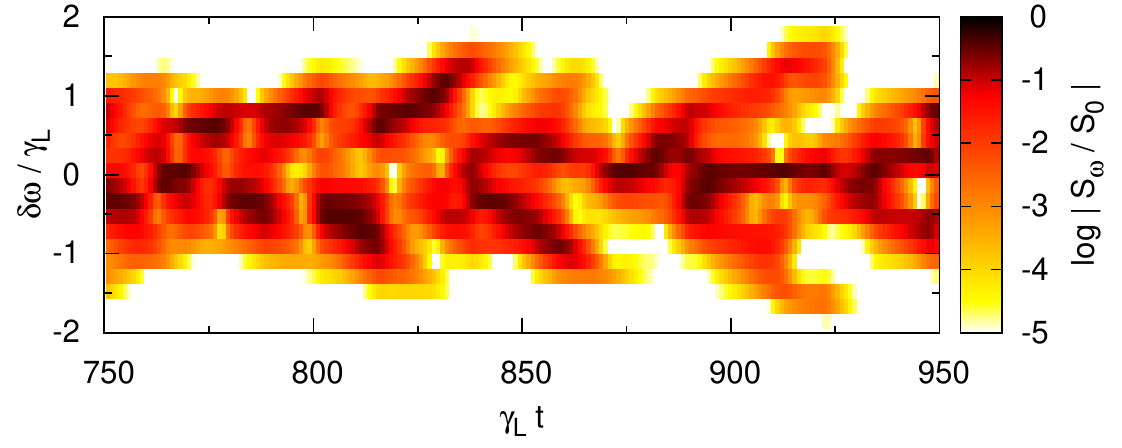}
\caption{Logarithmic scale plot of the time evolution of the frequency power spectrum for $\gamma _d = 0.2 \,\gamma _L $, $\nu _a = 0.02 \,\gamma _L$.}
\label{fig:SmallGamdChirping}
\end{center} \end{figure}

As we consider a single mode with a fixed wave number, sweeping frequencies correspond to evolving structures in velocity distribution. In Fig.~\ref{fig:chirping}(b), we observe formation and evolution of hole/clump pairs in phase-space, and they show clear correlation with peaks of the spectrogram shown in Fig.~\ref{fig:chirping}(a).
It should be noted that near marginal stability, $\gamma _d > 0.4 \gamma _L$ is given as a necessary condition for hole-clump pair creation in Ref.~\cite{berkbreizman97pop}. Although there is yet no theory in the opposite limit $\gamma \sim \gamma _L$, in our simulations we observe frequency sweeping even when $\gamma _d \ll \gamma _L$. A spectrogram is shown in Fig.~\ref{fig:SmallGamdChirping} for $\gamma _d / \gamma _L = 0.2$, $\nu _a / \gamma _L = 0.02$. Althought the chirping is not as pronounced as in Fig.~\ref{fig:chirping}(a), we observe that the dominant frequency sweeps $5\%$ of its initial value.

\subsubsection{Non-adiabatic chirping}

In Chap.~\ref{ch:experiment}, we consider a regime with relatively fast sweeping, $\dot{\delta \omega} / \omega _b^2 \approx 0.5$, which approaches the limit of validity of the above theory. $\dot{\delta \omega} / \omega _b^2$ can be seen as a measure of hole/clump adiabaticity, and is roughly proportional to $(\gamma _d \nu _a)^{1/2}/\gamma _{L0}$ in the Krook case. When $\dot{\delta \omega} / \omega _b^2 \approx 0.5$, $4\omega _b / \dot{\delta \omega} \approx 2\pi / \omega _b$, in other words a hole or a clump is shifted by its width in a bounce time of deeply trapped particles. In this regime, the previous analytic treatment is not relevant.

\begin{figure} \begin{center}
\includegraphics{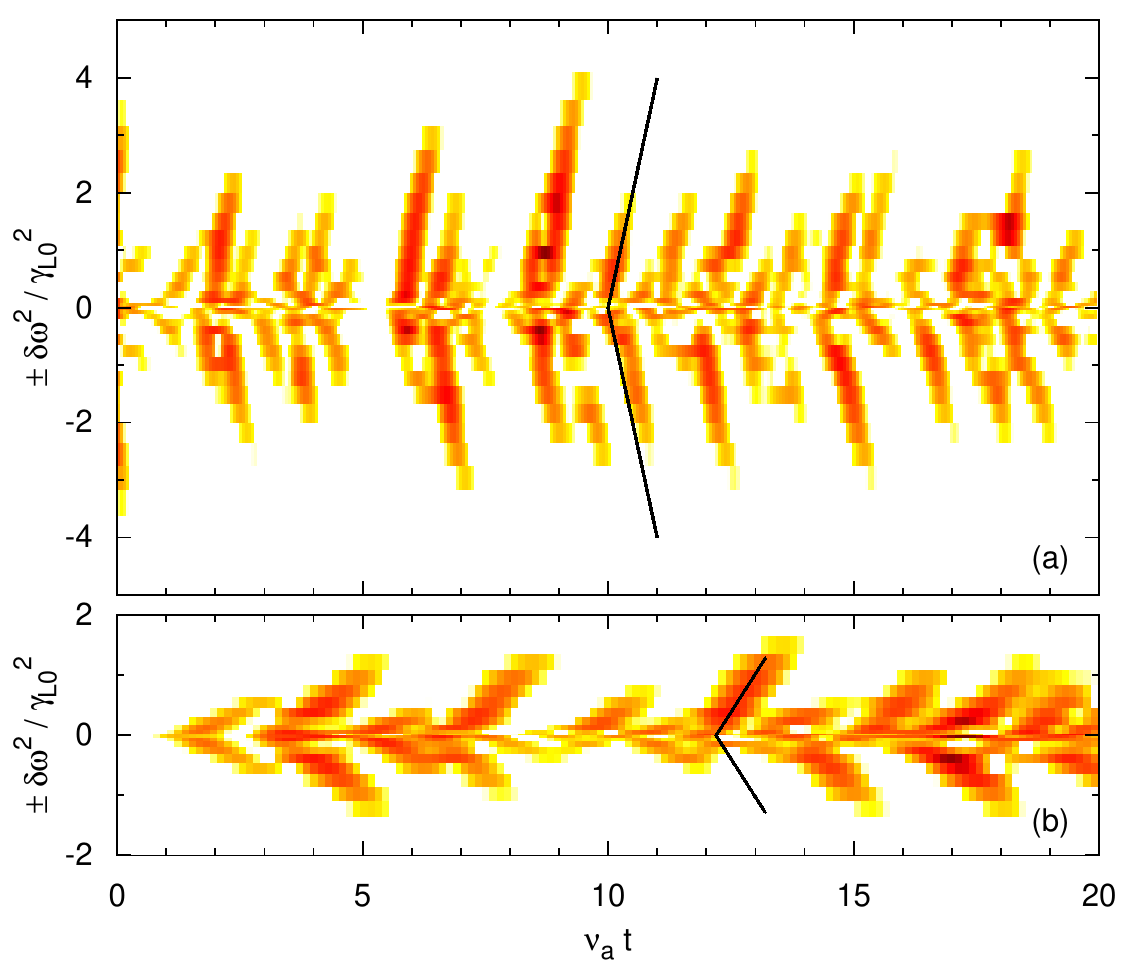}
\caption{Spectrogram of the electric field in $\delta f$ simulations, obtained with a moving Fourier window of size $\Delta =510$, for $\gamma _{L0}=0.1$ and (a) $\gamma _d/\gamma _{L0}=0.4$, $\gamma _d \nu _a / \gamma _{L0}^2 = 0.008$, $\beta=1.0$; (b) $\gamma _d/\gamma _{L0}=0.9$, $\gamma _d \nu _a / \gamma _{L0}^2 = 0.043$, $\beta=0.57$. Solid lines show the chirping velocity predicted by Eq.~(\ref{eq:chirping_velocity}), with correction coefficient $\beta$, and chirping lifetime predicted by Eq.~(\ref{eq:chirping_lifetime_krook}). In the remaining of this thesis, the logarithmic color scale for each spectrogram spans 3 orders of magnitude.}
\label{fig:exs_for_alpha}
\end{center} \end{figure}

\begin{figure} \begin{center}
\includegraphics{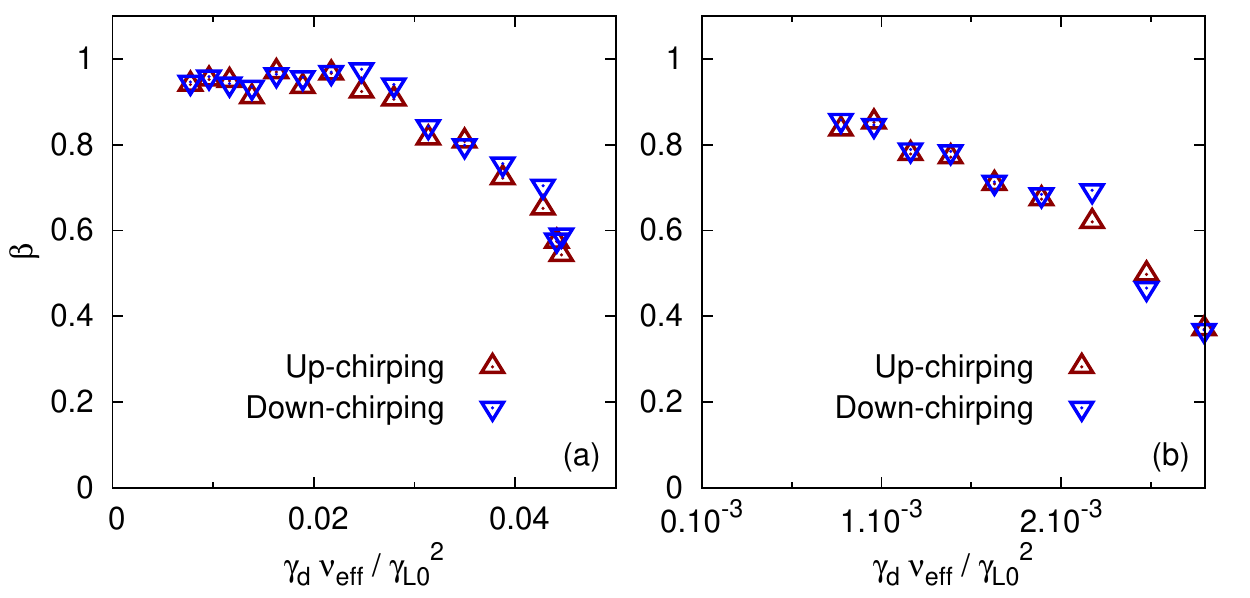}
\caption{Correction to Eq.~(\ref{eq:chirping_velocity}) when the time-scale of frequency shift is relatively short compared to the bounce period, measured in $\delta f$ simulations with $\gamma _{L0}=0.1$. The ratio $\gamma _d / \nu _\mathrm{eff}$ is such that $\delta \omega (1/\nu _\mathrm{eff})=0.2$ in Eq.~(\ref{eq:chirping_velocity}). (a) In the Krook case, spectrograms for two extreme points are shown in Fig.~\ref{fig:exs_for_alpha}. (b) With drag and diffusion, $\nu _f/\nu_d=0.1$.}
\label{fig:alpha_scan}
\end{center} \end{figure}

Clarifying this point is easier if we avoid effects of bulk evolution and of the shape of beam distribution. Thus we switch to the $\delta f$ model.
Fig.~\ref{fig:exs_for_alpha} shows spectrograms of chirping $\delta f$ COBBLES simulations, first in a regime which satisfies the assumptions of the above theory, then in a regime out of its scope. We observe similar square-root dependency of the frequency shift in time. This suggests that we can introduce the effect of non-adiabaticity on chirping velocity as a correction parameter $\beta$, defined as
\begin{equation}
\beta \;\equiv\; \dfrac{\delta \omega (t)}{\alpha \, \gamma _{L0} \, \sqrt{\gamma _d \, t}}, \label{eq:beta}
\end{equation}
which is a priori a function of all input parameters.
$\beta$ is obtained numerically for $\gamma _{L0}/\omega = 0.1$ in Fig.~\ref{fig:alpha_scan}. Results show that chirping velocity slows down compared to theory as we leave the adiabatic limit. 
We confirm that, inside the validity limit of the above theory, $\beta$ approaches unity. Even for relatively large values of $\dot{\delta \omega} / \omega _b^2$, chirping velocity has a smooth dependency on the kinetic parameters. The latter point is crucial with regard to the validity of the procedure described in Chap.~\ref{ch:experiment}.\\

\subsubsection{Effect of finite Krook collisions}

In Ref.~\cite{berkbreizman97pla}, Eq.~(\ref{eq:chirping_velocity}) is derived by changing variables to action-angle of the bounce-motion of particles trapped in a hole or a clump, and by noting that the unperturbed part $f_0(\omega)$ of $f$ does not contribute in the resonant power transfer of the power balance, Eq.~(\ref{result:power_transfer_W}). Thus, the wave equation involves only the deviation from the unperturbed distribution at the center of the evolving hole/clump, $g\equiv f - f_0(\omega+\delta \omega)$. Then, $g$ is expanded in powers of $\epsilon$,
\begin{equation}
g\;=\; g_0 \;+\; \epsilon g_1 \;+\; \ldots
\end{equation}
where $\epsilon \equiv \max (\ddot{\delta \omega}/\omega _b^3,\dot{\omega _b}/\omega_b^2,\omega _b/\delta \omega)\ll 1$, and it is shown that to zeroth order in $\epsilon$, the real and imaginary parts of the wave equation are reduced to
\begin{eqnarray}
\gamma _d &=& -\frac{2 \gamma _{L0}}{\pi \omega _b^2 \pds{f_0}{\omega}(\omega _0)}\, \int _0 ^{J_\mathrm{max}} \frac{\dot{\delta \omega}}{\omega _b^2} g_0 \, \mathrm{d} J, \label{eq:bounce_av_wave_1} \\
\delta \omega &=& -\frac{2 \gamma _{L0}}{\pi \omega _b^2 \pds{f_0}{\omega}(\omega _0)}\, \int _0 ^{J_\mathrm{max}} \left\langle \cos \psi \right\rangle g_0 \, \mathrm{d} J, \label{eq:bounce_av_wave_2}
\end{eqnarray}
where angle brackets indicates a bounce-average, $\psi$ is the spatial coordinate in a frame moving with the hole/clump, $J$ is the bounce-motion action, $J_\mathrm{max}=8\omega _b / \pi$, and $g_0$ is obtained by bounce-averaging the kinetic equation.
Then, in the reference, the collisionless limit is consider, in which case $g_0$ is simply
\begin{equation}
g_0(t)\;=\; f_0(\omega) \,-\, f_0(\omega + \delta \omega). 
\end{equation}
Eq.~(\ref{eq:chirping_velocity}) follows by assuming a constant gradient for $f_0$.

\begin{figure}
\begin{center}
\includegraphics{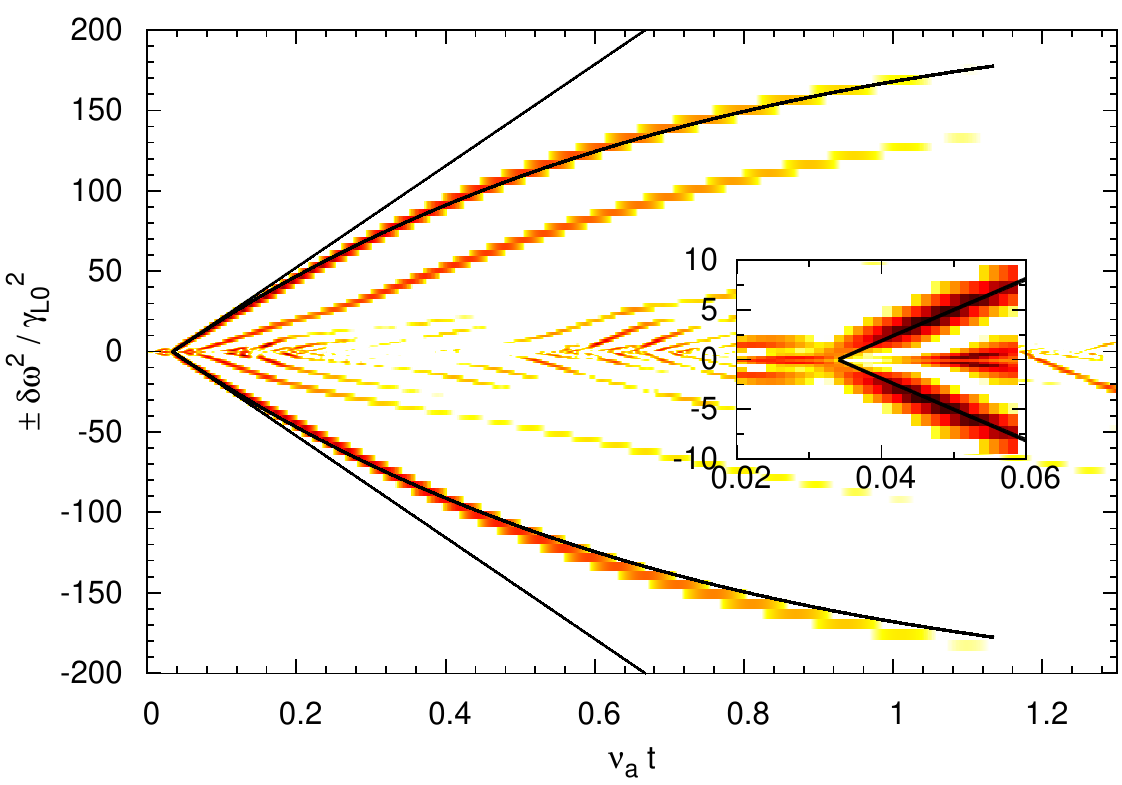}
\caption{Effect of finite Krook collisions on chirping velocity. Spectrogram of $\delta f$-COBBLES simulation with $\gamma _{L0}=0.05$, $\gamma _d = 0.045$, $\nu _a = 4\times 10^{-5}$, $N_x \times N_v = 128\times 4096$, and $\mathrm{CFL}=1.3$. Two straight lines correspond to Eq.~(\ref{eq:beta}). Two bended lines correspond to Eq.~(\ref{eq:chirping_velocity_with_krook}). In both cases, the lines span a time interval $1.1/\nu _a$ [See Eq.~(\ref{eq:chirping_lifetime_krook})], and we have included a correction coefficient $\beta=1.23$. Inset: zoom on the beginning of the first chirping event.}
\label{fig:chirping_finite_collision}
\end{center} \end{figure}

The above theory in the adiabatic limit, and the above numerical investigation of $\beta$ are valid on a timescale smaller than a collision time. In the following, we include the effect of finite collisions, to explain a deviation from square-root law in longer timescales. In the Krook case, the bounce-averaged kinetic equation for $g$ at lowest order is
\begin{equation}
\pd{g_0}{t} \,+\, \nu _a g_0 \;=\; - \pd{f_0}{t},
\end{equation}
whose solution is
\begin{equation}
g_0 (t) \;=\; e^{-\nu _a t}f_0(\omega) \,-\, f_0(\omega + \delta \omega) \,+\, \nu _a \int _0^t f_0[\omega+\delta \omega (t')] e^{\nu _a (t'-t)} \,\mathrm{d} t'.
\end{equation}
As in the reference, we assume a constant gradient for $f_0$,
\begin{equation}
f_0 [\omega + \delta \omega(t)] \;=\; f_0(\omega) \,+\, f_0'(\omega) \, \delta \omega (t),
\end{equation}
which yields
\begin{equation}
g_0 (t) \;=\; f_0'(\omega) \, \left[ \nu _a \, \int _0^t e^{\nu _a (t'-t)} \delta \omega (t') \,\mathrm{d} t' \,-\, \delta \omega (t) \right].
\end{equation}
We substitute the latter solution into Eqs.~(\ref{eq:bounce_av_wave_1}-\ref{eq:bounce_av_wave_2}) to find, in the limit $\dot{\delta \omega}/\omega _b^2 \ll 1$,
\begin{eqnarray}
\frac{\gamma _d}{\delta \omega} &=& 3\,\frac{\dot{\delta \omega}}{\omega _b^2}, \\
\omega _b (t) &=& \frac{16 \gamma _{L0}}{3 \pi ^2} \, \left[ 1\,-\,\frac{\nu _a}{\delta \omega (t)} \, \int _0^t e^{\nu _a (t'-t)} \delta \omega (t') \,\mathrm{d} t' \right].
\end{eqnarray}
We solve the latter equation system by expanding both $\omega _b$ and $\delta \omega$ in powers of $\nu _a t$. This lengthy but straightforward procedure yields
\begin{eqnarray}
\delta \omega (t) &=& \pm \alpha \, \beta \, \gamma _{L0} \,\sqrt{\gamma _d t} \, \left[ 1\,-\, \frac{1}{3} (\nu _a t) \,+\, \frac{7}{90} (\nu _a t)^2 \,-\, \frac{19}{1890} (\nu _a t)^3 \,+\, \ldots \right], \label{eq:chirping_velocity_with_krook}\\
\omega _b (t) &=& \frac{16 \gamma _{L0}}{3 \pi ^2} \, \left[ 1\,-\, \frac{2}{3} (\nu _a t) \,+\, \frac{8}{45} (\nu _a t)^2 \,-\, \frac{8}{315} (\nu _a t)^3 \,+\, \ldots \right],
\end{eqnarray}
where we have included the effect of non-adiabaticity.
Note that Eqs.~(\ref{eq:chirping_velocity})-(\ref{eq:chirping_steadystate}) are recovered in the collisionless limit. The effect of finite collision is to reduce chirping extent by bending shifting branches. This effect is not negligible since $\delta \omega$ is reduced by $27\%$ after a collision time, which is of the order of chirping lifetime as we will see in \ref{subs:lifetime}. This is illustrated in Fig.~\ref{fig:chirping_finite_collision}, which is the spectrogram for a $\delta f$ Krook simulation. The beginning of the first chirping event is fitted to the collisionless analytic prediction, Eq.~(\ref{eq:chirping_velocity}), to determine the correction parameter $\beta$, then we show the time-evolution of Eq.~(\ref{eq:chirping_velocity_with_krook}) with the same correction parameter, which is in good agreement with observed bended chirping.

\subsection{Chirping lifetime} \label{subs:lifetime}

\begin{figure} \begin{center}
\includegraphics{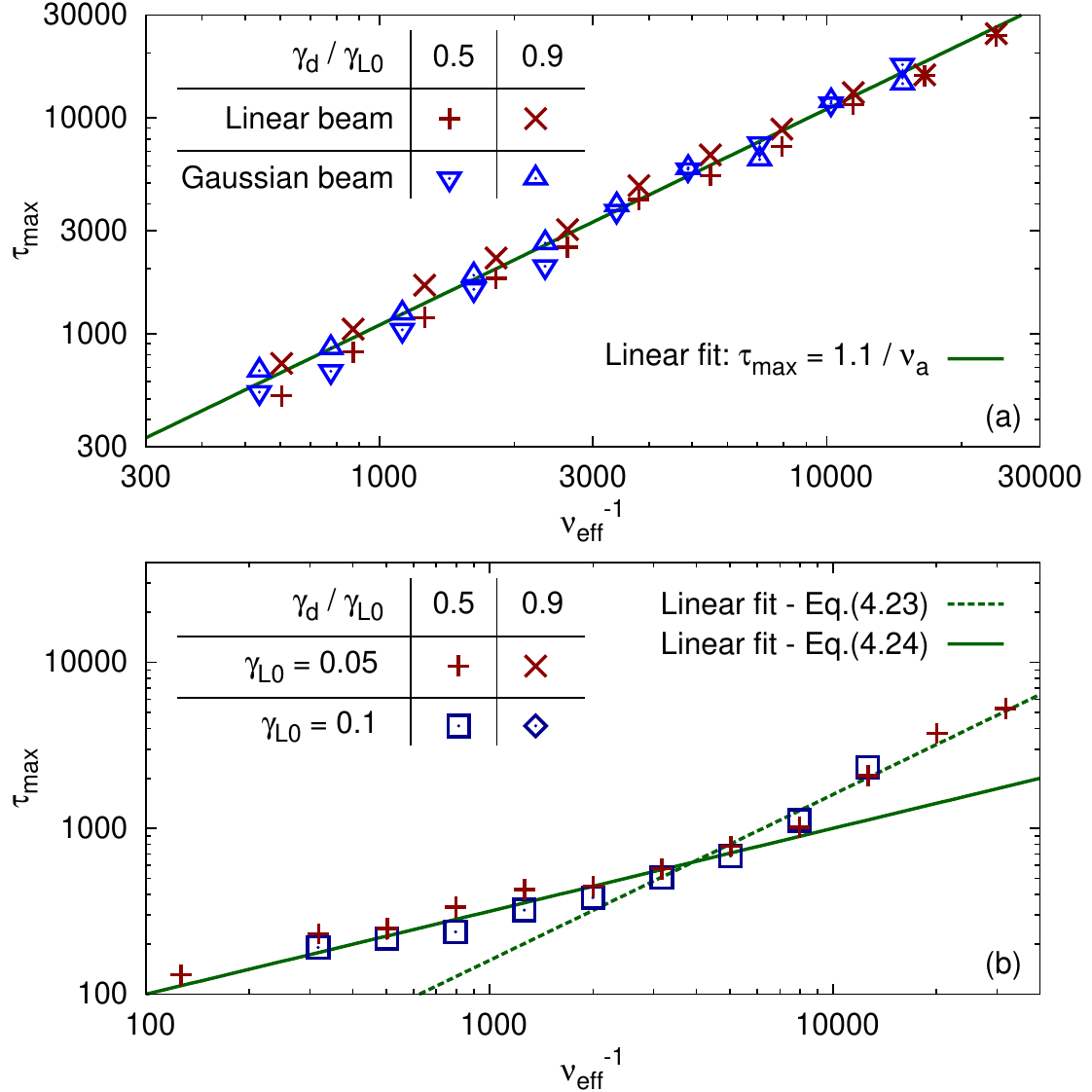}
\caption{Maximum lifetime of a hole or clump, far from marginal stability, $\gamma _d / \gamma _{L0} = 0.5$, and near marginal stability, $\gamma _d / \gamma _{L0} = 0.9$. (a) With a Krook collision operator. Crosses correspond to $\delta f$ simulations with initial distribution shown in Fig.~\ref{fig:df_f0}(a), while triangles correspond to full-$f$ simulations, where the initial distribution is a bump-on-tail with a Gaussian beam. In both cases, $\gamma _{L0}=0.05$. A solid line corresponds to Eq.~(\ref{eq:chirping_lifetime_krook}). (b) $\delta f$ model with drag/diffusion collision operator (for a linear distribution only), for two different values of linear drive. A dashed line corresponds to Eq.~(\ref{eq:chirping_lifetime_dragdiff_theory}) with $\iota _d=0.16$, a solid line to Eq.~(\ref{eq:chirping_lifetime_dragdiff}) with $\iota _d=10$. The drag is chosen so that it does not significantly alter chirping lifetime, $\nu _f/\nu _d=0.1$. An absence of point means that we do not observe repetitive chirping before the end of simulation ($t=100000$).}
\label{fig:chirplifetime}
\end{center} \end{figure}

The resonant velocity of a hole (a clump) does not increase (decrease) indefinitely. We define the lifetime $\tau$ of a chirping event as the time in which the corresponding power in the spectrogram decays below a fraction $e^{-2}$ of the maximum amplitude that is reached during this chirping event. The maximum lifetime $\tau _\mathrm{max}$ is the maximum reached by $\tau$ during a time-series, ignoring the first chirping event and any minor event.
It is reasonable to assume that the island structure is dissipated by collisional processes, in which case the maximum chirping lifetime should be of the form
\begin{equation}
\tau _\mathrm{max} \;=\; \dfrac{\iota _a}{\nu _a}, \label{eq:chirping_lifetime_krook}
\end{equation}
in the Krook case, and
\begin{equation}
\tau _\mathrm{max} \;=\; \iota _d\, \dfrac{\gamma _{L0}^2}{\nu _d^3}, \label{eq:chirping_lifetime_dragdiff_theory}
\end{equation}
in the case with drag and diffusion, when $\nu _f \ll \nu _d$, where $\iota _a$ and $\iota _d$ are constant parameters.
In Fig.~\ref{fig:chirplifetime}, we plot the maximum lifetime measured in $\delta f$-COBBLES simulations, where the ratio $\gamma _d / \gamma _{L0}$ is chosen as $0.5$ and $0.9$, i.e.~far from and close to marginal stability, respectively. A quantitative agreement is found with Eq.~(\ref{eq:chirping_lifetime_krook}), with $\iota _a=1.1$, for $\nu _a^{-1}$ spanning 2 orders of magnitude.
With the diffusive collision operator, chirping lifetime agrees with Eq.~(\ref{eq:chirping_lifetime_dragdiff_theory}) only for low collisionality. For high collisionality, diffusion affects the width of a hole or clump during the first phase of their evolution, namely drive by free-energy extraction, which in turn affects the decay by diffusion. Since chirping observed in experiments has a lifetime of the order of $\tau \sim 500$, we adopt a semi-empirical law obtained by a linear fit,
\begin{equation}
\tau _\mathrm{max} \;=\; \iota _d \, \left( \dfrac{\gamma _{L0}^2}{\nu _d^3} \right)^{0.5}, \label{eq:chirping_lifetime_dragdiff}
\end{equation}
with $\iota _d=10$. No repetitive chirping is found near marginal stability for $0.05\le \gamma _{L0}\le 0.1$, though longer computations may reveal this possibility.

In Chap.~\ref{ch:experiment}, Eqs.~(\ref{eq:chirping_lifetime_krook}) and (\ref{eq:chirping_lifetime_dragdiff}) are used as diagnostics for the effective collision frequency, thus it is important that these results are not too sensitive to the shape of fast-particle distribution, which is not measured accurately enough in experiments. To investigate this point, we repeat the previous analysis (in the Krook case), this time with an initial bump-on-tail distribution with a Gaussian beam (with full-$f$ COBBLES) instead of a constant gradient, or linear, beam (as is imposed in $\delta f$ COBBLES). Fig.~\ref{fig:chirplifetime}(a) shows that the agreement is kept, even if the shape of the distribution has a significant effect on the extent of chirping as can be seen for example in Fig.\ref{fig:chirping}(a).\\

\subsection{Chirping asymmetry}

It should be noted that, in Fig.~\ref{fig:chirping_finite_collision} for example, we observe a slight asymmetry between up- and down- shifting branches, which is in disagreement with theory, since the $\delta f$ BB problem with Krook collisions is symmetric around the resonant velocity $v_R$. We infer that the observed asymmetry is a numerical effect due to a difference in round-off errors between $v<v_R$ and $v>v_R$ domains, which is amplified in such chaotic regime. In a long-time averaged point-of-view, chirping observed in $\delta f$-COBBLES simulations should be symmetric, though.

In this section, we consider significant chirping asymmetry, rather than the above negligible numerical effect.

\subsubsection{Effects of drag}

\begin{figure}
\begin{center}
\includegraphics{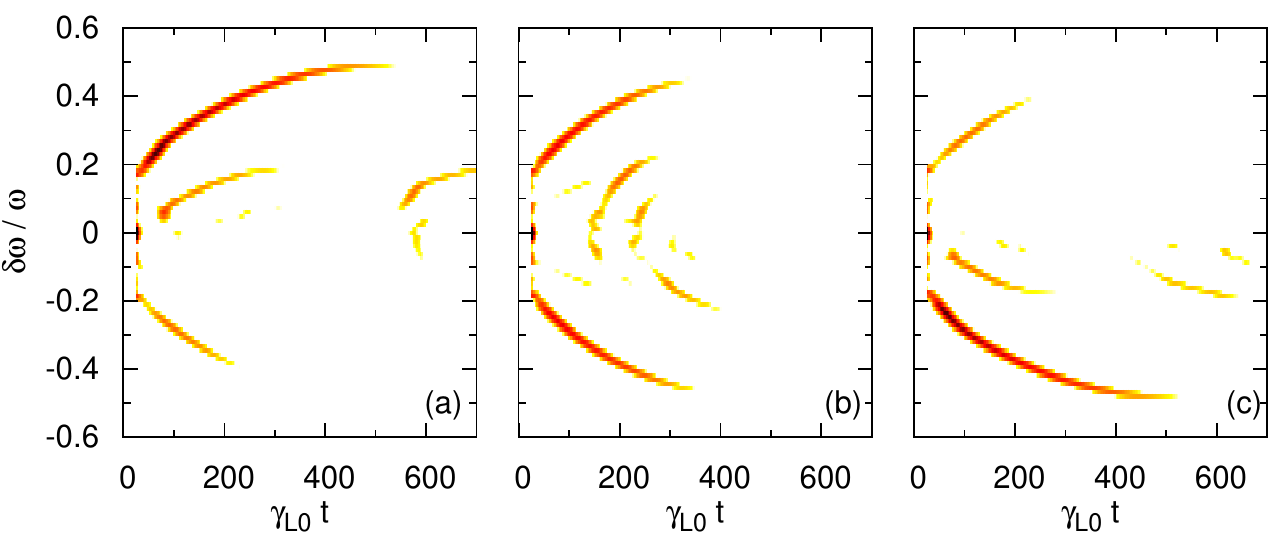}
\caption{Effect of drag on chirping asymmetry. Spectrograms of $\delta f$ COBBLES simulations, with input parameters $\gamma _{L0} = 0.099$, $\gamma _d=0.028$, $\nu _d=9.2\times 10^{-3}$. (a) Natural drag, $\nu _f=5.4\times 10^{-3}$. (b) Negligible drag, $\nu _f=8.0\times 10^{-4}$. (c) Inverted drag, $\nu _f=5.4\times 10^{-3}$.}
%\caption{Effect of drag on chirping asymmetry. $\gamma _L = 70$ms$^{-1}$, $\gamma _d=20$ms$^{-1}$, $\nu _D=6$ms$^{-1}$. Above: inverted drag, $\nu _f=4$ms$^{-1}$. Middle: negligible drag, $\nu _f=0.5$ms$^{-1}$. Below: natural drag, $\nu _f=4$ms$^{-1}$.}
\label{fig:opposite_drag}
\end{center}
\end{figure}

The drag term in the collision operator has a counter-intuitive effect on chirping asymmetry. Since the effect of drag on any phase-space structure is to advect it from large to small velocities, one could imagine that the presence of drag would make down-chirping dominant. The opposite is observed in our simulations. In Fig.~\ref{fig:opposite_drag}, we show a symmetric simulation obtained with negligible drag, and two asymmetric solutions with significant drag. For Fig.~\ref{fig:opposite_drag}(c), we changed the sign in front of $(\nu _f^2 / k) \partial _v f^B$ in Eq.~(\ref{model:diffdrag}).
An explanation for this paradox is proposed in Ref.~\cite{lilley10}.
% by splitting the drag term into a dynamic part $(\nu _f^2/k) \pds{f}{v}$, which slows particles down, and a sink part $-(\nu _f^2/k) \pds{f_0}{v}$, which acts to lower $f$ everywhere. The latter part reduces clumps and enhances holes.
%It is possible to explain this paradox in the following way. Shifting hole or clump is a way for the system to extract much more free energy than what is available in the fixed resonance situation \cite{berkbreizman95report}, and thus compensate for the loss of energy induced by external damping. In the ideal adiabatic and collisionless situation, a clump moves with its particles, thus its amplitude grows, since $f$ is conserved along particle orbits.

\subsection{Chirping quasi-period}

\begin{figure} \begin{center}
\includegraphics{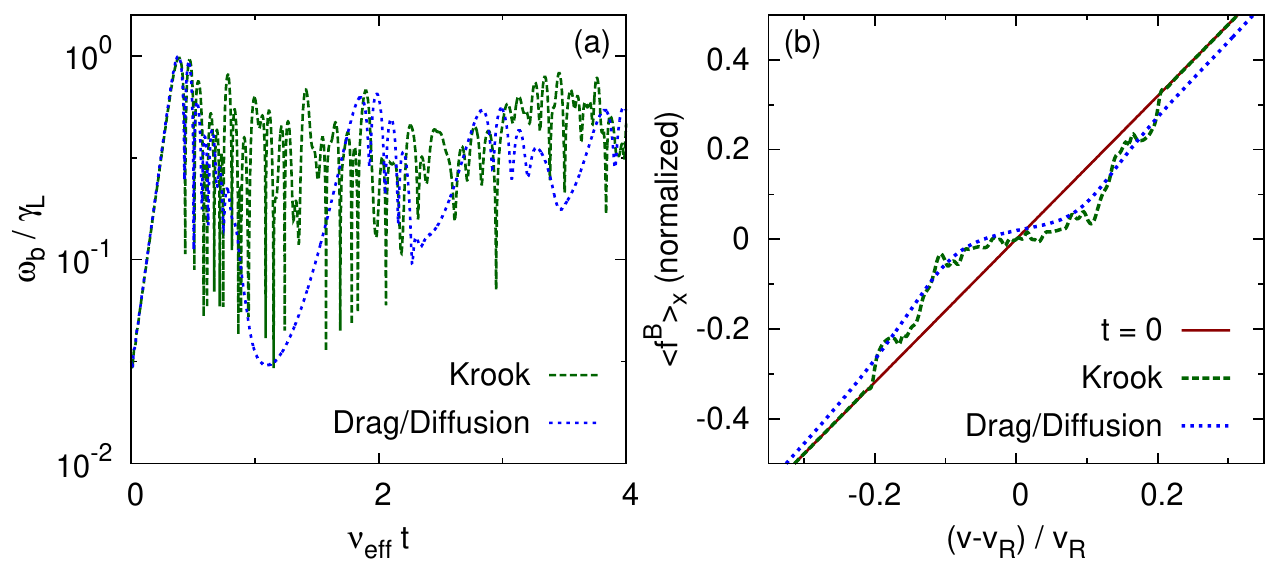}
\caption{(a) Time evolution of the bounce frequency. $\delta f$ simulations with $\gamma _{L0}=0.1$, $\gamma _d=0.05$, and $\nu _\mathrm{eff}=0.002$. In the drag/diffusion case, $\nu _f/\nu _d = 0.3$ (b) Distribution function, normalized to $\gamma _{L0} / \sqrt{2\pi}$. Solid line is the initial condition, dashed lines are for $\nu _\mathrm{eff} t=2$.}
\label{fig:periodic_chirping_amplitude_and_f0}
\end{center} \end{figure}

In some parameter regimes, chirping arising from the BB model with Krook collisions is quasi-periodic, although the phase between two major chirping events is generally not quiet, but filled with minor chirping events. In a regime where $\nu _f \ll \nu_d$, chirping arising from the BB model with drag/diffusion collisions is quasi-periodic too, but this time with clear quiescent phases in-between chirping events. In Fig.\ref{fig:periodic_chirping_amplitude_and_f0}(a), which shows periodic decay and recovery of perturbation amplitude, corresponding to major chirping events, we observe qualitatively different behavior between the two collision models. The velocity distribution after nonlinear saturation shown in Fig.\ref{fig:periodic_chirping_amplitude_and_f0}(b) illustrates the fact that several holes and clumps with different amplitudes co-exist in the Krook case, while diffusion smooths out fine-scale structures, which explains isolated chirping events we observe in the drag-diffusion case (See for example Fig.~\ref{fig:E32359}(c)). In both cases, no analytic theory has been developed that predicts the average time between two chirping events, $\Delta t_\mathrm{chirp}$. However, conceptually, there exists some relation with a subset of the input parameters. This concept is useful to develop diagnostic for TAEs in Chap.~\ref{ch:experiment}.

\begin{figure}
\begin{center}
\includegraphics{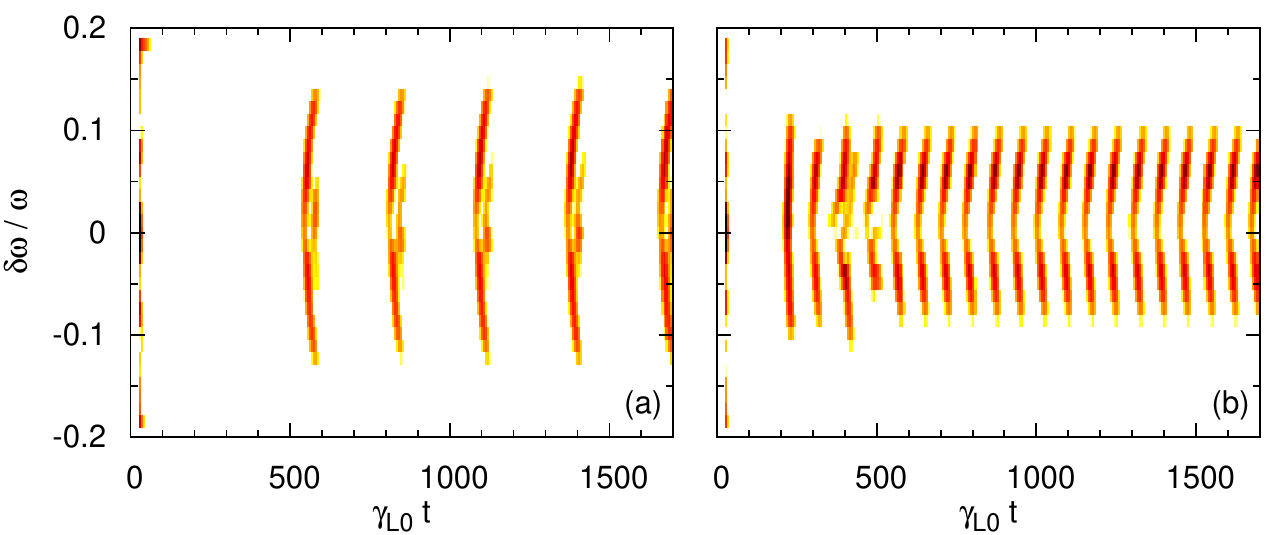}
\caption{Periodic chirping regime. Spectrograms of $\delta f$ COBBLES simulations, with input parameters $\gamma _{L0} = 0.099$, $\gamma _d=0.028$, and $\nu _f / \nu _d=0.2$. (a) Less collisions, $\nu _f=2.9\times 10^{-3}$, $\nu _d=1.5\times 10^{-2}$. (b) More collisions, $\nu _f=4.4\times 10^{-3}$, $\nu _d=2.2\times 10^{-2}$.}
\label{fig:periodic_chirping}
\end{center}
\end{figure}

%#Simulation started at : Date 07/07/2010 ; Time 10:33:31
%#nx = 00064 ; ny = 02048 ; single = 1
%#xk =   0.1000000E+01 ; vr =   0.1000000E+01 ; a =   0.1000000E-04 ; vmin =  -0.2
%#149668E+01 ; vmax =   0.4149668E+01 ; \mathrm{CFL} =   0.1056704E+01 ; gammalstar =   0.9
%#908883E-01 ; tfin =   0.3000000E+06 ; gamd =   0.2840426E-01 ; nua =   0.0000000
%#E+00 ; nuf =   0.2938514E-02 ; nuD =   0.1469257E-01 ; xl =   0.6283185E+01 ; dx
%# =   0.9817477E-01 ; dy =   0.3075848E-02 ; dt =   0.5000000E-01
%
%#Simulation started at : Date 07/07/2010 ; Time 10:33:31
%#nx = 00064 ; ny = 02048 ; single = 1
%#xk =   0.1000000E+01 ; vr =   0.1000000E+01 ; a =   0.1000000E-04 ; vmin =  -0.2
%#149668E+01 ; vmax =   0.4149668E+01 ; \mathrm{CFL} =   0.1056704E+01 ; gammalstar =   0.9
%#908883E-01 ; tfin =   0.3000000E+06 ; gamd =   0.2840426E-01 ; nua =   0.0000000
%#E+00 ; nuf =   0.4407771E-02 ; nuD =   0.2203886E-01 ; xl =   0.6283185E+01 ; dx
%# =   0.9817477E-01 ; dy =   0.3075848E-02 ; dt =   0.5000000E-01

\begin{figure}
\begin{center}
\includegraphics{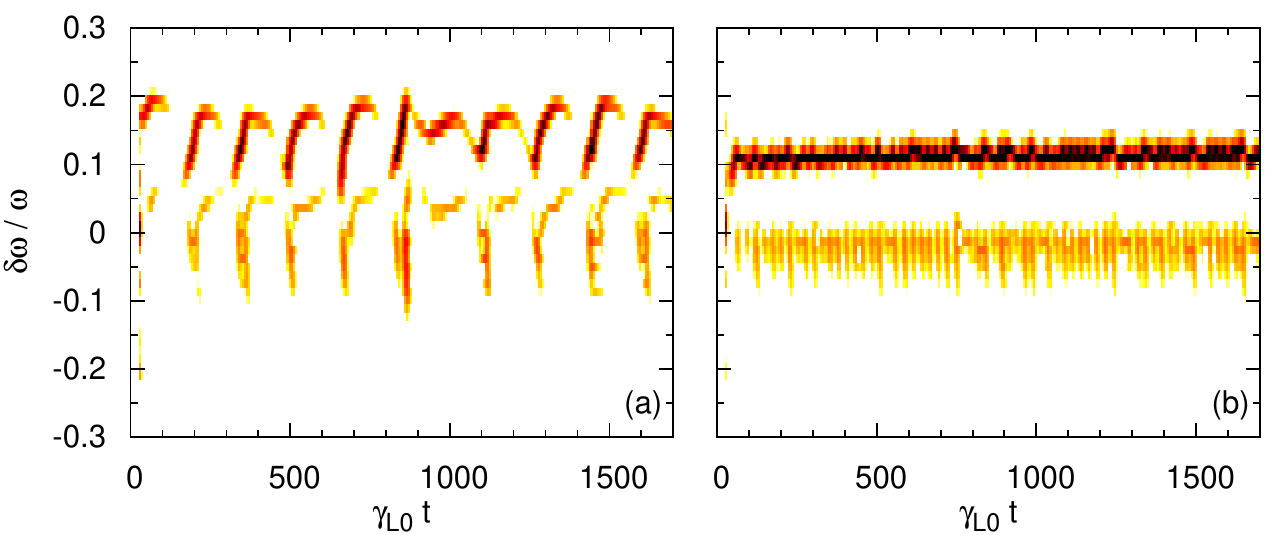}
\caption{Dragging regime. Spectrograms of $\delta f$ COBBLES simulations, with input parameters $\gamma _{L0} = 0.099$, $\gamma _d=0.028$, and $\nu _f / \nu _d=0.6$. (a) Less collisions, $\nu _f=9.7\times 10^{-3}$, $\nu _d=1.7\times 10^{-2}$. (b) More collisions, $\nu _f=1.5\times 10^{-2}$, $\nu _d=2.6\times 10^{-2}$.}
\label{fig:dragging_chirping}
\end{center}
\end{figure}

%#Simulation started at : Date 05/07/2010 ; Time 13:00:18
%#nx = 00064 ; ny = 02048 ; single = 1
%#xk =   0.1000000E+01 ; vr =   0.1000000E+01 ; a =   0.1000000E-04 ; vmin =  -0.2
%#149668E+01 ; vmax =   0.4149668E+01 ; \mathrm{CFL} =   0.1056704E+01 ; gammalstar =   0.9
%#908883E-01 ; tfin =   0.3000000E+06 ; gamd =   0.2840426E-01 ; nua =   0.0000000
%#E+00 ; nuf =   0.9723024E-02 ; nuD =   0.1652914E-01 ; xl =   0.6283185E+01 ; dx
%# =   0.9817477E-01 ; dy =   0.3075848E-02 ; dt =   0.5000000E-01
%
%#Simulation started at : Date 05/07/2010 ; Time 13:00:18
%#nx = 00064 ; ny = 02048 ; single = 1
%#xk =   0.1000000E+01 ; vr =   0.1000000E+01 ; a =   0.1000000E-04 ; vmin =  -0.2
%#149668E+01 ; vmax =   0.4149668E+01 ; \mathrm{CFL} =   0.1056704E+01 ; gammalstar =   0.9
%#908883E-01 ; tfin =   0.3000000E+06 ; gamd =   0.2840426E-01 ; nua =   0.0000000
%#E+00 ; nuf =   0.1512470E-01 ; nuD =   0.2571200E-01 ; xl =   0.6283185E+01 ; dx
%# =   0.9817477E-01 ; dy =   0.3075848E-02 ; dt =   0.5000000E-01

In the case with drag and diffusion, we found essentially two regimes, depending on the ratio $\nu _f / \nu _d$, with a threshold $r_{fd}$. When $\nu _f / \nu _d < r_{fd}$, the chirping instability is in a \textit{periodic chirping regime} as described above, with chirping events repeating at regular intervals and a clear quiescent phase between two successive chirping events. This behavior is illustrated in Fig.~\ref{fig:periodic_chirping}, which shows two spectrograms with $\nu _f / \nu _d=0.2$. In this regime, $\Delta t_\mathrm{chirp}$ decreases as drag is enhanced, as can be seen by comparing Fig.~\ref{fig:periodic_chirping}(a) with Fig.~\ref{fig:periodic_chirping}(b), since it accelerates the process of velocity-distribution recovery. However, $\nu _f$ is not the only factor that determines $\Delta t_\mathrm{chirp}$. If linear drive is decreased, more time is required to recover a gradient steep enough to overcome damping, hence $\Delta t_\mathrm{chirp}$ increases. Diffusion tends to recover the initial distribution, but much less efficiently than friction. It also tends to counteract formation of holes and clumps, hence $\Delta t_\mathrm{chirp}$ slightly increases with increasing $\nu _d$.
On the contrary, when $\nu _f / \nu _d > r_{fd}$, dynamical friction significantly modifies the shape of a chirping branch, up to a situation where the direction of a chirping hole is reversed. This is illustrated in Fig.~\ref{fig:dragging_chirping}, which shows two spectrograms with $\nu _f / \nu _d=0.6$. We refer to this latter regime as \textit{dragging regime}. Chirping such as in Fig.~\ref{fig:dragging_chirping}(a) has recently been observed and explained in Ref.~\cite{lilley10}. This is called as hooked chirping. In Fig.~\ref{fig:dragging_chirping}(b), drag and frequency sweeping contradicting effects seem to balance, creating a hole with a stable frequency shift.
Though we found a threshold $r_{fd}\sim 1$ in our range of investigation, more systematic scan remains to be done.

Note that in a case with diffusion only ($\nu _a=\nu _f=0$), after a first chirping event, the perturbation is damped to zero, and the gradient of $f^B$ at resonant velocity is significantly reduced from initial condition. The original gradient does not recover in a reasonable time for typical experimental values of $\nu _d$.
%If we use JT-60U diffusion coefficient, which is estimated in \ref{subs:analyse_e32359_dragdiff}, and normalize time with typical TAE frequency, a second chirping event in our simulations does not appear before $100$ ms.
Drag is an essential ingredient of repetitive chirping, since, by advecting $f-f_0$ from positive to negative velocities, it replaces, at the resonant velocity, a plateau by the large gradient which was formed at slightly larger velocity compared to the plateau.

%[Work needed: Estimation from power balance]

%\subsection{Chirping extent}
%
%\begin{itemize}
%\item Definition
%\item Theory (simply combining chirping velocity and lifetime results)
%\item Numerical validation
%\end{itemize}

%\subsection{Velocity-space avalanches}
%
%\begin{itemize}
%\item Is chirping an avalanche process ?
%\item Frequency of events vs amplitude
%\end{itemize}

\section{Subcritical instabilities\texorpdfstring{ ($\gamma < 0$)}{}\label{sec:subcritical}}

\begin{figure}
\begin{center}
\includegraphics{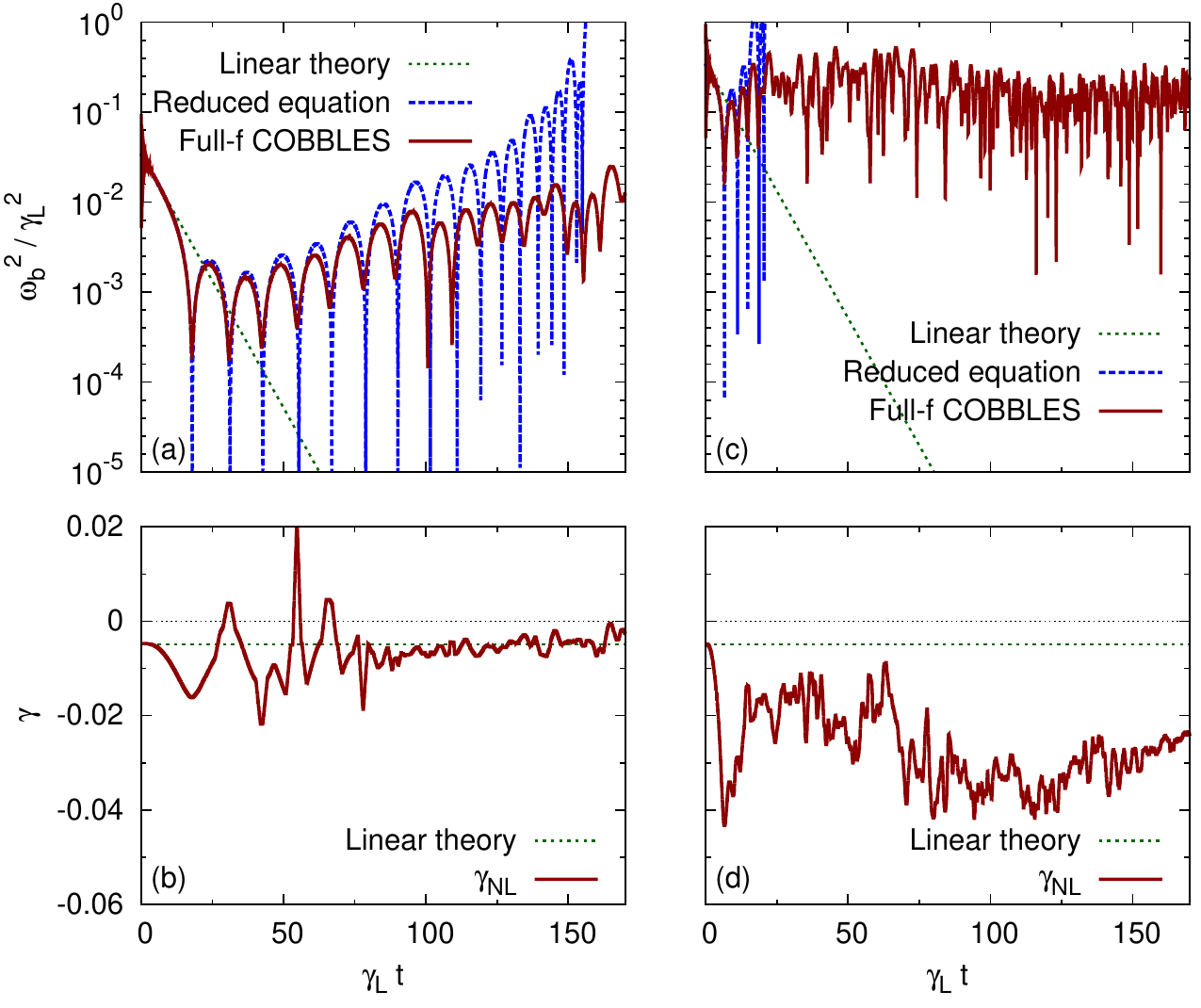}
\caption{(a) and (c) Time-evolution of electric field amplitude obtained by linear theory (dotted line), by solving Eq.~(\ref{eq:bb_integral_equation}) (dashed line), and by solving the full-$f$ initial value problem (solid line). (b) and (d) Dynamic estimation of \textit{nonlinear} growth rate (solid line). Dotted horizontal line is the value predicted by linear theory.}
\label{fig:subcritical_ff_and_reduced}
\end{center}
\end{figure}

Instabilities in a regime where the wave is linearly stable have been observed in $\delta f$ particle simulations \cite{berkbreizman99}. We recover such instabilities with both $\delta f$ and full-$f$ COBBLES. Fig.~\ref{fig:subcritical_ff_and_reduced}(a) and (c) show the time-evolution of electric field amplitude obtained for distribution B, with $\gamma _d = 0.042$ and $\nu _a = 2\times 10^{-4}$, and two different initial perturbation amplitude. Although at first the solution follows linear prediction, this trend reverses, consistently with Eq.~(\ref{eq:bb_integral_equation}), whose numerical solution is included in the figures. Such behavior seems to contradict linear theory, especially in Fig.~\ref{fig:subcritical_ff_and_reduced}(a), since the amplitude of perturbation at reversal clearly satisfies the linear ordering, $|f-f_0|/f_0\sim 10^{-3}$ in this case.

Subcritical solutions have also been identified with a different, two-species BB model, where thermal ion Landau damping is included in a self-consistent way rather than as a constant input parameter \cite{nguyen10prl}. In this case, both resonant drive and resonant damping are present, and subcriticality is explained in terms of a nonlinear reduction of damping by particle trapping of the second population. An interesting feature is the existence of metastable, or subcritical steady-state solutions, which were not observed in single-species BB simulations.
Since metastable modes exist for the two-species BB model in a regime consistent with BAEs in standard tokamak conditions \cite{nguyen10ppcf}, subcritical BAEs can reasonably be thought as possible. It is then natural to raise a similar question, that is the existence of TAEs in a subcritical regime. Besides, an experimental device must take a path that goes through the linearly stable region as chirping emerges. Thus it is important to understand the mechanism of nonlinear drive in this regime.

Here we show that, in our single-species BB model where external damping is treated as a fixed input parameter, subcriticality can be explained as an effect of holes and clumps.

\subsection{Mechanism}

Let us estimate the lowest-order correction to linear theory in the collisionless limit. Substituting $E(x,t)=\omega_b^2(t)\cos(\psi)/k$, where $\psi\equiv kx-\omega t + \varphi (t)$, into the power balance Eq.~(\ref{result:power_transfer_W}), yields
\begin{equation}
\frac{\mathrm{d} \omega _b^2}{\mathrm{d} t}\;+\; \omega \, \int f \,\cos(\psi) \frac{\mathrm{d} x}{L} \mathrm{d} v \;+\; \gamma _d \omega _b^2 \;=\;0,
\end{equation}
where the integral is taken over the resonant region. We decompose $f$ in Fourier space,
\begin{equation}
f(x,v,t) \;=\; \ov{f}(v,t) \;+\; \sum _{n=1}^{\infty} \left[ \hat{f}_n (v,t) e^{\imath n \psi} \,+\, \mathrm{c.c.} \right],
\end{equation}
which simplifies the power balance,
\begin{equation}
\frac{\mathrm{d} \omega _b^2}{\mathrm{d} t}\;=\; - \gamma _d \omega _b^2 \;-\; \omega \, \mathcal{R} \int \hat{f}_1 \, \mathrm{d} v, \label{eq:omegab_f1}
\end{equation}
where $\mathcal{R}$ denotes the real part.
Assuming the ordering $\ov{f}\gg \hat{f}_1 \gg\hat{f}_2$, the $n=1$ Fourier component of Vlasov equation yields
\begin{equation}
\frac{\partial \hat{f}_1}{\partial t} \;+\; \imath u \hat{f}_1 \;+\; \frac{\omega _b^2}{2} \, \frac{\partial \ov{f}}{\partial u} \;=\; 0,
\end{equation}
where $u\equiv kv-\omega$, whose solution is
\begin{equation}
\hat{f}_1 \;=\; - \int _0 ^t \frac{\omega _b^2(t')}{2} \, \frac{\partial \ov{f}}{\partial u}(u,t') e^{-\imath u (t-t')}  \mathrm{d} t'.
\end{equation}
Substituting the latter solution into Eq.~(\ref{eq:omegab_f1}) yields 
\begin{equation}
\frac{\mathrm{d} \omega _b^2}{\mathrm{d} t}\;=\; (\gamma _{L0}- \gamma _d) \omega _b^2 \;+\; \frac{\omega}{2k} \, \mathcal{R} \int _0 ^t \left[ \omega _b^2(t') \int \frac{\partial (\ov{f}-f_0) }{\partial u}(u,t') e^{-\imath u (t-t')} \mathrm{d} u \right] \mathrm{d} t' ,
\end{equation}
which is actually an intermediary step in the derivation of Eq.~(\ref{eq:bb_integral_equation}).
The latter equation formally shows that small deviations in the velocity distribution, which are not taken into account in linear theory, could build up in time and eventually significantly modifies linear prediction.
Physically, if we assume a fixed mode frequency, the drive is directly proportional to the velocity gradient in the resonant region. In Fig.~\ref{fig:subcritical_ff_and_reduced}(b) and (d), we estimate a \textit{nonlinear} growth rate $\gamma _\mathrm{NL}$ by measuring, at each time-step in the simulation, the slope of velocity distribution, averaged over a resonant region of width $\Delta v = 4\,\omega _b(t) / k$. We also average this value in time with a window $\Delta t=50$. We qualify $\gamma _\mathrm{NL}$ as nonlinear, in the sense that it takes into account modifications of distribution function, although it is not estimated in a self-consistent way. At initial time, $\gamma _\mathrm{NL}$ agrees with linear prediction, then it drops due to the flattening of the distribution. Afterwards, though there are cases where $\gamma _\mathrm{NL}$ becomes positive, as in Fig.~\ref{fig:subcritical_ff_and_reduced}(b), there are also cases where it stays negative, as in Fig.~\ref{fig:subcritical_ff_and_reduced}(d), which shows that we cannot assume a fixed mode frequency, since otherwise the wave would be stable in the latter case.

\begin{figure}
\begin{center}
\includegraphics{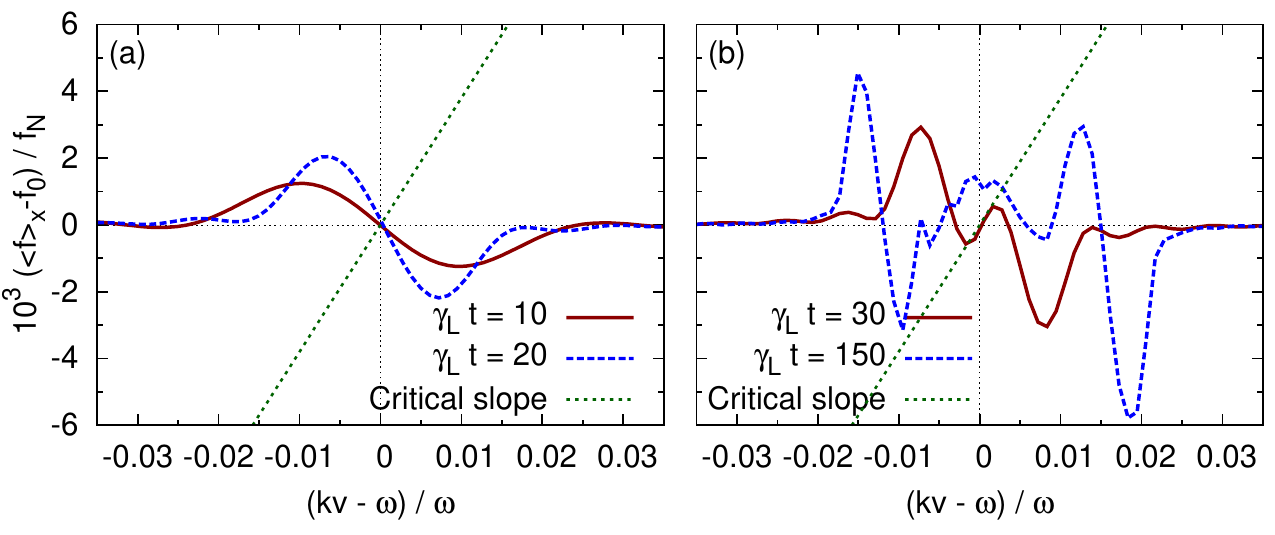}
\caption{Snapshots of deviation of velocity distribution from the initial one. Simulation parameters are the same as for Fig.~\ref{fig:subcritical_ff_and_reduced}(a) (we have increased $N_v$ to $8192$ in order to resolve small hole and clumps). Dotted line is a distribution such that $\gamma _{L0}=\gamma _d$.}
\label{fig:subcritical_ff_f0}
\end{center}
\end{figure}

Physically, subcritical instabilities can be explained by the following mechanism, which we illustrate in Fig.~\ref{fig:subcritical_ff_f0} by snapshots of $\ov{f}-f_0$ in the neighborhood of resonant velocity. Even if the initial perturbation is small, some particles are trapped and create a seed island ($\gamma _L t = 10 - 20$). As collisions try to recover the initial slope, small holes and clumps are created ($\gamma _L t = 30$). As holes and clumps shift their central velocity ($\gamma _L t = 150$), they enhance free-energy extraction compared to a situation where the mode frequency would be fixed \cite{berkbreizman95report}. This interpretation is consistent with the fact that we only observe subcritical instabilities with frequency sweeping. We do observe subcritical solutions categorized as chaotic, but even chaotic solutions actually feature slight frequency sweeping (See Fig.~\ref{fig:nonlinear_behaviors}(i)).
%if $\gamma _d=0$, then $\gamma<0$ means that the distribution slope is negative, which rules out the above mechanism

\subsection{Initial amplitude threshold}

\begin{figure}
\begin{center}
\includegraphics{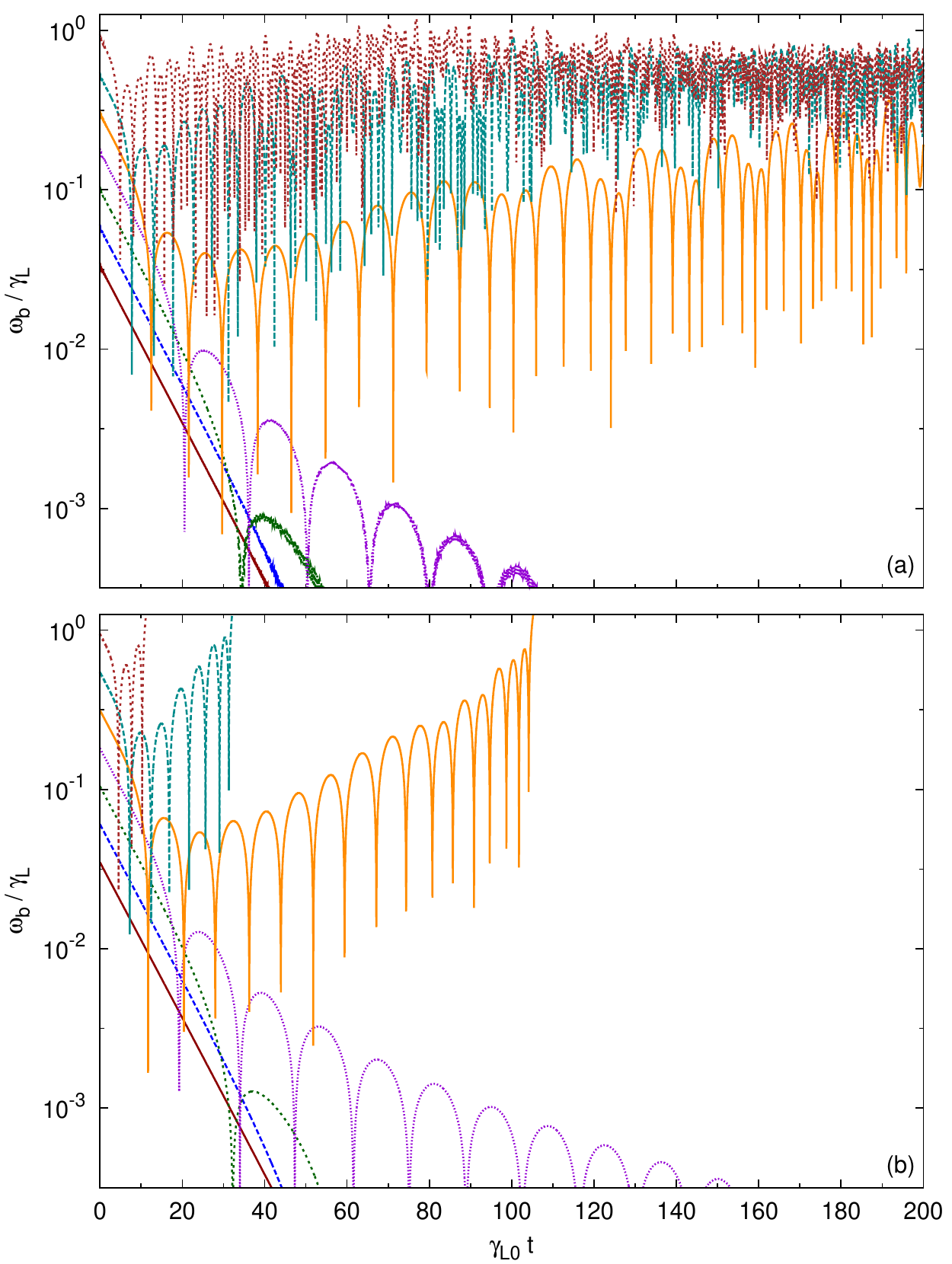}
\caption{Time-evolution of electric field amplitude in a subcritical regime, varying the initial perturbation amplitude. (a) $\delta f$-COBBLES simulations with $\gamma _{L0}=0.05$, $\gamma _d=0.062$, $\nu _a=10^{-4}$, $\mathrm{CFL}=0.65$. (b) Numerical solutions of Eq.~(\ref{eq:bb_integral_equation}).}
\label{fig:subcritical_threshold}
\end{center}
\end{figure}
%#nx = 00128 ; ny = 04096 ; single = 1
%#xk =   0.1000000E+01 ; vr =   0.1000000E+01 ; a =   0.2469150E-03 ; vmin =  -0.5
%#626196E+00 ; vmax =   0.2562620E+01 ; \mathrm{CFL} =   0.6525658E+00 ; gammalstar =   0.5
%#000000E-01 ; tfin =   0.3000000E+06 ; gamd =   0.6200000E-01 ; nua =   0.1000000
%#E-03 ; nuf =   0.0000000E+00 ; nuD =   0.0000000E+00 ; xl =   0.6283185E+01 ; dx
%# =   0.4908739E-01 ; dy =   0.7629979E-03 ; dt =   0.2500000E-01
%gaml = 0.04724

Ref.~\cite{berkbreizman99} gives a condition for subcritical instabilities to take off, as
\begin{equation}
\omega _b^2 \gtrsim \frac{(\nu _\mathrm{eff}-\gamma)^{5/2}}{\gamma _{L0}^{1/2}}, \label{eq:sub_thr}
\end{equation}
which is obtained by a dimensional analysis.
We investigate this threshold with a series of $\delta f$ simulations with different initial perturbation amplitudes. A large value is taken for $N_v$, so that recurrence time is long enough (here $T_R=412/\gamma _{L0}$).
In Fig.~\ref{fig:subcritical_threshold}(a), we observe a clear qualitative difference between solutions above and below $\omega _b(t=0)/\gamma _{L0} \sim 0.2$. This threshold is consistent with numerical solutions of Eq.~(\ref{eq:bb_integral_equation}) shown in Fig.~\ref{fig:subcritical_threshold}(b). It is also in qualitative agreement with Eq.~(\ref{eq:sub_thr}), which gives $\omega _b(t=0)/\gamma _{L0} \approx 0.15$, with $\gamma = -0.011$.
The agreement between Fig.~\ref{fig:subcritical_threshold}(a) and Fig.~\ref{fig:subcritical_threshold}(b) suggests that all information necessary to obtain an analytic prediction for the threshold, more precise than Eq.~(\ref{eq:sub_thr}), are included in the reduced equation, Eq.~(\ref{eq:bb_integral_equation}).

%We investigate this threshold by launching a series of simulations, with initial distribution B, $\gamma _d=0.042$, $\nu _a = 2\times 10^{-4}$, $N_x\times N_v=64\times 8192$ (so that recurrence time is $T_R=198/\gamma _L$), and $\mathrm{CFL}=2.8$, varying only the initial perturbation amplitude.
%In Fig.~\ref{fig:subcritical_threshold}(a), we observe that the instability threshold is $\omega _b(t=0)/\gamma _L \sim 0.1$ while Eq.~(\ref{eq:sub_thr}) gives $\omega _b(t=0)/\gamma _L \approx 0.015$, since $\gamma = -0.0042$. This discrepancy suggests that a more rigorous analytic treatment is needed.

\chapter{Spectroscopic analysis of chirping TAEs}
\label{ch:experiment}

As long as the background plasma parameters are not significantly changed, chirping events in most tokamak experiments are quasi-periodic, with a quiescent phase between two chirping branches, which lasts a few milliseconds. It should be noted that this statement does not seem to apply to DIII-D \cite{heidbrink95}.
The spectrogram of a quasi-periodic chirping solution contains many pieces of information, which include chirping velocity, lifetime, quasi-period and asymmetry. This chapter shows how to use these information as diagnostics. We apply the BB model to AEs with frequency that sweeps only $\sim 10\%$ of the linear frequency, so that we can reasonably assume that no phase-space structure interacts with the bulk plasma. This situation is consistent with the $\delta f$ approach, where thermal populations are assumed adiabatic. In addition, the choice of $\delta f$ model removes complications of the full-$f$ approach that are due to effects of bulk.

We consider TAEs in a time interval during which quasi-periodic, perturbative chirping is observed, and during which background plasma parameters are not significantly changed. We assume a weak drive by energetic particles, such that chirping velocity is determined by a natural evolution of phase-space structures, rather than by a forcing from turbulent transport. We further assume that frequency shifting occurs well within the gap of Alfv\'en continuum, so that chirping lifetime is determined by collision processes, rather than by continuum damping.

In the spectrogram of magnetic field perturbations measured by a Mirnov coil at plasma edge, we extract the linear mode frequency $f_\mathrm{A}$, the average chirping velocity, $d\delta \omega ^2 / dt$, the maximum chirping lifetime, $\tau _\mathrm{max}$, and the average chirping period, $\Delta t_\mathrm{chirp}$. The goal is to find the input parameters for $\delta f$ COBBLES simulations that are such that chirping shows similar features than those observed in experiments. When we compare simulation and experimental results, we simply re-normalize time by $\omega _\mathrm{A} = 2\pi f_\mathrm{A}$, where $f_\mathrm{A}$ is the TAE linear frequency. In this chapter, we adopt a convention of the EP literature, where growth rates and collision frequencies are given in percentage of $\omega _\mathrm{A}$. In the remaining of this work, all simulations are performed with $N_x \times N_v = 64 \times 2048$ grid points, and a time-step width $\Delta t=0.05$, unless stated otherwise. A large number of grid points in velocity-space is necessary to avoid recurrence effect during quiescent phases between chirping events.

\section{Fitting procedure\label{subsec:analysis_method}}

In the previous chapter, we saw that if we normalize time with the frequency of the mode, then chirping velocity, lifetime and period are dictated by the input parameters of the model, $\gamma _{L0}$, $\gamma _d$, and, either $\nu _a$, either $\nu _f$ and $\nu _d$. In the Krook case, we have a 3-variables, 3-equations system, which we solve by a fitting procedure described in \ref{subs:fitting_krook}. With drag and diffusion, there is one additional degree of freedom, hence the solution is not unique, but the boundaries of chirping regime limit the possible range of input parameters.

Eq.~(\ref{eq:beta}) gives a relation between linear drive and external damping,
\begin{equation}
\gamma _{L0} ^2 \, \gamma _d \;=\; \dfrac{1}{\alpha ^2 \beta ^2} \, \dfrac{d\delta \omega ^2}{dt}. \label{eq:gams_from_spectro}
\end{equation}
With the Krook model, chirping is limited to a range where $0.2 < \gamma _d / \gamma _{L0} < 1.1$ \cite{lesur09}. We found a similar constraint in our simulations with drag and diffusion, although a full scan of parameter space remains to be done. From this observation, in both cases, $\gamma _{L0}$ is given within roughly $30\%$ error, and $\gamma _d$ within $50\%$ error, by
\begin{eqnarray}
\gamma _{L0} &\approx & 1.3\, \left(\dfrac{1}{\alpha ^2 \beta ^2} \,\dfrac{d\delta \omega ^2}{dt}\right)^{\frac{1}{3}}, \label{eq:approx_gaml}\\
\gamma _{d} &\approx & 0.7\, \left(\dfrac{1}{\alpha ^2 \beta ^2} \,\dfrac{d\delta \omega ^2}{dt}\right)^{\frac{1}{3}}.\label{eq:approx_gamd}
\end{eqnarray}
We refine these estimations in a manner that depends on the collision model we adopt.

% More precisely:
% 1.34
% 0.70

% $V$, defined as
%\begin{equation}
%$V \;\equiv\; \sqrt{\dfrac{2 \norm{\delta \omega}}{\alpha ^2} \, \dfrac{d\norm{\delta \omega}}{d\norm{t}}$.
%\end{equation}

\subsection{With Krook collisions\label{subs:fitting_krook}}

The analysis described here aims at estimating the values of $\gamma _{L0}$, $\gamma _d$, and $\nu _a$ for which the $\delta f$ BB model fits experimental observations in terms of chirping characteristics.
Eq.~(\ref{eq:chirping_lifetime_krook}) yields the effective collision frequency,
\begin{equation}
\nu _a \;=\; \dfrac{\iota _a}{\tau _\mathrm{max}}. \label{eq:nua_from_spectro}
\end{equation}
Note that this effective collision frequency is meaningful only in the framework of a modelisation by the simple Krook operator of all dissipative processes, namely particle collisions, particle source and particle sink. Thus, $\nu _a$ can not be quantitatively compared with experimental measurements of collision frequency, unless particle source and sink terms are fully identified as well.
Eqs.~(\ref{eq:gams_from_spectro}) and (\ref{eq:nua_from_spectro}) form a system of two equations with three unknowns. The remaining unknown is found by fitting the chirping period. In our simulations, the chirping period is estimated by searching for the dominant frequency in the Fourier spectrum of electric field amplitude. To ensure a reasonable accuracy, simulations are performed for a time $t \gg \Delta t_\mathrm{chirp}$.
If the experiment belongs to a regime where $\beta=1$, the above procedure is systematic. However, if $\beta$ is significantly smaller than unity, an iterative procedure is needed, with a feedback between $\beta$ and $\gamma _{L0} ^2 \, \gamma _d$.\\

\subsection{With drag and diffusion}

\begin{figure} \begin{center}
\includegraphics[scale=0.6]{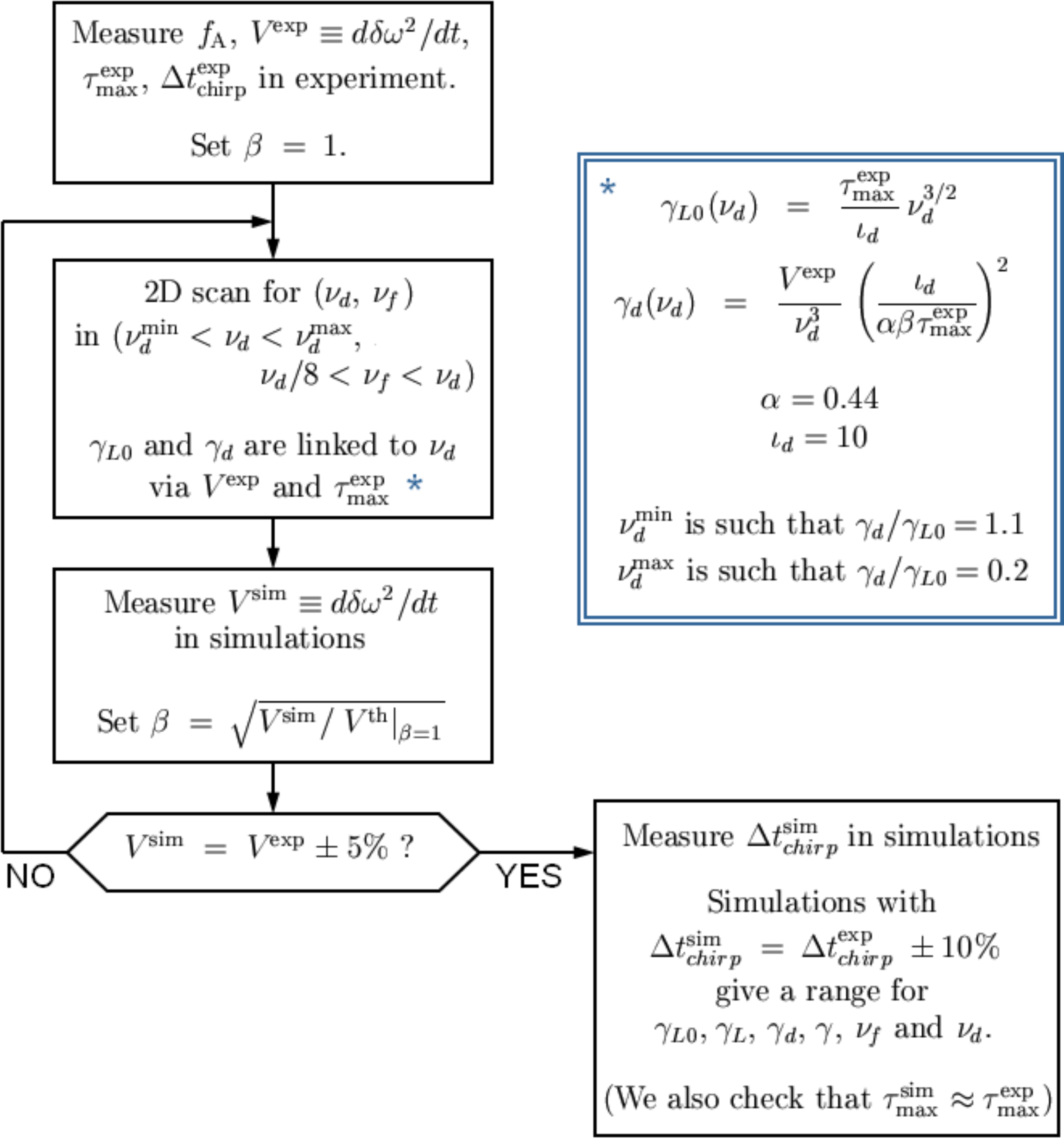}
\caption{Flowchart of chirping features fitting procedure in the drag/diffusion case.}
\label{fig:fitting}
\end{center} \end{figure}

The analysis described here aims at estimating the values of $\gamma _{L0}$, $\gamma _d$, $\nu _f$ and $\nu _d$ for which the $\delta f$ BB model fits experimental observations.
Eqs.~(\ref{eq:chirping_lifetime_dragdiff}) and (\ref{eq:gams_from_spectro}) form a system of two equations with four unknowns. 
The boundaries of chirping regime yield an estimation of $\nu _d$ within $20\%$ error,
\begin{equation}
\nu _d \;\approx \; 1.2 \,\left( \dfrac{\iota _d}{\tau _\mathrm{max}} \right) ^{\frac{2}{3}} \, \left(\dfrac{1}{\alpha ^2 \beta ^2} \,\dfrac{d\delta \omega ^2}{dt}\right)^{\frac{2}{9}}.\label{eq:approx_nud}
\end{equation}
On the one hand, it is shown in Ref.~\cite{lilley09} that for typical neutral beam-heated experiments, the ratio $\nu _d / \nu _f$ is of the order of unity. On the other hand, when $\nu _f \ge \nu _d$, we leave the periodic chirping regime for the dragging regime. Thus the relevant range for friction is $\nu _f \lesssim \nu _d$. In this regime, $\Delta t_\mathrm{chirp}$ increases with decreasing $\nu _f$, $\gamma$, and increasing $\nu _d$.
To obtain $\nu _f$ and refine the above estimations, we need a two dimensional scan in ($\nu _f$, $\nu _d$), where we search for solutions that fit the chirping period. In general, $\beta\neq 1$, and trial-and-errors are required to adjust chirping velocity to the experimental value. This procedure is summarized as a flowchart in Fig.~\ref{fig:fitting}.

% More precisely:
% 1.20

\section{Application to MAST\label{sec:analysis_mast}}

\begin{figure} \begin{center}
\includegraphics{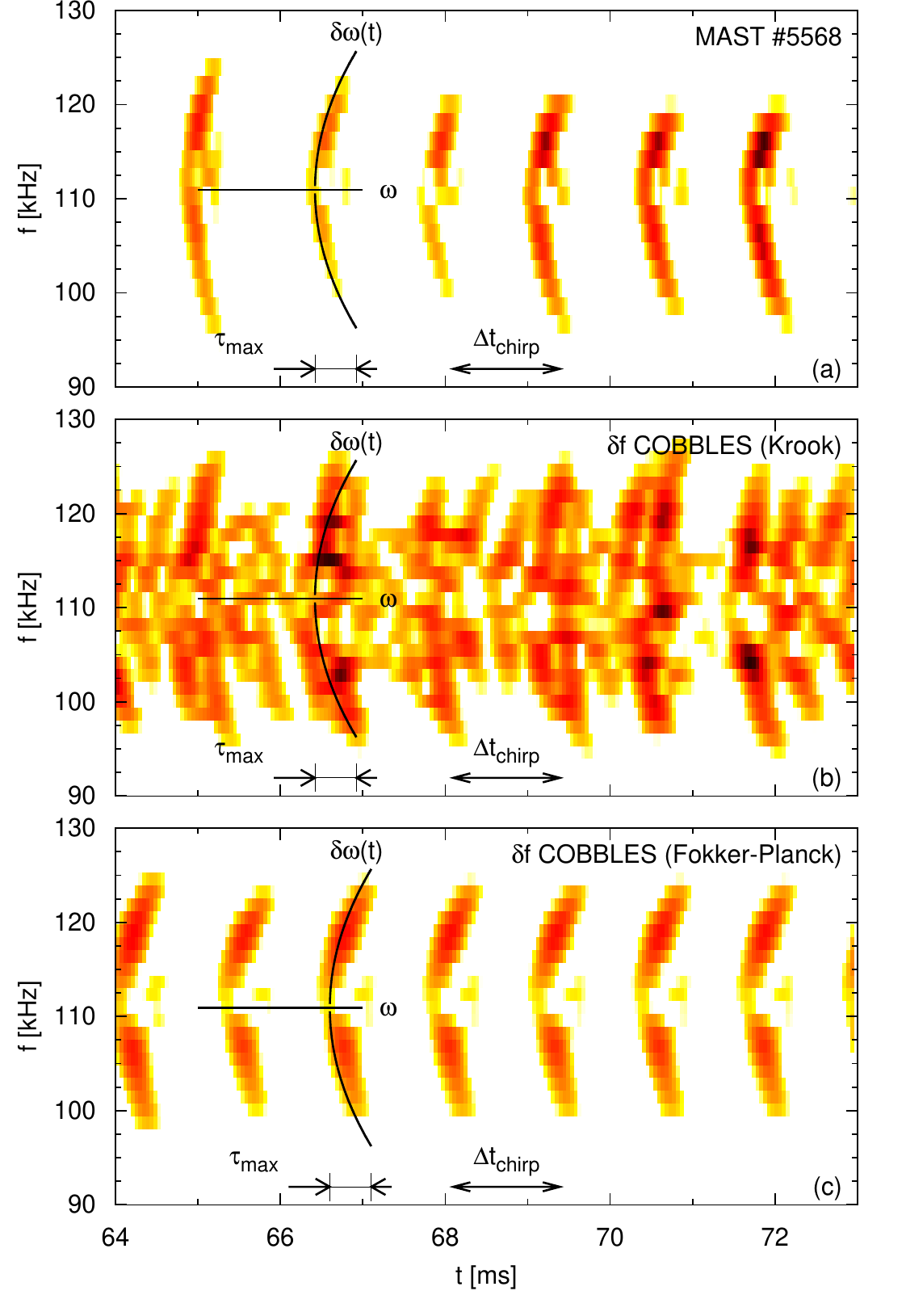}
\caption{(a) Spectrogram of magnetic fluctuations in MAST discharge $\#$5568, obtained with a moving Fourier window of size $2$ ms. (b) and (c) Spectrogram of the electric field, obtained with a moving Fourier window of size $0.6$ ms, where the input parameters of the $\delta f$ BB model were chosen to fit the magnetic spectrogram for MAST discharge $\#$5568. A solid curve shows the analytic prediction for chirping velocity and lifetime. (b) Krook collisions, correction parameter $\beta=0.85$. (b) Friction-diffusion collisions, correction parameter $\beta=0.65$.}
\label{fig:MAST5568}
\end{center} \end{figure}

The frequency sweeping mode in the MAST discharge $\#$5568 between $64$ and $73$ ms has been identified as a global $n=1$ TAE \cite{pinches04S47}. In the magnetic spectrogram shown in Fig.~\ref{fig:MAST5568}(a), we measure $f_\mathrm{A} = 111$ kHz, $d\delta \omega^2 / dt = 4.6 \times 10^{-5}$, $\tau _\mathrm{max} = 0.35\times 10^{3}$, and $\Delta t_\mathrm{chirp} = 0.9\times 10^{3}$ (on average).

\subsection{With Krook collisions}

Eqs.~(\ref{eq:gams_from_spectro}) and (\ref{eq:nua_from_spectro}) yield $\nu_a = 3.1 \times 10^{-3}$, and $\gamma _{L0} \sqrt{\gamma _d} = 1.5\times 10^{-2}$. 
However, the results of our analysis suggest that the plasma belongs to a regime where $\beta = 0.85$, hence we adjust the value of the product $\gamma _{L0} \sqrt{\gamma _d}$ to $1.5\times 10^{-2}/0.85\,=\,1.8\times 10^{-2}$.

\begin{figure} \begin{center}
\includegraphics{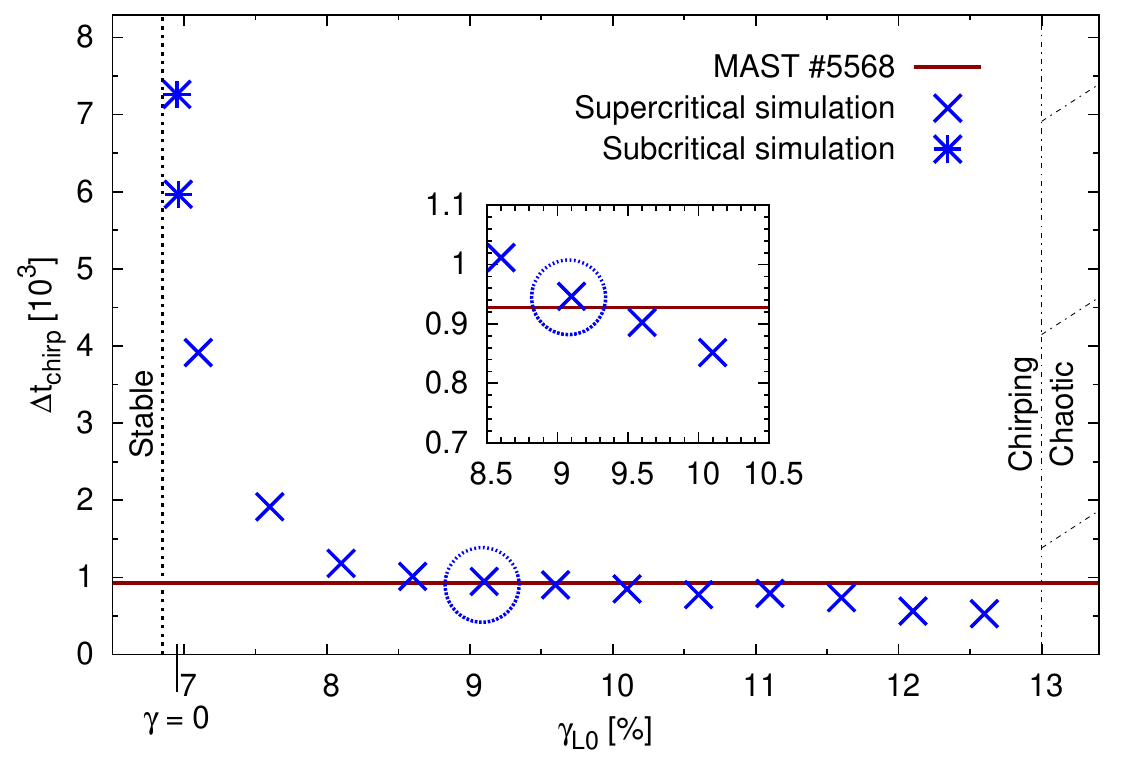}
\caption{Scan of chirping quasi-period for the set of parameters corresponding to MAST discharge $\#$5568. The hatched region correspond to a chaotic regime without frequency sweeping. The linear stability threshold is indicated as $\gamma = 0$, and the nonlinear stability threshold by a vertical dashed line. The chirping quasi-period agrees with the experiment for $\gamma _L \approx 7.7\%$. Inset: zoom of the most relevant region. The spectrogram for the circled simulation is shown in Fig.~\ref{fig:MAST5568}(b).}
\label{fig:scanMAST}
\end{center} \end{figure}

A scan for this set of parameters is performed by changing the slope of the distribution. Fig.~\ref{fig:scanMAST} shows that the chirping quasi-period depends on $\gamma _L$ in a roughly monotonous way. Note that the above scan needs to be performed on a relatively narrow range of $\gamma _L$, since the limits of subcritical regime and non-chirping (chaotic) regime yield a first estimate as $\gamma _L \sim 6-10 \%$ , $\gamma _d \sim 2-7 \%$. Here, the nonlinear stability threshold is defined as the largest value of $\gamma _L$ for which the electric field amplitude vanishes in the time-asymptotic limit, independently of the initial perturbation amplitude; the chaotic regime is defined and categorized in a way described in Sec.~\ref{subs:benchmark}. We observe that the two-points correlation of electric field amplitude decreases as the system approaches marginality.

Fig.~\ref{fig:MAST5568}(b) is the spectrogram for the simulation which is emphasized by a circle in Fig.~\ref{fig:scanMAST}. The features of main chirping events agree with the experimental observation. However, we observe a series of minor chirping events in between, which are absent in the experimental spectrogram. Another caveat is that only symmetric chirping is observed with the $\delta f$ BB model with Krook collisions and a linear velocity distribution. Thus, the application of this method with Krook collisions is restricted to symmetric or nearly-symmetric chirping experiments.
Linear parameters estimated from this analysis are shown in Tab.~\ref{tab:params}. Our analysis suggests that the TAE in this discharge is above marginal stability, with $\gamma \sim \gamma _L$.

Eq.~(\ref{eq:chirping_steadystate}) yields a saturated bounce frequency as $\omega _b \approx 4.9\%$, which is satisfied in this simulation. In Ref.~\cite{pinches04S47}, the peak amplitude of magnetic perturbation was determined with edge magnetic perturbation amplitudes and linear eigenfunctions calculated by the MISHKA code \cite{mikhailovskii97}, and a scaling between $\omega _b$ and the mode amplitude was obtained with the HAGIS code. According to the latter analysis, $\omega _b \approx 5.4\%$, which is consistent with our result. However, this agreement is not enough to validate our procedure.

\subsection{With drag and diffusion}

\begin{figure} \begin{center}
\includegraphics{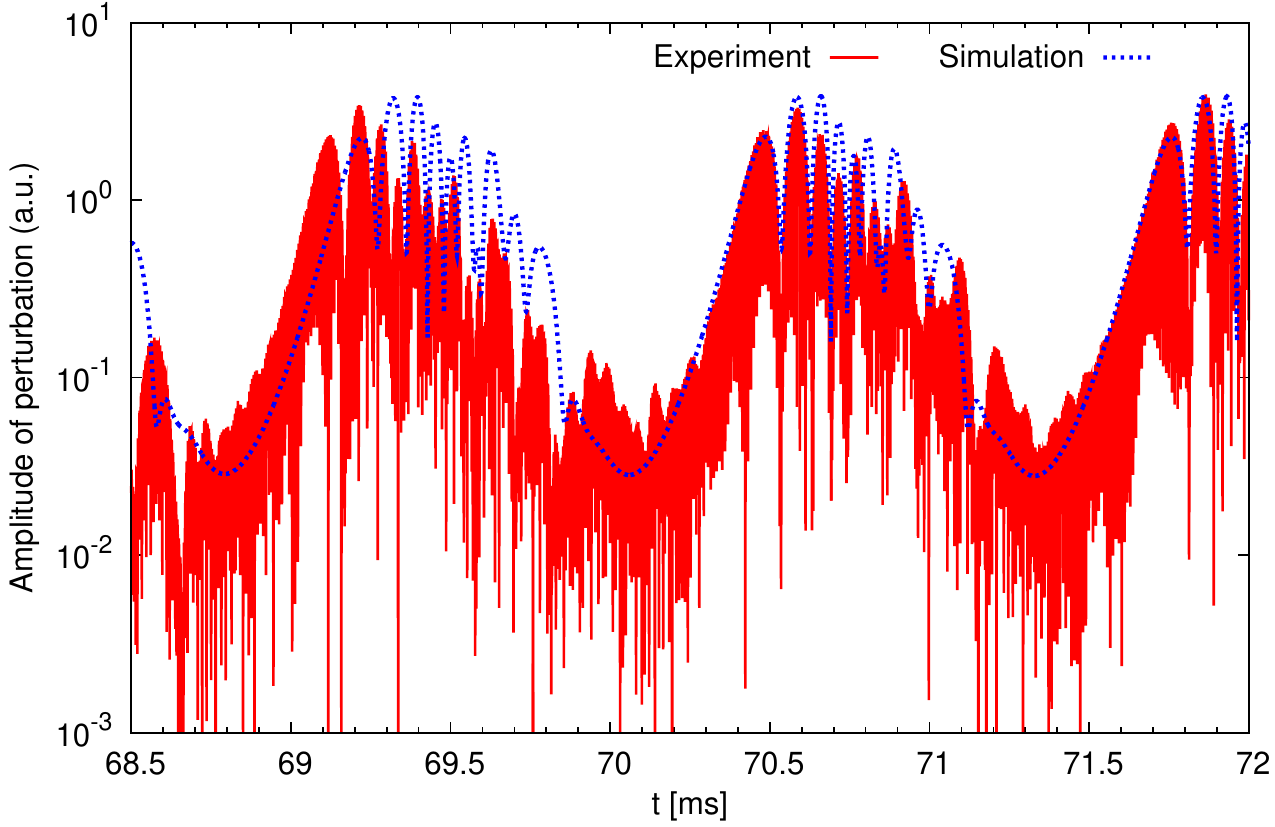}
\caption{Time-evolution of the perturbation. The signal is filtered between $90$ and $130$ kHz. The parameters of the simulation are shown in the second line of Tab.~\ref{tab:params}. For the simulation, to avoid hiding experimental data, we show the amplitude of perturbations (the envelope) instead of the perturbations themselves. Note the use of arbitrary units (we compare normalized quantities only).}
\label{fig:MAST5568_amplitude_dragdiff}
\end{center} \end{figure}

We perform a first, rough scan in ($\nu _f$, $\nu_d$) parameter space, assuming $\beta=1$. Measuring average chirping velocity in repetitive chirping solutions yields an estimation of the correction parameter, $\beta = 0.65$. Then Eqs.~(\ref{eq:approx_gaml}), (\ref{eq:approx_gamd}) and (\ref{eq:approx_nud}) yield $\gamma _{L0}=7.1^{\pm 2}  \%$, $\gamma _d=3.8^{\pm 2} \%$, and $\nu _d=1.6^{\pm 0.3} \%$.
We perform a second, more careful scan, which consists of a series of $4\times 8$ simulations in the domain ($1.3 \%\le \nu _d \le 1.9\%$, $1 \le \nu _d /\nu _f \le 8$), where $\gamma _{L0}$ and $\gamma _d$ are constrained by Eqs.~(\ref{eq:chirping_lifetime_dragdiff}) and (\ref{eq:gams_from_spectro}). The only repetitive chirping solution with $800 < \Delta t_\mathrm{chirp} < 1000$ we found is shown in Fig.~\ref{fig:MAST5568}(c). We verify that chirping features measured in this simulation fit the experiment, $d\delta \omega^2 / dt = 4.7\times 10^{-5}$ ($4.6\times 10^{-5}$ for up-chirping, $4.8\times 10^{-5}$ for down-chirping), $\tau _\mathrm{max} = 0.35\times 10^{3}$ ($0.36\times 10^{3}$ for up-chirping, $0.34\times 10^{3}$ for down-chirping), and $\Delta t_\mathrm{chirp} = 0.9\times 10^{3}$. The estimated linear parameters are shown in Tab.~\ref{tab:params}. Other solutions could be found with additional simulations in the same domain, but the latter estimations are quite accurate because of the narrow range of periodic chirping regime. To validate this analysis, we compare the amplitude of perturbations in Fig.~\ref{fig:MAST5568_amplitude_dragdiff}. Since the growth rate of chirping structure is neither $\gamma$ nor $\gamma _L$, and the decay rate not simply $\gamma _d$, but a function of several linear parameters, the agreement we obtain is not trivial (We measure a growth rate of $2.3\%$, and a decay rate of $0.6\%$).

Eq.~(\ref{eq:chirping_steadystate}) yields a saturated bounce frequency as $\omega _b \approx 6.0\%$, which is also in agreement with the value of $5.4\%$ estimated in Ref.~\cite{pinches04S47}.

In principle it should be possible to perform an independant estimation of collision frequencies using equilibrium parameters. However, some ambiguities in experimental data prevents such verification until more information are obtained.

%\begin{table*} \begin{center}
%\caption{Frequencies and growth rates estimated from the magnetic spectrogram of chirping TAEs.}
%\begin{tabular}{l l c c c c c c c}
%\hline \hline
%Experiment & Collision model & $\gamma _{L0}$ [$\%$] & $\gamma _L$ [$\%$] & $\gamma _d$ [$\%$] & $\nu_a$ [$\%$] & $\nu_f$ [$\%$] & $\nu_d$ [$\%$] & $\gamma$ [$\%$]  \\
%\hline
%MAST $\#$5568 & Krook & 8.1 & 7.7 & 3.6 & 0.31 & $\ldots$ & $\ldots$ & 4.8 \\ 
%MAST $\#$5568 & Friction-diffusion & 11.2 & 10.5 & 4.4 & $\ldots$ & 0.57 & 2.2 & 6.0 \\ 
%JT-60U E32359 & Krook & 9.4 & 8.5 & 8.6 & 0.25 & $\ldots$ & $\ldots$ & 0.7 \\ 
%JT-60U E32359 & Friction-diffusion & 9.8 & 9.2 & 4.7 & $\ldots$ & 0.36 & 1.7 & 4.6 \\ 
%\hline  \hline
%\end{tabular} 
%\label{tab:params}
%\end{center} \end{table*}

\newlength{\collwidth}
\settowidth{\collwidth}{riction-diffusi}
\newlength{\fcollwidth}
\settowidth{\fcollwidth}{Experimen}

\begin{table*} \begin{center}
\caption{Frequencies and growth rates estimated from the magnetic spectrogram of chirping TAEs, in percentage of the linear frequency $\omega _\mathrm{A}$.}
\begin{tabular}{l c c c c}
\hline \hline
\makebox[\fcollwidth]{Experiment} & \multicolumn{2}{c}{MAST $\#$5568} & \multicolumn{2}{c}{JT-60U E32359} \\
\makebox[\fcollwidth]{Coll.~model} & \makebox[\collwidth]{Krook} & Friction-diffusion & \makebox[\collwidth]{Krook} & Friction-diffusion \\
\hline
$\gamma _{L0}$ & 8.1 & 11.2 & 9.4 & 9.8 \\
$\gamma _L$ & 7.7 & 10.5 & 8.5 & 9.2 \\
$\gamma _d$ & 3.6 & 4.4 & 8.6 & 4.7 \\
$\nu_a$ & 0.31 & $\ldots$ & 0.25 & $\ldots$ \\
$\nu_f$ & $\ldots$ & 0.57 & $\ldots$ & 0.36 \\
$\nu_d$ & $\ldots$ & 2.2 & $\ldots$ & 1.7 \\
$\gamma$ & 4.8 & 6.0 & 0.7 & 4.6 \\ 
\hline  \hline
\end{tabular} 
\label{tab:params}
\end{center} \end{table*}

\section{Application to JT-60U\label{sec:analysis_jt60u}}

\begin{figure} \begin{center}
\includegraphics[width=\textwidth]{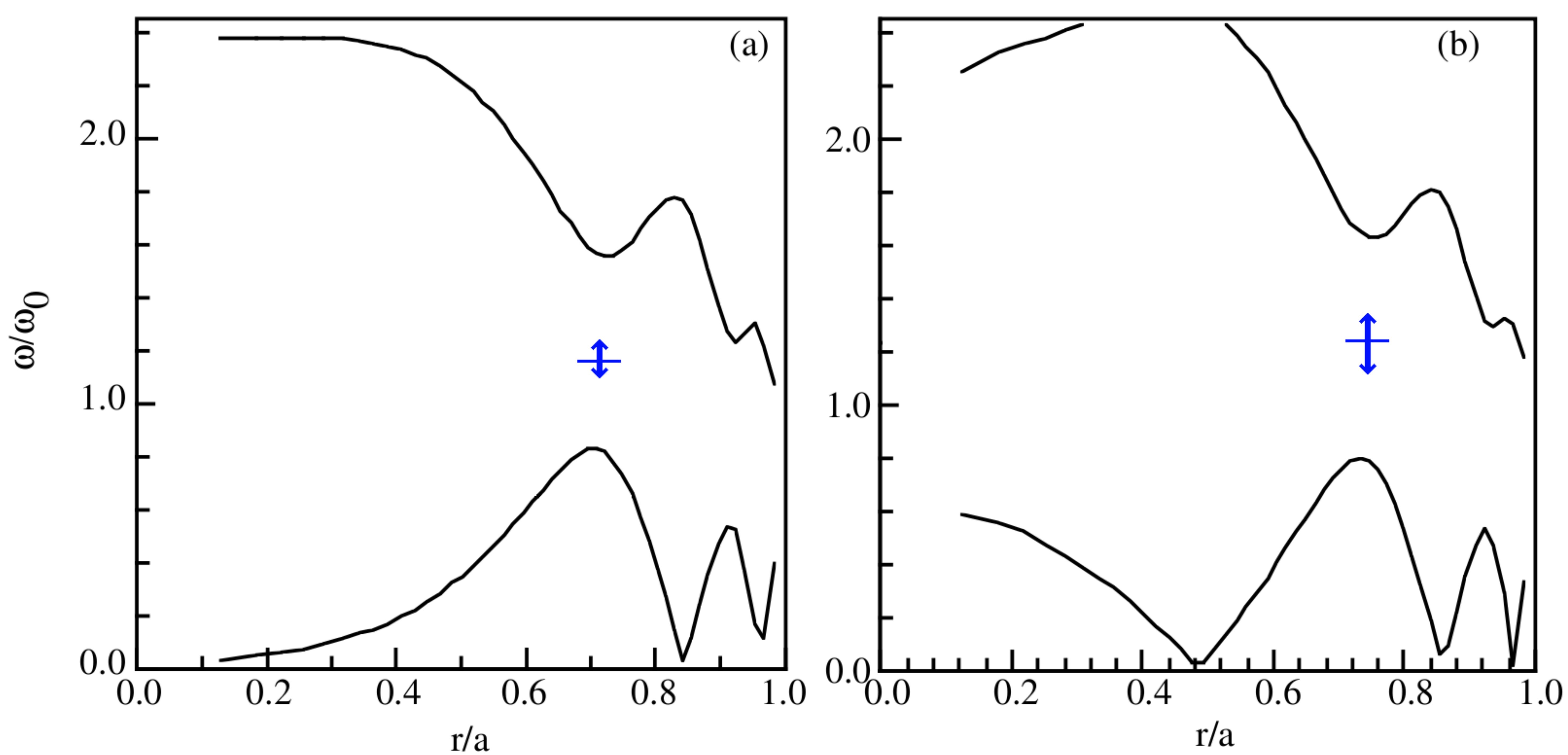}
\caption{SAW continuum for $n=1$, reproduced from Figs.~3 and 4 of Ref.~\cite{shinohara01}, normalized to $\omega _0 = 2\pi f_0$, to which we superpose a horizontal segment at the TAE linear frequency, and a vertical double-arrow, which represents the extent of chirping. (a) At $t=3.65$, $f_0=49$kHz. (b) At $t=3.75$, $f_0=39$kHz.}
\label{fig:continuum_E36378}
\end{center} \end{figure}

In JT-60U, TAEs are destabilized by a negative ion-based neutral beam (N-NB), which injects deuterons at $E_b=360$ keV.
%In the spectrogram shown in Fig.~\ref{fig:E32359}(a), we measure $\norm{\delta \omega _\mathrm{max}} \,=\, 0.16$, $\norm{\Delta t_\mathrm{chirp}} \,=\, 3100$, and $d\norm{\delta \omega}^2 / d\norm{t} \,=\, 63\times 10^{-6}$.
In the discharge E32359, around $t=4.2$ s, fast-FS modes have been identified as $m/n=2/1$ and $3/1$ TAEs \cite{kusama99}. In the spectrogram shown in Fig.~\ref{fig:E32359}(a), we measure $f_\mathrm{A} = 53$ kHz, $d\delta \omega^2 / dt = 6.3\times 10^{-5}$, $\tau _\mathrm{max} = 0.44\times 10^{3}$, and $\Delta t_\mathrm{chirp} = 3\times 10^{3}$ (on average).
Frequency sweeps between $43$ and $62$ kHz. By comparing these frequencies with the SAW continuum gap in Fig~\ref{fig:continuum}, it looks as if up-chirping was extending slightly into the continuum. However, the above continuum was estimated by assuming concentric circular flux surfaces, and a radial gradient of Shafranov shift can significantly increase the gap width. Unfortunately, current profiles are not available for this shot, forbidding more precise estimation of the continuous spectrum. However, there is a shot with similar background plasma parameters, E36378, for which current profiles are available and the continuum has been calculated \cite{shinohara01}. In Fig.~\ref{fig:continuum_E36378}, we superpose the extent of chirping, measured in a magnetic spectrogram, to the SAW continuum calculated in the reference. At $t=3.65$s, frequency sweeps between $53$ and $61$ kHz, while the continuum is broken between $41$ and $76$ kHz. At $t=3.75$s, frequency sweeps between $43$ and $52$ kHz, while the continuum is broken between $31$ and $63$ kHz. In both cases, the extent of chirping TAEs is well within the gap. Thus we can reasonably assume that chirping lifetime is determined by collision processes, rather than by continuum damping.

\subsection{With Krook collisions}

\begin{figure} \begin{center}
\includegraphics{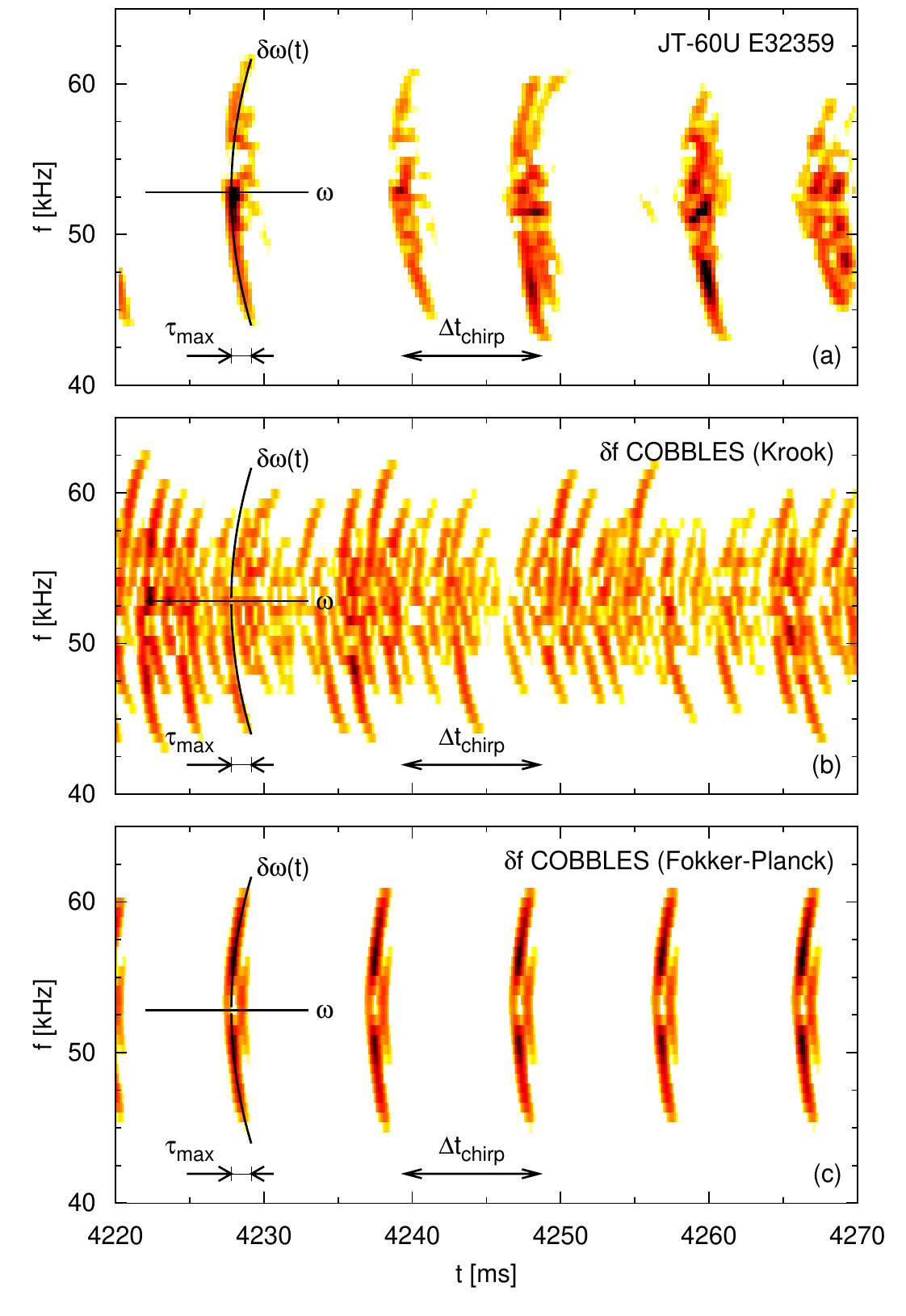}
\caption{(a) Spectrogram of magnetic fluctuations during fast-FS modes in JT-60U discharge E32359, obtained with a moving Fourier window of size $2$ ms. (b) and (c) Spectrogram of the electric field, where the input parameters of the $\delta f$ BB model were chosen to fit the magnetic spectrogram for JT-60U discharge E32359. A solid curve shows the analytic prediction for chirping velocity and lifetime. (b) Krook collisions, correction parameter $\beta=0.65$. (c) Friction-diffusion collisions, correction parameter $\beta=0.85$.}
\label{fig:E32359}
\end{center} \end{figure}

Eqs.~(\ref{eq:gams_from_spectro}) and (\ref{eq:nua_from_spectro}) yield $\nu_a = 2.5 \times 10^{-3}$, and $\gamma _{L0} \sqrt{\gamma _d} = 1.8\times 10^{-2}$. 
However, a preliminary analysis suggests that the plasma belongs to a regime where $\beta = 0.65$, hence we adjust the value of the product $\gamma _{L0} \sqrt{\gamma _d}$ to $1.8\times 10^{-2}/0.65\,=\,2.8\times 10^{-2}$.

\begin{figure} \begin{center}
\includegraphics{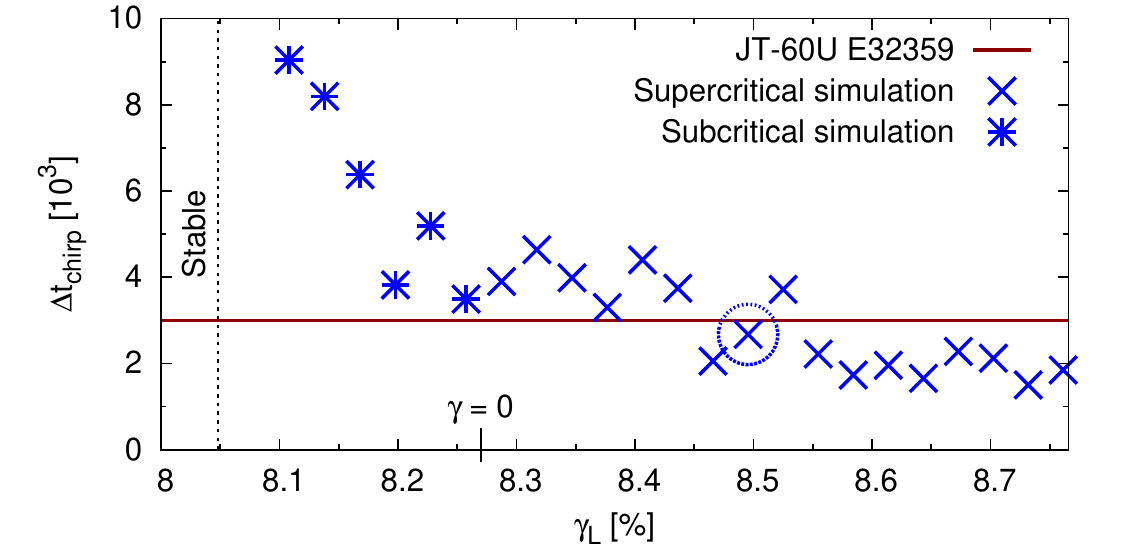}
\caption{Scan of the chirping quasi-period for the set of parameters corresponding to JT-60U discharge E32359. The region where $\gamma _L > 8.8 \%$, where $\Delta t_\mathrm{chirp} < 2\times 10^{3}$, is not included in this plot. Both linear and nonlinear stability threshold are indicated. The chirping quasi-period agrees with the experiment for $\gamma _L \approx 8.5\%$. The spectrogram for the circled simulation is shown in Fig.~\ref{fig:E32359}(b).}
\label{fig:scanE32359}
\end{center} \end{figure}

%A scan for this set of parameters is performed by changing the slope of the distribution. Fig.~\ref{fig:scanE32359} shows that the chirping quasi-period depends on $\gamma _L$ in a roughly monotonous way. Note that the scan needs to be performed on a relatively narrow range of the kinetic parameters, since the limits of subcritical regime and non-chirping (chaotic) regime yield a first estimate as $\gamma _L \sim 8-12 \%$ , $\gamma _d \sim 4-10 \%$. Here, the nonlinear stability threshold is defined as the largest value of $\gamma _L$ for which the electric field amplitude tends to zero in the time-asymptotic limit, independently of the initial perturbation amplitude; the chaotic regime is defined and categorized in a way described in Ref.~\cite{lesur09}. We observe that the two-points correlation of electric field amplitude decreases as the system approaches marginality.\\
A scan for this set of parameters is given in Fig.~\ref{fig:scanE32359}. The limits of subcritical regime and non-chirping regime yield $\gamma _L \sim 8-12 \%$, $\gamma _d \sim 4-10 \%$.
Fig.~\ref{fig:E32359}(b) is the spectrogram for the simulation which is emphasized by a circle in Fig.~\ref{fig:scanE32359}. The features of main chirping events agree with the experimental observation. However, quiescent phases are replaced by many chirping events with amplitude slightly smaller than the major ones.
Linear parameters estimated from this analysis are shown in Tab.~\ref{tab:params}. Our analysis suggests that the TAE in this discharge is marginally unstable, with $\gamma / \gamma _L \sim 0.1$, even though $\gamma _L < \gamma _d$, which is not inconsistent with Eq.~(\ref{eq:gaml_minus_gamd}) since $\gamma _{L0} > \gamma _d$. However, these values are inconsistent with estimations that take into account drag and diffusion processes. Since the following analysis with drag and diffusion shows much better agreement with the experiment, we imply that the Krook model is insufficient to describe nonlinear features related to chirping repetition.\\

\subsection{With drag and diffusion\label{subs:analyse_e32359_dragdiff}}

\begin{figure} \begin{center}
\includegraphics{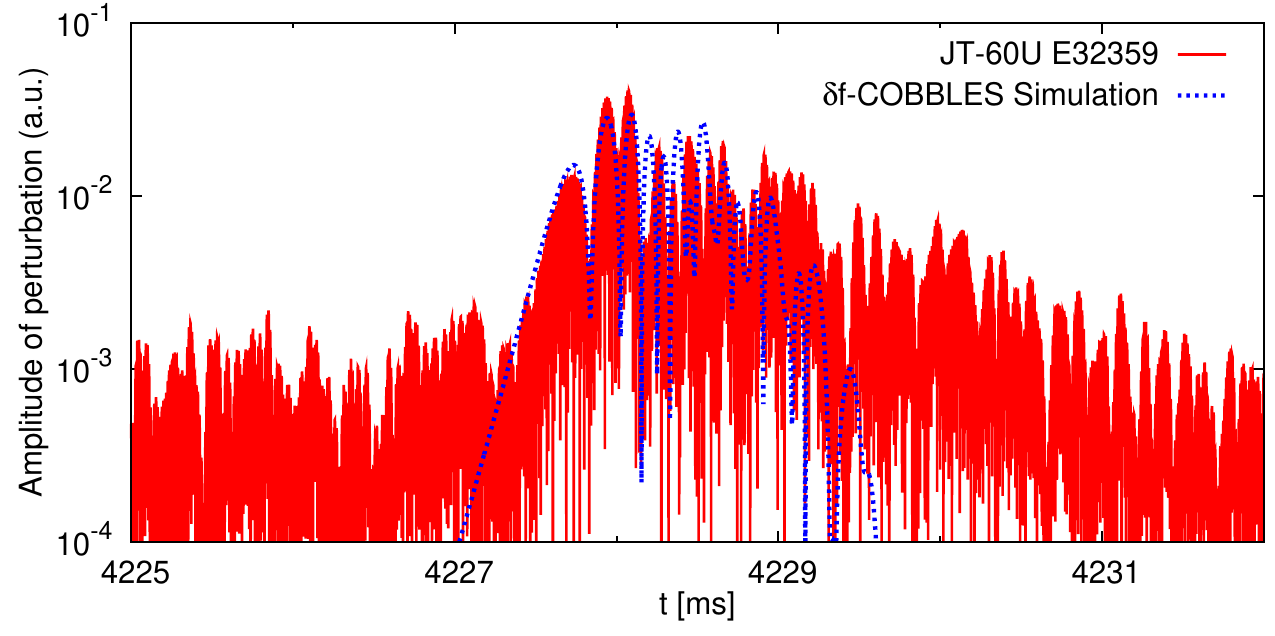}
\caption{Evolution of the perturbation during a single chirping event. The signal is filtered between $40$ and $65$ kHz. In these arbitrary units, $10^{-3}$ roughly corresponds to a noise level. The parameters of the simulation are shown in the fourth line of Tab.~\ref{tab:params}. For the simulation, to avoid hiding experimental data, we show the amplitude of perturbations (the enveloppe) instead of the perturbations themselves.}
\label{fig:E32359_amplitude_dragdiff}
\end{center} \end{figure}

A first scan in ($\nu _f$, $\nu_d$) yields an estimation of the correction parameter, $\beta = 0.85$. Then Eqs.~(\ref{eq:approx_gaml}), (\ref{eq:approx_gamd}) and (\ref{eq:approx_nud}) yield $\gamma _{L0}=10^{\pm 3}  \%$, $\gamma _d=5^{\pm 3} \%$, and $\nu _d=1.7^{\pm 0.3} \%$.
We perform a second scan, which consists of a series of $4\times 8$ simulations in the domain ($1.5 \%\le \nu _d \le 2.2\%$, $1 \le \nu _d /\nu _f \le 8$), where $\gamma _{L0}$ and $\gamma _d$ are constrained by Eqs.~(\ref{eq:chirping_lifetime_dragdiff}) and (\ref{eq:gams_from_spectro}). The only repetitive chirping solution with $2500 < \Delta t_\mathrm{chirp} < 3500$ we found is shown in Fig.~\ref{fig:E32359}(c). We verify that chirping features measured in this simulation fit the experiment, $d\delta \omega^2 / dt = 7.1\times 10^{-5}$ ($6.2\times 10^{-5}$ for up-chirping, $7.9\times 10^{-5}$ for down-chirping), $\tau _\mathrm{max} = 0.45\times 10^{3}$ ($0.47\times 10^{3}$ for up-chirping, $0.43\times 10^{3}$ for down-chirping), and $\Delta t_\mathrm{chirp} = 3.1\times 10^{3}$. Estimated linear parameters are shown in Tab.~\ref{tab:params}. To validate this analysis, we compare the amplitude of perturbations in Fig.~\ref{fig:E32359_amplitude_dragdiff}. We obtain a quantitative agreement, with a growth rate of $2.3\%$, and a decay rate of $0.3\%$.

As an independent validation, we now estimate the values of $\nu _f$ and $\nu _d$ from background plasma parameters.
%\begin{figure} \begin{center}
%\includegraphics{SpeciesSNB}
%\caption{Assumed distributions in parallel velocity at the resonant surface in JT-60U, E32359. The Alfv\'en velocity is indicated by a tic. Note that at constant $P_{\zeta}$, beam ions can increase their parallel velocity from $v_{b\|}=4.6 \cdot 10^6$ m.s$^{-1}$ at the magnetic axis, up to $8 \cdot 10^6$ m.s$^{-1}$ at the resonant surface.}
%\label{fig:species}
%\end{center} \end{figure}
%The assumed distributions in parallel velocity at the resonant surface are illustrated in Fig.~\ref{fig:species}. 
In the discharge E32359 around $t=4.2$ s, the resonant surface of the $m/n=2/1$ and $3/1$ TAE is located around $r=0.7$ m. The magnetic shear is estimated from the $q$ profile \cite{gorelenkov00}, $S=0.8$. The deuteron plasma has the following characteristics, $B_0=1.2$ T, $R_0=3.3$ m, and the tangential radius of the N-NB is $R_T=2.6$ m. At $r=0.7$ m, $n_e=1.4\cdot 10^{-19}$ m$^{-3}$, and $T_0=0.75$ keV. We take into account carbon impurities with $Z_{\mathrm{eff}}=2.7$.
With these equilibrium measurements, Eqs.~(\ref{eq:nuf}-\ref{eq:nud}) yield $\nu _f/\omega _\mathrm{A} = 1.2\%$ and $\nu _d/\omega _\mathrm{A} = 1.7\%$.
We obtain a quantitative agreement for $\nu _d$, which further validates our fitting procedure. However, $\nu _f$ was underestimated by $70\%$ compared to the latter estimation from background plasma parameters. Though error bars in experimental data may account for this discrepancy, it is also possible that our model misses some mechanism that would enhance friction.

Note that electrons account for $99\%$ of $\nu _f^2$, which reflects a high Alfv\'en velocity, while impurities account for $57\%$ of $\nu _d^3$, which is consistent with the fact that pitch-angle scattering is more effective with heavier particles. This observation suggests that impurities tend to reduce the ratio $\nu _f / \nu _d$, which reduces the frequency of AE activity. 
%consistently with the fact that $v_\mathrm{A}$ is above the critical velocity $v_c\equiv (3\pi ^{1/2} m_e n_i Z_i^2 / 4 m_i n_e)^{1/3} v_{Te}= 10^6$ m.s$^{-1}$

\section{Validity of fitting procedure}

Overall, our fitting procedure with drag and diffusion shows quantitative agreement for the time-evolution of perturbation, for an independent estimation of the linear drive in the case of MAST, and for an independent estimation of collision frequencies in the case of JT-60U. On the other hand, our fitting procedure with Krook collisions give similar linear rates only in MAST case, and in general we could not clearly reproduce periodic chirping with quiescent phases. Thus we infer that a fitting procedure of chirping features to estimate linear properties of background plasma is valid, provided that drag and diffusion are taken into account.

\chapter{Conclusion}
\label{ch:conclusion}

Resonant interactions between NBI-induced far passing energetic ions and TAEs were described by a Hamiltonian perturbation in guiding-center angle-action variables. By reducing the latter Hamiltonian, a Fokker-Planck collision operator, and external damping processes to a new, 2D phase-space, we showed how the input parameters of the BB model are quantitatively related to plasma parameters and characteristics of the TAE.

We developed a kinetic code to solve the initial-value BB model. COBBLES was verified by checking conservation properties, benchmarked against numerical simulations in former works, and validated against well-known linear and nonlinear theories. The feasibility of long-time simulations for a low-$\gamma _L$ distribution hinged upon quick convergence and strong numerical stability of the CIP-CSL scheme. 

With both $\delta f$ and full-$f$ versions of COBBLES, we explored theory in several parameter regimes, namely steady-state, periodic, chaotic, chirping, and subcritical. We performed a scan of the nonlinear evolution of electric field in the whole ($\gamma _d$,\, $\nu _a$) parameter space. A new diagnostics allowed us to identify the chirping region. Although holes and clumps were not expected to appear when $\gamma _d < 0.4 \, \gamma _L$, we found that the frequency sweeping region can expand to a low external dissipation regime, around $\gamma _d \approx 0.2 \, \gamma _L$.

Numerical results display good quantitative agreements with theory in several regimes. The limits of validity correspond to the assumptions used in theory. A perturbative numerical approach which doesn't take into account the kinetic response of the bulk may feature spurious agreement outside of the validity limits when the resonant region reaches a non negligible portion of bulk particles. 

We found a regime of quasi-periodic chirping with both Krook and drag/diffusion collision operators.
Since quantitative agreement with theory suggests the predictability of nonlinear chirping characteristics based on fundamental linear kinetic parameters, the latter may be estimated in the opposite way from chirping data in experiments. More precisely, chirping velocity and lifetime yield two relations among $\gamma _L$, $\gamma _d$, and collision frequencies; and a fitting of $\Delta t_\mathrm{chirp}$ yields an estimation of remaining unknowns. Note that major advantages of this technique are 1.~kinetic parameters in the core of the plasma estimated only from a spectrogram of magnetic fluctuations measured at the edge, without expensive kinetic/MHD calculations nor detailed core diagnostics, and 2.~unified treatments of supercritical and subcritical AEs.
We showed that diffusion is essential to reproduce quiescent phases observed in experiments between chirping events, and drag is required to observe repetitive chirping, since drag allows the distribution to recover in a reasonable time.
We confronted this procedure by analyzing AEs on MAST and JT-60U. We found quantitative agreement with measured magnetic fluctuations for the growth and decay of chirping structures, quantitative agreement with an estimation of the linear drive based on measured magnetic fluctuation amplitude in the case of MAST, and qualitative agreement with collision frequencies estimated from experimental background measurements in the case of JT-60U. In the latter estimations, impurities, which were not included in estimations of Ref.~\cite{lilley09}, account for the main part of velocity diffusion.
In both cases, our analysis suggests that TAEs belong to a regime which is relatively far from marginal stability, where total growth rate is of the same order as linear drive.

Accurate estimations of linear drive, damping rate, and overall growth rate, open the way to promising advanced numerical and experimental investigations. 
\begin{enumerate}
\item Applying the same procedure to dozens of available chirping TAE data would produce a database from which useful trends could be extracted.
\item The estimated growth rates and collision frequencies could be translated in terms of spatial gradient of energetic-particle distribution, assuming a slowing-down distribution in energy, and in terms of coefficients of the 3D Fokker-Planck collision operator. This work would yield all input parameters of drift-kinetic perturbative 3D simulations that assume a fixed and arbitrary external damping, such as in the HAGIS code, where drag and diffusion is being implemented.
\item Since $\gamma$ is roughly proportional to $\omega \pds{f}{W}+n \pds{f}{P_\zeta}$, where $W$ is the energy, it should be possible to control the growth rate by varying the total number of injected high-energy ions. A possible scenario is to start from a NBI experiment with chirping TAEs, and reproduce the same conditions except for different NBI power $P_\mathrm{NBI}$ at the time of chirping. With a scan in $P_\mathrm{NBI}$, dependency of $\gamma$ on $P_\mathrm{NBI}$ and existence of subcritical TAEs could be investigated.
\end{enumerate}

%\chapter*{Publications and orals}

\subsection*{List of publications}

\begin{itemize}
\item \textit{Spectroscopic determination of kinetic parameters for frequency sweeping Alfv{\'e}n Eigenmodes}, M.~Lesur, Y.~Idomura, K.~Shinohara, X.~Garbet, and the JT-60U team, Phys.~Plasmas 17, 122311 (2010) \cite{lesur10}.

\item \textit{Estimation of kinetic parameters based on chirping Alfv{\'e}n eigenmodes}, M.~Lesur, Y.~Idomura, K.~Shinohara, and X.~Garbet, Proc.~of 23rd IAEA Fusion Energy Conference (to appear).

\item \textit{Nonlinear modification of the stability of fast particle driven modes in tokamaks}, C.~Nguyen, X.~Garbet, V.~Grandgirard, J.~Decker, Z. Guimar{\~a}es-Filho, M.~Lesur, H.~L{\"u}tjens, A.~Merle, and R.~Sabot, Plasma Phys.~and Cont.~Fus.~52, 124034 (2010) \cite{nguyen10ppcf}.

\item \textit{Existence of metastable kinetically driven modes}, C.~Nguyen, H.~L{\"u}tjens, X.~Garbet, V.~Grandgirard, and M.~Lesur, Phys.~Rev.~Lett.~105, 205002 (2010) \cite{nguyen10prl}.

\item \textit{Reduced model analysis of supercritical and subcritical chirping Alfv{\'e}n eigenmodes}, M.~Lesur, Y.~Idomura, K.~Shinohara, and X.~Garbet, Proc.~of 8th BPSI annual meeting, Reports of RIAM, Kyushu University, S-5, 37 (2010).

\item \textit{Fully nonlinear features of the energy beam-driven instability}, M.~Lesur, Y.~Idomura, and X.~Garbet, Phys.~Plasmas 16, 092305 (2009) \cite{lesur09}.

\end{itemize}

\subsection*{List of oral presentations}

\begin{itemize}

\item \textit{Estimation of Kinetic Parameters based on Chirping Alfv{\'e}n Eigenmodes - Effect of drag and diffusion}, 4th ITPA Topical Meeting on Energetic Particles, Seoul (October 2010).

\item \textit{Spectroscopic determination of kinetic parameters for frequency sweeping Alfv{\'e}n eigenmodes}, 4th ITPA Topical Meeting on Energetic Particles, Garching (April 2010).

\item \textit{Estimation of Kinetic Parameters based on Chirping Alfv{\'e}n Eigenmodes - Application to JT-60U}, 15th Numerical EXperiment of Tokamak (NEXT) Workshop, Kyoto (March 2010).

\item \textit{Reduced model analysis of chirping Alfv{\'e}n Eigenmodes - Application to MAST}, 8th Burning Plasma Simulation Initiative (BPSI), Kyushu (December 2009).

\item \textit{The energetic beam-driven instability - Kinetic nonlinearities and subcritical instability in JT-60U}, 7th General Scientific Assembly of the Asia Plasma and Fusion Association (APFA2009) and the Asia-Pacific Plasma Theory Conference (APPTC2009), Aomori (October 2009).

\item \textit{Nonlinear evolution of frequency sweeping (chirping)}, 5th Festival de Théorie, Aix-en-Provence (July 2009).

\item \textit{Fully nonlinear features of the energetic beam-driven instability}, 14th Numerical EXperiment of Tokamak (NEXT) Workshop, Kyoto (March 2009).

\end{itemize}

\appendix

\chapter{Validity limit of Lie transform perturbation theory}\label{app:lie}

\def\Action{J}
\def\Angle{\zeta}
\def\NewAction{I}
\def\NewAngle{\xi}
\def\Hamiltonian{h}
\def\NewHamiltonian{H}
\def\PhaseSpace{z}
\def\NewPhaseSpace{Z}
\def\FundamentalOneForm#1{\gamma _{#1}}
\def\NewFundamentalOneForm#1{\Gamma _{#1}}

\section{Test problem}

To illustrate the improvements of Lie transform perturbation theory compared to classical perturbation theory, let us apply both formulations to a simple mathematical problem,
\begin{eqnarray}
\dfrac{\mathrm{d} x}{\mathrm{d} t} & = & y,  \\
\dfrac{\mathrm{d} y}{\mathrm{d} t} & = & -x \;-\; \epsilon \, x^3,
\end{eqnarray}
where $\epsilon$ is a small parameter, with initial conditions $\left. x \right| _{t=0}=0$ and $\left. y \right| _{t=0}=K$.
These are the equations of motion of an Hamiltonian system, with Hamiltonian
\begin{equation}
\Hamiltonian(x,y) \;=\;  \dfrac{x^2}{2} \;+\; \dfrac{y^2}{2} \;+\; \dfrac{\epsilon}{4} \, x^4 \;=\; \Hamiltonian_0(x,y) \;+\; \epsilon \, \Hamiltonian_1(x).
\end{equation}
In action-angle variables $\PhaseSpace\equiv(\Angle ,\, \Action)$, which are defined by
\begin{eqnarray}
x & = & \sqrt{2\Action}\, \sin \Angle,  \\
y & = & \sqrt{2\Action}\, \cos \Angle,
\end{eqnarray}
the unperturbed motion is described by
\begin{eqnarray}
\Angle(t) & = & t,  \\
\Action(t) & = & K,
\end{eqnarray}
and the perturbed system by the total Hamiltonian,
\begin{equation}
\Hamiltonian(\Angle ,\,\Action) \;=\;  \Action \;+\; \epsilon \, \Action^2 \, \sin^4(\Angle).
\end{equation}

\section{Classical perturbation theory}

\begin{figure}
\begin{center}
\includegraphics{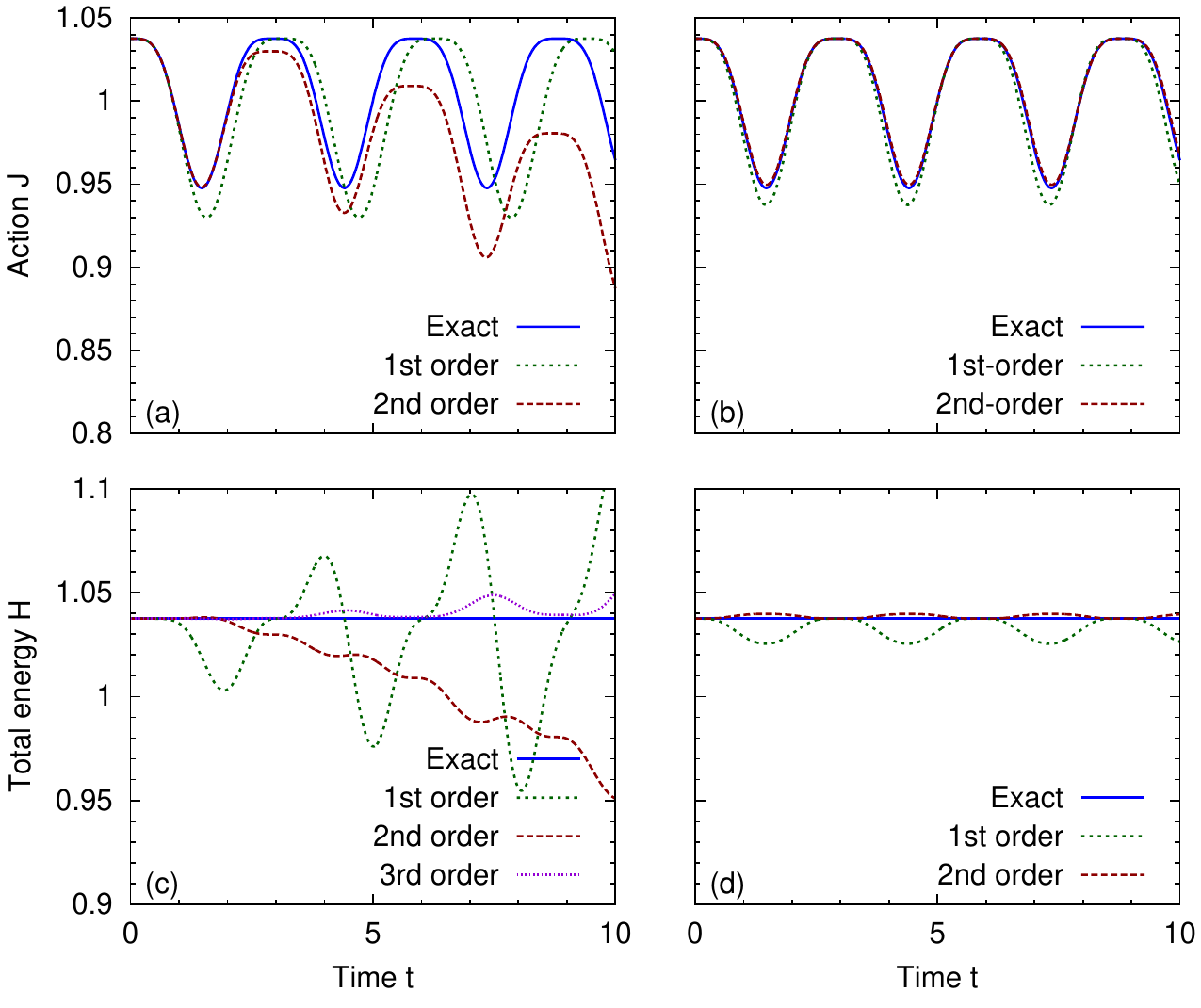}
\caption{Evolution of action and energy. (a) and (c), classic perturbation theory. (b) and (d), Lie transform theory. In both cases, $\epsilon = 0.1$.}
\label{fig:Lie_Action_Energy}
\end{center}
\end{figure}

Classical perturbation theory yields a $n^{\mathrm{th}}$-order solution of Hamilton's equations,
\begin{equation}
\dfrac{\mathrm{d} \PhaseSpace_0}{\mathrm{d} t} \;+\; \epsilon \, \dfrac{\mathrm{d} \PhaseSpace_1}{\mathrm{d} t} \;+\; \ldots \;+\; \epsilon ^n \, \dfrac{\mathrm{d} \PhaseSpace_n}{\mathrm{d} t} \;=\; \left( 
\begin{array}{cc}
	0 & 1 \\
	-1 & 0
\end{array}
 \right) \cdot \, \left. \pd{\Hamiltonian}{\PhaseSpace} \right| _{\PhaseSpace=\PhaseSpace_0 + \ldots + \epsilon ^{n-1} \, \PhaseSpace_{n-1} }, \label{eq:classic_motion}
\end{equation}
after an expansion in the small parameter $\epsilon$, $\PhaseSpace = \PhaseSpace_0 + \epsilon \PhaseSpace_1 + \epsilon ^2 \PhaseSpace_2 + \ldots$.
Fig.~\ref{fig:Lie_Action_Energy} (a) and (c) show the evolution of action and total energy given by the application Eq.~(\ref{eq:classic_motion}) at successive orders. These results are compared with the exact solution obtained by solving the original Hamilton's equations with a fourth-order Runge-Kutta scheme \cite{NumericalRecipes} with a very small time-step width. At small times, the solution given by this method is close to the exact solution, but we rapidly observe a strong secularity effect that makes the average action spuriously shrink or expand.

\section{Lie perturbation theory}

Let us now treat the same problem, this time with Lie perturbation theory. The fundamental one-form, $\FundamentalOneForm{ } = \FundamentalOneForm{0} + \epsilon \FundamentalOneForm{1} + \epsilon ^2 \FundamentalOneForm{2} + \ldots $, can be separated into its equilibrium part and the perturbation,
\begin{eqnarray}
\FundamentalOneForm{0} & = & \Action\, d\Angle \;-\; \Hamiltonian_0(\Action)\, \mathrm{d} t,  \\
\FundamentalOneForm{1} & = &  -\; \Hamiltonian_1(\Angle ,\,\Action)\, \mathrm{d} t,
\end{eqnarray}
and $\gamma _n = 0$ for $n \geq 2$.

To simplify the fundamental one-form, we change variables from $\PhaseSpace$ to $\NewPhaseSpace\equiv(\NewAngle ,\, \NewAction)$. Following Lie perturbation theory, the zeroth-order one-form and the time coordinate remain unchanged, $\NewFundamentalOneForm{0}=\FundamentalOneForm{0}$, and $T\equiv t$. Applying Eqs.~(\ref{eq:GiveNewOneForm}), (\ref{eq:GiveGaugeScalar}), and (\ref{eq:GiveGeneratingFunctions}), we find the new one-form, gauge scalar, and generating functions, which yield a solution which is valid up to $n^{\mathrm{th}}$-order in $\epsilon$. We choose initial conditions such that orbits have a same energy level.
At first order,
\begin{eqnarray}
\NewAngle & = & \left( 1\,+\, \dfrac{3}{4}\epsilon \, \NewAction \right) \, t, \label{eq:firstOrderAnalyticSol1}\\
\NewAction & = &  K. \label{eq:firstOrderAnalyticSol2}
\end{eqnarray}
This solution can be obtained in terms of the original coordinates by using the pull-back operator,
\begin{eqnarray}
\Angle &=& \NewAngle \;-\; \epsilon \dfrac{\NewAction}{4} \, \left[ 2 \sin(2\NewAngle) \,-\, \dfrac{1}{4} \sin(4\NewAngle) \right] \;+\; o(\epsilon ^2), \label{eq:firstOrderAnalyticSol3}\\
\Action &=& \NewAction \;-\; \epsilon \dfrac{\NewAction^2}{4} \, \left[ \dfrac{1}{2} \cos(4\NewAngle) \,-\, 2 \cos(2\NewAngle) \right] \;+\; o(\epsilon ^2). \label{eq:firstOrderAnalyticSol4}
\end{eqnarray}
At second order,
\begin{eqnarray}
\NewAngle & = & \left( 1\,+\, \dfrac{3}{4}\epsilon \, \NewAction \,-\, \dfrac{51}{64}\epsilon ^2 \, \NewAction^2 \right) \, t, \label{eq:secondOrderAnalyticSol1}\\
\NewAction & = &  K \,+\, \dfrac{17}{64}\epsilon ^2 \, K^3 \,-\, \dfrac{51}{256}\epsilon ^3 \, K^4, \label{eq:secondOrderAnalyticSol2}
\end{eqnarray}
and the pull-back operator is,
\begin{eqnarray}
\Angle &=& \NewAngle \;-\; \epsilon \dfrac{\NewAction}{4} \, \left[ 2 \sin(2\NewAngle) \,-\, \dfrac{1}{4} \sin(4\NewAngle) \right] \nonumber\\
& & +\; \epsilon ^2 \dfrac{\NewAction^2}{512} \left[ \sin(8\NewAngle) \,-\, 16 \sin(6\NewAngle) \,-\, 4\sin(4\NewAngle) \,+\, 400 \sin(2\NewAngle)\right] \;+\; o(\epsilon ^3), \label{eq:secondOrderAnalyticSol3}
\end{eqnarray}
\begin{eqnarray}
\Action &=& \NewAction \;-\; \epsilon \dfrac{\NewAction^2}{4} \, \left[ \dfrac{1}{2} \cos(4\NewAngle) \,-\, 2 \cos(2\NewAngle) \right]\nonumber\\
& &+\; \epsilon ^2 \dfrac{\NewAction^3}{64} \left[ 2\cos(6\NewAngle) \,+\, 6 \cos(4\NewAngle) \,-\, 42\cos(2\NewAngle) \,+\, 17 \right] \;+\; o(\epsilon ^3). \label{eq:secondOrderAnalyticSol4}
\end{eqnarray}
Fig.~\ref{fig:Lie_Action_Energy} (b) and (d) show action and total energy for analytic solutions, Eq.~(\ref{eq:firstOrderAnalyticSol1}-\ref{eq:firstOrderAnalyticSol4}) and (\ref{eq:secondOrderAnalyticSol1}-\ref{eq:secondOrderAnalyticSol4}), with $\epsilon = 0.1$. There is good agreement with the exact solution, without any secularity effect.

\begin{figure}
\begin{center}
\includegraphics{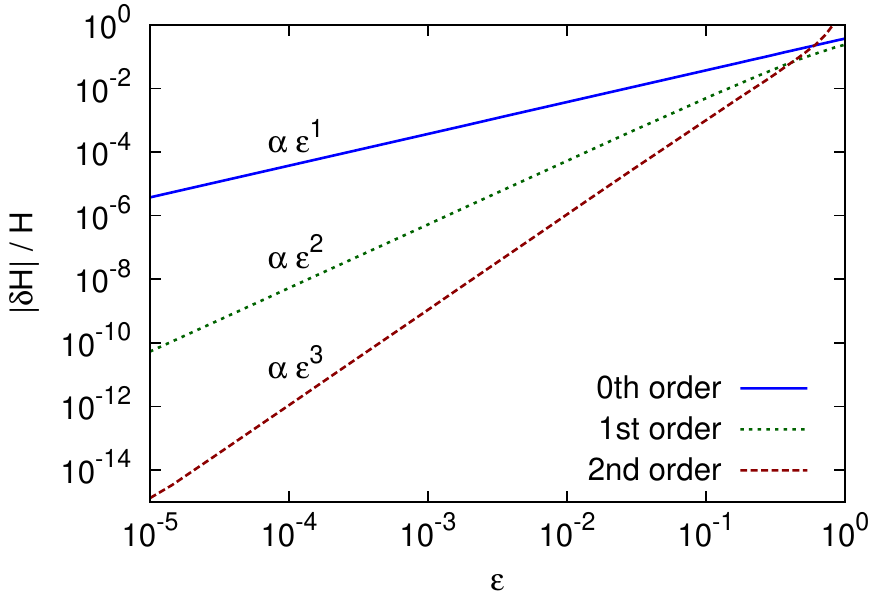}
\caption{Relative error in total energy conservation for solutions obtained with Lie perturbation theory at $t=12.5\pi$. For small $\epsilon$, solid, dotted, and dashed lines scale as $\epsilon$, $\epsilon ^2$, and $\epsilon ^3$, respectively.}
\label{fig:Lie_Error_vs_Epsilon}
\end{center}
\end{figure}

%To investigate the limit of validity of Lie transform theory, we 
Fig.~\ref{fig:Lie_Error_vs_Epsilon} shows the relative error in total energy conservation against the small parameter $\epsilon$. It is confirmed that the error is one order higher than the order of perturbation analysis.

\chapter{Validity limit of gyrokinetics in slab geometry} \label{app:slab_gyrokinetics}

Most of linear and nonlinear numerical investigations of TAEs are based on a hybrid model where energetic particles follow gyrokinetic equations, while background thermal populations are described by a MHD model \cite{todo98}. Such simulations can exhibit chirping that is qualitatively similar to experiments \cite{todojpfr03}. Ultimately, the results we obtain in this thesis with our 1D approach should be compared with global gyrokinetic simulations - actually, this should have been an intermediary step before carrying out comparisons with experiments.

%Recently, advances in numerical techniques and computational power allowed to include thermal ion kinetic effects by treating them with a gyrokinetic model \cite{lang09}.

This appendix summarizes some results of gyrokinetic theory, which could be used in future works. Note that the following results are not used within the core of this manuscript.
The goal is to investigate a break-down of gyrokinetics when time-scale and length-scale of the perturbation approach the gyromotion time-scale and length-scale, and when the amplitude of perturbation approaches the equilibrium. Since we used Lie-transform perturbation theory to derive the set of gyrokinetic equations, this provides a second illustration of this technique.

\section{Review of gyrokinetics}

\begin{figure}
\begin{center}
\includegraphics[angle=-90]{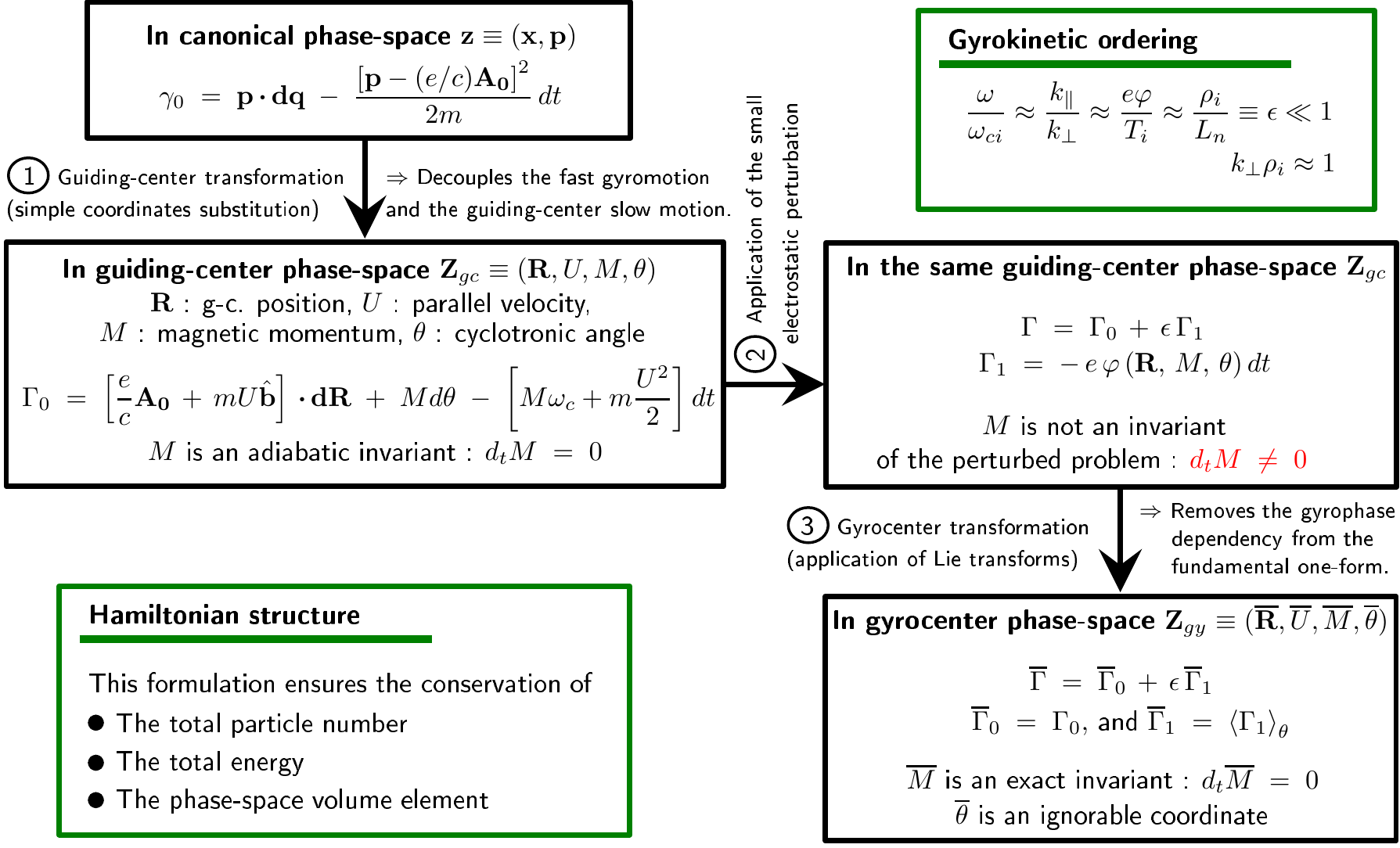}
\caption{Schematic summary of the modern derivation of gyrokinetic equations for an electrostatic perturbation.}
\label{fig:gyrokinetics}
\end{center}
\end{figure}

Let us consider the evolution of a distribution $f$ in 6D (3D position, 3D velocity) phase-space $z$, given by a Vlasov equation,
\begin{equation}
\partial _t f \,-\, \left\{ h,\,f \right\}_z \;=\; 0,
\end{equation}
where $h$ is the Hamiltonian, and $\{\}_z$ are Poisson brackets.
This description can be simplified by removing the irrelevant fast gyromotion from the complete motion of charged particles, thus reducing dimensionality to 5D. There are two ways to proceed,
\begin{itemize}
\item averaging Vlasov equation over a gyroperiod,
\begin{equation}
\partial _t \left< f \right> _{\xi} \,-\, \left< \left\{ h,\,f \right\}_z \right> _{\xi} \;=\; 0,
\end{equation}
where $\xi$ is the cyclotronic angle, or
\item finding a coordinate transformation from $z$ to $\ov{Z}$, where $f$ and $h$ are changed to $\ov{F}$ and $\ov{H}$ such that
\begin{equation}
\partial _t \ov{F} \,-\, \left\{ \ov{H},\,\ov{F} \right\}_{\ov{Z}} \;=\; 0,
\end{equation}
\end{itemize}
and $\ov{H}$ does not depend on $\xi$.
Lie transform provides a modern way of deriving gyrokinetic equations in the latter fashion, which preserves the Hamiltonian structure of the initial problem. These conservation properties are essential for robust long-time computations. Fig.~\ref{fig:gyrokinetics} is a schematic summary of the modern derivation of gyrokinetic equations.

\section{Single particle orbit in slab geometry}

We develop a numerical simulation to solve the equations of motion for a single ion in both particle coordinates, guiding-center coordinates, and gyrocenter coordinates. The aim is to compare 3 ways of computing the orbit of a single particle.

We consider an electrostatic perturbation $\varphi$ of frequency and wave vector $\omega$ and $\boldsymbol{k}$, respectively.
For the gyrokinetic treatment to be valid, the gyrokinetic ordering \cite{hahm88},
\begin{eqnarray} \label{eq:GKOrdering}
\dfrac{\omega}{\Omega _{ci}} \approx \dfrac{k_{\|}}{k_{\bot}} \approx \dfrac{e\varphi}{T_i} \approx \dfrac{\rho _i}{L_n} \equiv \epsilon & \ll & 1 ,\\
k_{\bot} \, \rho _i & \approx & 1,
\end{eqnarray}
has to be satisfied. Here $\Omega _{ci}$ is the ion cyclotronic frequency, $T_i$ the ion temperature, $\rho _i$ the ion Larmor radius, and $L_n$ a scale-length of variation of equilibrium plasma.

As a test problem, we assume an homogeneous magnetic field $\boldsymbol{B_0} = B_0 \boldsymbol{b}$ and a time-independant, single-wave electrostatic perturbation,
\begin{equation}
\varphi(\boldsymbol{x}) \;=\; \varphi _0 \cos (\dotp{k}{x}).
\end{equation}
Then, the gyrokinetic ordering is reduced to $\dfrac{k_{\|}}{k_{\bot}} \approx \dfrac{e\varphi}{T_i} \ll 1$, and $k_{\bot} \rho _i \approx 1$.

\subsection{Equations of motion}

We normalize length, time, velocity, and electric potential, to Debye length $\lambda _D$, inverse ion plasma frequency $\omega _{pi}^{-1}$, ion thermal velocity $v_{Ti}$, and the ratio $T_i / e$ respectively.

\subsubsection{In particle position coordinates}

The equations of motion expressed in particle position coordinates are given by Newton's equations with Lorentz's force,
\begin{eqnarray}
\dot{\boldsymbol{x}} & = & \boldsymbol{v},  \\
\dot{\boldsymbol{v}} & = & \boldsymbol{E} + \Omega _{ci} \crop{v}{b}.
\end{eqnarray}

\subsubsection{In guiding-center coordinates}

We recall the guiding-center variables $\boldsymbol{Z} \equiv (\boldsymbol{R},U,M,\xi)$,
\begin{eqnarray}
\boldsymbol{R} & \equiv & \boldsymbol{x} \; - \; \epsilon \dfrac{v_{\bot}}{\Omega _{ci}} \boldsymbol{a},\\
U & \equiv & \dotp{v}{b},\\
M & \equiv & \dfrac{v_{\bot}^2}{2\Omega _{ci}},\\
\xi & \equiv & \tan ^{-1} \left( \dfrac{\dotp{v}{e_1}}{\dotp{v}{e_2}} \right).
\end{eqnarray}
The equations of motion expressed in guiding-center coordinates \cite{littlejohn83} are given by
\begin{eqnarray}
\dot{\boldsymbol{R}} & = & v_\| \boldsymbol{b} + \Omega _{ci} ^{-1} \crop{b}{\grad} \varphi , \\
\dot{U} & = & - \dotp{b}{\grad} \varphi ,\\
\dot{M} & = & - \Omega _{ci} \pds{\varphi}{\xi} ,\\
\dot{\xi} & = & \Omega _{ci} \left( 1 + \pds{\varphi}{M} \right).
\end{eqnarray}

\subsubsection{In gyrocenter coordinates}

In Ref.~\cite{hahm88}, a complete set of gyrokinetic equations is derived in the homogeneous magnetic field case, based on Lie-transform perturbation theory. In the gyrocenter variables $\ov{\boldsymbol{Z}} \equiv (\ov{\boldsymbol{R}},\ov{U},\ov{M},\ov{\xi})$, the equations of motion are 
\begin{eqnarray}
\dot{\ov{\boldsymbol{R}}} & = & \ov{U} \boldsymbol{b} + \Omega _{ci} ^{-1} \crop{b}{\ov{\grad}} \left< \varphi \right> ,\\
\dot{\ov{U}} & = & - \dotp{b}{\ov{\grad}} \left< \varphi \right> ,\\
\dot{\ov{M}} & = & 0 ,\\
\dot{\ov{\xi}} & = & \Omega _{ci} \left( 1 + \pds{\left< \varphi \right>}{\ov{M}} \right).
\end{eqnarray}

The first equation yields the velocity of the gyrocenter as a sum of parallel velocity and $\crop{E}{B}$ velocity. The second one shows the effect on parallel acceleration of the electric field. The third one confirms that the new magnetic momentum is an exact invariant. The last one can be ignored in practical gyrokinetic simulations. However, it is evolved in our simulation since it is needed to retrieve guiding-center coordinates from gyrocenter coordinates and compare physical quantities expressed in different sets of coordinates.

\subsection{Numerical results}

\subsubsection{Test problem}

Let the plasma be such that $\omega _{pi}=10$ GHz, $v_{Ti}=10^5$ m.s$^{-1}$ and $\lambda _D = 10$ $\mu$m, with a magnetic field amplitude $B=1$T. Then $\Omega _{ci} = 96$ MHz and $\rho _i = 1.04$ mm, or $\Omega _{ci} = 9.6 \times 10^{-3}$ and $\rho _i = 104$ in normalized units.
We define two small parameters $\epsilon _E \;\equiv\; \varphi _0$ and $\epsilon _k \;\equiv\; k_\| / k_{\bot}$.
In our test problem, we choose $\epsilon _E = 0.1$, $k_\bot = 1/\rho _i$, $k_\|=0.1k_\bot$.
The initial conditions are $\xi=0$, and $v_\bot=v_\|=1$, so that $M(t=0)=0.5$.

This set of parameters satisfies the gyrokinetic ordering,
\begin{eqnarray}
\epsilon _E \sim \epsilon _k &\ll & 1 ,\\
k_\bot \rho _i &\sim & 1 .
\end{eqnarray}

\subsubsection{Comparison of numerical results in several coordinate systems}

To compare numerical results between themselves, we transform each coordinate system into guiding-center coordinates. For the particle coordinates we define
\begin{eqnarray}
\boldsymbol{R}_{xv} &\equiv& \boldsymbol{x} - \dfrac{v_{\bot}}{\Omega _{ci}} \boldsymbol{a}(\xi ^*)  \\
U_{xv} &\equiv& \dotp{v}{b} \\
M_{xv} &\equiv& \dfrac{v_{\bot}^2}{2} \\
\xi _{xv} &\equiv&  \xi ^* ,
\end{eqnarray}
where
\begin{eqnarray}
v_{\bot} &\equiv & \sqrt{\dotp{v}{e_1}^2+\dotp{v}{e_2}^2},\\
\xi ^* &\equiv & \tan ^{-1} \left( \dotp{v}{e_1} , \dotp{v}{e_2} \right) .
\end{eqnarray}

Guiding-center coordinates are taken as the reference, $\boldsymbol{Z_{gc}}\equiv \boldsymbol{Z}$.

Finally the guiding-center coordinates are related with the gyrocenter coordinates $\ov{\boldsymbol{Z}}$ by the pull-back transform,
\begin{eqnarray}
\boldsymbol{R}_{gy} &\equiv& \ov{\boldsymbol{R}} - \epsilon _E \dfrac{\ov{\grad} \tilde{\Phi} \times \boldsymbol{b}}{\Omega _{ci} ^2} ,\\
U_{gy} &\equiv& \ov{U} - \epsilon _E \epsilon _k \dfrac{\ov{\nabla _\|} \tilde{\Phi}}{\Omega _{ci}} ,\\
M_{gy} &\equiv& \ov{M} - \epsilon _E \tilde{\varphi} ,\\
\xi _{gy} &\equiv&  \ov{\xi} + \epsilon _E \pd{\tilde{\Phi}}{\ov{M}},
\end{eqnarray}
where we have explicitly written the small parameters as reminders of the order of each term. Note that the generating vector for the parallel velocity is one order higher than the others.

\begin{figure}
\begin{center}
\includegraphics{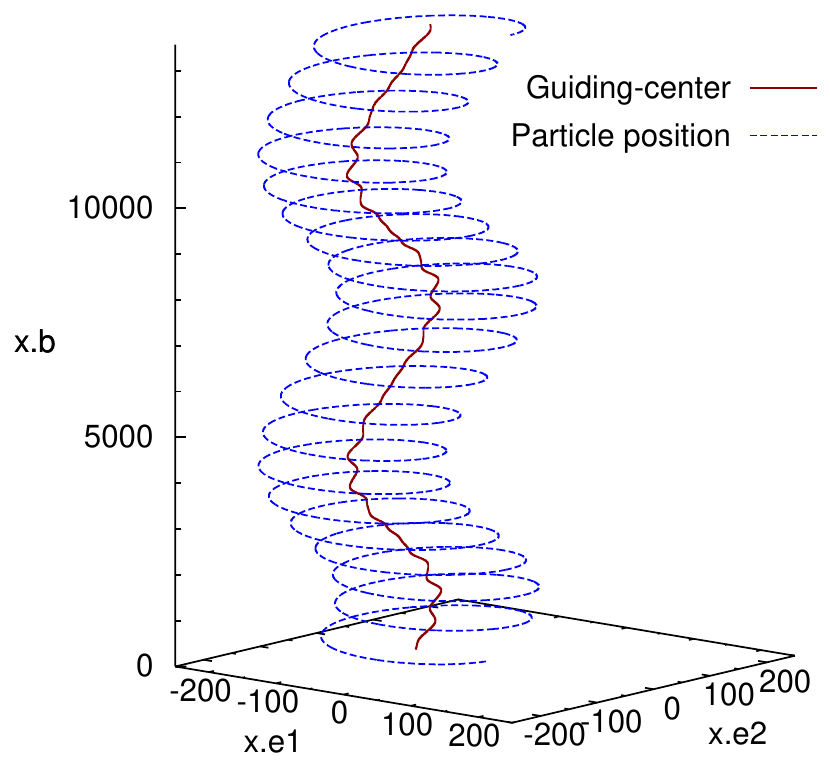}
\caption{Time-evolution of a particle position in an uniform magnetic field with an electrostatic perturbation. Here, guiding-center trajectory overlaps the trajectories of both pull-back of gyrocenter, and push-forward of particle position. Simulation with $\epsilon = 0.1$ and $\Delta t = 10^{-3}\, T_{ci}$.}
\label{fig:position}
\end{center}
\end{figure}

\begin{figure}
\begin{center}
\includegraphics{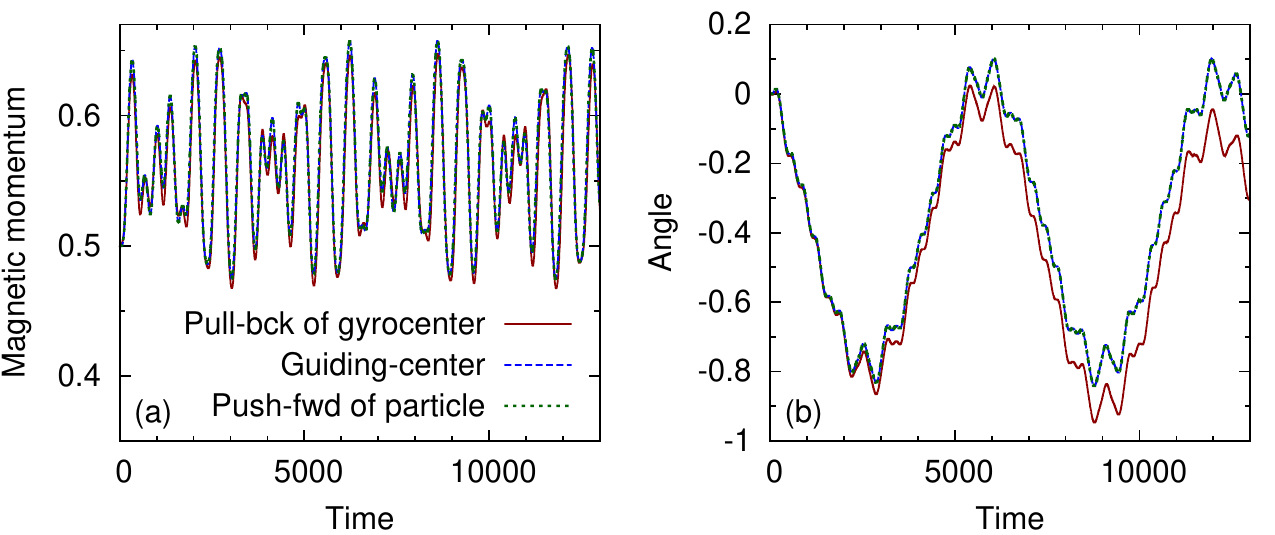}
\caption{(a) Magnetic momentum. (b) Gyroangle perturbation (difference between $\xi$ and the unperturbed solution $\xi _0 = \Omega _{ci} t$). Simulation with $\epsilon = 0.1$ and $\Delta t = 10^{-3}\, T_{ci}$.}
\label{fig:momentandangle}
\end{center}
\end{figure}

\begin{figure}
\begin{center}
\includegraphics{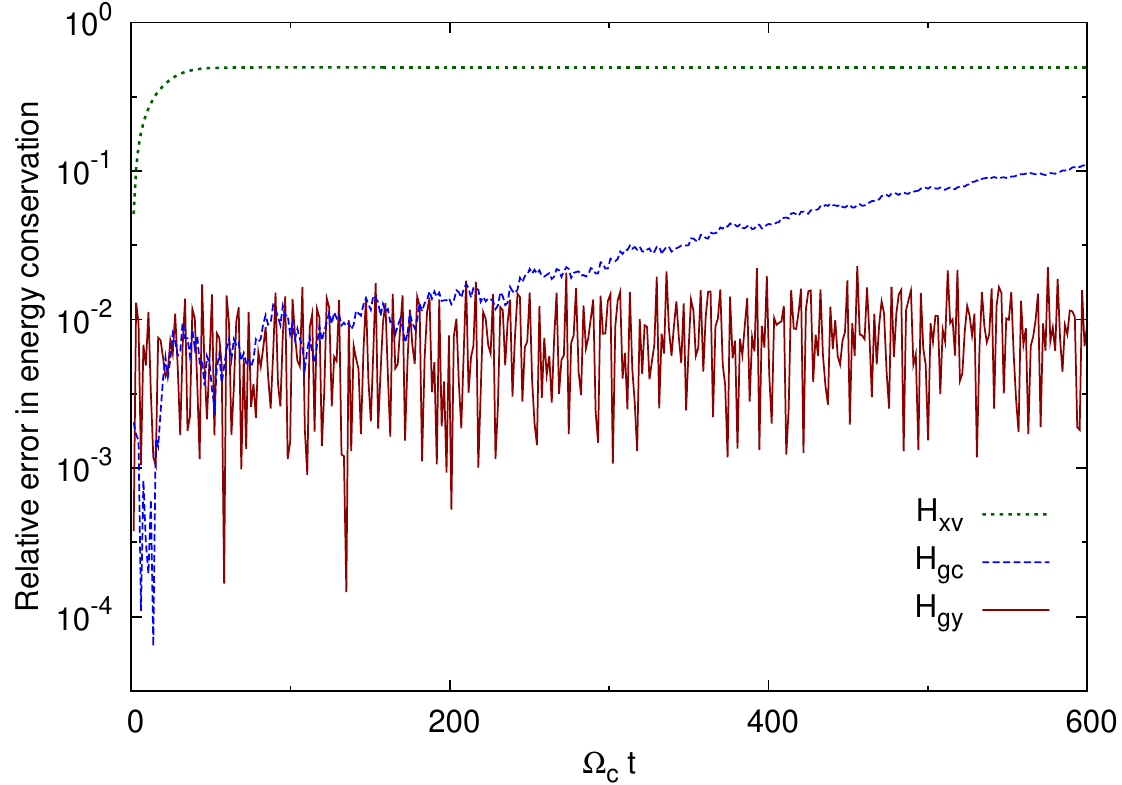}
\caption{Error in energy conservation. The time-step width of these simulations is $\Delta t = T_{ci}/2$.}
\label{fig:errnrj}
\end{center}
\end{figure}

In order to compare results obtained by the gyrokinetic code with the true orbit of the particle and its guiding-center, we firstly run a numerical simulation with a very small time-step, $\Delta t = 10^{-3}\, T_{ci}$, where $T_{ci}\equiv 2\pi / \Omega _{ci}$ is a gyration period. Fig.~\ref{fig:position} shows the orbit of the particle calculated in position/velocity coordinates, and the orbit of the guiding-center, which overlaps with orbits of both the push-forward of the particle position and the pull-back of the gyrocenter.
Then we choose a practical time-step width, and compare, in Fig.~\ref{fig:errnrj}, the guiding center Hamiltonian calculated in the three coordinates system,
\begin{eqnarray}
H_{xv} & = & M_{xv} \,+\, \half U_{xv}^2 \,+\, \epsilon _E \, \varphi (\boldsymbol{R_{xv}},M_{xv},\xi _{xv}), \\
H_{gc} & = & M_{gc} \,+\, \half U_{gc}^2 \,+\, \epsilon _E \, \varphi (\boldsymbol{R_{gc}},M_{gc},\xi _{gc}), \\
H_{gy} & = & M_{gy} \,+\, \half U_{gy}^2 \,+\, \epsilon _E \, \varphi (\boldsymbol{R_{gy}},M_{gy},\xi _{gy}). \label{eq:gcHamgy}
\end{eqnarray}
We observe a secularity effect in the numerical error of particle position. The gyrocenter Hamiltonian,
\begin{equation}
\ov{H} \;=\; \ov{M} \,+\, \half \ov{U}^2 \,+\, \epsilon _E \, \left< \varphi \right> _{\ov{\xi}} (\ov{\boldsymbol{R}},\ov{M}) ,
\end{equation}
is trivially perfectly conserved.

\subsubsection{Analysis of the error}

\begin{figure}
\begin{center}
\includegraphics{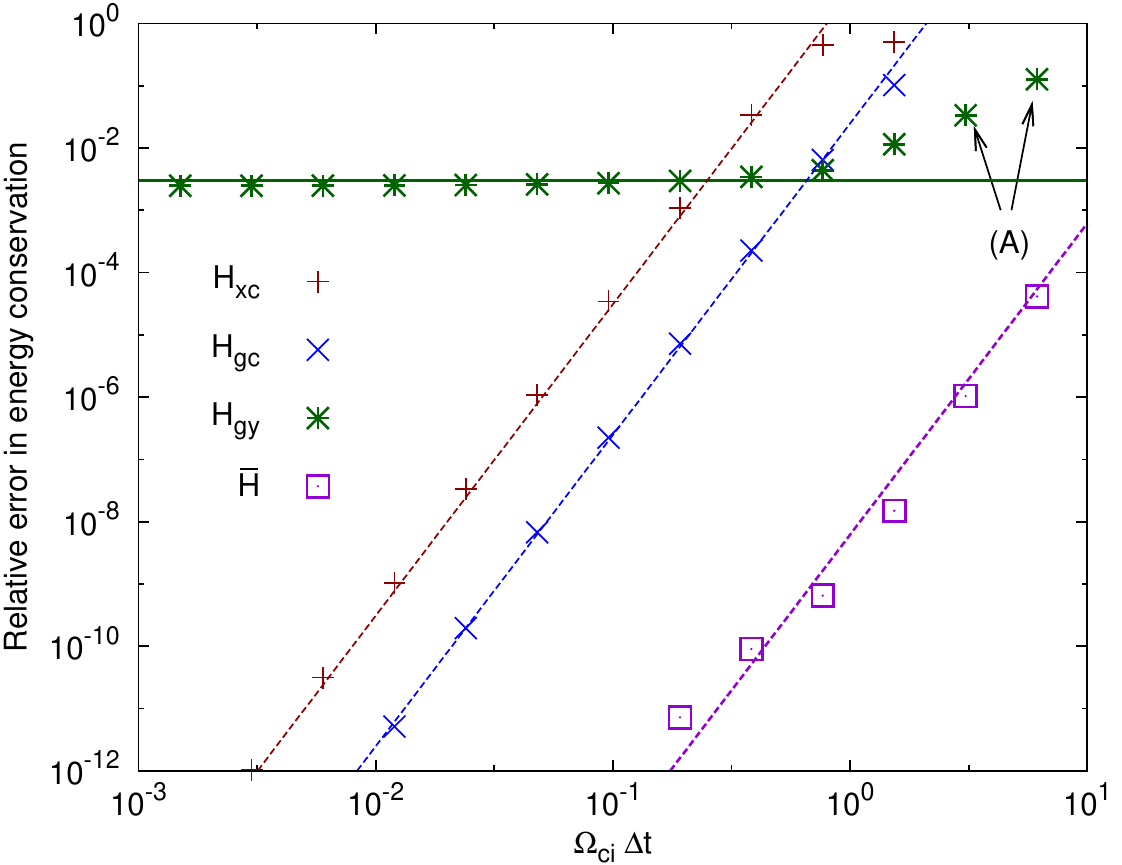}
\caption{Error in energy conservation at $t=100 \, T_{ci}$, against time-step width. A solid line follows a $\Delta t ^0$ law, and dashed lines follow a $\Delta t ^5$ law. Two points marked $(A)$ show a break-up of gyrokinetics close to cyclotronic period.}
\label{fig:errordt5}
\end{center}
\end{figure}

Firstly, we analyse the numerical error. Our code uses a fourth-order Runge-Kutta scheme \cite{NumericalRecipes}. As a consequence, the error between numerical and analytic solutions is proportionnal to $\Delta t^5$.
Fig.~\ref{fig:errordt5} shows the variation of the relative error in the conservation of guiding-center and gyrocenter Hamiltonians, calculated after a fixed, long time (hundred gyration periods), with increasing time-step.
This plot is consistent with the prediction of numerical error for this numerical scheme. Note that the error for each coordinate is proportional to $\Delta t^5$, save the magnetic momentum calculated in gyrocenter coordinates, which is trivially exactly conserved for any time-step width.
The curve of gyrokinetic-calculated error crosses the other ones at a fraction of the cyclotronic period, meaning that a gyrokinetic computation becomes rapidly more efficient. However, the points marked $(A)$ show that even when the gyrocenter Hamiltonian is precisely conserved, the conservation of guiding-center Hamiltonian breaks-down when the time-step width is close to the cyclotronic period, $\Delta t \approx \, T_{ci}$. This is caused by a lack of accuracy of the pull-back transformation of coordinates, coming from a lack of accuracy for large time-step width in the computation of $\tilde{\Phi}$.

\begin{figure}
\begin{center}
\includegraphics{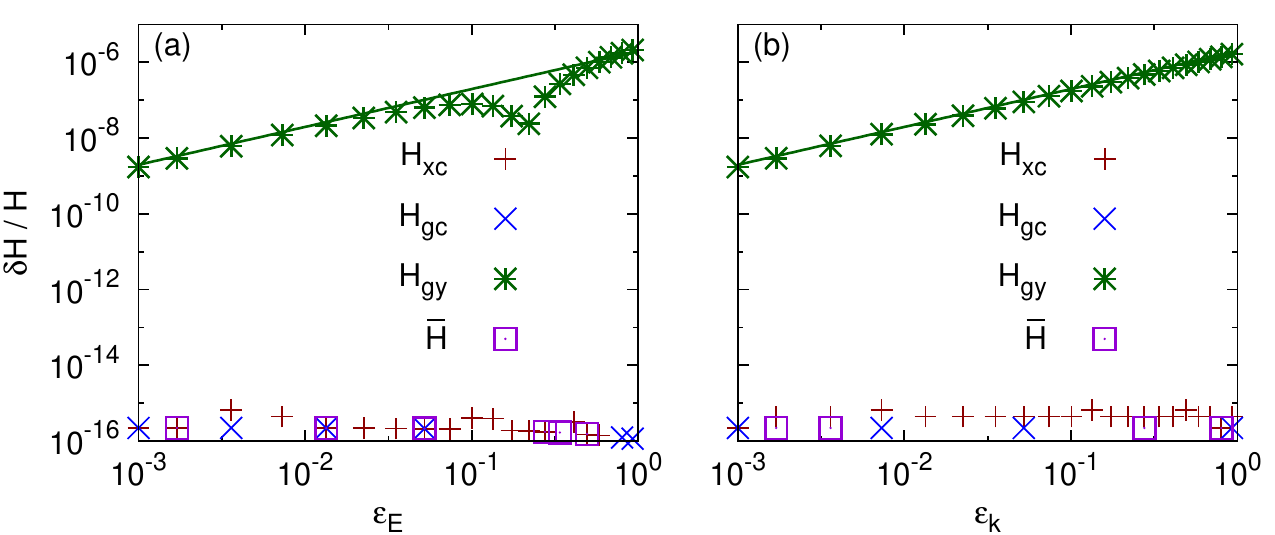}
\caption{Error in total energy conservation after one time-step, when $\epsilon _E$ and $\epsilon _k$ are different. (a) Scan in electric field amplitude $\epsilon _E$. A solid line follows $\epsilon _E ^1$ law. Simulation parameters are $\Delta t = 0.01 (2\pi/\Omega _{ci})$, and $\epsilon _k = 10^{-3}$. (b) Scan in $\epsilon _k=k_\| / k_{\bot}$. A solid line follows $\epsilon _k ^1$ law. Simulation parameters are $\Delta t = 0.01 \, T_{ci}$, and $\epsilon _E = 10^{-3}$.}
\label{fig:errorepEk}
\end{center}
\end{figure}

\begin{figure}
\begin{center}
\includegraphics{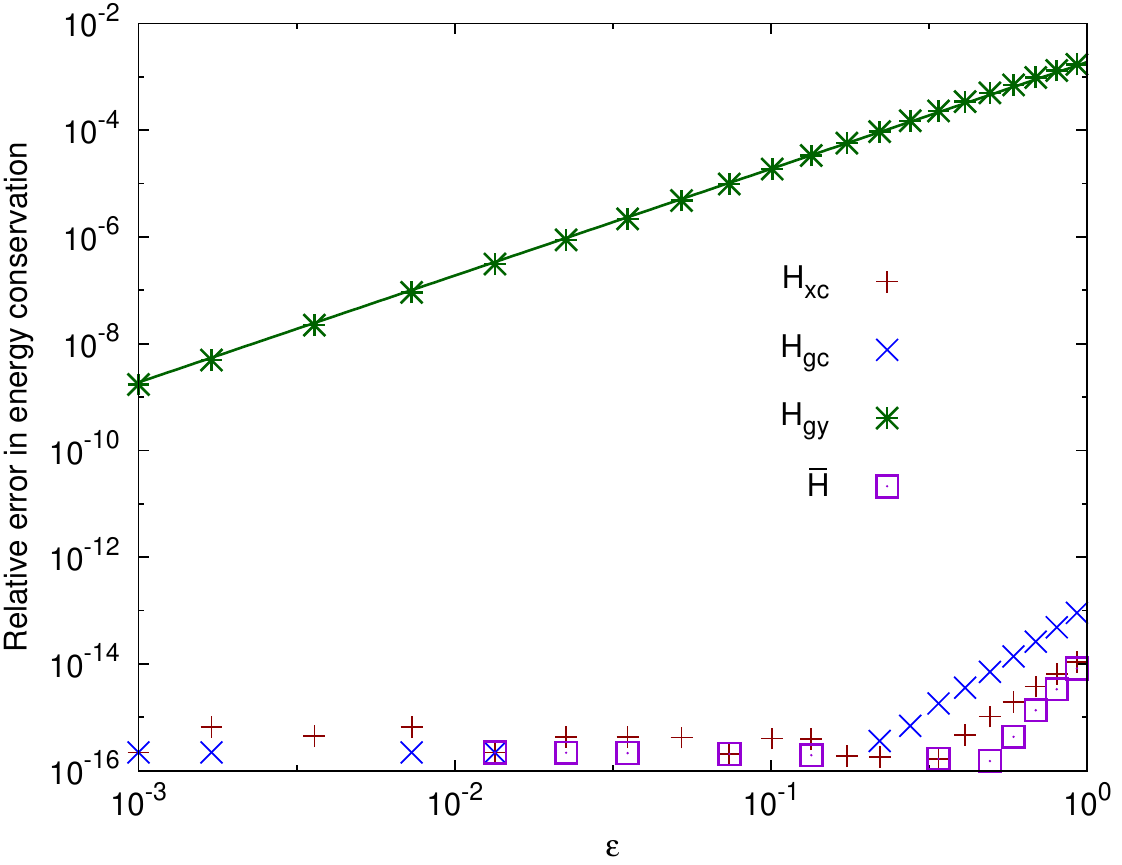}
\caption{Error in total energy conservation after one time-step, against the small parameter of gyrokinetic expansion $\epsilon = \epsilon _k = \epsilon _E$. Time-step width of simulations is $\Delta t = 0.01 \, T_{ci}$.}
\label{fig:erroreps}
\end{center}
\end{figure}

Fig.~\ref{fig:errorepEk} shows the variation of the relative error in the conservation of guiding-center and gyrocenter Hamiltonians with the small parameters $\epsilon _E$ and $\epsilon _k$, for a very small time-step width. Fig.~\ref{fig:erroreps} shows the variation of the relative error in the conservation of guiding-center and gyrocenter Hamiltonian with the small parameter $\epsilon$, when $\epsilon _k = \epsilon _E = \epsilon$.
To predict the variation of the relative error in the conservation of the Hamiltonian (\ref{eq:gcHamgy}), let us calculate its difference with the gyrocenter Hamiltonian, whose error in conservation does not depend on the small parameters,
\begin{eqnarray}
H_{gy} - \ov{H} & = & M_{gy} \,-\, \ov{M} \,+\, \half \left( U_{gy}^2 - \ov{U}^2 \right) \,+\, \epsilon _E \, \varphi (\boldsymbol{Z_{gy}}) \,-\, \epsilon _E \, \left< \varphi \right> (\boldsymbol{\ov{Z}}), \\
   & = & - \epsilon _E \, \tilde{\varphi}(\boldsymbol{\ov{Z}}) \,-\, \half \left( U_{gy}+\ov{U} \right) \epsilon _E \epsilon _k \dfrac{\ov{\nabla _\|} \tilde{\Phi}}{\Omega _{ci}} \,+\, \epsilon_E \, \tilde{\varphi}(\boldsymbol{\ov{Z}}) \nonumber \\
 & & \qquad \qquad \qquad \qquad \qquad \qquad \qquad \qquad \qquad + \epsilon _E ^2 \boldsymbol{g_1^Z}\bdot \pd{\varphi}{\boldsymbol{Z}}\\
   & \sim & \epsilon _E \, \left( 2 \epsilon _k \,+\, \epsilon _E \right).
\end{eqnarray}
The latter result is consistent with both Fig.~\ref{fig:errorepEk} and Fig.~\ref{fig:erroreps}.

\section{Conclusion}

Using a full set of gyrokinetic equations, which is valid up to first order in the small parameter $\epsilon$, in the case of an homogeneous magnetic field and a small electrostatic perturbation that satisfies the gyrokinetic ordering, we computed a single charged-particle orbit. We confirmed that the error due to the gyrokinetic reduction is of the order of $\epsilon ^2$.
This simple test problem shows the accuracy of gyrokinetic equations compared to a direct computation of the particle orbit in position/velocity or guiding-center coordinates. We note that in practical gyrokinetic codes, information about the gyroangle is discarded, because of its inaccuracy and its irrelevancy.

%\begin{itemize}
%\item Description of the test problem and of the small parameters $\epsilon _E \equiv \left| \frac{e\varphi}{m v_{Ti}^2} \right|$ and $\epsilon _k \equiv k_\| / k_\perp$
%\item Particle position coordinates $z$ and gyrocenter coordinates $\bar{Z}$ expressed as functions of the guiding-center coordinates $Z$ (Since all comparisons are made in $Z$)
%\item Error in energy conservation for the orbit calculated in $z$, $Z$, and $\bar{Z}$
%\item Dependency of this error with the time-step width
%\end{itemize}

\chapter{The CIP-CSL scheme in 2D phase-space} \label{app:numerical_implementation}
%Numerical implementation of COBBLES
%\section{} \label{app_sec:cipcsl}

To solve the general 1D advection equation,
\begin{equation}
\pds{F}{t} \;+\; u \, \pds{F}{\lambda} \;=\; 0,
\end{equation}
in a conservative form, we evolve both $F$ and its integrated value $R$. $\tilde{F}^n$ ($\tilde{R}^n$) is the continuous interpolated solution for $F$ ($R$) at a time $t_n=n\Delta t$. $F^n_{m}$ is the discrete value of $\tilde{F}^n$ at the grid point of coordinate $\lambda_m \,=\, m\,\Delta \lambda$, and 
\begin{equation}
R^n_{m} \;=\; \int _{\lambda _m} ^{\lambda _{m+1}} \tilde{F}^n(\lambda)\,\mathrm{d}\lambda.
\end{equation}
$F^n$ is an array which contains the values $F^n_{m}$ for every point $m$ on a grid point of coordinate $\lambda _{m}$. We define the 1D algorithm $CIPCSL1D$($u$, $F^n$, $F^{n+1}$, $R^n$, $R^{n+1}$, $\lambda$, $\Delta\lambda$, $\Delta t$) as follows.

For each $m$,
\begin{itemize}%[-]
\item Define $\lambda _{p_m} \;\equiv\; \lambda _{m} \,-\, u \, \Delta t$ ;
\item Find the grid point $k_m$ satisfying $\lambda _{k_m} \leq \lambda _{p_m} \leq \lambda _{k_m + 1}$, and define $\left\langle \lambda _{m} \right\rangle \;\equiv\; \lambda _{p_m} \,-\, \lambda _{k_m}$ ;
\item Define the coefficients \begin{equation} \phi _{k_m} \;\equiv\; \dfrac{F^n_{k_m} + F^n_{k_m+1}}{\Delta\lambda ^2} \;-\; \dfrac{2 R^n_{k_m}}{\Delta\lambda ^3}, \end{equation} \begin{equation} \eta _{k_m} \;\equiv\; -\dfrac{2 F^n_{k_m} + F^n_{k_m+1}}{\Delta\lambda} \;+\; \dfrac{3 R^n_{k_m}}{\Delta\lambda ^2}; \end{equation}
\item Advect $R$, \begin{equation} R^{n+1}_{m} \;=\; R^n_{k_m} \;+\; D^n_{m+1} \;-\; D^n_{m}, \end{equation} where \begin{equation} D^n_{m} \;=\; \phi _{k_m} \, \left\langle \lambda _{m} \right\rangle ^3 \;+\; \eta _{k_m} \, \left\langle \lambda _{m} \right\rangle ^2 \;+\; F^n_{k_m} \, \left\langle \lambda _{m} \right\rangle ; \end{equation}
\item Advect $F$, \begin{equation} F^{n+1}_{m} \;=\; 3 \, \phi _{k_m} \, \left\langle \lambda _{m} \right\rangle ^2 \;+\; 2 \, \eta _{k_m} \, \left\langle \lambda _{m} \right\rangle \;+\; F^n_{k_m}. \end{equation}
\end{itemize}
%All quantities like $f$ are sampled on an uniform Eulerian grid of width $\Delta\lambda$. $F_m$ represents the value of $f$ at the grid point $\lambda _{m} = m \, \Delta\lambda$. $R_m$ represents the integral of $f$ between two neighbouring grid points $\lambda _{m}$ and $\lambda _{m+1}$. We define the 1D algorithm $CSL1D$($u$, $F$, $F^n$, $R$, $R^n$, $\lambda$, $\Delta\lambda$, $\Delta t$) as follows.

This 1D algorithm is extended to the 2D phase-space $(x,v)$ in the following way. $f^n_{i,j}$ is the value of the distribution $f$ at the grid point of coordinates $x_i \,=\, i\,\Delta x$, $v_j \,=\, v_{\mathrm{min}} \,+\, j\,\Delta v$, at a time $t_n=n\Delta t$. We define the density within a cell and the line densities,
\begin{equation} \rho ^n_{i,j} \;=\; \int _{x_i} ^{x_{i+1}} \int _{v_j} ^{v_{j+1}} \tilde{f}^n(x,v)\,\mathrm{d}x\,\mathrm{d}v, \end{equation}
\begin{equation} \sigma ^n_{x\,i,j} \;=\; \int _{x_i} ^{x_{i+1}} \tilde{f}^n(x,v_j)\,\mathrm{d}x, \end{equation}
\begin{equation} \sigma ^n_{v\,i,j} \;=\; \int _{v_j} ^{v_{j+1}} \tilde{f}^n(x_i,v)\,\mathrm{d}v. \end{equation}
The first advection in the $x$-direction is performed by calling successively, for each $j$,
\[ \begin{array}{c@{\ }c@{\ }c@{\ }c@{\ }c@{\ }c@{\ }c@{\ }c}
CIPCSL1D(\ov{v}_{j}, & \sigma _v^n, & \sigma _v^*, & \rho ^n,     & \rho ^*,     & x, & \Delta x, & \Delta t / 2),\\
CIPCSL1D(v_{j},      & f^n,         & f^*,         & \sigma _x^n, & \sigma _x^*, & x, & \Delta x, & \Delta t / 2),
\end{array} \]
with $\ov{v}_{j}\,\equiv\, (v_{j}+v_{j+1})/2$. Similarly, the advection in the $v$-direction is performed by calling successively, for each $i$,
\[ \begin{array}{c@{\ }c@{\ }c@{\ }c@{\ }c@{\ }c@{\ }c@{\ }c}
CIPCSL1D(q \ov{E}_i / m, & \sigma _x^*, & \sigma _x^{**}, & \rho ^*,     & \rho ^{**},     & v, & \Delta v, & \Delta t),\\
CIPCSL1D(q E_i / m,      & f^*,         & f^{**},         & \sigma _v^*, & \sigma _v^{**}, & v, & \Delta v, & \Delta t),
\end{array} \]
with $\ov{E}_{i}\,\equiv\, (E_{i}+E_{i+1})/2$. Then we repeat the advection in the $x$-direction,
\[ \begin{array}{c@{\ }c@{\ }c@{\ }c@{\ }c@{\ }c@{\ }c@{\ }c}
CIPCSL1D(\ov{v}_{j}, & \sigma _v^{**}, & \sigma _v^{n+1}, & \rho ^{**},     & \rho ^{n+1},     & x, & \Delta x, & \Delta t / 2),\\
CIPCSL1D(v_{j},      & f^{**},         & f^{n+1},         & \sigma _x^{**}, & \sigma _x^{n+1}, & x, & \Delta x, & \Delta t / 2).
\end{array} \]
Here, subscripts $*$ and $**$ were used to designate intermediate steps between $t_n$ and $t_{n+1}$.
To avoid spurious leakage of particles, we impose a zero flux at the velocity boundaries by setting
\begin{equation}
\left\langle \lambda _{j=1} \right\rangle \,=\, \left\langle \lambda _{j=N_v} \right\rangle \,=\, 0.
\end{equation}
\\

\begin{figure}
\begin{center}
\includegraphics{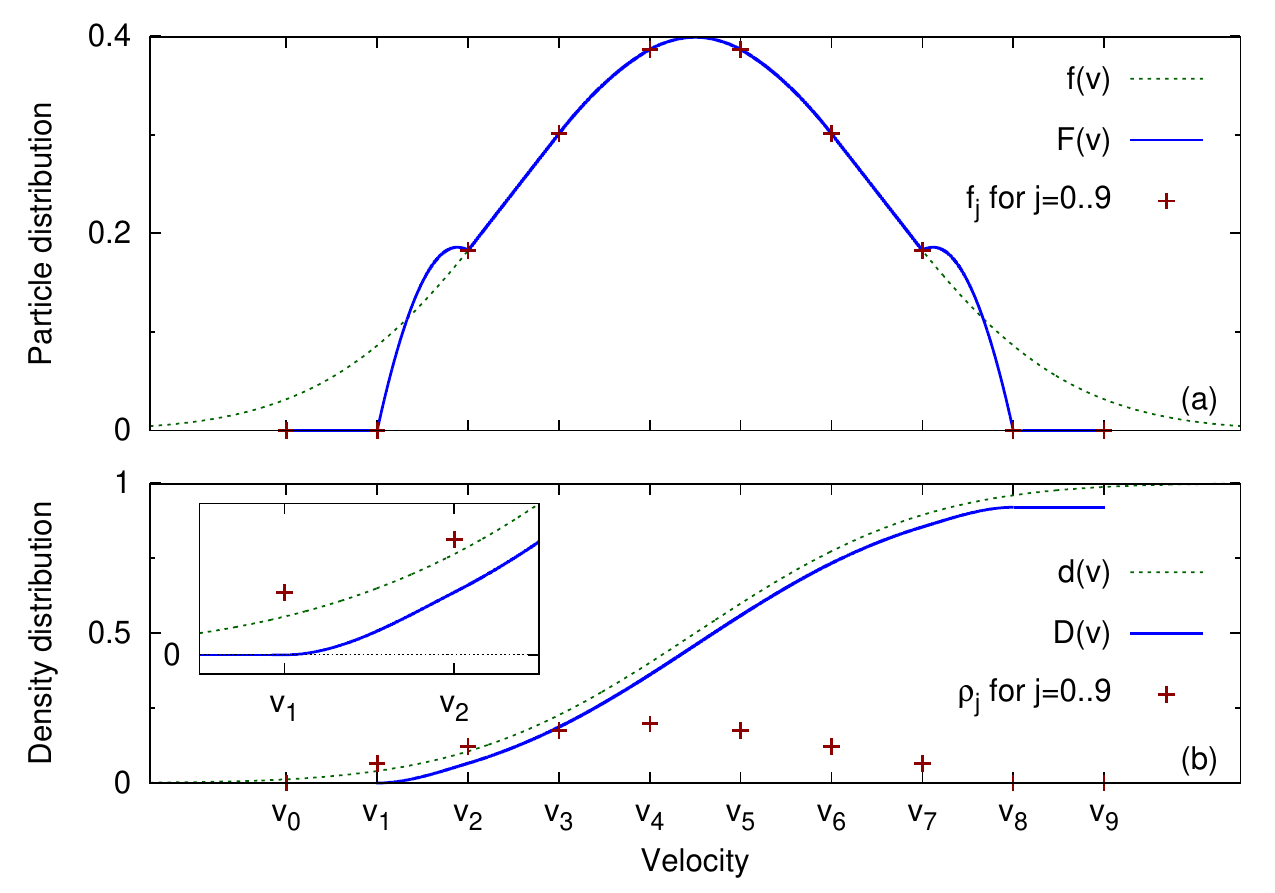}
\caption{Initial Maxwellian distribution discretized on $N_v=8$ grid points. (a) Particle distribution $f$, interpolation polynomial $F$, and discretized values $f_j$. (b) Density distribution $d$, interpolation polynomial $D$, and discretized values of grid-point density $\rho _j$. Inset: zoom on the first points.}
\label{fig:cipcsl_f_d}
\end{center}
\end{figure}

Boundary conditions in the velocity distribution are illustrated in Fig.~\ref{fig:cipcsl_f_d}(a), which shows the discretization on $N_v=8$ grid points of a Maxwellian distribution $f(v)$, and its interpolation polynomial $F$. Notice two buffer points $j=0$ and $j=N_v+1$. Fig.~\ref{fig:cipcsl_f_d}(b) shows the integral function,
\begin{equation}
d(v)\;\equiv\; \int _{-\infty}^v f(v') \mathrm{d} v',
\end{equation}
and its interpolation polynomial $D$. Discretization of $\rho$, which is defined as
\begin{equation}
\rho(v_j)\;\equiv\; \int _{v_j}^{v_{j+1}} f(v') \mathrm{d} v',
\end{equation}
is included. Here we chose a small number of grid points and small cut-off velocities to emphasize border effects. For realistic parameters, the discrepancies between $f$ and $F$, and between $d$ and $D$ are negligible.

%\section{Splitting schemes}
%
%[Redaction needed: describe study of different ways of splitting DCE]

%\section{Parallelization}
%
%COBBLES was parallelized in the velocity direction using MPI library.
%
%[Redaction needed: \begin{itemize} \item implementation using MPI \item week scaling, with and without OpenMP \item strong scaling, with and without OpenMP \end{itemize}]

\chapter{Nonlinear regime categorization algorithm} \label{app:categorization_algo}

The algorithm we developed to categorize the behavior of each numerical solution is an improved version of the algorithm designed by Vann \cite{vann03}, with an other way of sorting out chaotic from periodic behavior, a new diagnostics to distinguish chirping from merely chaotic solutions, and where we take into account numerical issues that appear when $\gamma _L$ is smaller.

The categorization is based on an analysis of the time-series of normalized electric field energy, $A(t)\equiv\mathcal{E}(t)/\mathcal{E}_0$, with $\mathcal{E}_0\,=\,v_{th}^2\,/\,2$. Firstly, we define the global minimum as $A_\mathrm{gm}\equiv \min \{A(t)\}_{0<t<t_\mathrm{max}}$, where $t_{\mathrm{max}}$ is the time-duration of the simulation. Then we drop the initial transient phase, and extract a time window $t_{\mathrm{min}}<t<t_{\mathrm{max}}$ over which $A(t)$ is sampled at a rate $\Delta t _s$. Here, $t_{\mathrm{min}}$ is an estimation of the time-duration of the transient phase. Within this window, we define the mean value $\left\langle A \right\rangle$, maximum $A_{\mathrm{max}} \equiv \max \{A(t)\}$ and minimum $A_{\mathrm{min}} \equiv \min \{A(t)\}$, the oscillation amplitude $\Delta A \equiv A_{\mathrm{max}}-A_{\mathrm{min}}$, and the local minima (maxima) as points where $A(t)$ is smaller (larger) than $A(t - \Delta t _s)$ and $A(t + \Delta t _s)$. As a measure of the periodicity, we compute the two-point correlation function as
\begin{equation}
R(\tau) \; = \; \dfrac{1}{m} \sum _{j=0}^{m-1} \dfrac{ \left\langle \tilde{A}_j(\tau , \tau ') \tilde{A}_{j+1}(\tau , \tau ') \right\rangle }{\left\langle \tilde{A}_j(\tau , \tau ') ^2 \right\rangle ^{1/2}\,  \left\langle \tilde{A}_{j+1}(\tau , \tau ') ^2 \right\rangle ^{1/2}},
\end{equation}
where, for each correlation window size $\tau$,
\begin{equation}
\tilde{A}_j(\tau , \tau ') \,\equiv\, A(t_{\mathrm{max}}\,-\,j\,\tau\,-\,\tau ') - \left\langle A (t_{\mathrm{max}}\,-\,j\,\tau) \right\rangle,
\end{equation}
$m=(t_{\mathrm{max}}-t_{\mathrm{min}})/\tau$ is the number of periods included inside the total time-window, and angular brackets represent a time-average over a period,
\begin{equation}
\left\langle \tilde{A}(\tau , \tau ') \right\rangle = \dfrac{1}{\tau} \int _0 ^{\tau} \tilde{A}(\tau , \tau ') \mathrm{d}\tau '.
\end{equation}
The overall correlation $R_0$ is defined as the maximum of $R(\tau)$ for $\tau > \tau _c$, where $\tau _c$ is the shortest period such that $R(\tau _c)\leq 0$. In other words, $R_0$ is the normalized amplitude of the peak in the two-point correlation function corresponding to the dominant frequency. This measure, which is used to sort out chaotic from periodic behavior, is different from the one given in Ref.~\onlinecite{vann03}.
We also provide a criterion to sort out chirping solutions. We compute the frequency power spectrum of the time-evolution of the electric field at some position, $E(x=0,t)$, and normalize the amplitude of the spectrum to its maximum value. At each time, we extract the largest and smallest frequency for which the amplitude in the power spectrum is significant, i.e. larger than a threshold $\epsilon _6$. Then, we define $\Delta \omega _{\mathrm{max}}$ as the maximum difference between the largest and smallest frequency, normalized to the plasma frequency.
We then proceed to the following decision tree:
\begin{enumerate}%[1.]
\item \label{test0} IF $A_\mathrm{gm} \,<\, \epsilon _0$ THEN damped
\item ELSE IF $\Delta \omega _{\mathrm{max}}\,>\,\epsilon _7$ THEN chirping
\item ELSE IF $\left\langle A \right\rangle \,<\, \epsilon _1$ THEN damped
\item ELSE IF $A(t)$ is monotonic AND $\Delta A / \left\langle A \right\rangle \,<\, \epsilon _2$ THEN steady-state
\item ELSE IF $A(t)$ is monotonically decreasing (zero local extrema) THEN damped
\item ELSE IF each minima (maxima) is larger (smaller) than the former OR $\Delta A / \left\langle A \right\rangle \,<\, \epsilon _3$ OR $\Delta A \,<\, \epsilon _4$ THEN steady-state
\item ELSE IF $R_0 > 1-\epsilon_5$ THEN periodic
\item ELSE IF the number of extrema is not less than four THEN chaotic,
\end{enumerate}
where $\epsilon _i$ are thresholds that must be carefully adjusted. Special care is taken in adjusting $\epsilon _7$ empirically so that frequency splitting is not mistaken for frequency sweeping.

In this decision tree, the logical test \ref{test0} is an addition to the decision tree given in Ref.~\cite{vann03}. For damped solutions, as the electric field becomes small, the particles experience free streaming, leading to spurious recurrence effects after half a recurrence time $T_R/2$ $=$ $\pi / k\Delta v$.
Free streaming occurs when $E_0 \,T_R  < \Delta v$, and $\epsilon _0$ is chosen to reflect this condition. For the benchmark in \ref{subs:benchmark}, this logical step is switched off as the recurrence effect is less problematic for shorter-time simulations.

\chapter{Chirping features analysis algorithm} \label{app:chirping_algo}

Numerical validation of chirping theory, and development of semi-empirical laws for chirping features, both hinge upon systematic quantification of chirping velocity, lifetime, and period or quasi-period.

We developed an algorithm to measure features of major chirping events in both full-$f$ and $\delta f$ COBBLES simulations, for both Krook and drag/diffusion collisions, except in the dragging regime. The starting point is a spectrogram $P(t_i,\omega _j)$, such as plotted in Fig.~\ref{fig:E32359}(b) or (c), which is obtained from time-series of the electric field at $x=0$ with a moving Fourier window of size $\Delta t$, and the linear frequency $\omega$, which is obtained from one of our linear analysis tools described in Sec.~\ref{sec:bblin}. Here, $i$ is an index of the median time of a Fourier window, and $j$ is an index of the frequency, where $\omega _{j+1}-\omega _j = 2\pi / \Delta t$.
Let us consider upward chirping only. The first step is to identify major chirping branches. A simple approach would be to extract from the spectrogram, at each time step, the local maximum, and try to fit a square root function following maxima whose amplitude exceeds some threshold. However, two or more chirping branches including minor events can easily be confused as one (Minor and major events are not well separated categories, it is difficult to set a threshold to isolate major events). Trying to start from the tip of a branch and follow it down by jumping to the closest maximum fails for the same reason. The problem with these procedures is that, when we extract maxima, we loose useful information, namely the amplitude of these maxima.
The following approach makes use of some additional information. It requires a rough estimation of chirping velocity $V^0 \approx (d\delta \omega ^2/dt)^{1/2}$, and a rough estimation of chirping lifetime $\tau _{\mathrm{max}}^0$, both of which are easily obtained from theory using input parameters of the simulation.
First, we initialize a test chirping velocity $V$ to $V^0$. For each time-step $t_{i_0}$ of the spectrogram, the algorithm takes the sum $S_{i_0}(V)$ of $P(t_i,\omega _j)$, where $i$ starts from $i_0$, and $j(i)$ is such that $\omega _j (t_i)$ roughly follows a square-root law with a coefficient given by $V$. We repeat this procedure for each $i_0$, and for test chirping velocities $0.1 V^0 < V < 10 V^0$. A maximum of $S_{i_0}(V)$ should indicate the initial time $t_{i_0}$ and velocity $V$ of a major chirping event. We found this is the case most of the time for data that are analyzed in this work. Simultaneously, we extract the maximum $P_\mathrm{max}$ reached by $P$ during this chirping event.
The second step is to identify the tip of chirping. Using the estimated chirping velocity and $\tau _{\mathrm{max}}^0$, we deduce its approximate location $i^0_\mathrm{tip}$ and $j^0_\mathrm{tip}$. The algorithm, starting from well above $j^0_\mathrm{tip}$ and decrementing $j$, searches around $i^0_\mathrm{tip}$ for the first value of $P$ above a threshold which is given by $e^{-2}P_\mathrm{max}$. This should correspond to the tip of the considered chirping event, $i_\mathrm{tip}$ and $j_\mathrm{tip}$. Since the algorithm is sometimes mistaken, we go on decrementing $j$, while following the first local maximum to the left (left meaning smaller time here). When we reach a $j$ such that $\omega _j\approx \omega$, we check that $i$ is close to $i^0$.
Then the lifetime is given by $t_{i_\mathrm{tip}}-t_{i_0}$. Otherwise the chirping event is discarded. Repeating the above steps while avoiding the time region $t_{i_0}<t<t_{i_\mathrm{tip}}$ yields velocity and lifetime of a significant number of major chirping events.
Finally the chirping period is simply obtained by extracting the maximum in the Fourier spectrum of electric field time-series. The whole procedure requires only a few seconds on one typical laptop processor, even for long-time simulations ($t\sim 10^5$).

The same algorithm is used to measure chirping features in experimental data. In this case, the starting point is a spectrogram of magnetic perturbations measured by a Mirnov coil at the edge of the plasma, such as Fig.~\ref{fig:E32359}(a), the linear frequency $\omega$ and a rough estimation of chirping velocity, and of chirping lifetime, which are directly measured on the spectrogram.

%\subsection*{List of posters}
%
%\begin{itemize}
%
%\item \textit{Estimation of Kinetic Parameters based on Chirping Alfv{\'e}n Eigenmodes}, 23rd IAEA Fusion Energy Conference, Daejeon (October 2010).
%
%\end{itemize}

\bibliographystyle{alpha}

\newcommand{\etalchar}[1]{$^{#1}$}

\end{document}